\documentclass[12pt,paper=b5,pagesize=auto,twoside]{scrbook}
\usepackage[utf8]{inputenc}
\usepackage{ngerman,amsmath,amssymb,amstext,latexsym,graphicx,color}
\usepackage{subfig}
\usepackage[pdfborder={0 0 0}]{hyperref}
\usepackage{bibgerm}
\usepackage{pdfpages} 
\usepackage{chngcntr}
\usepackage[compress,square,numbers]{natbib}
\counterwithout{footnote}{chapter}

\numberwithin{equation}{section}

\begin{document}

\newcommand{\figtxt}[1]{{\footnotesize\sffamily #1}}
\newcommand{\figsubtxt}[1]{{\scriptsize\sffamily #1}}
\newcommand{\todo}[1]{\,\!}
\definecolor{gray}{rgb}{0.4, 0.3, 0.4}
\newcommand{\remark}[1]{\,\!}
\newcommand{\wh}[1]{{{\color{white}#1}}}
\newcommand{\Ito}{It\={o}}
\newcommand*{\tscale}[2][1]{\scalebox{#1}{#2}}
\newcommand*{\mscale}[2][1]{\scalebox{#1}{$#2$}}
\newcommand{\di}{\;\mathrm{d}} 
\newcommand{\ldi}{\!\!\mathrm{d}} 
\newcommand{\dd}{\mathrm{d}} 
\newcommand{\Di}{\;\mathcal{D}} 
\newcommand{\lDi}{\!\!\!\!\mathcal{D}} 
\newcommand{\pt}{\partial}
\newcommand{\pst}{p^{{\textrm{\scriptsize st}}}}
\newcommand{\pstnn}{p^{{\textrm{\scriptsize st}}}_\mr{nn}}
\newcommand{\Jst}{j^{{\textrm{\scriptsize st}}}}
\newcommand{\vJst}{\vek{j}^{{\textrm{\scriptsize st}}}}
\newcommand{\peq}{p^{{\textrm{\scriptsize eq}}}}
\newcommand{\Jeq}{j^{{\textrm{\scriptsize eq}}}}
\newcommand{\pch}{p^{{\textrm{\scriptsize ch}}}}
\newcommand{\ddpst}{\dot p^{{\textrm{\scriptsize st}}}}
\newcommand{\bSS}{\bar\SS}
\newcommand{\bp}{\bar p}
\newcommand{\bP}{\xbar{P}}
\newcommand{\tp}{\widetilde p}
\newcommand{\tP}{\widetilde P}
\newcommand{\ddp}{\dot{p}}
\newcommand{\Pasy}{P_\mr{asy}}
\newcommand{\xc}{x(\cdot)}
\newcommand{\sx}{x^*}
\newcommand{\st}{t^*}
\newcommand{\vxc}{\vek{x}(\cdot)}
\newcommand{\bxc}{\bar x(\cdot)}
\newcommand{\bbxc}{\bar{\bar x}(\cdot)}
\newcommand{\vbxc}{\vek{\bar x}(\cdot)}
\newcommand{\sxc}{x^*(\cdot)}
\newcommand{\ddx}{\dot{x}}
\newcommand{\dddx}{\ddot{x}}
\newcommand{\bx}{\bar x}
\newcommand{\bddx}{\dot{\bar x}}
\newcommand{\Dt}{\D t}
\newcommand{\Dx}{\D x}
\newcommand{\uc}{u(\cdot)}
\newcommand{\tuc}{\tilde u(\cdot)}
\newcommand{\uM}{u_\mr{M}}
\newcommand{\buc}{\bar u(\cdot)}
\newcommand{\bu}{\bar u}
\newcommand{\ddu}{\dot{u}}
\newcommand{\dddu}{\ddot{u}}
\newcommand{\bucOm}{\bar u(\cdot,\Om)}
\newcommand{\bucSm}{\bar u(\cdot,\Sm)}
\newcommand{\bucW}{\bar u(\cdot,W)}
\newcommand{\Du}{\D u}
\newcommand{\Dr}{\D r}
\newcommand{\tr}{\tilde r}
\newcommand{\wc}{u(\cdot)} 
\newcommand{\ddw}{\dot{u}} 
\newcommand{\trh}{\tilde\r}
\newcommand{\bt}{{\bar\tau}}
\newcommand{\kc}{\k(\cdot)}
\newcommand{\bk}{\bar\k}
\newcommand{\bkc}{\bar\k(\cdot)}
\newcommand{\bbkc}{\bar{\bar\k}(\cdot)}
\newcommand{\bd}{{\bar d}}
\newcommand{\vchar}{v_\mr{ch}}
\newcommand{\lchar}{\ell_\mr{ch}}
\newcommand{\Kchar}{k_\mr{ch}}
\newcommand{\Knu}{k_\nu}
\newcommand{\Rey}{\mr{R\hspace{-0.9pt}e}}
\newcommand{\Reych}{\Rey_\mr{ch}}
\newcommand{\Reycr}{\Rey_\mr{cr}}
\newcommand{\dfr}{d_\mr{fr}}
\newcommand{\vrms}{v_\mr{rms}}
\newcommand{\vek}[1]{\boldsymbol{#1}}
\newcommand{\mat}[1]{{\boldsymbol{\mr{#1}}}}
\newcommand{\op}[1]{\widehat{\mr{#1}}}
\newcommand{\ddD}{\dot D}
\newcommand{\ddF}{\dot F}
\newcommand{\Sr}{S_\mr{r}}
\newcommand{\Sd}{S_\mr{d}}
\newcommand{\Sb}{S_\mr{b}}
\newcommand{\Si}{S_\mr{i}}
\newcommand{\Sna}{S_\mr{na}}
\newcommand{\Sa}{S_\mr{a}}
\newcommand{\Se}{S_\e}
\newcommand{\Sm}{S_{\mr{m}}}
\newcommand{\Sre}{S_{\mr{r}\e}}
\newcommand{\Sirr}{S_\mr{irr}}
\newcommand{\Sirrna}{S_\mr{ex}^{\mscale[0.65]{\geq}}}
\newcommand{\Stot}{S_{\mr{tot}}}
\newcommand{\bStot}{\bar S_{\mr{tot}}}
\newcommand{\DS}{\D S}
\newcommand{\DSeq}{\D S^\mr{eq}}
\newcommand{\DSst}{\D S^\mr{st}}
\newcommand{\Ds}{\D S}
\newcommand{\Qrev}{Q_\mr{rev}}
\newcommand{\Qirr}{Q_\mr{irr}}
\newcommand{\Wdiss}{W_\mr{diss}}
\newcommand{\Rdiss}{R_\mr{diss}}
\newcommand{\Qhk}{Q_\mr{hk}}
\newcommand{\Qex}{Q_\mr{ex}}
\newcommand{\Shk}{S_\mr{hk}}
\newcommand{\Sex}{S_\mr{ex}}
\newcommand{\si}{s_\mr{i}}
\newcommand{\sm}{s_\mr{m}}
\newcommand{\stot}{s_\mr{tot}}
\newcommand{\se}{s_\mr{e}}
\newcommand{\sd}{s_\mr{d}}
\renewcommand{\sb}{s_\mr{b}}
\newcommand{\seq}{s^\mr{eq}}
\newcommand{\sst}{s^\mr{st}}
\newcommand{\Seq}{S^\mr{eq}}
\newcommand{\Sst}{S^\mr{st}}
\newcommand{\ddS}{\dot S}
\newcommand{\ddSi}{\dot S_\mr{i}}
\newcommand{\ddSe}{\dot S_\mr{e}}
\newcommand{\ddSna}{\dot S_\mr{na}}
\newcommand{\ddSa}{\dot S_\mr{a}}
\newcommand{\dds}{\dot s}
\newcommand{\ddsi}{\dot s_\mr{i}}
\newcommand{\ddsm}{\dot s_\mr{m}}
\newcommand{\ddstot}{\dot s_\mr{tot}}
\newcommand{\ddse}{\dot s_\mr{e}}
\newcommand{\ddsd}{\dot s_\mr{d}}
\newcommand{\ddsb}{\dot s_\mr{b}}
\newcommand{\ddseq}{\dot s^\mr{eq}}
\newcommand{\ddsst}{\dot s^\mr{st}}
\newcommand{\ddWdiss}{\dot W_\mr{diss}}
\newcommand{\ddRdiss}{\dot R_\mr{diss}}
\newcommand{\ddQirr}{\dot Q_\mr{irr}}
\newcommand{\ddQhk}{\dot Q_\mr{hk}}
\newcommand{\ddQex}{\dot Q_\mr{ex}}
\newcommand{\ddshk}{\dot s_\mr{hk}}
\newcommand{\ddsex}{\dot s_\mr{ex}}
\newcommand{\Ekin}{E_\mr{kin}}
\newcommand{\DF}{\D\FF}
\newcommand{\DG}{\D\GG}
\newcommand{\DI}{\D\II}
\newcommand{\Dp}{\D\p}
\newcommand{\Dseq}{\D s^\mr{eq}}
\newcommand{\Dsst}{\D s^\mr{st}}
\newcommand{\tME}{t_\mr{ME}}
\newcommand{\rME}{r_\mr{ME}}
\newcommand{\DtME}{\Dt_\mr{ME}}
\newcommand{\kB}{k_\mr{B}}
\newcommand{\tx}{z}
\newcommand{\x}{\tilde x}
\newcommand{\Str}{S_{\!u}}
\newcommand{\Df}{D^{^{\text{\tiny (1)\!\!}}}}
\newcommand{\df}{d^{^{\text{\tiny (1)\!\!}}}}
\newcommand{\tDf}{\tilde D^{^{\text{\tiny (1)\!\!}}}}
\newcommand{\tdf}{\tilde d^{^{\text{\tiny (1)\!\!}}}}
\newcommand{\bDf}{\bar D^{^{\text{\tiny (1)\!\!}}}}
\newcommand{\Dfsq}{D^{^{\text{\tiny (1)}^2\!\!}}}
\newcommand{\Dfx}{{D^{^{\text{\tiny (1)\!\!}}}}}
\newcommand{\Dg}{D^{^{\text{\tiny (2)\!\!}}}}
\newcommand{\dg}{d^{^{\text{\tiny (2)\!\!}}}}
\newcommand{\tDg}{\tilde D^{^{\text{\tiny (2)\!\!}}}}
\newcommand{\bDg}{\bar D^{^{\text{\tiny (2)\!\!}}}}
\newcommand{\Dgsq}{D^{^{\text{\tiny (2)}^2\!\!}}}
\newcommand{\Dgx}{{D^{^{\text{\tiny (2)}\prime\!\!}}}}
\newcommand{\Dgxx}{{D^{^{\text{\tiny (2)}\prime\prime\!\!}}}}
\newcommand{\bDgx}{{\bar D^{^{\text{\tiny (2)}\prime\!\!}}}}
\newcommand{\Dgxsq}{{D^{^{\text{\tiny (2)}\prime^2\!\!}}}}
\newcommand{\Dk}{D^{^{\text{\tiny ($k$)\!\!}}}}
\newcommand{\dk}{d^{^{\text{\tiny ($k$)\!\!}}}}
\newcommand{\tDk}{\tilde D^{^{\text{\tiny ($k$)\!\!}}}}
\newcommand{\Dth}{D^{^{\text{\tiny (3)\!\!}}}}
\newcommand{\Dfr}{D^{^{\text{\tiny (4)\!\!}}}}
\newcommand{\Dgg}{D^{^{\text{\tiny ($\geq\!\!3$)\!\!}}}}
\newcommand{\Dggg}{D^{^{\text{\tiny ($\geq\!\!4$)\!\!}}}}
\newcommand{\Dfg}{D^{^{\text{\tiny (1,2)\!\!}}}}
\newcommand{\dfg}{d^{^{\text{\tiny (1,2)\!\!}}}}
\newcommand{\Dtf}{D^{^{\text{\tiny (3,4)\!\!}}}}
\newcommand{\Dff}{D^{^{\text{\tiny (1-4)\!\!}}}}
\newcommand{\Ak}{\varPsi^{^{\text{\tiny ($k$)\!\!}}}}
\newcommand{\Akk}{\varPsi^{^{\text{\tiny ($2k$)\!\!}}}}
\newcommand{\Af}{\varPsi^{^{\text{\tiny (1)\!\!}}}}
\newcommand{\Ag}{\varPsi^{^{\text{\tiny (2)\!\!}}}}
\newcommand{\Afr}{\varPsi^{^{\text{\tiny (4)\!\!}}}}
\newcommand{\Agg}{\varPsi^{^{\text{\tiny ($\geq\!\!3$)\!\!}}}}
\newcommand{\Afg}{\varPsi^{^{\text{\tiny (1,2)\!\!}}}}
\newcommand{\tAk}{D^{^{\text{\tiny ($k$)\!\!}}}}
\newcommand{\tAkk}{D^{^{\text{\tiny ($2k$)\!\!}}}}
\newcommand{\tAf}{D^{^{\text{\tiny (1)\!\!}}}}
\newcommand{\tAg}{D^{^{\text{\tiny (2)\!\!}}}}
\newcommand{\tAfr}{D^{^{\text{\tiny (4)\!\!}}}}
\newcommand{\tAgg}{D^{^{\text{\tiny ($\geq\!\!3$)\!\!}}}}
\newcommand{\tAfg}{D^{^{\text{\tiny (1,2)\!\!}}}}
\newcommand{\Mk}{M^{^{\text{\tiny ($k$)\!\!}}}}
\newcommand{\Mf}{M^{^{\text{\tiny (1)\!\!}}}}
\newcommand{\Mg}{M^{^{\text{\tiny (2)\!\!}}}}
\newcommand{\vDf}{\vek{D}^{^{\text{\tiny (1)\!\!}}}}
\newcommand{\mDg}{\mat{D}^{^{\text{\tiny (2)\!\!}}}}
\newcommand{\dta}{\dot{\tilde{a}}}
\newcommand{\tf}{\tilde{f}}
\newcommand{\tg}{\tilde{g}}
\newcommand{\txi}{\tilde{\xi}}
\newcommand{\tZ}{\tilde{Z}}
\newcommand{\fa}{f_{i_\a}}
\newcommand{\ga}{g_{i_\a}}
\newcommand{\fax}{f_{i_\a}'}
\newcommand{\gax}{g_{i_\a}'}
\newcommand{\Sc}{S_{u_L}}
\newcommand{\ee}[1]{\mathrm{e}^{#1}}
\newcommand{\eee}[1]{\text{\large e}^{\,\displaystyle #1}}
\newcommand{\mr}[1]{\mathrm{#1}}
\newcommand{\const}{\mr{const}}
\newcommand{\bs}[1]{\boldsymbol{#1}}
\newcommand{\HH}{\mscale[0.9]{\mathcal{H}}}
\renewcommand{\SS}{\mathcal{S}}
\newcommand{\LL}{\mathcal{L}}
\newcommand{\hLL}{\hat{\mathcal{L}}}
\newcommand{\hLLFP}{\hat{\mathcal{L}}_\mathrm{FP}}
\newcommand{\hLLFPd}{\hat{\mathcal{L}}^{\,\dagger}_\mathrm{FP}}
\newcommand{\hLLFPdd}{\hat{\mathcal{L}}^{\,\dagger\dagger}_\mathrm{FP}}
\newcommand{\FF}{\mathcal{F}}
\newcommand{\GG}{\mathcal{G}}
\newcommand{\ZZ}{\mathcal{Z}}
\newcommand{\II}{\mathcal{I}}
\newcommand{\JJ}{\mathcal{J}}
\newcommand{\NN}{\mathcal{N}}
\newcommand{\YY}{\mathcal{Y}}
\newcommand{\WW}{\mathcal{W}}
\newcommand{\EE}{\mathcal{E}}
\newcommand{\OO}{\mathcal{O}}
\newcommand{\tY}{\widetilde\YY}
\newcommand{\g}{\gamma}
\newcommand{\G}{\Gamma}
\renewcommand{\b}{\beta}
\renewcommand{\a}{\alpha}
\renewcommand{\d}{\delta}
\newcommand{\e}{\epsilon}
\newcommand{\eps}{\varepsilon}
\newcommand{\beps}{\bar\varepsilon}
\newcommand{\p}{\varphi}
\renewcommand{\l}{\lambda}
\renewcommand{\L}{\varLambda}
\renewcommand{\t}{\tau}
\newcommand{\D}{\varDelta}
\newcommand{\z}{\zeta}
\renewcommand{\k}{\kappa}
\renewcommand{\u}{\upsilon}
\newcommand{\om}{\omega}
\newcommand{\OmO}{\varOmega} 
\newcommand{\OmI}{\Omega} 
\renewcommand{\j}{\chi} 
\newcommand{\jd}{\theta} 
\renewcommand{\r}{\varrho}
\newcommand{\s}{\sigma}
\newcommand{\sigi}{\s_\infty}
\newcommand{\laplace}{\Delta}
\newcommand{\ra}{\rightarrow}
\newcommand{\la}{\leftarrow}
\newcommand{\Ra}{\Rightarrow}
\newcommand{\lolra}{\longleftrightarrow}
\newcommand{\lora}{\longrightarrow}
\newcommand{\loRa}{\Longrightarrow}
\newcommand{\lola}{\longleftarrow}
\newcommand{\lra}{\leftrightarrow}
\newcommand{\Lra}{\Leftrightarrow}
\newcommand{\lla}{\left\langle}
\newcommand{\rra}{\right\rangle}
\newcommand{\lb}{\left|}
\newcommand{\rb}{\right|}
\newcommand{\+}{\!+\!}
\renewcommand{\-}{\!-\!}
\newcommand{\eq}{\!=\!}
\newcommand{\meq}{\stackrel{!}{=}}
\newcommand{\cneq}[1]{\stackrel{\mr{#1}}{\neq}}
\newcommand{\ceq}[1]{\stackrel{\mr{#1}}{=}}
\renewcommand{\leadsto}{\tscale[1.3]{$\bs{\dashrightarrow}$}}
\newcommand{\sep}{\;,\quad}
\newcommand{\dfn}{\mathrel{\mathop:}=}
\newcommand{\dfnrv}{=\mathrel{\mathop:}}
\newcommand{\dfns}{\!\mathrel{\mathop:}=\!}
\newcommand{\dfnsrv}{\!=\mathrel{\mathop:}\!}
\newcommand{\mfrac}[2]{\mscale[0.83]{\dfrac{#1}{#2}}}
\newcommand{\mbinom}[2]{\mscale[0.83]{\dbinom{#1}{#2}}}
\newcommand{\nn}{\nonumber\\}
\newcommand{\nnn}{\nonumber\\\nonumber}
\newcommand{\intev}{\sum\limits_\pm\int\limits_{-\infty}^{\a_\mr{max}}}
\newcommand{\ka}{\kappa_{\!\pm}\!(\a)}
\newcommand{\rha}{\varrho_{\!\pm}\!(\a)}
\newcommand{\trha}{\tilde\varrho_{\!\pm}\!(\a)}
\newcommand{\abs}{\\[5pt]}
\newcommand{\crs}{\addcontentsline{toc}{section}{cursor}}
\newcommand{\figsize}{\footnotesize}
\newcommand{\<}{\hspace{-4pt}}
\newcommand{\dint}{\displaystyle\int}
\newcommand{\dsum}{\displaystyle\sum}
\newcommand{\dprod}{\displaystyle\prod}
\newcommand{\tint}{\textstyle\int}
\newcommand{\tsum}{\textstyle\sum}
\newcommand{\tprod}{\textstyle\prod}

\newcommand*\xbar[1]{%
  \hbox{%
    \vbox{%
      \hrule height 0.5pt 
      \kern0.4ex
      \hbox{%
        \kern-0.15em
        \ensuremath{#1}%
        \kern-0.1em
      }%
    }%
  }%
}

\renewcommand{\figurename}{Figure}
\renewcommand{\bibname}{References}

\pagestyle{empty}

\begin{minipage}{0.38\textwidth}
  \wh{Unter dem Titel}
\end{minipage}
\begin{minipage}{0.6\textwidth}
  \begin{flushright}
    \includegraphics[width=0.7\textwidth]{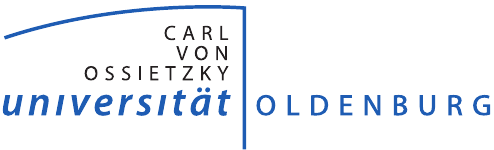}
  \end{flushright}
\end{minipage}
\vspace{70pt}
\begin{center}
	\hrule \vspace{7pt}
	{\Large {\bf Markov Processes linking\\[7pt] Thermodynamics and Turbulence} } \\[7pt]
	\hrule \vspace{70pt}
\end{center}

\begin{center}
{\large {\bf Dissertation } } \\[120pt]
\end{center}

\noindent Institut für Physik\\
Fakultät für Mathematik und Naturwissenschaften\\
Carl von Ossietzky Universität Oldenburg\\[40pt]

\noindent {\bf Daniel Nickelsen }\\[5pt]
geboren am 15.01.1983 in Bredstedt\\[53pt]

\cleardoublepage

\vspace*{0.8\textheight}
\begin{flushleft}
	\noindent
	Erstgutachter:\; {\bf Prof. Dr. Andreas Engel }\\[2pt]
	Zweitgutachter:\; {\bf Prof. Dr. Joachim Peinke }\\[5pt]
	Tag der Disputation:\; {\bf 05.08.2014 }
\end{flushleft}

\pagestyle{plain}
\pagenumbering{roman}
\setcounter{page}{1}

\section*{Abstract}
\figtxt{This thesis deals with Markov processes in stochastic thermodynamics and fully developed turbulence.}\\[7pt]
\figtxt{In the first part of the thesis, a detailed account on the theory of Markov processes is given, forming the mathematical fundament. In the course of developing the theory of continuous Markov processes, stochastic differential equations, the Fokker-Planck equation and Wiener path integrals are introduced and embedded into the class of discontinuous Markov processes. Special attention is paid to the difficulties that arise in the case of multiplicative noise.\\
Referring to the paradigm of Brownian motion, the thermodynamic quantities work, heat and entropy, and the accompanied first law and second law, are formulated on the level of individual trajectories using stochastic differential equations.\\
One of the prominent results of stochastic thermodynamics are so-called fluctuation theorems which reveal an intimate relation between entropy production and irreversibility. Using the path integral formulation, fluctuations theorems are derived in a formal setting.\\
Applications of fluctuation theorems to thermodynamic systems are dependent on a reliable statistics of rare events. To access the statistics of rare events, an asymptotic method is developed. The first order asymptotics is derived for the general case of multiplicative noise, the second order asymptotics for the simpler case of additive noise. The application of the asymptotic method is demonstrated for work distributions in physically relevant models}.\\[7pt]
\figtxt{The second part of the thesis carefully introduces the concept of fully developed turbulence and gives an account on established theories. These theories can be reformulated in terms of Markov processes. An overview of these Markov representations is compiled, including a few new approaches.\\
In view of the results of stochastic thermodynamics, implications for the Markov representation are exploited. Fluctuation theorems are derived and applied to experimental data. In an experimental as well as theoretical analysis, fluctuation theorems are found to be sensitive to intermittent small-scale fluctuations. The above mentioned asymptotic method is used to assess these fluctuations.\\
In closing the thesis, the discussed approaches to fully developed turbulence are contrasted with each other using their Markov representations, and an interpretation of the respective Markov processes is offered.}

\newpage
\section*{Zusammenfassung}
\figtxt{Diese Dissertation beschäftigt sich mit Markov Prozessen in stochastischer Thermodynamik und voll entwickelter Turbulenz.}\\[7pt]
\figtxt{Der erste Teil der Arbeit führt ausführlich in die Theorie der Markov Prozesse ein. Die Darstellung dieses mathematischen Grundgerüstes schließt stochastische Differentialgleichungen, die Fokker-Planck Gleichung und Wiener Pfadintegrale ein. Diese dadurch beschriebenen stetigen Markov Prozesse werden in die Formulierung unstetiger Markov Prozesse eingebettet. Ein besonderes Augenmerk ist dabei auf die Komplikationen gerichtet, die sich für den Fall multiplikativen Rauschens ergeben.\\
In Bezug auf das Musterbeispiel Brown'scher Bewegung werden aus stochastischen Differentialgleichungen Arbeit, Wärme und Entropie, und die damit einhergehenden ersten beiden Hauptsätze der Thermodynamik auf der Ebene individueller Trajektorien eingeführt.\\
Herausragende Ergebnisse stochastischer Thermodynamik sind sogenannte Fluktuationstheoreme, die die enge Beziehung zwischen Entropieproduktion und Irreversibilität deutlich machen. Die Fluktuationstheoreme werden formal aus der Pfadintegral-Formulierung hergeleitet.\\
Die Anwendungen der Fluktuationstheoreme auf thermodynamische Systeme ist abhängig von einer zuverlässigen Statistik seltener Ereignisse. Um die Statistik seltener Ereignisse zugänglich zu machen, wird eine asymptotische Methode entwickelt. Die Asymptotik erster Ordnung wird für den allgemeinen Fall multiplikativen Rauschens hergeleitet, die Asymptotik zweiter Ordnung für additives Rauschen. Die Anwendung der asymptotischen Methode wird für Arbeitsverteilung in physikalisch relevanten Modellen demonstriert.}\\[7pt]
\figtxt{Der zweite Teil der Arbeit führt sorgfältig in das Konzept voll entwickelter Turbulenz ein und verschafft einen Überblick über etablierte Theorien. Diese Theorien können auch als Markov Prozesse formuliert werden. Es wird eine Zusammenstellung dieser Markov Darstellungen angefertigt, die auch ein paar neue Ansätze einschließt.\\
In Hinblick auf Ergebnisse der stochastischen Thermodynamik wird ihre Bedeutung für die Markov Darstellung ausgewertet. Fluktuationstheoreme werden berechnet und auf experimentelle Daten angewandt. Sowohl hinsichtlich einer experimentellen als auch einer theoretischen Untersuchung stellt sich heraus, dass Fluktuationstheoreme dem Auftreten intermittenter klein-skaliger Fluktuationen Rechnung tragen. Unter Benutzung der oben genannten asymptotischen Methode werden diese Fluktuationen beurteilt.\\
Abschließend werden die besprochenen Theorien voll entwickelter Turbulenz anhand ihrer Markov Darstellung gegenübergestellt, und eine Interpretation der entsprechenden Markov Prozesse wird unterbreitet.}

\renewcommand{\contentsname}{Contents}

\makeatletter
\@openrightfalse
\makeatother

\tableofcontents

\makeatletter
\@openrighttrue
\makeatother

\cleardoublepage

\pagestyle{headings}
\pagenumbering{arabic}
\setcounter{page}{1}
\chapter*{Introduction}
\addcontentsline{toc}{chapter}{Introduction}
In the preceding century, two branches of physics\footnote{And many others, of course.} have been advanced considerable.
\\ One branch is the theory of Brownian motion, pushed forward by Einstein, Smoluchovski and Langevin in the beginning of the preceding century, and from which in the last 20 years the developing field of stochastic thermodynamics emerged.\\
The other branch is studying universal features of turbulent flows, initiated by Kolmogorov and Obhukov who addressed scaling laws for the statistics of fluctuations of flow velocity and coined the conception of cascading turbulent structures.\\

The motion of a Brownian particle suspended in a fluid is a continuous Markov process, which means that the current state of the particle only depends on the most recent event. The events in Brownian motion are collisions between particle and fluid molecules, provoking an irregular and random motion of the particle. Brownian motion is observable for tiny particles for which it is not unlikely that fluid molecules collide predominantly with only one side of the particle, kicking the particle to the other side. For large particles, too many molecules are involved in the collisions, such that a possible excess of collisions on one side of the particle is negligible.\\
The incidents in which the small particles are kicked forward constitute to a local rectification of thermal noise, cooling down the fluid molecules involved in the collision, and thus {\it consuming} entropy. On the other hand, fast particles collect significant more collisions on its front side than on its backside, provoking a friction that slows down the particle. Due to this friction, decelerating particles entail a local heating of the fluid, and thus {\it produce} entropy. The balance between entropy consuming and entropy producing collisions are quantified by so-called {\it fluctuation theorems}, which are prominent results of {\it stochastic thermodynamics}.\\
Stochastic thermodynamics is the thermodynamics of nanoscopic systems which may be arbitrarily far from equilibrium. The smallness of the systems is crucial in order to observe the entropy consuming events as exemplified for the Brownian particle, and non-equilibrium can be imposed by driving the particle with an external force or preparing a non-equilibrium initial state. Stochastic thermodynamics extents the first law and the second law to the level of individual trajectories of nanoscopic systems with non-equilibrium dynamics.\\
The fluctuation theorems are used to extract equilibrium information from non-equilibrium measurements, a procedure being dependent on a sufficient occurrence of the entropy consuming events in the system. Methods to assess the entropy consuming events, or to correct for an insufficient sampling of the rare events, are focus of current research.\\

Universal features of turbulent flows are traditionally sought in the scaling properties of the moments of velocity fluctuations in the flow. Less than 20 years ago, a new approach to characterise turbulent flows has been suggested, in which spatial fluctuations of flow velocity are modelled by a Markov process. These Markov processes address the repeated break-up of turbulent structures due to the non-linear interactions in the flow. The individual trajectories are probes of the spatial structures of the flow, or more specific, velocity fluctuations on different spatial scales. The evolution of these structures towards smaller scales is captured by the conception of {\it turbulent cascades}.\\
Though valid on a macroscopic scale, stochastic thermodynamics does usually not provide more insights when applied to macroscopic systems. Therefore, since turbulent flows are of rather macroscopic dimension, the implications of the results of stochastic thermodynamics for the Markovian approach to turbulent flows are rather unexplored. However, it has proved recently that fluctuation theorems implied by the Markov representation of turbulent cascades do apply for a surprisingly small ensemble of probes.\\
The formulation of turbulent cascades as Markov processes raise intriguing questions. Acquire the formal expressions for entropy productions a meaningful form for turbulent cascades? What is the statement of the associated second law? What do the entropy consuming trajectories look like? Under what conditions do entropy consuming trajectories arise sufficiently often to allow a practical use of fluctuation theorems? What are possible applications of fluctuation theorems for characterisations of turbulent flows? Can the rare events in a turbulent flow be assessed?\\
The aim of this thesis is to find answers to such questions.\\

The first part of the thesis introduces the relevant aspects of the theory of Markov processes, and elucidates the thermodynamic interpretation of Markov processes in the framework of stochastic thermodynamics.\\
The second part introduces to the basic theory of turbulence and compiles the Markov representation of established approaches to turbulence, combining a survey of existing results and a report of novel Markov representations. On the basis of the introduced Markov representations, the above questions are tackled.\\
Each part is divided into chapters, which are in turn divided into sections.\\

To be able to transfer results of stochastic thermodynamics to turbulence, the involved concepts have to be well understood. In contradiction to the usual thermodynamic setting, the Markov representations of turbulent cascades are always characterised by a diffusion that depends on the current state of the process, which entails ambiguities and technical difficulties. A special focus of the first part of the thesis is therefore the implications of state dependent diffusion, and consequently, the theory of stochastic thermodynamics is introduced on the abstract level of arbitrary diffusion. To maintain intuition, the paradigm of Brownian motion serves as an intuitive example whenever appropriate.\\

This thesis is publication based and includes three peer-reviewed publications \cite{Nickelsen2011,Nickelsen2012,Nickelsen2013}. Some aspects elucidated in the included publications are also discussed in the main text to allow a fluent introduction to stochastic thermodynamics and the theory of turbulence, and to maintain the central theme of the thesis.\\
Bringing together two rather distinct fields of research, each of which treated in a separate part of the thesis, the publications are included into the respective parts. The central results are given by the publications, a couple of minor, unpublished results are included in the main text.

\cleardoublepage
\renewcommand{\thechapter}{\Roman{chapter}}

\chapter{Stochastic Thermodynamics}
\vspace*{30pt}
The properties of nanoscopic systems can not be described by equilibrium thermodynamics. Examples of nanoscopic systems include colloidal particles, biological cells and nanoscopic devices. Characteristic for nanoscopic systems is the stochasticity in their degrees of freedom. The developing field of stochastic thermodynamics addresses the thermodynamic properties of these systems, taking into account the stochasticity and, in addition, remains valid in situations arbitrarily far from equilibrium. In the last 20 years, stochastic thermodynamics has put forth intriguing results, a prominent example being the relation between entropy production and irreversibility on the level of individual trajectories.\\[30pt]

\noindent The mathematical fundament of stochastic thermodynamics is the theory of Markov processes which will be introduced in the first chapter of this part. The second chapter explicates the thermodynamic interpretation of Markov processes, including the first and second law of thermodynamics, and an account on applications of the so-called fluctuation theorems. The third chapter introduces an asymptotic method and exemplifies its use in models relevant for the application of fluctuation theorems.

\cleardoublepage

\section{Markov Processes (MPs)} \label{s_Markov_processes}
\newcommand{\din}{d_0}
In this chapter we introduce the theory of Markov processes (MPs) which forms the mathematical basis of this thesis. The basic material will mainly build on the books by Gardiner \cite{Gardiner2009} and van Kampen \cite{vanKampen2007} on stochastic processes, the book by Chaichian and Demichev on path integrals \cite{ChaichianDemichev01}, and the article by Lau and Lubensky on state dependent diffusion \cite{LauLubensky07PRE}.\\
A Markov process is a subclass of stochastic processes which involve a time dependent random variable $X(t)$. The random variable is not restricted to one dimension, but we will primarily consider Markov processes with one degree of freedom. A realisation of a stochastic process $X(t)$ is a sequence of measured values $x_1, x_2, x_3, \dots$ at times $t_1>t_2>t_3>\dots$, completely described by the joint probability density function (PDF)
\begin{align} \label{eq:MPs_jointPDF}
	p(x_1,t_1;\,x_2,t_2;\,x_3,t_3;\,\dots) \;.
\end{align}
If the stochastic process is a MP, the necessary information to define the process uniquely reduces to univariate PDFs.\\
A stochastic process is qualified as a MP by the {\it Markov assumption}. The Markov assumption states that conditioned on having measured values $y_1,\,y_2,\,\dots$ at times $\t_1 \geq \t_2 \geq \dots$, the probability of measuring $x_1,\,x_2,\,\dots$ at later times $t_1 \geq t_2 \geq \dots \t_1 \geq \t_2$ will only depend on the most recent measured value $y_1$:
\begin{align} \label{eq:MarCond}
  p(x_1,t_1;\,x_2,t_2;\,\dots|y_1,\t_1;\,y_2,\t_2;\,\dots) = p(x_1,t_1;\,x_2,t_2;\,\dots|y_1,\t_1) \;.
\end{align}
This definition of a MP implies that we can write for the joint PDF
\begin{align}
	&p(x_1,t_1;\,x_2,t_2;\,x_3,t_3;\,\dots\,x_n,t_n) \\
	&\qquad= p(x_1,t_1|x_2,t_2)p(x_2,t_2|x_3,t_3)\cdots p(x_{n-1},t_{n-1}|x_n,t_n)p(x_n,t_n) \nonumber \;,
\end{align}
now defined solely by the conditional PDF $p(x_{i-1},t_{i-1}|x_i,t_i)$ and the univariate PDF $p(x_n,t_n)$ with $t_{i-1}>t_i$.\\
A direct consequence of the Markov assumption is the {\it Chapman-Kolmogorov equation} (CKR)
\begin{align} \label{eq:CKR}
  p(x_1,t_1|x_3,t_3) = \int p(x_1,t_1|x_2,t_2)\,p(x_2,t_2|x_3,t_3) \di x_2 \;.
\end{align}
The CKR states that the transition probability from time $t_3$ to $t_1$ can be subdivided into transition probabilities from $t_3$ to $t_2$ and then from $t_2$ to $t_1$. Such a sequence of realisations $(x_1, x_2, x_3, \dots)$, in which each realisation $x_i$ determines the probability to observe the next realisation $x_{i+1}$, is commonly referred to as a Markov chain.\\

Markov processes can by subdivided into continuous and discontinuous components. This distinction addresses the values of $X(t)$ at infinitesimal time steps which will be rendered more precisely in the course of this chapter.\\
Regarding the time variable $t$, the above introduction suggests that time is discrete. In fact, the model equations for Markov processes used in this thesis assume that time is continuous. This assumption does not hold for realistic MPs, since any real stochastic process will have a time scale $\tME$ below which the Markov assumption does not hold. We will call $\tME$ the {\it Markov-Einstein time scale}, for reasons which will be discussed in the following section.\\
We will start in section \ref{ss_LE} with Brownian motion, from which intuitively a stochastic differential equation (SDE) arises, the {\it Langevin equation} (LE). The thus motivated SDEs will then be formally defined and discussed. In sections \ref{ss_FPE} and \ref{ss_WPI}, we introduce two equivalent formulations of MPs, the Fokker-Planck equation (FPE) and Wiener path integrals (WPI). Until then, only continuous MPs are considered. In section \ref{ss_ME} we will come back to the CKR and show how continuous MPs fit into the greater class of MPs that also involve discontinuous realisations. In the appendix, \ref{AA_Itocalc}-\ref{AA_overview}, we provide additional material regarding continuous Markov processes.

\subsection{Stochastic differential equations (SDEs)} \label{ss_LE}
\paragraph{The Langevin equation (LE)} In 1908, Paul Langevin\remark{wenn historisch dann mit vornamen} was the first to set up a SDE, in an attempt to simplify Albert Einstein's theory of Brownian motion \cite{Lemons1997}, which Einstein formulated in 1905 \cite{Einstein1905}. Langevin considered spherical particles suspended in a medium (fluid), on which an external force $F_\mr{ex}$ acts (e.g. gravity). Denoting the centre of mass by $x$ and ignoring effects of the medium, Newton's equation of motion reads
\begin{align} \label{eq:SDEs_Newton}
	m\dddx(t) = F_\mr{ex}\big(x(t)\big) \;,
\end{align}
where dots denote derivatives with respect to time $t$, and $m$ is the mass of the particle.\\
For spherical particles in a fluid, it is known that the particles experience a frictional force $F_\mr{fr}$ proportional to the velocity of the particles (\cite{landau1987fluid} p.\,70ff),
\begin{align} \label{eq:SDEs_Stokes}
	F_\mr{fr} = -\,\g\,\ddx \sep \g \dfn 6\pi\nu R \;. 
\end{align}
Here, $\nu$ is the dynamic viscosity and $R$ is the radius of the particle. The above equation is known as {\it Stoke's law}, and $\g$ is referred to as friction coefficient. Stoke's law holds at laminar flow conditions (that is, small Reynolds numbers $\Rey=R\ddx/\nu$).\\
It is clear that due to this friction, the particles will eventually come to rest. Sufficient small particles, however, exhibit an ongoing irregular motion, observed by Robert Brown in 1827, and is hitherto known as Brownian motion. The origin of this motion is the thermal energy of the fluid, due to which the fluid molecules hold a kinetic energy of the order of magnitude $\kB T$ where $\kB$ is Boltzmann's constant and $T$ is the temperature of the fluid. If the particles are small enough, the kinetic energy transferred by collisions from the fluid molecules to the particles results into an observable irregular motion.\\
In his approach to Brownian motion, Langevin considered a thermal force 
\begin{align} \label{eq:SDEs_thermal force}
	F_\mr{th}=\sqrt{2\g \kB T}\xi(t)
\end{align}
acting on the particles, where $\xi(t)$ is a $\d$-correlated random variable sampled from a Gaussian distribution with zero mean and infinite variance.\remark{einzusehen, wenn man in die korrelation ($\d$-funktion) $t=t'$ einsetzt. Dass die Summe (integral) $Z$ trotzdem Gauss ist (obwohl CLT eigentlich nicht gilt), folgt aus $\lla\dd W^n\rra=0$ für $n\geq3$.} Due to its $\d$-correlation, $\xi(t)$ is also called Gaussian white noise. The pathological properties of $\xi(t)$ are elucidated in the appendix, and the main consequences will be discussed shortly.\\
Including the frictional and the thermal force into Newton's equation (\ref{eq:SDEs_Newton}), we arrive at the {\it Langevin equation} (LE)
\begin{align} \label{eq:Langevin}
	\dddx(t) + \g\,\ddx(t) = -V'\big(x(t)\big) + \sqrt{2\g \kB T}\xi(t) \;.
\end{align}
Here we assumed that $F_\mr{ex}(x)$ arises from a potential $V(x)$,\linebreak $F_\mr{ex}(x)=-V'(x)$, which can always be achieved in one dimension, and we set $m\dfn 1$ (force is measured in units of acceleration). Note that the unit of $\xi(t)$ is $1/\sqrt{\mr{s}}$\remark{weil delta-funktion ne pdf ist, also density, und die haben immer als einheit 1/einheit(argument), damit wenn drüber integriert ne dimensionslose eins rauskommt (normalisierung, bzw bei $\d$-fkt definition). und hier ist die varianz von $\xi$ ne $\d$-funktion, also hat $\xi^2$ die einheit 1/s und $\xi$ damit die einheit $1/\sqrt s$.}. \\
The LE illustrates the interplay between damping of the particles due to frictional effects and energy injection to the particles by thermal kicks: The frictional effects withdraw kinetic energy from the particle, energy that is received by the fluid and in turn passed back to the particles as thermal fluctuations. Hence, even in the absence of an external force, Brownian particles exhibit an ongoing irregular motion.\\
The thermal energy $\kB T$ entering the thermal force is intuitive, whereas the occurrence of the friction coefficient $\g$, proportional to $R$ and $\nu$, see (\ref{eq:SDEs_Stokes}), is counter-intuitive. Whereas it is clear that for larger particles both the friction coefficient and the ability to absorb thermal fluctuations increase, it appears dubious that the fluid molecules should exert stronger forces on the Brownian particle for increased viscosity. We unravel this paradox\footnote{In Langevin's derivation the magnitude of the thermal force arises from satisfying the equipartition theorem \cite{Lemons1997}\remark{siehe auch Risken p.33f}.} by arguing that with higher viscosity, the collisions between fluid molecules and Brownian particle become more inelastic, and consequently, more heat is transferred to the fluid, which, in order to maintain the constant temperature $T$, is in return received by the Brownian particle through an increased strength of kicks by the fluid molecules. This fast equilibration mechanism is coarse grained as Gaussian white noise, and $\g$ should in that context merely be thought of a coupling coefficient between fluid and particle.\remark{wenn ich die temperatur erhöhe, erhöhe ich gleichmaßen die stöße in und gegen bewegungsrichtung, die reibung wird also nicht beeinflusst. Wenn ich die Viskosität, also die Reibung $\g$ erhöhe, und die Stärke der Fluktuationen sich nicht ändern würde, dann erwärmt sich auch das Fluid und die thermischen Stöße werden stärker. Ich will aber die Temperatur konstant halten, also muss diese extra Stoßkraft separat eingehen. Deswegen $\g\kB T$ in LE.}\remark{Anders ausgedrückt, stärkere Reibung bedeutet inelastischere Stöße, bedeutet mehr Wärme im Fluid welche gleich wieder ans Teilchen abgegeben wird ($T=\const$, ich kann ja unabhängig an $\g$ drehen), also stärkere Fluktuationen. Diese Kollisions-Dynamik wird in einen einzigen Schritt zusammengefasst und als weißes Rauschen modelliert, in dessen Stärke dann also auch gamma eingehen muss.}\\
The assumption that $\xi(t)$ is on average zero, accounts for the fact that a resting particle experiences decelerating and accelerating collisions in equal measures. The excess of decelerating collisions for a moving particle is taken into consideration by the frictional force. As a rising temperature increases the decelerating and accelerating collisions in equal shares, the temperature does not enter the frictional force.\\

The above considerations suggest that fluctuating forces and dissipative forces have the same origin - the collisions with fluid molecules. This interrelation of fluctuation and dissipation is the essence of fluctuation-dissipation theorems, which in its simplest form is the Einstein relation
\begin{align} \label{eq:D_einstein}
  D = \frac{\kB T}{\g} \;,
\end{align}
where $D$ is the diffusion constant \cite{Einstein1905}. The Einstein relation expresses that the diffusion of a particle is a result of both thermal fluctuations and frictional dissipation.\\
The diffusion of particles was subject of research before Einstein derived (\ref{eq:D_einstein}) from his theory of Brownian motion. In the course of that research, Adolf Fick derived in 1855 (\cite{Gardiner2009} p.\,336)
\begin{align} \label{eq:Fick}
	j(x,t) = -D\,\pt_xp(x,t) \;,
\end{align}
where $j(x,t)$ measures the flux of particles per area and time, and $p(x,t)$ is the particle density. The determination of the diffusion constant by Einstein in 1905 given by (\ref{eq:D_einstein}) revealed the rather unexpected relation between friction and diffusion and was confirmed by Smoluchovski (1906) and by Langevin (1908). The more general fluctuation-dissipation theorems are part of response theory, we direct the interested reader to the recent overview article by Marconi et al. \cite{MARCONI2008}.\\
Fick's law (\ref{eq:Fick}) can be generalised to diffusing particles subject to external forces, which leads to an evolution equation for $p(x,t)$ where only force and diffusion enter.\remark{kontinuitätsgleichung führt auf diffusionsgleichung ohne drift, zusammen mit drift wird ne FPE draus.} This equation is known as Fokker-Planck equation, to which we will come back in the next section.\\

We return to the LE. For particles with small mass, frictional effects prevail over inertial effects. In this so-called {\it overdamped limit}, the inertial term is negligible, and we obtain the {\it overdamped} LE
\begin{align} \label{eq:LE_newton_force}
	\g\ddx(t) = -V'\big(x(t)\big) +  \sqrt{2\g \kB T}\xi(t) \;.
\end{align}
The overdamped LE is often given in terms of velocities,
\begin{align} 
	\ddx(t) &= -\G V'\big(x(t)\big) + \sqrt{2\kB T\G}\xi(t) \nn
	&= -\G V'\big(x(t)\big) + \sqrt{2D}\xi(t) \sep D=\kB T\G \;, \label{eq:LE_newton}
\end{align}
where $\G=1/\g$ is the mobility of the particle, and $-\G V'(x)$ is now the {\it drift velocity} and $\sqrt{2\kB T\G}\xi(t)$ accounts for abrupt changes in velocity after collisions with fluid molecules. In this form, the occurrence of the mobility $\G$ in both the drift and diffusion term is intuitive, which in retrospect justifies the dependency of the thermal force $\sqrt{2\g\kB T}\xi(t)$ in (\ref{eq:LE_newton_force}) on $\g$.\remark{The crux is that $\g$ enters the thermal force as square root of $\g$, such that $D=\kB T\G$. In a sense, by the square root, $\kB T$ and $\g$ share the contribution to the thermal force $\sqrt{2\g \kB T}\xi(t)$.}\\
It can be shown that the solutions of the LE (\ref{eq:LE_newton}) are continuous Markov processes (\cite{Gardiner2009} p.\,92). Brownian motion itself, however, is not a Markov process on arbitrary time scales, since the Markov assumption does not hold for the detailed dynamics of the collisions. But the outcome of each collision does only depend on the initial condition of this collision, that is on only the most recent collision, which is precisely the Markov assumption. The time scale $\tME$ above which Brownian motion can be assumed to be Markovian is therefore of the same order of magnitude as the mean time between collisions. Referring to Einsteins theory of Brownian motion, the time scale $\tME$ above which a process is Markovian is often called {\it Markov-Einstein scale}.

\begin{figure}[t] 
	\begin{center}
		\includegraphics[width=0.95\textwidth]{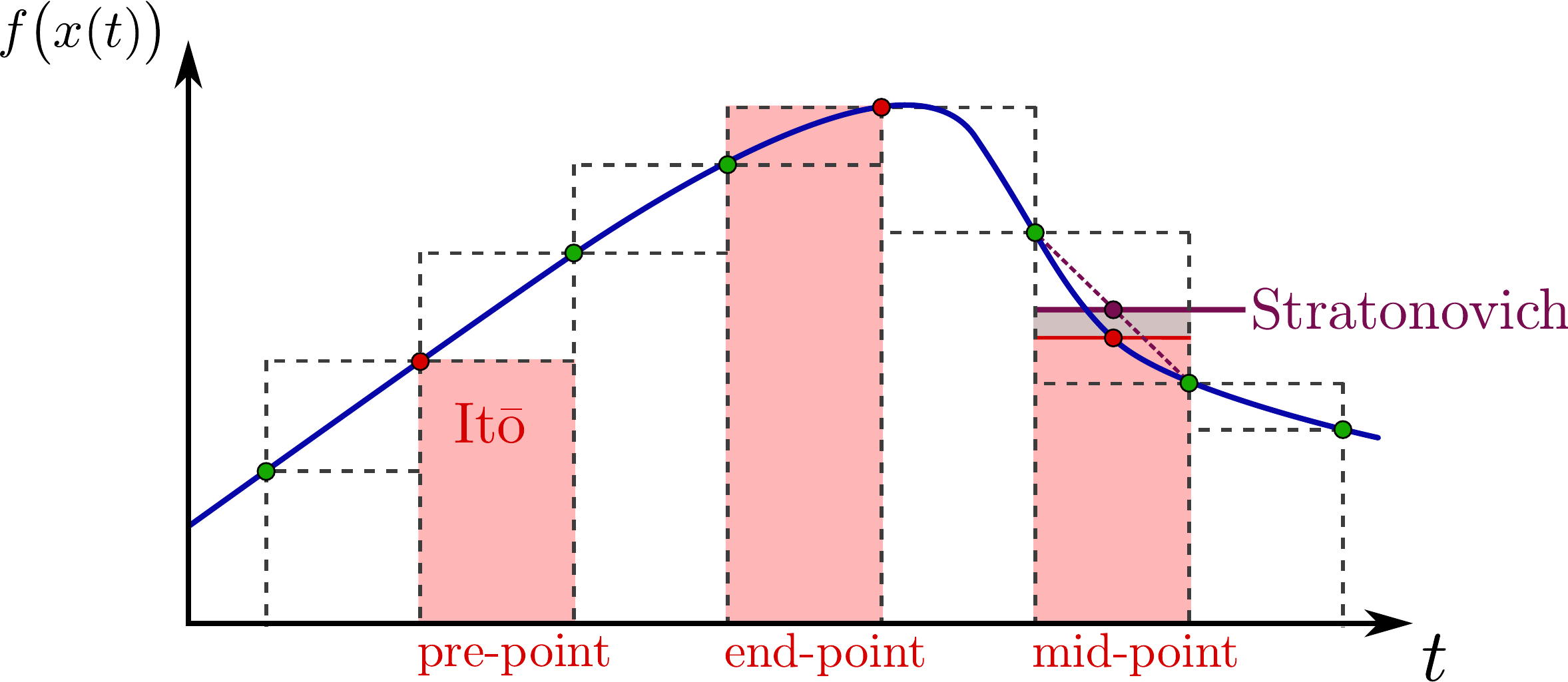}
		\caption{\label{ff:a-point}\figtxt{Discretisation rules and conventions. The integrand of a stochastic integral is given as the blue line (for clarity, as a smooth function). The vertical and horizontal dashed lines indicate lower and upper sums. The coloured circles mark the discretisation intervals. The red circles exemplary designate points taken for pre-point, mid-point and end-point, the light-red coloured areas are the resulting rectangles to approximate the integral. The \Ito~convention is identical with the pre-point rule, and the Stratonovich convention, added as purple symbols, is identical with the trapezoidal rule.}}
	\end{center}
\end{figure}

\paragraph{Stochastic calculus} Bridging Brownian motion to the theory of Markov processes brings us back to the formal level. We will return to the intuitive picture in chapter \ref{s_td_interpretation}.\\
On the formal level, $x(t)$ is now a continuous degree of freedom in a MP, and the LE is a {\it stochastic differential equation} (SDE). The general form of a SDE reads
\begin{subequations} \label{eq:LE-no-alpha}
	\begin{align} \label{eq:LE-no-alpha_ode}
		\ddx_t=f(x_t,t)+g(x_t,t)\,\xi(t) \sep x(t=t_0)=x_0 \;,
	\end{align}
	\begin{align} \label{eq:LE-no-alpha_xi}
		\lla\xi(t)\rra=0 \sep \lla\xi(t)\xi(t')\rra=\d(t-t') \;,
	\end{align}
\end{subequations}
where for clarity $x(t)$ is denoted by $x_t$. In view of the inherent time scale separation of realistic Markov processes, the white noise $\xi(t)$ is not exactly $\d$-correlated, but rather of the form
\begin{align} \label{eq:SDEs_xi_not-delta}
	\lla\xi(t)\xi(t')\rra \sim \mfrac{1}{\tME}\, \ee{-\frac{|t-t'|}{\tME}} \;.
\end{align}
In this case, the noise is called {\it coloured} noise.\remark{fourier trafo von $\d$-korreliertem $\xi(t)$ is constant, also weiß, ansonsten eben bunt.} In the limit $\tME\to0$, the {\it white-noise limit}, the exponential function becomes the $\d$-function $\d(t-t')$.
Solving SDEs for time-steps $\Dt<\tME$ is therefore an interpolation of the real process to non-Markovian time scales.\\
By the attempt to solve a SDE, however, we encounter a problem. Naive integration yields
\begin{align} \label{eq:SDEs_int}
	x(t)-x(t_0) = \int_{t_0}^{t}f(x_\t,\t)\di \t + \int_{t_0}^{t}g(x_\t,\t)\,\xi(\t) \di \t \;.
\end{align}
Due to the stochastic variable $x(t)$, the above integrals are {\it stochastic integrals} for which the usual Riemann definition of an integral is not suitable. The consequence is that, in the continuous limit, the value of the stochastic integral is not unique, depending on whether we define the integral as a lower sum or upper sum. We will therefore parametrise the value of stochastic integrals with $\a$, where $\a=0$ corresponds to taking the value of the integrand at the beginning of each discretisation interval, $\a=1$ to the end of each interval, and $0<\a<1$ to somewhere in-between. The two mostly used definitions of stochastic integrals are the {\it pre-point} rule ($\a=0$) and the {\it mid-point} rule ($\a=1/2$). See figure \ref{ff:a-point} for an illustration.\\
\\
It is clear that, since the stochastic integrals in (\ref{eq:SDEs_int}) depend on $\a$, also the solution of the SDE will depend on $\a$. If we take, for instance, the SDE
\begin{align} \label{eq:SDEs_GBM}
	\ddx_t=a\,x(t)+\sqrt{2b}\,x(t)\,\xi(t) \sep x(t=t_0)=x_0 \,
\end{align}
which is known as {\it geometric Brownian motion} (GBM), we find that the solution reads for $x_0x(t)>0$ (cf. (\ref{eq:A1_GBM_Ito_xsol}), (\ref{eq:A5_a2a}))\remark{$\tilde\a=0$, dann kann ich in pre-point lösen und hab noch $\a$ drin.}
\begin{align} \label{eq:SDEs_GBM_sol}
	x(t) = x_0\,\exp\big[\big(a+(2\a-1)\,b\big)\,(t-t_0)+\sqrt{2b}\,Z(t-t_0)\big] 
\end{align}
where $Z(t)$ is a Gaussian random variable with zero mean and variance $t$. We see that for mid-point, $\a=1/2$, the deterministic parameter $a$ of the SDE determines the mean, and the stochastic parameter $b$ fixes the variance of $\ln x(t)/x_0$. On the other hand, for $a=0$ but $\a\neq1/2$, the mean of $\ln x(t)/x_0$ does not vanish, despite the vanishing deterministic parameter. Such a noise induced drift is generally referred to as {\it spurious drift}, in the sense that it does not arise physically but from the discretisation rule applied to the SDE.\\
The immediate consequence of the ambiguity of SDEs is that generally, along with the SDE itself, also the discretisation rule needs to be defined, otherwise the solution of the SDE is not unique. The definition of the discretisation rule is also referred to as an {\it interpretation} of the SDE, or as {\it convention}.\\
A crucial point is that for $\a\neq1/2$ the usual rules of calculus are not applicable. A change of variable in the SDE or the integration rules for polynomials in $x(t)$, for instance, need to be modified. For $\a=0$, such a modified calculus was developed by the mathematician Kiyoshi \Ito, now known as {\it \Ito~calculus}. For an introduction to \Ito~calculus see appendix \ref{AA_Itocalc}, for further readings we recommend the books by Gardiner \cite{Gardiner2009} or van Kampen \cite{vanKampen2007}. The charm of the \Ito~convention is the mathematical feasibility of the rigorous theory. For practical purposes, the mid-point rule is often preferable, since then the ordinary rules of calculus apply. Defining the mid-point rule along with a SDE is usually referred to as {\it Stratonovich convention}, after the physicist Ruslan L. Stratonovich.\footnote{More specific, contrary to the mid-point rule, the Stratonovich stochastic integral takes the average of beginning and end of the discretisation intervals (\cite{Gardiner2009} p.\,96), see also figure \ref{ff:a-point}. A distinction that becomes irrelevant for small discretisation intervals.}\\

The ambiguity of SDEs implies a dilemma. Suppose we want to describe a stochastic system with a SDE, and identify from the physics of the system the deterministic and stochastic influences as $f(x,t)$ and $g(x,t)$. So far so good, but how can we decide from physical arguments which convention is the correct one, such that we can expect that the SDE correctly predicts the outcome of experimental measurements? This dilemma is known as {\it \Ito-Stratonovich dilemma}, and still debated in current research \cite{Tupper2012,Pesce2012,Yang2013,Durang2013,Kuroiwa2013,Farago2014,Rubin2014}. To deal with this dilemma, we will follow the argumentation by van Kampen \cite{Kampen1981} (see also \cite{vanKampen2007} p.\,232 ), in which he distinguishes between {\it external} and {\it internal} noise. External noise refers to the case where the source of stochasticity is separate from the source of deterministic dynamics, whereas internal noise arises from an intertwined source of both stochastic and deterministic dynamics. In other words, external noise comes on top of a pure deterministic process and vanishes on average, internal noise is intrinsically present in the system and implies a non-vanishing average tendency.\label{a-dilemma}\\
Van Kampen claims that in the case of external noise, a SDE can be meaningfully set up, and Stratonovich convention has to be used. This is in accord with our previous observation regarding the example of GBM that $\lla x(t)\rra\!\equiv\! x_0$ for $f(x,t)\!\equiv\!0$ and $\a\!=\!1/2$. Additional support for van Kampen's claim follows from reverting to coloured noise as in (\ref{eq:SDEs_xi_not-delta}). In this case, solutions of the SDE do not depend on $\a$, and the \Ito-Stratonovich dilemma is baseless. In the white noise limit, the solutions of the SDE with coloured noise reproduce the solutions of the SDE with white noise in the Stratonovich interpretation.\\
If the stochasticity of the processes arises from internal noise, van Kampen argues that a SDE is not the natural equation to describe the process. Instead a {\it master equation}, which we will discuss in chapter \ref{ss_ME}, is advisable. Under certain conditions, a master equation can be recast as a Fokker-Planck equation which is formally equivalent to a SDE. The interpretation of the SDE is then simply a matter of taste, as long as the equivalence to the Fokker-Planck equation is retained. The defining functions $f(x,t)$ and $g(x,t)$ of the SDE will then depend on $\a$, and it is tedious to attribute a meaning to $f(x,t)$ and $g(x,t)$ for various $\a$. For thermodynamic diffusion processes, Lau and Lubensky argued that the choice $\a=1$ is the most convenient one \cite{LauLubensky07PRE}.\\

In closing, we mention that the practical significance of SDEs is to generate realisations $x(t)$ of a stochastic process $X(t)$. This can be achieved by solving the SDE explicitly and trace the randomness of $x(t)$ back to random variables of known PDFs as we did in (\ref{eq:SDEs_GBM_sol}). The analytic solution of a SDE, however, can rarely be achieved. The numerical solution of SDEs is therefore the predominant practice when dealing with SDEs. The numerical results of this thesis are achieved by the first order integration scheme given by (\ref{eq:A3_SDEint_alpha}).\\
In the appendix \ref{AA_overview}, we provide an overview of equivalent descriptions of continuous Markov processes that are defined via a SDE. We also demonstrate how a SDE in various interpretations can be reformulated to an equivalent SDE in a different interpretation.

\subsection{The Fokker-Planck equations (FPE)} \label{ss_FPE}
In the previous chapter we discussed SDEs which are the evolution equations for a stochastic variable $x(t)$. Instead of considering realisations $x(t)$, we can also describe the process by PDFs $p(x,t)$ of $x$ for each instant of time $t$. The evolution equation for these PDFs is the {\it Fokker-Planck equation} (FPE), which we will discuss in this section.\\

\paragraph{Equivalence to SDEs} In the appendix \ref{AA_SDE2FPE}, we demonstrate that the equivalent FPE of a SDE of the form (\ref{eq:LE-no-alpha}) reads
\begin{align} 
  \ddp(x,t) &= \pt_x\big[-f(x,t) - \a g'(x,t)g(x,t) + \pt_x\tfrac{1}{2}g(x,t)^2\big]p(x,t) \;, \nn
  p(x,t=t_0)) &= p_0(x) \label{eq:FPE_fg} \;,
\end{align}
with the initial PDF $p_0(x)$. Note that the FPE depends on $\a$, as was to be expected in view of the discussion regarding the interpretation of SDEs in the previous section. However, a FPE alone does not involve stochastic integrals and should as such not be plagued by the interpretation ambiguity. We therefore rewrite the FPE by defining
\begin{subequations} \label{eq:def_D12}
	\begin{align}
		\Df(x,t) &\dfn f(x,t) +\a g'(x,t)g(x,t) \;, \label{eq:def_D1} \\
		\Dg(x,t) &\dfn \tfrac{1}{2}g(x,t)^2 \label{eq:def_D2}
	\end{align}
\end{subequations}
as suggested by Lau and Lubensky \cite{LauLubensky07PRE}. The deterministic coefficient\linebreak $\Df(x,t)$ is known as {\it drift} coefficient, and the {\it diffusion} coefficient $\Dg(x,t)$ is the state and time dependent equivalent of the diffusion constant we met in Fick's law (\ref{eq:Fick}) and the Einstein relation (\ref{eq:D_einstein}).\footnote{The enumeration of drift as (1) and diffusion as (2) will become clear when discussing in \ref{ss_ME} how a FPE arises from a master equation.} In terms of $\Dfg(x,t)$, the FPE reads
\begin{align} \label{eq:FPE_D1D2}
  \ddp(x,t) = \pt_x\big[\-\Df(x,t) \+ \pt_x\Dg(x,t)\big]p(x,t) \sep p(x,t_0) = p_0(x)
\end{align}
which is sometimes also referred to as the \Ito~form of the FPE, as for $\a=0$ we have $\Df(x,t)=f(x,t)$.\\
By defining the probability current density
\begin{align} \label{eq:def_J}
  j(x,t) = -\big[-\Df(x,t) + \pt_x\Dg(x,t)\big]p(x,t) \;,
\end{align}
the FPE obtains the form of a continuity equation \remark{(van Kampen p. 193)}
\begin{align} \label{eq:FPE_J}
  \pt_tp(x,t) + \pt_xj(x,t) = 0 \;.
\end{align}
As objected by Lau and Lubensky \cite{LauLubensky07PRE}, however, setting $\Df(x,t)\equiv0$ does not reproduce the correct generalisation of Fick's law (\ref{eq:Fick}) for state dependent diffusion. If we instead define
\begin{subequations} \label{eq:def_FD}
\begin{align} 
	F(x,t) &\dfn \Df(x,t) - \pt_x\Dg(x,t) \\
	D(x,t) &\dfn \Dg(x,t)\footnotemark
\end{align}
\end{subequations}
\footnotetext{The superfluous definition of $D(x,t)$ is for convenience and a better distinction from $\Dfg(x,t)$.}and perform the derivative of $\pt_x\Dg(x,t)$ in (\ref{eq:FPE_D1D2}), the FPE and the current take the form
\begin{align} 
  \ddp(x,t) &= \pt_x\big[\-F(x,t) \+ D(x,t)\pt_x\big]p(x,t) \sep p(x,t_0) = p_0(x) \;, \label{eq:FPE_FD} \\
  j(x,t) &= F(x,t)\,p(x,t) - D(x,t)\,p'(x,t) \;, \label{eq:def_J_FD}
\end{align}
which is sometimes referred to as the Stratonovich form, although $F(x,t)=f(x,t)$ for $\a=1$ (instead of $\a=1/2$). If we now set $F(x,t)\equiv0$, we obtain the correct generalisation of Fick's law (\ref{eq:Fick}). In \ref{ss_td-interpration_energy}, we will see that $F(x,t)$ is the drift velocity in reaction to an external force, and the stationary distribution of the Stratonovich FPE coincides with the thermal equilibrium distribution. The connection to thermal equilibrium and the correct generalisation of Fick's law led Lau and Lubensky to call the choice $\a=1$ the thermodynamic consistent convention.\footnote{More specific, they argued that the definition $F(x,t)=f(x,t) + (\a-1) g'(x,t)g(x,t)$ is thermodynamic consistent, in which $\a=1$ is the most convenient choice.}\remark{diss p.29} An overview of equivalent descriptions of continuous Markov processes defined by FPEs is provided in the appendix (\ref{AA_overview}).\\
We note that an useful quantity is the local average velocity $\lla\ddx|x,t\rra$ which allows to write the probability density current as the product
\begin{align} \label{eq:def_avg-xt}
	j(x,t) = \lla\ddx|x,t\rra p(x,t) \;.
\end{align}
We also note that for the initial PDF $p_0(x)=\d(x-x_0)$, the solution of the FPE will be the conditional PDF $p(x,t|x_0,t_0)$, from which by
\begin{align}
  p(x,t) = \int p(x,t|x_0,t_0)\,p_0(x_0) \di x_0
\end{align}
the solution of the FPE for any other initial PDF can be determined. Therefore, $p(x,t|x_0,t_0)$ is the Green's function of a FPE.\\

\paragraph{Moment equation} Often of interest are the moments of a PDF instead of the PDF itself. We define the moments of $p(u,r)$ as
\begin{align}
  S^{\,n}_x(t) \dfn \lla\, x^n \,\rra_{p(x,t)} = \int x^n\,p(x,t) \di x \;.
\end{align}
If $p(u,r)$ is the solution of a FPE, we find by multiplying the FPE with $x^n$ and integrating with respect to $x$,
\begin{align} 
  \int x^n\,\ddp(x,t) \di x &= \int x^n\,\pt_x\big[-\Df(x,t) + \pt_x\Dg(x,t)\big]p(x,t) \di x \label{eq:FPE_moments_pre1} \\
  &= -\int x^n\pt_x\Df(x,t)p(x,t) \di x + \int x^n\pt_x^2\Dg(x,t)p(x,t) \di x \nn
  &= n\,\lla\,x^{n\-1}\Df(x,t)\,\rra_{p(x,t)} + n(n\-1)\,\lla\,x^{n\-2}\Dg(x,t)\,\rra_{p(x,t)} \nonumber
\end{align}
where the last line follows from repeated integration by parts and vanishing boundary terms due to the normalisation condition of $p(x,t)$.\\
Since $x$ is in (\ref{eq:FPE_moments_pre1}) a independent variable and not to be confused with the time-dependent path $x(t)$ as we had in the context of SDEs, we immediately obtain for the $n$-th moment $S^{\,n}_x(t)$ the ODE
\begin{align} \label{eq:FPE_moments}
  \dot S^{\,n}_x(t) = n\lla x^{n\-1}\Df(x,t)\rra_{p(x,t)} + n(n\-1)\lla x^{n\-2}\Dg(x,t)\rra_{p(x,t)} \,.
\end{align}
For polynomial $\Dfg$ in $u$, the averages on the right hand side of (\ref{eq:FPE_moments_pre1}) reduce to a combination of moments $S^{\,m}_x(t)$ of different order $m$, and consequently, the integration of (\ref{eq:FPE_moments_pre1}) to obtain $S^{\,n}_x(t)$ requires in general the knowledge of moments of order $m\neq n$.\\
For the special case $\Df(x,t)=d_1(t)x$ and $\Dg(x,t)=d_2(t)x^2$, however, we get
\begin{align} 
  \dot S^{\,n}_x(t) &= n\,\lla\,x^{n\-1}d_1(t)x\,\rra_{p(x,t)} + n(n\-1)\,\lla\,x^{n\-2}d_2(t)x^2\,\rra_{p(x,t)} \nn
  &= \big[nd_1(t)+n(n\-1)d_2(t)\big]\,S^{\,n}_x(t) \;, \label{eq:FPE_moments_GBM}
\end{align}
with the solution
\begin{align}
  S^{\,n}_x(t) = S^{\,n}_x(t_0)\,\exp\bigg[\int_{t_0}^{t} (d_1(t)-d_2(t))n+d_2(t)n^2\di t\bigg] \;.
\end{align}
The $S^{\,n}_x(t)$ are the moments of a log-normal distribution with mean $\mu=\int_{t_0}^{t} (d_1(t)-d_2(t))\di t$ and variance $\s^2=2\int_{t_0}^{t} d_2(t))\di t$.\\
For the simple choice $d_1(t)\equiv a+2\a b$ and $d_2(t)\equiv b$, which, as we see from (\ref{eq:FPE_fg}), is equivalent to GBM defined in (\ref{eq:SDEs_GBM}) for arbitrary convention $\a$, we obtain $\mu=\big(a+(2\a-1)b\big)(t-t_0)$ and $\s^2=2b(t-t_0)$, which are indeed mean and variance of $\ln x(t)$, where $x(t)$ is the solution (\ref{eq:SDEs_GBM_sol}) of the SDE (\ref{eq:SDEs_GBM}) of GBM. This simple example demonstrates the equivalence of SDEs and FPEs for arbitrary $\a$.\\

\paragraph{Stationary solution} An important characteristic of a FPE is its stationary solution $\pst(x)$. For a stationary process, that is $F(x,t)=F(x)$ and $D(x,t)=D(x)$ are time independent, the stationary distribution $\pst(x)$ is invariant under the dynamics defined by the FPE and can be determined by
\begin{align}
  0\meq\pt_x\big[\-F(x) + D(x)\pt_x\big]\pst(x) \,.
\end{align}
By writing the above equation in terms of the Fokker-Planck operator
\begin{align} \label{eq:FPE_OP}
  \hLLFP=\pt_x\big[\-F(x) + D(x)\pt_x\big] \sep \hLLFP\pst=0\,\pst \;,
\end{align}
it becomes apparent that $\pst(x)$ is the eigenfunction corresponding to zero eigenvalue, and hence the time evolution of every initial distribution $p_0(x)$ will in most cases converge to that stationary distribution $\pst(x)$\todo{(hat irgendwas damit zu tun, dass ev=0 der niedrigste ev ist, also dem der niedrigsten energie entspricht, siehe nochmal gardiner, risken. wahrscheinlich hat dann gbm kein wirkliches pst, da es nicht ergodisch ist? muss doch wohl in gardiner, risken, wiki, google, ... stehen)}.\\
For the case of time dependent $F(x,t)$ and $D(x,t)$, an {\it instantaneous} stationary distribution $\pst(x,t)$ can be defined. With 'instantaneous' is meant that the time variable $t$ in $F(x,t)$, $D(x,t)$ and $\pst(x,t)$ is treated as a parameter which can be fixed to a certain value, whereas the time $t$ in $p(x,t)$ and $\ddp(x,t)$ is the evolving time. To clarify this point, we temporarily introduce the constant parameter $\k$ and write for the FPE
\begin{align} \label{eq:FPE_protocol}
  \ddp(x,t) = \pt_x\big[-F(x;\k) + D(x;\k)\pt_x\big]p(x,t)
\end{align}
with now $F(x;\k)$ and $D(x;\k)$ being constant in time. The stationary distribution $\pst(x;\k)$ also depends on $\k$ and by setting $\ddpst(x;\k)=0$ in the FPE (\ref{eq:FPE_D1D2}), we arrive at the conditional equation for $\pst(x;\k)$,
\begin{align}
  0 = \pt_x\big[F(x;\k) - D(x;\k)\pt_x\big]\pst(x;\k) \;.
\end{align}
Integration with respect to $x$ yields
\begin{align}
  c(\k) = F(x;\k)\,\pst(x;\k) - D(x;\k)\,\pt_x\pst(x;\k)
\end{align}
with a constant $c(\k)$. Referring to (\ref{eq:def_J}), we identify the integration constant to be the stationary probability current, $c(\k)=\Jst(\k)$. In one dimension and for no periodic boundary conditions, we demand that $\Jst(\k)\equiv0$ to ensure $\ddpst(x;\k)=0$.\\
Upon isolation of $\pst(x;\k)$ and integration, we obtain
\begin{align} \label{eq:def_pst_FD}
	\pst(x;\k) = \exp\Big[-\big[\p(x;\k) - \GG(\k)\big]\Big] \;,
\end{align}
defining
\begin{subequations} \label{eq:def_pst_FD}
  \begin{align}
		\p(x;\k) &= -\int_{-\infty}^x \frac{F(x';\k)}{D(x';\k)} \di x' \;, \label{eq:defs_pst_phi_FD} \\
		\GG(\k) &= -\ln Z(\k) = -\ln \int \exp\big[-\p(x;\k)\big]\di x \;, \label{eq:defs_pst_FF+Z_FD}
	\end{align}
\end{subequations}
where $Z(\k)$ is the normalisation constant included into the exponential of $\pst(x;\k)$ in terms of $\GG(\k)$.\\
For each value of $\k$ and any initial distribution $p_0(x)$, the solution of the FPE will in general become $\pst(x;\k)$ in the limit $t\to\infty$. By writing again time $t$ instead of parameter $\k$, $\pst(x,t)$ hence is the stationary distribution for any instance of time $t$ at which we keep $F(x,t)$ and $D(x,t)$ fixed and let the dynamics evolve. Instead of simply $\k=t$, we could also outsource the time-dependency of $F(x,t)$ and $D(x,t)$ by leaving $\k(t)$ unspecified. The {\it protocol} $\kc$ can take any functional form, of which $\k(t)=t$ is just the simplest choice. We will come back to the notion of protocols in the context of fluctuation theorems in section \ref{ss_td-interpration_FTs}.\\

If the system is in a steady state, i.e. $p(x,t)=\pst(x)$, the appendant probability current $\Jst(x)$ must necessarily be constant in time which is a direct consequence of the continuity equation (\ref{eq:FPE_J}). Except for periodic boundary conditions, the current $\Jst(x)$ must vanish at its boundaries to preserve probability. Hence the probability current must vanish in a steady state, $\Jst(x)\equiv0$.\\
If, however, the MP takes place in higher dimensions, i.e. $\vek{X}(t)\in\mathbb{R}^n$ with $n\geq2$, the continuity equation becomes
\begin{align} \label{eq:highdim_conti-equ}
	\dot p(\vek{x},t) + \vek{\nabla} \vek{j}(\vek{x})=0 \;,
\end{align}
from which it is evident that it now is sufficient to require that the stationary probability current is divergence free, i.e. $\vek{\nabla}\vJst(\vek{x})=0$. This requirement implies that the vector field $\vJst$ is a pure curl, it is therefore natural to divide $\vek{F}(\vek{x})$ into a conservative part deriving from the scalar potential $\p(\vek{x},t)$ and an additional non-conservative vector-field $\vek{A}(\vek{x},t)$,
\begin{align} \label{eq:highdim_Fcnc}
	\vek{F}(\vek{x},t) =  \mat{D}(\vek{x},t)\,\big[\-\vek{\nabla}\p(\vek{x},t) + \vek{A}(\vek{x},t)\big] \,,
\end{align}
where we now have a diffusion matrix $\mat{D}(\vek{x},t)$ instead of a scalar diffusion coefficient.\\
The FPE and the probability current now read
\begin{align} 
	\dot p(\vek{x},t) &= -\vek{\nabla}\big[\vek{F}(\vek{x},t) - \mat{D}(\vek{x},t)\vek{\nabla}\big]p(\vek{x},t) \;, \label{eq:highdim_FPE}\\ 
	\vek{j}(\vek{x},t) &= \vek{F}(\vek{x},t)p(\vek{x},t) - \mat{D}(\vek{x},t)\vek{\nabla}p(\vek{x},t) \;. \label{eq:highdim_J}
\end{align}
We further write the stationary distribution as
\begin{align} \label{eq:highdim_pst}
	\pst(\vek{x},t)=\exp\big[-\phi(\vek{x},t)\big] \;,
\end{align}
where it has to be stressed that the non-equilibrium potential $\phi(\vek{x},t)$ is distinct from $\p(x,t)$ defined in (\ref{eq:defs_pst_phi_FD}) in the sense that\linebreak $\vek{F}(\vek{x},t) \neq \mat{D}(\vek{x},t) \vek{\nabla} \phi(\vek{x},t)$ due to the non-conservative part of $\vek{F}(\vek{x},t)$ in (\ref{eq:highdim_Fcnc}). In fact, the analytic determination of $\phi(\vek{x},t)$ is only in particularly simple cases possible. \\
Substituting $\pst(\vek{x},t)$ from (\ref{eq:highdim_pst}) and $\vek{F}(\vek{x},t)$ from (\ref{eq:highdim_Fcnc}) into (\ref{eq:highdim_J}), we get for the stationary current 
\begin{align} \label{eq:highdim_Jst}
	\vJst(\vek{x},t) = \mat{D}(\vek{x},t)\vek{A}(\vek{x},t)\,\pst(\vek{x},t)
\end{align}
and consequently for the stationary local mean velocity
\begin{align} \label{eq:highdim_mean-ddx}
	\lla\vek{\ddx}|\vek{x},t)\rra_\mr{st} = \mat{D}(\vek{x},t)\vek{A}(\vek{x},t) \;.
\end{align}
Only for the case of no non-conservative forces, i.e. $\vek{A}(\vek{x},t)\equiv0$, we have a vanishing stationary current.\\

It is insightful to express the FPE in terms of the stationary distribution and stationary current. To this end, we plug $\pst(\vek{x},t)$ from (\ref{eq:highdim_pst}) into the current (\ref{eq:highdim_J})\,
\begin{align}
	\vJst(\vek{x},t) = \big[\vek{F}(\vek{x},t)+\mat{D}(\vek{x},t)\vek{\nabla}\phi(\vek{x},t)\big]\,\ee{-\phi(\vek{x},t)} \;,
\end{align}
and solve for $\vek{F}(\vek{x},t)$,
\begin{align}
	\vek{F}(\vek{x},t) = \ee{\phi(\vek{x},t)}\vJst(\vek{x},t)-\mat{D}(\vek{x},t)\vek{\nabla}\phi(\vek{x},t) \;,
\end{align}
to arrive at
\begin{align}
	\dot p(\vek{x},t) &= -\vek{\nabla}\Big[\ee{\phi(\vek{x},t)}\vJst(\vek{x},t)-\mat{D}(\vek{x},t)\vek{\nabla}\phi(\vek{x},t) - \mat{D}(\vek{x},t)\vek{\nabla}\Big] p(\vek{x},t) \nn
	&= -\vek{\nabla}\,\Big[\ee{\phi(\vek{x},t)}\vJst(\vek{x},t) -\mat{D}(\vek{x},t)\ee{-\phi(\vek{x},t)}\vek{\nabla}\ee{\phi(\vek{x},t)}\Big]\,p(\vek{x},t) \nn
	&= -\vek{\nabla}\,\bigg[\lla\dot{\vek{x}}|\vek{x},t\rra p(\vek{x},t) -\mat{D}(\vek{x},t)\pst(\vek{x},t)\vek{\nabla}\frac{p(\vek{x},t)}{\pst(\vek{x},t)}\bigg] \;. \label{eq:highdim_FPEst}
\end{align}
The Fokker-Planck operator can then be written as
\begin{align} \label{eq:FPE_OP_st}
	\hLLFP(\vek{x},t) = -\vek{\nabla}\,\Big[\ee{\phi(\vek{x},t)}\vJst(\vek{x},t) - \mat{D}(\vek{x},t)\ee{-\phi(\vek{x},t)}\vek{\nabla}\ee{\phi(\vek{x},t)}\Big] \;,
\end{align}
which shows that the dynamics can independently be fixed by choosing $\{\phi,\,\mat{D},\,\vJst\}$. In particular, fixing $\phi$ and $\mat{D}$ but varying the current $\vJst$ (making sure it is still divergence free), we can define a family of dynamics with identical $\phi(\vek{x},t)$ and $\mat{D}(\vek{x},t)$ but distinct currents $\vJst(\vek{x})$. We will pick up this notion in the context of entropies in section \ref{ss_td-interpration_entropy} and fluctuation theorems in section \ref{ss_td-interpration_FTs}.\\

More details regarding the FPE and its various applications can be found in the book by Risken \cite{Risken89}.

\subsection{Wiener path integrals (WPIs)} \label{ss_WPI}
So far, we addressed continuous Markov processes $X(t)$ by evolution equations for realisations $x(t)$, SDEs, and an evolution equation for the PDF $p(x,t)$ of the random variable $x$ at time $t$, the FPE. In this section, we introduce a third notion, which allocates to a certain realisation $x(t)$ a probability measure $P[\xc]$. To emphasise that we do not evaluate a function $x(t)$ at a certain time $t$ but rather take the whole path from initial time $t_0$ to final time $t$, we will write $\xc$ for realisations of the continuous MP. The probabilistic description of sample paths $\xc$ is given by {\it Wiener path integrals} (WPIs), which were introduced by Norbert Wiener in the 1920s for Brownian motion (and related diffusion processes), and which we will recast here on a formal level. In doing so, we build on the book by Chaichian and Demichev \cite{ChaichianDemichev01}, and the article by Lau and Lubensky \cite{LauLubensky07PRE}.\\

The point of origin of WPIs is the Chapman-Kolmogorov relation (CKR) (\ref{eq:CKR}) which is of the form
\begin{align}
  p(x_3,t_3|x_1,t_1) = \int p(x_3,t_3|x_2,t_2)\,p(x_2,t_2|x_1,t_1) \di x_2 \;.
\end{align}
Here, $(x_1,x_2,x_3)$ are subsequent measurements of a continuous MP at times $t_1>t_2<t_3$. In order to capture not only three points but the whole path $\xc$, we combine $N\-2$ CKRs and write the Green's function of a FPE as
\begin{align} \label{eq:WPIs_CKRmulti}
	p(x,t|x_0,t_0) = \int\left[\prod\limits_{i=2}^{N-1}\dd x_i\right]\;\prod\limits_{i=1}^{N-1}\,p(x_{i+1},t_{i+1}|x_i,t_i)
\end{align}
where $(x_1,t_1) \eq (x_0,t_0)$ and  $(x_N,t_N) \eq (x,t)$. This is the prototype of a WPI.\\ 
The conditional probability $p(x_{i+1},t_{i+1}|x_i,t_i)$ is called the {\it propagator}, in the sense that it propagates upon integration the probability of $(x_i,t_i)$ to $(x_{i+1},t_{i+1})$.\\
Based on the SDE $\ddx_t=f(x_t,t)+g(x_t,t)\xi(t)$ in $\a$-point and for small times $\Dt=t_{i+1}\-t_i$, the propagator can be well approximated by
\begin{subequations}
	\begin{align}
		p(x_{i+1},t_{i+1}|x_i,t_i) &\simeq \frac{1}{\sqrt{2\pi\Dt \ga^2}}\,\eee{-\Dt\,s_i(x_i,\,x_{i+1})} \;, \\
		s_i(x_i,\,x_{i+1}) &\dfn \frac{1}{2\ga^2}\bigg(\frac{x_{i+1}\-x_i}{\Dt}
			+\fa+\a \ga \gax\bigg)^2-\a \fax \;.
	\end{align}
\end{subequations}
Here, the index $i_\a$ denotes that $\fa$ and $\ga$ have to be evaluated in $\a$-point. The detailed calculation that led to this result can be found appendix \ref{AA_SDE2WPI}, in essence, it arises from a probability transformation from $\xi(t)$ to $x(t)$.\\
Substituting the above propagator into the repeated CKR (\ref{eq:WPIs_CKRmulti}) yields
\begin{align}
	p(x,t|x_0,t_0) \simeq \frac{1}{\sqrt{2\pi\Dt g_{N_\a}^2}} \int\left[\prod\limits_{i=2}^{N\-1}\!\frac{\dd x_i}{\sqrt{2\pi\Dt g_{i_\a}^2}}\right]\exp\left[-\e\sum\limits_{j=1}^{N\-1} s_j(x_j,x_{j\+1})\right] \label{eq:WPI_discr}
\end{align}
In the continuous limit, $\Dt\to0$ and $N\to\infty$, the sum becomes the stochastic integral
\begin{align} 
	\SS[\xc] &\dfn \int_{t_0}^{t} \frac{\big[\ddx_\t - F(x_\t,\t) + (2\a\-1)D'(x_\t,\t)\big]^2}{4D(x_\t,\t)}-J(x_\t,\t)\di \t \;, \nn
	J(x,t) &\dfn \a F'(x,t) + \mfrac{1}{2} D''(x,t) \label{eq:WPI_action}	
\end{align}
and the remaining integration is symbolically abbreviated as
\begin{align}\label{eq:WPI_int-measure}	
	\int\limits_{(x_0,t_0)}^{(x,t)} \lDi\xc \dfn \frac{1}{\sqrt{4\pi D_N\Dt}} \int\left[\prod\limits_{i=2}^{N\-1}\!\!\frac{\dd x_i}{\sqrt{4\pi D_i \Dt}}\right] \;.
\end{align}
Here, we already substituted $F(x,t)=f(x,t)+(\a\-1)g'(x,t)g(x,t)$ and $D(x,t)=\frac{1}{2}g(x,t)^2$ as defined in the context of a FPE, see (\ref{eq:def_D12}) and (\ref{eq:def_FD}). The functional $\SS[\xc]$ is usually referred to as the {\it action} of the Wiener process, and $J(x,t)$ can be identified as a Jacobian of the underlying transformation from $\xi(t)$ to $x(t)$.\\
Using the above definitions for $\SS[\xc]$ and $\Di\xc$, we finally arrive at the WPI
\begin{subequations} \label{eq:WPI}
\begin{align} 
	p(x,t|x_0,t_0) &= \int\limits_{(x_0,t_0)}^{(x,t)} \Di\xc\,P[\xc|x_0] \;, \label{eq:WPI_P} \\[5pt]
	P[\xc|x_0] &= \exp\big[-\SS[\xc]\,\big] \label{eq:WPI_S} \;,
\end{align}
\end{subequations}
with the probability density functional $P[\xc|x_0]$, which we will simply call {\it path probability}.\\
In appendix \ref{AA_overview}, the action of WPIs is listed for continuous MPs defined via a SDE or a FPE.\\

The formulation of continuous MPs in terms of WPIs might seem cumbersome, but despite its formal complexity, it offers a great deal of intuition in practical applications which will partially unfold in section \ref{ss_td-interpration_FTs} and \ref{s_asymp} in the context of fluctuation theorems and asymptotic approximations.\\
For now we note that a WPI is the sum of the probabilities of all possible paths $\xc$ that connect $(x_0,t_0)$ with $(x,t)$.\remark{These paths are continuous but non-differentiable. In fact, also differentiable paths are allowed, but it can be shown that they do not contribute to the path integral (\cite{ChaichianDemichev01} p.\,27). The marginal role of differentiable paths can be traced back to their fractal dimension of one, whereas trajectories of Brownian motion are known to have a fractal dimension greater than one.} The functional $P[\xc|x_0]$ is to be understood as a {\it density} in function space, whereas the {\it probability} of a single path is rather $P[\xc|x_0]\!\Di\xc$, bearing resemblance to a random variable $x$ with PDF $p(x,t)$, for which the probability $p(x)\dd x$ is the product of the density $p(x)$ and an infinitesimal volume $\dd x$. In that sense, loosely speaking, the measure $\Di\xc$ can be thought of as an infinitesimal tube that encloses the path $\xc$. However, $\Di\xc$ remains to be a highly symbolic object.\\
The form of the action $S[\xc]$ can also be furnished with intuition. For this purpose, we take the Stratonovich convention $\a=1/2$ in which we have no spurious drift. We observe that $S[\xc]$ is larger, the more $\xc$ deviates from the solution of the deterministic equation $\ddx_\t - F(x_\t,\t)$, that is when the actual velocity $\ddx$ deviates from the drift velocity $F(x,t)$. On the other hand, if $D(x,t)$ is large along $\xc$, the penalty for deviating from the deterministic motion, $(\ddx_\t - F(x_\t,\t))^2$, gets mitigated; in other words, a high diffusivity implies a higher probability for large excursions. Thus, we retrieve the interplay between deterministic and stochastic dynamics.\\
The occurrence of the diffusion coefficient in the denominator of $S[\xc]$ allows for another intuitive interpretation. Let us assume that we can factor out an overall diffusion magnitude $\din$ and write $D(x,t)=\din \tilde D(x,t)$, where $\tilde D(x,t)$ is now dimensionless. For moderate values of $\din$, an ensemble of trajectories fluctuating around the deterministic solution contribute to the path integral. For small values of $\din$, the variety of contributing trajectories narrows down to slightly fluctuating trajectories close to the deterministic solution. In the limit $\din\to0$, only the deterministic solution will survive\remark{bin mir hier nicht sicher, ob tatsächlich für state-dependent noise der markoskopische pfad überlebt, denn mode$\neq$average.}. This limit is known as the {\it weak noise limit}, and approximations with $\din$ as small parameter are called {\it weak noise approximations} (\cite{Gardiner2009} p.\,169, \cite{ChaichianDemichev01} p.\,27). We note that the weak noise approximation is closely related to the stationary-phase approximation of Feynman path integrals in quantum mechanics (\cite{ChaichianDemichev01} p.\,169).\\ 
Quite analogous to WPIs, a Feynman path integral considers all paths that are {\it simultaneously} realised in the quantum mechanical regime, for instance in the double-slit experiment. This correspondence is not a coincidence, since the Schrödinger equation is a Fokker-Planck equation with imaginary diffusion coefficient proportional to $\hslash$. In the limit $\hslash\to0$, only the classical path survives. The stationary-phase approximation is a semi-classical method in which small quantum fluctuations are considered that are characterised by actions large compared $\hslash$. \\

We return to the WPI. Per construction, the path probability $P[\xc|x_0]$ is still conditioned on the initial value $x_0$. Augmenting $P[\xc]|x_0]$ with the initial distribution $p_0(x)$
\begin{align}
	P[\xc] \dfn p_0(x_0)\,P[\xc|x_0] \;,
\end{align}
and integration with respect to the initial value $x_0$ yields 
\begin{align} \label{eq:WPI_int-x0}
  p(x,t) = \int \dd x_0\!\! \int\limits_{(x_0,t_0)}^{(x_t,t)} \lDi\xc\,P[\xc] \;.
\end{align}
The path probability $P[\xc]$ is normalised upon integration with respect to all paths connecting the initial and final points, and integration with respect to these points,
\begin{align} \label{eq:WPI_norm}
	1 = \int \dd x_t \int \dd x_0\!\! \int\limits_{(x_0,t_0)}^{(x_t,t)} \lDi\xc\,P[\xc] \;.
\end{align}
Building on this normalisation condition, we can write for the average of an integral observable $Y=\YY[\xc]$,
\begin{align}
	\lla Y\rra = \int \dd x_t \int \dd x_0\!\! \int\limits_{(x_0,t_0)}^{(x_t,t)} \lDi\xc\,P[\xc]\,\YY[\xc] \;,
\end{align}
as the analogue of $\lla y(x)\rra_{p(x,t)}=\int y(x)p(x)\di x$. Here, the path probability $P[\xc]$ weights each path according to the chance of its realisation. Note that the above average $\lla\rra$ defined in terms of a path integral is equivalent to the {\it ensemble average} of values for $Y$ resulting from a representative (in principle infinite large) ensemble of realisations $\xc$.\\
We can also write the PDF of the observable $Y$ as a WPI,
\begin{align} \label{eq:WPI_pdf_int-val}
	P(Y) &= \int \dd x_t \int \dd x_0\, p_0(x_0)\!\! \int\limits_{(x_0,t_0)}^{(x_t,t)} \lDi\xc\,P[\xc|x_0]\,\d(Y-\YY[\xc]) \;.
\end{align}
Due to the $\d$-function, the integration space is restricted to paths that obey $Y-\YY[\xc]$, thus the WPI collects the probabilities of all paths that give rise to the desired value $Y$. We will come back to the WPI representation of a PDF in chapter \ref{s_asymp}, in which we will approximate the above WPI with an asymptotic method formally equivalent to the weak noise approximation.\\

Finally, we note that the formalism of WPIs can be generalised to higher dimensions. We refrain from giving an account on this generalisation and refer the interested reader to \cite{ChernyaChertkoJarzyn06JoSMTaE}.\\

\subsection{Discontinuous Markov processes} \label{ss_ME}
In the previous three sections, we were concerned with continuous MPs. In this section, we will embed the continuous subclass of MPs into the larger context of MPs that also allow discontinuous realisations (\cite{Gardiner2009} p.\,45ff).\\

We begin with formulating the continuity condition
\begin{align} \label{eq:ContinCond_Markov}
  \lim\limits_{\Dt\to0}\frac{1}{\Dt} \int\limits_{|\tx-x|>\e} p(\tx,t\+\Dt\,|\,x,t) \di \tx = 0 \quad\forall\,\e>0
\end{align}
to define the borderline between continuous and discontinuous MPs. In words, the continuity condition requires that the probability for arbitrary small jumps in the process, $|\tx-x|>\e$, has to approach zero {\it faster} than the time step $\Dt$.\remark{the limit is uniformly in $y$, $t$ and $\Dt$} Building on the above continuity condition, we distinguish continuous from discontinuous MPs by formulating the following three defining conditions,
\begin{subequations} \label{eq:CKR_3conds}
  \begin{align}
		\label{eq:CKR_3conds_jump}
    &\lim\limits_{\Dt\to0}\,\frac{1}{\Dt}\,p(\tx,t\+\Dt\,|\,x,t) \dfnrv \j(\tx\,|\,x,t) \;, \\[5pt] 
    \label{eq:CKR_3conds_drift}
    &\lim\limits_{\Dt\to0}\frac{1}{\Dt} \int\limits_{|\tx-x|<\e} (\tx-x)\,p(\tx,t\+\Dt\,|\,x,t) \di \tx \dfnrv A(x,t) + \OO(\e) \;, \\[5pt]
    \label{eq:CKR_3conds_diff}
    &\lim\limits_{\Dt\to0}\frac{1}{2\Dt} \int\limits_{|\tx-x|<\e} (\tx-x)^2\,p(\tx,t\+\Dt\,|\,x,t) \di \tx \dfnrv B(x,t) + \OO(\e) \;.
  \end{align}
\end{subequations}
Here, $\j(\tx\,|\,x,t)$ measures the violation of the continuity condition (\ref{eq:ContinCond_Markov}) via jumps $x$ to $\tx$ at time $t$ and is referred to as the {\it jump measure}. The coefficients $A(x,t)$ and $B(x,t)$ are related to the mean and variance of the MP. For that reason, a non-zero $\j(\tx\,|\,x,t)$ is responsible to the generation of discontinuous realisations, whereas $A(x,t)$ and $B(x,t)$ are connected with continuous realisations\remark{(letzteres hängt damit zusammen, dass $\xi$ gaussian ist, und nur dann die lösungen der SDE stetig sind. andere (heavy tailed) verteilungen für $\xi$ führen auf jumps, und müssen durch $\j$ berücksichtigt werden)}. $B$ and $\j$ must be positive.\\
The crucial question is, why are $A(x,t)$ and $B(x,t)$ sufficient to capture the continuous part of the MP. The question is equivalent to asking why the noise $\xi(t)$ in SDEs must be Gaussian in order to have continuous solutions. The answer is that one can indeed prove that (\cite{Gardiner2009} p.\,47f)
\begin{align} \label{eq:only_drift+diff}
  \lim\limits_{\Dt\to0}\frac{1}{\Dt} \int\limits_{|\tx-x|<\e} (\tx-x)^n\,p(\tx,t\+\Dt\,|\,x,t) \di \tx = \OO(\e) \quad\forall\,n\geq3 \;,
\end{align}
which follows from estimating the cubic equivalent of (\ref{eq:CKR_3conds_diff}) against $B(x,t)$, the quartic form against the cubic, and so on.\\

\paragraph{The master equation} Having set up the above classification, the next step is to pursue an evolution equation for $p(x,t)$ that, in contrast to the FPE, also includes discontinuous realisation of the MP. This equation is the differential form of the CKR. The starting point to derive the differential CKR is the time derivative of the expectation of a test function $h(x)$, $\pt_t \int h(x) p(x,t) \di x$. By writing the time derivative as the limit of a difference quotient, substituting the CKR, making use of conditions (\ref{eq:CKR_3conds}) and upon integration by parts, we can read off the differential CKR as the effect of the CKR on the test function $h(x)$ as
\begin{align} \label{eq:diffCKR}
  \pt_t p(x,t) = &-\pt_x A(x,t) p(x,t) + \pt_x^{\,2} B(x,t) p(x,t) \nn
  &+ \int \j(x|\tx,t) p(\tx,t) - \j(\tx|x,t) p(x,t) \di \tx \;.
\end{align}
We note that all solutions of the differential CKR satisfy the CKR, and all MPs obey the CKR, but not all solutions of the differential CKR are necessarily Markovian.\remark{i remember that every MP can be written as a ME} However, the cases in which the solutions of the differential CKR are not Markovian build on pathological choices for $\j(\tx\,|\,x,t)$ which we do not consider here.\\
By setting one or two of the coefficients $A(x,t)$, $B(x,t)$ and $\j(x|\tx,t)$ to zero, different subclasses of MPs are described. By assuming, e.g., a vanishing probability of jumps in the process, $\j(x|\tx,t)\equiv0$, we recover the Fokker-Planck equation (\ref{eq:FPE_D1D2}) with drift $A(x,t)=\Df(x,t)$ and diffusion $B(x,t)=\Dg(x,t)$ describing a continuous MP. Also the deterministic case $B=0$ and $\j(x|\tx,t)\equiv0$ can be considered as a MP (\cite{Gardiner2009} p.\,54,353), and is then referred to as a Liouville process, in which the only randomness arises from sampling the initial values of the process from an initial PDF $p_0(x)$.\\
The simplest discontinuous case is setting drift $A(x,t)\equiv0$ and diffusion $B(x,t)\equiv0$, for which we find
\begin{align} \label{eq:ME}
  \pt_t p(x,t) &= \int \j(x|\tx,t) p(\tx,t) - \j(\tx|x,t) p(x,t) \di \tx \nn 
  p_0(x) &= p(x,t=t_0) \;.
\end{align}
The above evolution equation is known as {\it master equation}.\\
The master equation can be interpreted: The probability $p(x,t)$ of a state $x$ increases by jumps from states $\tx$ to $x$ and decreases by jumps from state $x$ to states $\tx$.\\
Since $p(x,t)$ in the second term does not depend on $\tx$, we can rewrite the master equation as
\begin{align} \label{eq:ME_escape}
  \pt_t p(x,t) = \int \j(x|\tx,t) p(\tx,t)\di \tx  - \r(x,t) p(x,t) \;,
\end{align}
where we have defined the {\it escape rate}
\begin{align} \label{eq:def:escape}
  \r(x,t) = \int \j(\tx|x,t) \di \tx \;.
\end{align}
Note that using an appropriate $\j(x|\tx,t)$, a pure jump process can be set up in which only discrete jumps occur, although the MP $X(t)$ does not need to be restricted to discrete values. If, however, the state-space is discrete, we have instead of $X(t)$ the time-dependent random variable $N(t)\in\mathbb{Z}$ with realisations $n(t)$, and the master equation for the PDF of measuring $n$ at time $t$ becomes
\begin{subequations} \label{eq:ME_discr}
  \begin{align}
		\pt_t p(n,t) &= \sum\limits_m\big[ \j(n|m,t) p(m,t)  - \j(m|n,t) p(n,t)\big] \label{eq:ME_discr_a} \\
		&= \sum\limits_m\big[ \j(n|m,t) p(m,t)\big] - \r(n,t) p(n,t) \;, \label{eq:ME_discr_escape} 
	\end{align}
	\begin{align}
		\r(n,t) &= \sum\limits_m \j(m|x,t) \di \tx \sep p_0(n)=p(n,t=t_0) \;.
	\end{align}
\end{subequations}
In this case, we are bound to have a pure jump process.\\
We mention that a useful property of jump processes is that for time independent jump measures $\j(\tx|x)$, the time $\Dt$ between jumps is exponentially distributed, $p(\Dt,x)=\exp[-\r(x)\Dt]$ (\cite{Gardiner2009} p.\,52). This fact allows for simple simulation algorithms to generate realisations of jump processes, for instance the Gillespie algorithm \cite{Gillespie1977}.\\
\remark{Note, that for the somewhat peculiar jump measure (eigentlich ja auch nicht erlaubt, da das keine pdf ist, und $\j>0$ unklar ist)
\begin{align} \label{eq:jump2FPE}
  \j(x|\tx,t) = \big[ -\pt_x\Df(x,t) + \pt_x^{\,2}\Dg(x,t) \big]\,\d(x-\tx)
\end{align}
the Fokker-Planck equation can be reproduced (\cite{Risken89} p.\,51) (auf die idee das einzusetzen komme ich ja aber nur, wenn ich vorher die fCKR angeschaut habe, und eine aussage über stetige und unstetige prozesse erlaubt so eine blinde zuordnung der springdichte auch nicht). The usage of $\d(x-\tx)$ in the jump measure (\ref{eq:jump2FPE}) assigns a vanishing probability to jumps $\tx-x\neq0$, aber wenn ich zum beispiel in $\j$ oben noch einen $D^3$ term einfügen würde, dann wäre der prozess nicht mehr stetig, das ginge aus definition eines solchen $\j$ nicht hervor.}\\

\paragraph{The Kramers-Moyal expansion} The master equation for a pure jump process can be expanded in the moments of the jump measure. This expansion is known as the {\it Kramers-Moyal expansion} (KME) (\cite{Gardiner2009} p.\,275). To derive it, we define the jump density 
\begin{align} \label{eq:def_jump-distr}
  \jd(y;x,t) \dfn \j(x\+y|x,t) \;.
\end{align}
In terms of $\jd(y;x,t)$, the master equation (\ref{eq:ME}) reads
\begin{align} \label{eq:ME_jumps}
  \pt_t p(x,t) = \int \jd(y;x\-y,t)\,p(x\-y,t) - \jd(y;x,t)\,p(x,t) \di y \;.
\end{align}
By expanding the integrand
\begin{align}
  \begin{split}
		\jd(y;x\-y,t) p(x\-y,t) &= \jd(y;x,t)p(x,t) \\
		&\quad+ \sum\limits_{k=1}^{\infty}\frac{(-1)^k}{k!}\frac{\pt^k}{\pt x^k}\,\big[\jd(y;x,t)p(x,t)\big]\,y^k
  \end{split} \label{eq:KME_pre}
\end{align}
and substitution into the master equation (\ref{eq:ME_jumps}), we get the KME
\begin{align}
	\pt_t p(x,t) &= \int \sum\limits_{k=1}^{\infty}\frac{(-1)^k}{k!}\frac{\pt^k}{\pt x^k}\Big[\jd(y;x,t)p(x,t)\Big]\,y^k \di y \nn
	&= \sum\limits_{k=1}^{\infty}\frac{(-1)^k}{k!}\frac{\pt^k}{\pt x^k}\Big[\Ak(x,t)p(x,t)\Big] \;, \label{eq:KME}
\end{align}
where we defined the moments of the jump density as
\begin{align}
	\Ak(x,t) &\dfn \int y^k\,\jd(y;x,t) \di y \nn
	&= \int (\tx\-x)^k\,\j(\tx|x,t) \di \tx \label{eq:def_Ak} \;.
\end{align}
Truncating the KME after the second term, we obtain a FPE with drift $\Df(x,t)=\Af(x,t)$ and diffusion $\Dg(x,t)=\frac{1}{2}\Ag(x,t)$ which is often used to approximate the jump process defined by the jump density $\jd(y;x,t)$ as a continuous process (\cite{Risken89} p.\,70ff).\\
The KME in the form (\ref{eq:ME_jumps}) rules a pure jump process. To include an underlying continuous MP into the expansion, we substitute the KME (\ref{eq:KME_pre}) into the full differential CKR (\ref{eq:diffCKR}) and find the expansion
\begin{align} 
  \pt_t p(x,t) &= \;-\,\pt_x \big[A(x,t)+\Af(x,t)\big] p(x,t) \nn
  &\;\,\quad+ \pt_x^{\,2} \big[B(x,t)+\tfrac{1}{2}\Ag(x,t)\big] p(x,t) \nn
  &\;\,\quad+  \sum\limits_{k=3}^{\infty}\frac{(-\pt_x)^k}{k!}\big[\Ak(x,t)p(x,t)\big] \nn
  &= \sum\limits_{k=1}^{\infty}(-\pt_x)^k\big[\Dk(x,t)p(x,t)\big]\label{eq:KME_contin}
\end{align}
with
\begin{subequations} \label{eq:def_KMC}
\begin{align} 
	\tAf(x,t) &\dfn A(x,t)+\Af(x,t)) \;, \\
	\tAg(x,t) &\dfn B(x,t)+\mfrac{1}{2}\Ag(x,t) \;, \\
	\tAk(x,t) &\dfn \mfrac{1}{k!}\Ak(x,t) \quad \text{for } k\geq3 \;.
\end{align}
\end{subequations}
This expansion is also known as a KME, and we will refer to the $\Dk(x,t)$ as {\it Kramers-Moyal coefficients} (KMCs).\remark{Denoting the KMCs by $\Dk$ is ambiguous since we denote drift and diffusion coefficients also by $\Dfg$. However, in the context of continuous MPs, the first two KMCs are identical to drift and diffusion coefficients. And for discontinuous MPs, drift and diffusion are the approximated coefficients for the FPE. Hier unterscheide ich also $A$, $B$ von den $\tAk$, eine unterscheidung die in der restlichen arbeit überwiegend fallengelassen wird. siehe discussion unter (\ref{eq:tAk-vs-Ak})}\\
If the conditional PDFs of a MP are known, we see by (\ref{eq:CKR_3conds}) that the KMCs can be obtained from
\begin{subequations} \label{eq:Dk_esti}
  \begin{align} 
		\tAk(x,t) = \lim\limits_{\Dt\to0}\,\frac{1}{k!\Dt}\Mk(x,t;\,\Dt) \label{eq:Dk_esti_D} \;, \\
		\Mk(x,t;\,\Dt) = \int (\tx\-x)^k\,p(\tx,t\+\Dt\,|\,x,t) \di \tx \label{eq:Dk_esti_M} \;,
	\end{align}
\end{subequations}
where we defined the conditional moments $\Mk(x,t;\,\Dt)$. The above prescription is often used to estimate $\tAk(x,t)$ from experimental measurements by estimating the conditional moments $\Mk(x,t;\,\Dt)$ for various $\Dt$. In this context, the knowledge of the Markov-Einstein time scale $\DtME$, which we discussed in section \ref{ss_LE}, is of relevance, since the formulae of prescription (\ref{eq:Dk_esti}) rely on the Markov assumption which only holds for $\Dt>\DtME$. To obtain an estimate for $\DtME$ from experimental data, the time-constant of the auto-correlation function of the measured $x(t)$ may serve. In the simplest approach, the limit in (\ref{eq:Dk_esti_D}) can be substituted by $\tAk(x,t)=\Mk(x,t;\,\DtME)$ (\cite{vanKampen2007} p.\,195), but many more sophisticated procedures have been established \cite{Stresing2011,Honisch2011,Honisch2012,Friedrich2011a,Tang2013}, commonly regarded to as {\it Markov analysis}.\\

It is important to keep in mind that the $\Ak(x,t)$ in (\ref{eq:def_Ak}) are defined as the moments of the jump measure $\j(\tx|x,t)$, whereas using the prescription (\ref{eq:Dk_esti}) to estimate $\tAk(x,t)$ from experimental data rests upon the conditional distribution $p(\tx,t\+\Dt\,|\,x,t)$ being insensitive to the formal distinction between the continuous and discontinuous components of the process $X(t)$.\\
To clarify our point, let us discuss the correspondence from (\ref{eq:def_KMC}), 
\begin{subequations} \label{eq:tAk-vs-Ak}
	\begin{align} 
		\tAf(x,t) &= A(x,t)+\Af(x,t) \label{eq:tAk-vs-Ak_1} \;, \\
		\tAg(x,t) &= B(x,t)+\mfrac{1}{2}\Ag(x,t) \label{eq:tAk-vs-Ak_2} \;, \\
		\tAgg(x,t) &= \mfrac{1}{k!}\Agg(x,t) \label{eq:tAk-vs-Ak_g3} \;.
	\end{align}
\end{subequations}
In the case of a continuous process $X(t)$, i.e. $\Ak(x,t)\equiv0$, it follows from (\ref{eq:tAk-vs-Ak}) that by estimating $\tAfg(x,t)$ we find the drift and diffusion coefficients of the FPE ruling $X(t)$. In the case of an underlying jump process, however, $\tAfg(x,t)$ includes also the first two moments of the jump density, and the FPE is a mere approximation of the true process. Realistic MPs taken to be continuous will always have an underlying jump process, the $\Dfg(x,t)$ as above will therefore never equal $A(x,t)$ and $B(x,t)$ exactly. However, in many applications, the discontinuous component of a MP can be neglected (\cite{Friedrich2011a}, \cite{Risken89} p.\,77ff). In the major part of this thesis, when considering continuous MPs, we will therefore drop the distinction between $A(x,t)$, $B(x,t)$ and $\Dfg(x,t)$.\\
If we were to decide whether $X(t)$ is approximately continuous, we could in principle estimate $\tAk(x,t)$ for an even $k\geq4$ (a symmetric distribution has vanishing odd moments) and, if $\tAk(x,t)\equiv0$, argue that $\j(\tx|x,t)\equiv\d(\tx-x)$ which assigns zero probability to non-zero jump widths. We can also formulate this finding as that ``for a positive $p(\tx,t\+\Dt\,|\,x,t)$, the KME may stop after the first or the second term, and if it does not stop after the second term, it must contain an infinite number of terms'' (Risken, p. 70). This statement is known as the {\it theorem of Pawula}, originally proved by applying a generalised Schwarz inequality to a combination of conditional moments \cite{Pawula1967}.\\
However, restricting ourselves to the fourth coefficient as being, due to experimental and statistical limitations, the statistically easiest accessible, it is not possible to show that exactly $\tAfr(x,t)\equiv0$. Instead, it can only be found that $\tAfr(x,t)\approx0$ within a certain error margin, or that $\tAfr(x,t)$ is negligible small compared to $\tAg(x,t)$, both of which does not mean that the $\tAk(x,t)$ for $k>4$ are negligible too. In that sense, the KME is not a systematic expansion in a small parameter (\cite{Gardiner2009} p.\,276, \cite{vanKampen2007} p.\,199). An improved expansion is the {\it system size expansion} suggested by van Kampen, which coincides with the KME in the weak noise limit (\cite{Gardiner2009} p.\,276ff, \cite{vanKampen2007} p.\,199ff).\\

From the FPE we derived an equation for the moments $S_x^{\,n}(t)$ of the PDF $p(x,t)$. Integrating the KME (\ref{eq:KME_contin}), we can also obtain a moment equation for MPs involving both continuous and discontinuous components,
\begin{align} 
	\dot S_x^{\,n}(t) &= \sum\limits_{k=1}^{\infty}\int x^n (-\pt_x)^k\,\big[\Dk(x,t)p(x,t)\big] \di x \nn
	&= \sum\limits_{k=1}^{n}\int \mfrac{n!}{(n-k)!}\,\big[x^{n-k}\Dk(x,t)p(x,t)\big] \di x \nn
	&= \sum\limits_{k=1}^{n}\mfrac{n!}{(n-k)!}\,\lla x^{n-k}\Dk(x,t)\rra  \nn
	\begin{split}
	&= n\lla x^{n-1}A(x,t)\rra + n(n\-1)\,\lla x^{n-2}B(x,t)\rra \\
	&\quad+ \sum\limits_{k=1}^{n}\!\mbinom{n}{k}\!\lla x^{n-k}\Ak(x,t)\rra \;. 
	\end{split} \label{eq:KME_moms}
\end{align}
Note that now the sum involves a finite number of terms, in contrast to the infinite sum in the KME. The consequence is that, apart from drift and diffusion, only the $n$ first moments of the jump density contribute to the $n$-th moment of $p(u,r)$.\\
For the special case that $A(x,t)=a(t)\,x$, $B(x,t)=b(t)\,x^2$ and \linebreak $\Ak(x,t)=\dk\!(t)\,x^k$, we obtain a closed equation for the moments,
\begin{align} \label{eq:KME_moms_special}
	\dot S_x^{\,n}(t) = \bigg[na(t) + n(n\-1)b(t) + \sum\limits_{k=1}^{n}\mbinom{n}{k}\dk\!(t)\bigg]\,S_x^{\,n}(t) \;.
\end{align}
The solution of this equation reads
\begin{align}  \label{eq:KME_moms_special_sol}
	S^{\,n}_x(t) = S^{\,n}_x(t_0)\,\exp\bigg[\int_{t_0}^{t} na(t)+n(n\-1)b(t) + \sum\limits_{k=1}^{n}\mbinom{n}{k}\dk\!(t)\di t\bigg] \;.
\end{align}
\remark{The moments equation from the KME, (\ref{eq:KME_moms}), implies that the $n$-th moment of $p(x,t)$ is fixed by the first $n$ KMCs $\Dk$. Therefore, in the context of continuous MPs, using $\Df$ and $\Dg$ in a FPE has the advantage that the {\it drift} $\Df$ alone governs the first moment. On the other hand, position-dependent diffusion induces a drift $\Dgx$. Using therefore the {\it force} $F=\Df-\Dgx$ is appropriate in physical considerations, as then $F$ may derive from a physical potential, as we will see in the next chapter.}\remark{dieser noise induced drift muss dann doch rein physikalisch betrachtet non-conservative sein. ich mein, natürlich kann ich rein rechnerisch ein potential aufstellen, dass durch ableiten die externe kraft und die noise induced kraft reproduziert. aber das potential ist dann nicht physikalisch, also gravitation, laserfeld, em-feld, ... interessant wäre, eine LE ohne $V$, also $F$ zu probieren, aber mit orts-abhängiger mobilität, wo dann ja ne noice induced drift zu sehen müsste. dann gibts hier ja erstmal die stationäre verteilung $1/\Dg$, und damit ja auch ein FT. was allerdings passiert, wenn ich dagegen eine kraft $F$ setze? der unterschied der dynamik in $F$ richtung und der dynamik in $\Dgx$ richtung sorgt vielleicht dafür, dass bei umkehr der beiden kräfte zwar das $\pst$ erhalten bleibt, aber die dynamik sich umkehrt, und ich 3faces kriege.}

\newpage\paragraph{Summary}
We close this chapter with a summary of MPs.\\
MPs are stochastic processes where each event depends solely on the most recent one. Realistic MPs are only Markovian above a certain time scale, the Markov-Einstein time scale $\tME$. A paradigmatic example is Brownian motion which arises from collisions of nanoscopic particles with fluid molecules, and the outcome of each collision depends only on the outcome of the previous collision. Here, $\tME$ is the average collision time. An intuitive approach the Brownian motion is the Langevin equation.\\
The Langevin equation is a SDE. We have seen that solutions of SDEs are realisations $x(t)$ of continuous MPs, determined by a deterministic and by a stochastic component, where the stochastic component arises from a Gaussian white noise $\xi(t)$. Due to the Markov assumption, $\xi(t)$ is $\d$-correlated. Instead of addressing the realisations $x(t)$ itself, a FPE is an evolution equation for the PDF $p(x,t)$ and defined by drift and diffusion coefficients, $\Df(x,t)$ and $\Dg(x,t)$. The FPE is advantageous with regard to the calculation of moments $\lla x(t)^n\rra$ and allows to determine the instantaneous stationary distribution $\pst(x,t)$, to which the process tries to relax at each instant of time. In the representation of WPIs, we defined the probability density functional $P[\xc]$ for paths $\xc$. The advantage of path integrals will unfold in the following chapter.\\
If the white noise $\xi(t)$ is not Gaussian, MPs are discontinuous, characterised by a jump density $\jd(y;x,t)$. In this case, a description via a FPE or WPI is not possible, or in some cases only an approximation. The evolution equation for $p(x,t)$ is the master equation, formally equivalent to the KME involving the KMCs $\Dk(x,t)$. For a discontinuous MPs, all even $\Dk(x,t)$ are non-zero, and for continuous MPs, the first two $\Dk(x,t)$ are drift and diffusion of the FPE, the remaining $\Dk(x,t)$ vanish. From the definition of the KMCs, an estimation of drift and diffusion from experimental data is possible.\\
The drift and diffusion coefficients $\Dfg(x,t)$ arise formally from the KME and are directly related to the moments of the jump density. In a thermodynamic setting, more relevant is the drift velocity $F(x,t)=\Df(x,t)-\pt_x\Dg(x,t)$, as it arises from the physical potential that defines thermal equilibrium. The thermodynamic interpretation of continuous MPs is the subject of the following chapter.

\cleardoublepage
\section{Thermodynamic interpretation of continuous MPs} \label{s_td_interpretation}
In the previous chapter we were mainly concerned with the formal definition and implications of Markov processes with barely physical reference. In this chapter, we will imbue the theory of continuous Markov processes with life by offering a thermodynamic interpretation. This interpretation will build on the equivalence between the stationary solution of the Fokker-Planck equation and the canonical equilibrium distribution.\\
In the first section of this chapter, \ref{ss_td-interpration_energy}, we remark on thermodynamic consistency in the case of multiplicative noise, and will then identify expressions for heat and work from the Langevin equation and discuss the appendant first law. In section \ref{ss_td-interpration_entropy}, we use the FPE to establish the bridge to the Gibbs entropy together with the second law. In section \ref{ss_td-interpration_FTs}, we will approach the irreversibility of non-equilibrium processes using Wiener path integrals and elucidate the relation to entropy.\\

The distinctive feature of the thermodynamic interpretation will be its formulation on the level of individual realisations $x(t)$ of processes arbitrarily far from equilibrium. Non-equilibrium can be imposed by a time dependent external force, or by preparing the initial state of the process off equilibrium\footnote{Also non-equilibrium constraints, such as multiple noise-terms (reservoirs), impose non-equilibrium, which will play a minor role in what follows.}. The former are {\it driven processes}, the latter are {\it relaxation processes}. The signature of non-equilibrium is in both cases that the solution of the FPE, $p(x,t)$, does not coincide with the stationary solution $\pst(x,t)$. The combination of stochasticity and non-equilibrium is the scope of {\it stochastic thermodynamics}.\\
In order to be transferable to other fields of application, we will leave the defining coefficients of the Markov process, e.g. $F(x,t)$ and $D(x,t)$, unspecified whenever possible. To maintain an intuitive level, however, we will occasionally come back to Brownian motion in which $F(x,t)=-\G V'\big(x(t)$ arises from an external potential and $D=\kB T\G$ is defined by the Einstein relation (\ref{eq:D_einstein}), and we will also on the formal level refer to the picture of particle and fluid.\\
More details to the material presented in this chapter can be found in a recent review article by Seifert \cite{Seifert2012}. For an appealing introduction to the field of stochastic thermodynamics we recommend the overview article \cite{Seifert08TEPJB} by the same author.

\subsection{Energy balance and the first law} \label{ss_td-interpration_energy}
Before we begin with identifying heat and work from the Langevin equation, we recap the stationary solution of the FPE and discuss its relation to equilibrium for various forms of the diffusion coefficient.\\

\paragraph{Thermodynamic consistency} We start with ordinary Brownian motion subject to an external force, for which we derived in (\ref{ss_LE}) the Langevin equation from Newton's equation of motion, (\ref{eq:LE_newton_force}), which involved a frictional force, an external force and a thermal force,
\begin{align} \label{eq:LE_td-interpretation}
	\g\ddx(t) = -V'\big(x(t)\big) +  \sqrt{2\g \kB T}\xi(t) \;.
\end{align}
Here, $V(x,t)$ is a potential that gives rise to the external force. The attendant canonical equilibrium distribution to $V(x,t)$ (the Boltzman distribution) reads
\begin{align} \label{eq:def_peq}
	\peq(x,t) = \frac{1}{Z(t)}\,\exp\big[-\b V(x,t)\big]
\end{align}
with $\b\dfns1/\kB T$ and partition sum $Z(t)=\int \exp\big[-\b V(x,t)\big] \di x$.\\
On the other hand, recall that the FPE (\ref{eq:FPE_FD})
\begin{align} \label{eq:FPE_td-interpretation}
	\ddp(x,t) &= \pt_x\big[\-F(x,t) \+ D(x,t)\pt_x\big]p(x,t) \sep p(x,t_0) = p_0(x) \;, 
\end{align}
implies a stationary distribution (\ref{eq:def_pst_FD}) of the form 
\begin{align} \label{eq:FPE_td-interpretation_pst}
	\pst(x,t) = \exp\big[-\p(x,t)+\GG(t)\big]
\end{align}
with
\begin{subequations}
  \begin{align}
		\p(x,t) &= -\int_{-\infty}^x \frac{F(\x,t)}{D(\x,t)} \di\x   \label{eq:FPE_td-interpretation_phi} \\
		\GG(t) &= -\ln \int \exp\big[-\p(x,t)\big] \di x  \label{eq:FPE_td-interpretation_GG}
	\end{align}
\end{subequations}
By taking drift velocity $F(x,t)$ and diffusion coefficient $D(x,t)$ to be
\begin{subequations}
\begin{align}
	F(x,t) &= -\G V'(x,t) \;, \\
	D &= \kB T\G \;,
\end{align}
\end{subequations}
where $\G=1/\g$ is the mobility, we establish the equivalency of the FPE to the LE from Newton's equation of motion as stated above in (\ref{eq:LE_td-interpretation}) (cf. also (\ref{eq:A5_overviewFPE-FD})). The stationary distribution then coincides with the equilibrium distribution (\ref{eq:def_peq}),
\begin{align}
	\p(x) &= -\int_{-\infty}^x \frac{-\G V'(x')}{\kB T\,\G} \di x'  = \b V(x) \;.
\end{align}
\vspace{7pt}\\
In the above example, in which the diffusion coefficient is constant,\linebreak $D=\kB T\G$, the interpretation ambiguity of the LE with regard to the rule of discretisation is irrelevant. In cases in which the diffusion coefficient does not depend on $x$, the noise is called {\it additive noise}, as $\xi(t)$ enters the LE for $x(t)$ additively. In contrast, a state dependent diffusion coefficient implies a multiplication of $\xi(t)$ with $x(t)$, accordingly, in this case the noise is referred to as {\it multiplicative noise}. Therefore, the interpretation ambiguity of the LE arises only for multiplicative noise.\\
The ambiguity for multiplicative noise raises the question how the above discussed correspondence between the stationary distribution and thermal equilibrium is affected. In view of the Einstein relation, $D=\kB T\G$, multiplicative noise may occur for position dependent mobility $\G(x)$ or in the presence of a temperature gradient $T(x)$. We discuss both cases shortly.\\
The mobility of Brownian particles typically depend on their position in the presence of geometrical confinement, see the article by Lau and Lubensky \cite{LauLubensky07PRE}. The LE (\ref{eq:LE_newton}) derived from Newton's equation of motion becomes
\begin{align} \label{eq:td-inter_multi_Gamma_LE}
	\ddx_t = -\G(x_t)V'(x_t,t) + \sqrt{2\kB T\,\G(x_t)}\,\xi(t) \;.
\end{align}
This is now a typical case of internal noise as discussed by van Kampen \cite{Kampen1981}, since $\G(x_t)$ influences both the deterministic and the stochastic component. Or in other words, in absence of the external force, $V'(x,t)\equiv0$, the deterministic component of the process does not vanish, which can immediately be seen from drift velocity and diffusion coefficient of the equivalent FPE (\ref{eq:A5_overviewSDE_FPE}),
\begin{subequations} \label{eq:td-inter_multi_Gamma_FD}
\begin{align}
	F(x,t) &= -\G(x)V'(x,t) - (1-\a)\kB T\,\G'(x_t) \;, \\
	D(x,t) &= \kB T\,\G(x) \;.
\end{align}
\end{subequations}
If we dealt with external noise, we could take the above LE in the Stratonovich convention, $a=1/2$. Instead, we choose $\a=1$, since we then retain thermal equilibrium,
\begin{align} \label{eq:td-inter_multi_Gamma_pst}
	\p(x) &= -\int_{-\infty}^x \frac{-\G('x)V'(x')}{\kB T\,\G(x')} \di x'  = \b V(x) \;,
\end{align}
as $\G(x)$ cancels for arbitrary dependency on $x$.\\
The starting point of this consideration was a naive generalisation of the LE. To avoid the resulting interpretation ambiguity, it is advisable to base the considerations on the FPE. The appropriate $F(x,t)$ and $D(x,t)$ follow then from a comparison of the FPE with the generalised Fick's law which indeed results into (\ref{eq:td-inter_multi_Gamma_FD}) for $\a=1$ \cite{LauLubensky07PRE}. The equivalent LE to this FPE reads (\ref{eq:A5_overviewFPE-FD_SDE})
\begin{align} \label{eq:td-inter_multi_Gamma_LE_alpha}
	\ddx_t = -\G(x_t)V'(x_t,t) + (1-\a)\kB T\,\G'(x_t) + \sqrt{2\kB T\,\G(x_t)}\,\xi(t) \;,
\end{align}
for which no interpretation ambiguity arises since the dynamics is the same for any choice of $\a$. The accordance with Fick's law and thermal equilibrium of the above LE (\ref{eq:td-inter_multi_Gamma_LE_alpha}) for any $\a$ (or the LE (\ref{eq:td-inter_multi_Gamma_LE}) for $\a=1$) is what Lau and Lubensky mean by thermodynamic consistency \cite{LauLubensky07PRE}.\\
The case where a temperature gradient causes multiplicative noise is intricate. If we again naively modify the LE as above with now constant mobility $\G$ and position dependent temperature $T(x)$, the equivalent FPE is defined by (\ref{eq:A5_overviewSDE_FPE}),
\begin{subequations} \label{eq:td-inter_multi_Tgrad_FD}
\begin{align}
	F(x,t) &= -\G V'(x) - (1-\a)\G\kB\,T'(x_t) \;, \\
	D(x,t) &= \G \kB T(x) \;,
\end{align}
\end{subequations}
and instead of $\p(x)=\b V(x,t)$, the stationary distribution is now defined by
\begin{align}
	\p(x) &= \int_{-\infty}^x \frac{V'(x')+(1-\a)\kB\,T'(x')}{\kB T(x')} \di x' \nn
	&= (1-\a)\ln T(x) + \int_{-\infty}^x \frac{V'(x')}{\kB T(x')}\;.
\end{align}
\remark{This is because the friction affects both the dissipation and the coupling between particles and thermal fluctuations such that both effects cancel in equilibrium (FDT), whereas the temperature only influences the strength of thermal fluctuations.} At this point it is not clear which $\a$ to choose, since the dynamics of suspended particles in fluids under the influence of a temperature gradient, known as thermophoresis, is still subject to current research \cite{Duhr2006,Duhr2006a,Wurger2007,Piazza2008,Hottovy2012,Yang2013}. It is debatable whether a naively modified LE constitutes a fruitful approach to thermophoresis.

\paragraph{Heat, work and the first law}
We now turn to the energy balance implied by the LE due to its derivation from Newton's equation of motion. We start again with ordinary Brownian motion, and then mention the difficulties that arise in the case of multiplicative noise.\\
To obtain the energy balance of Brownian motion, we multiply the LE with an infinitesimal piece of trajectory $\dd x_t$ and obtain
\begin{align} \label{eq:LE_en-bal_add}
	\Big[\g\,\ddx_t - \sqrt{2\kB T\g}\,\xi(t)\Big]\,\dd x_t = -V'(x_t,t)\,\dd x_t \;.
\end{align}
On the left hand side of (\ref{eq:LE_en-bal_add}), we find the energy loss due to motion against the friction force minus the energy received from collisions with fluid molecules (which may also be negative), which in total is the heat transferred into the fluid along $\dd x_t$. If the heat capacity of the fluid is much larger than the Brownian particles, we can assume that the fluid maintains a constant temperatur $T$ and serves as an ideal heat bath. Denoting the infinitesimal amount of dissipated heat by $\dot Q(t)$, the total heat dissipated by a particle along the trajectory $\xc$ in a time-interval $\t=t_0\dots t$ reads
\begin{align} \label{eq:def_Q}
	Q[\xc] = \int_{t_0}^{t} \dot Q(\t) \di \t = -\int_{t_0}^{t} V'(x_\t,\t)\,\ddx_\t \di \t \;.
\end{align}
Substitution of the total derivative of $V(x,t)$, that is\linebreak $\dd_\t V(x_\t,\t) = \dot V(x_\t,\t) + V'(x_\t,\t)\ddx_\t$, reveals that $Q[\xc]$ can be split into an integral and a boundary term,
\begin{align}
	Q[\xc] &= \int_{t_0}^{t} \dot V(x_\t,\t) - \dd_\t V(x_\t,\t) \di \t \nn
	&= \int_{t_0}^{t} \dot V(x_\t,\t) \di \t - \big[V(x_t,t) - V(x_0,t_0)\big] \;. \label{eq:1st_law_pre}
\end{align}
It is reasonable to identify the integral term as the work done on the system, as it constitutes a transfer of energy to a resting particle by externally lifting its potential energy, in contrast to heat dissipation which takes place only for the moving particle. In view of the canonical equilibrium distribution (\ref{eq:def_peq}), we identify the boundary term as the equilibrium difference in energy.\\
By denoting the work done on the particle as
\begin{align} \label{eq:def_W}
	W[\xc] = \int_{t_0}^{t} \dot V(x_\t,\t) \di\t 
\end{align}
and the equilibrium energy difference as
\begin{align}
	\D U = V(x_t,t) - V(x_0,t_0) \;,
\end{align}
the first law on the level of individual trajectories follows from (\ref{eq:1st_law_pre}) as
\begin{align} \label{eq:1st_law}
  \D U = W[\xc] - Q[\xc] \;.
\end{align}
Hence, the equilibrium difference of energy between initial point and final point of the particle is the work done on the particle minus the dissipated heat. This derivation of the first law from Langevin dynamics was first noted by Sekimoto \cite{Sekimoto1998}.\\
From the canonical point of view, we identify the Brownian particles as a {\it system} coupled to an {\it equilibrium heat bath} at constant temperature. We assume this heat bath to be the ambient medium, that is the fluid into which the particles are suspended. The assumption that the fluid is an ideal heat bath implies an instantaneous thermal equilibration for any value of $x(t)$. In fact, this equilibration takes place in finite time and is the fast dynamics on time scales $t<\tME$ at which the Markov assumption does not hold. Hence, the inclusion of only the outcomes of collisions as white noise into the LE constitutes a coarse-graining of the fast dynamics to the slow dynamics modelled explicitly by the LE.\remark{$x(t)$ is sozusagen die slow dynamics, und zu jedem zeitpunkt wabert die schnelle (als instantan angenommene) equilibrierung des fluids im hintergrund.} Identifying the fast dynamics as the dynamics of the fluid molecules, and the slow dynamics as the dynamics of the particle, we may also regard $Q[\xc]$ in (\ref{eq:def_Q}) as the transfer of heat from slow dynamics to fast dynamics, and the work $W[\xc]$ as the energy externally injected into the system by influencing the slow dynamics directly.\remark{Note that $Q[\xc]$ and $W[\xc]$ may also be negative.}\\

Due to the friction force $\g\,\ddx_t$, the derivation of the first law (\ref{eq:1st_law}) takes explicitly non-equilibrium effects into account. To separate equilibrium from non-equilibrium contributions to the first law (\ref{eq:1st_law}), consider the canonical equilibrium distribution \label{eq:def_peq} in the form
\begin{align}
	-\ln \peq(x,t) &= \b V(x,t) + \ln Z(t) \nn
	&= \frac{1}{\kB T}\Big[ V(x,t) - \FF(t) \Big]
\end{align}
with Helmholtz free energy 
\begin{align} \label{eq:def_FF}
	\FF(t) = -\kB T \ln Z(t) = -\kB T\ln \int \eee{-\b V(x,t)} \di x \;.
\end{align}
For reversible transition from equilibrium states $x_0$ at time $t_0$ to states $x$ at time $t$, the reversible work $\DF=\FF(t)-\FF(t_0)$ is the only work done on the system, and hence $-\D \Qrev = (\D U - \DF)$ is the heat reversibly received from the heat bath.\footnote{In the context of stochastic thermodynamics it is common to denote by $Q$ the heat delivered to the heat bath, in contrast to classical thermodynamics where $Q$ is the heat received from the heat bath.} Denoting by $\Wdiss[\xc]$ the work done in addition to the equilibrium work $\DF$, and by $\Qirr[\xc]$ the heat transferred to the medium in addition to the reversible heat $\Qrev$, we rewrite the first law as (\ref{eq:1st_law})
\begin{align}
	W[\xc] &= \DF + \Wdiss[\xc] \nn
	&= \D U + \Qrev + \Qirr[\xc] \label{eq:1st_law_irr} \;.
\end{align}
Limiting ourselves at first to quasi-steady process control, we see that spending the work $W[\xc]$, the internal energy of the system is increased by $\D U$, and heat $\Qrev$ is generated that is instantaneously transferred to the heat bath. Reversing the process, still in the case of quasi-steady process control, we can gain the free energy difference $\DF$ as useful work by making direct use of lowering the internal energy by $\D U$ and retrieving $\Qrev$ from the heat bath.\\
In the general (not quasi-steady) case, the additional heat $\Qirr[\xc]$ is dissipated, which in contrast to the quasi-steady case can not be retrieved from the heat bath in the reversed process, and the extra work $\Wdiss[\xc]$ has to be spent in order to compensate that heat loss\remark{and maintain $\D U$.}. We will refer to $\Qirr[\xc]$ as irreversible heat, and to $\Wdiss[\xc]$ as dissipative work. Note that in contrast to equilibrium state variables which only depend on initial and final state, the process functionals $\Qirr[\xc]$ and $\Wdiss[\xc]$ depend explicitly on the trajectory $\xc$ that connects initial and final state.\\
Depending on the process control, $\Qrev$ can be retrieved or dissipated by the system, and the associated equilibrium entropy difference $\DSeq=-\Qrev/T$ of the system can be positive or negative. Taking system and heat bath together, however, $\DSeq$ of the system cancels with $-\DSeq$ in the heat bath, and we are left with the irreversible entropy production (EP) $\Sirr[\xc]=\Qirr[\xc]/T$ which is, according to the second law, on average non-negative. We will demonstrate in the next section, how $\Sirr[\xc]$ arises along with other entropies from the FPE.\\

But before we turn to entropies arising from the FPE, let us briefly discuss the first law derived from the LE for multiplicative noise. For the case of state dependent mobility $\G(x)$, or friction $\g(x)=1/\G(x)$, the energy balance based on the LE derived from Newton's equation of motion now reads
\begin{align} \label{eq:LE_en-bal_mobi}
	\Big[\g(x_t)\ddx_t \- \sqrt{2\kB T\g(x_t)}\xi(t)\Big]\dd x_t = \bigg[\-V'(x_t,t)\-\frac{\kB T}{2}\frac{\g'(x_t)}{\g(x_t)}\bigg]\dd x_t \,. \!\!
\end{align}
Here, we have taken the thermodynamic consistent LE (\ref{eq:td-inter_multi_Gamma_LE_alpha}) which reproduces for any convention $\a$ the canonical equilibrium distribution in the stationary state and chosen $\a=1/2$ to resort to ordinary calculus. By following the same lines as for additive noise, we arrive at the modified first law
\begin{align} \label{eq:1st_law_mobi}
	\D U + \D G = W[\xc] - Q[\xc]
\end{align}
with the extra boundary term
\begin{align}
	\D G \dfn \kB T\,\ln\sqrt{\mfrac{\G(x_t)}{\G(x_0)}} \;.
\end{align}
We note that the quantity $T\D G$ is reminiscent of an entropy difference.\remark{dafür müsste $\G(x)$ sowas wie ne gleichgewichts-verteilung sein, und geht ja auch tatsächlich als $\ln\G(x)$ in $\p(x)$ ein. related to entropic barrier? [Zwanzig92]}\\
For the case of a temperature gradient, which may even be time-dependent, we have the energy balance 
\begin{align} \label{eq:LE_en-bal_Tgrad}
	\Big[\g\ddx_t \- \sqrt{2\g\kB T(x_t,t)}\xi(t)\Big]\dd x_t = \Big[\-V'(x_t,t)\-\mfrac{\kB T}{2}T'(x_t,t)\Big]\dd x_t \,. \!\! 
\end{align}
In this case, all ingredients of the first law are augmented by thermophoretic terms
\begin{subequations} \label{eq:1st_law_Tgrad}
  \begin{align}
		\D U &\quad\mapsto\quad \Big[V(x_\t,\t)+\mfrac{\kB}{2}T(x_\t,\t)\Big]_{t_0}^{t} \;,\\
		W[\xc] &\quad\mapsto\quad \int_{t_0}^{t} \dot V(x_\t,\t)+\mfrac{\kB}{2}\dot T(x_\t,\t) \di\t \;,\\
		Q[\xc] &\quad\mapsto\quad \int_{t_0}^{t} \Big[\-V'(x_\t,\t)-\mfrac{\kB}{2}T'(x_\t,\t)\Big]\,\ddx(\t) \di\t	\;,	
	\end{align}
\end{subequations}
where the second term in $\D U$ accounts for the difference in thermal energy between initial and final position, the extra term in $W[\xc]$ is due to lifting the thermal energy level of the particle externally (work-like), and for $Q[\xc]$ also the thermophoretic force proportional to the temperature gradient $T'(x)$ is considered.\\
We refrain from attempting a further interpretation of these results for the multiplicative case, and refer the interested reader to \cite{Lopez2007,Benjamin2008,Yang2013}.\remark{siehe auch 'genLE' notizen}

\subsection{Entropies and the second law} \label{ss_td-interpration_entropy}
In the previous section, we formulated the first law on the level of single stochastic trajectories in non-equilibrium. In doing so, we separated equilibrium state variables from non-equilibrium process functionals, the former only dependent on initial and final state, the latter on the specific trajectory. The introduced non-equilibrium functionals are dissipative work $\Wdiss[\xc]$ and irreversible heat dissipation $\Qirr[\xc]$, were $\Wdiss[\xc]$ has to be spent to compensate $\Qirr[\xc]$, hence, $\Wdiss[\xc]=\Qirr[\xc]$. The irreversibly dissipated heat is energy that can not be transformed into useful work. As a measure for useless energy, {\it entropy} is defined as irreversibly dissipated heat divided by the temperature of the heat bath that receives the heat. We will refer to the entropy change as {\it entropy production} (EP), which may also include equilibrium entropy differences. We denote the {\it total} EP of the compound of system and heat bath, that is the irreversible EP associated with $\Qirr[\xc]$, by $\Sirr[\xc]=\Qirr[\xc]/T$. Perceiving the compound of system and bath as an isolated system, the second law applies for the total EP, $\Sirr[\xc]>0$.\\
The aim of this section is to identify $\Sirr[\xc]$ from the FPE, along with equilibrium EP, and two distinct EPs for non-equilibrium steady states (NESSs) that separately obey the second law. We will also demonstrate the equivalence of two established notions of EP.\remark{doppelt: Although formulated for general $F(x,t)$, $D(x,t)$, to maintain intuition on this formal level, we will expediently revert to the paradigm of the particle suspended in a medium with drift velocity $F(x,t)=-\G V'(x,t)$ und diffusion $D=\G\kB T$, and we will also on the formal level refer to the picture of particle and fluid.} \\

\remark{We begin with defining the equilibrium entropy of the system as the Gibbs entropy of the equilibrium distribution ,
\begin{align}
	\seq(x,t) \dfn -\kB\ln \peq(x,t)
\end{align}
and obtain the reversible EP of the system between states $x_0$ and $x$ as
\begin{align} 
	\D \seq = \seq(x_t,t)\Big|_{t_0}^{t} = \frac{\D U - \DF}{T} = \frac{-\D \Qrev}{T} \;,
\end{align}
which will be produced by the process and received by the medium for a quasi-steady process that connects initial and final state. Note that that the above equation is in agreement with $F=U-TS$ known from macroscopic equilibrium thermodynamics. (doppelt: By reversing the process, we retrieve the work $\DF$ and the systems receives the heat $-\Qrev$.) (denk an piston, quasi-stationäre komprimierung verrichtet arbeit und gibt wärme an umgebung ab, und das ist genau die arbeit die quasi-stationären expandieren wieder gewonnen wird und die wärme die aufgenommen werden muss (abkühlung!), damit original-zustand wieder erreicht.)\\
Using the first law (\ref{eq:1st_law}) and the definitions of heat (\ref{eq:def_Q}) and work (\ref{eq:def_W}), the equilibrium EP $\D \seq$ can also be stated for non-reversible processes control on the level of individual trajectories. To this end, consider the EP rate along a trajectory, 
\begin{align} 
	\dd_\t \seq(x(\t),\t) = \kB\b(\dot V(x(\t),\t) + V'(x(\t),\t)\ddx(\t) - \dot\FF(\t))
\end{align}
and integrate it to
\begin{align} 
	\DSeq &= \int_{t_0}^{t} \dd_\t \seq(x(\t),\t) \nn
	&= \frac{W[\xc]-Q[\xc]-\DF}{T} \;.
\end{align}
(doppelt: For a reversible transition from $x(t_0)$ to $x(t)$, the work $W[\xc]$ equals the reversible work $\DF$, and we find again $T\D \seq=-\Qrev[\xc]$. If the process is done in finite time, the extra work $\Wdiss[\xc]=W[\xc]-\DF$ has to be performed on the particle, which must equal the irreversible heat dissipation $\Qirr[\xc]=Q[\xc]-\Qrev$.) We hence recover the equilibrium entropy
\begin{align} 
	T\,\DSeq = \Wdiss[\xc]-(\Qirr[\xc]+\Qrev) = \Qrev
\end{align}
for {\it arbitrary} individual realisations $\xc$. We thus can determine the equilibrium entropy difference $\DSeq$ between the two states at time $t_0$ and $t$ from non-equilibrium measurements, provided, of course, that we know the free energy difference $\DF$.\\}

As a preliminary, consider the  probability current density $j(x,t)\eq F(x,t)p(x,t) \- D(x,t)p'(x,t)$ defined in (\ref{eq:def_J_FD}).
In equilibrium, the current must vanish, \mbox{$\Jeq(x,t)\equiv0$}, which implies a basic condition to be satisfied in order for a equilibrium state to exist. Equivalent to demanding $\Jeq(x,t)\equiv0$ is the condition
\begin{align} \label{eq:detbal_pre}
	F(x,t) - D(x,t)\frac{{\peq}'(x,t)}{\peq(x,t)} =  \frac{\Jeq(x,t)}{\peq(x,t)} = \lla\ddx|x,t\rra_{\mr{eq}} \meq 0 \;.
\end{align}
Substituting $\peq(x,t)=\exp(-\p(x,t)-\GG(t))$, this condition becomes
\begin{align} \label{eq:detbal}
	F(x,t) + D(x,t)\p'(x,t) = 0 \;,
\end{align}
known as the {\it detailed balance condition} \cite{VandenBroeck2010}.\\
To satisfy the detailed balance condition, the external force must arise from a potential, $F(x,t)=-D(x,t)\p'(x,t)$, which can always be achieved in one dimensional processes. For the Brownian particle, i.e. $\p(x,t)=\b V(x,t)$ and $F(x,t)=-\G V'x,t)$, the detailed balance equation becomes the Einstein relation (\ref{eq:D_einstein}), $\b D=\G$. Note that the two generalisations for multiplicative noise discussed in the previous section (position-dependent mobility and temperature gradient) also respect the detailed balance condition.\\
The general form (\ref{eq:detbal_pre}) of the detailed balance condition, along with the local average of velocity $\lla\ddx|x,t\rra$, will be important in the considerations of this section.\\

\paragraph{Trajectory based EP}
We now turn to entropic terms in a FPE\remark{genLE p.9ff}. A reasonable definition for the system entropy on the formal level of a FPE is the Gibbs entropy\remark{(ist ja auch eine td interpretation von $p(x,t)$, mehr ist auf FPE-Ebene ja ersmtal nicht möglich, um weiter zu interpretieren müsste ich auf die Physik der LE zurückgreifen und $F=-\G V'$ und $D=\G\kB T$ einsetzen.)}\remark{closely related to Shannon entropy from information theory, where no $\kB$ and $\ln=\log_2$.}
\begin{subequations} \label{eq:gibbs}
  \begin{align}
		S(t) &= \int p(x,t)\,s(x,t) \di x = \lla s(x,t)\rra \;, \label{eq:gibbs_S} \\ 
		s(x,t) &= -\kB\ln p(x,t) \;, \label{eq:gibbs_s}
	\end{align}
\end{subequations}
where $p(x,t)$ is the solution of the FPE. In the following we will follow common practice and set $\kB\dfn1$, for which entropy becomes dimensionless and temperature is measured in units of energy\remark{multiply subsequent entropies with $\kB$ to retrieve physical entropies}.\\ \renewcommand{\kB}{}
In contrast to the thermodynamic quantities defined from the LE in the previous section, $S(t)$ and $s(x,t)$ are a-priori not defined on the level of individual trajectories $\xc$. To establish the formulation for individual $\xc$, we simply evaluate $s(x,t)$ along the trajectory $x(\t)$, that is $s\big(x(\t),\t\big)$. Implicit and explicit differentiation with respect to $\t$ yields
\begin{align} \label{eq:EPR_s_traj_pre1}
	\dds(\t) = - \kB\frac{p'(x,t)}{p(x,t)}\bigg|_{x=x(\t)}\,\ddx(\t) - \kB\frac{\dot p(x,t)}{p(x,t)}\bigg|_{x=x(\t)} \;,
\end{align}
where we have written for short $\dd_\t s\big(x(\t),\t\big)=\dot s(\t)$. Integration of the EP {\it rate} $\dds(\t)$ with respect to $\t$ recovers the notion of EPs as in the previous section.\\
To include the FPE (\ref{eq:FPE_FD}) into our considerations, we use the current $j(x,t)$ defined in (\ref{eq:def_J_FD}) and rewrite the FPE as
\begin{align}
	\frac{p'(x,t)}{p(x,t)} =  - \frac{j(x,t)}{D(x,t)p(x,t)} + \frac{F(x,t)}{D(x,t)} \;.
\end{align}
Substitution of the above form of the FPE into (\ref{eq:EPR_s_traj_pre1}) yields $\dds(\t)$ in terms of $F(x,t)$ and $D(x,t)$,
\begin{align}
	\dds(\t) &= \kB\frac{j(\t)}{D(\t)p(\t)}\,\ddx(\t) - \kB\frac{F(\t)}{D(\t)}\,\ddx(\t) - \kB\frac{\dot p(\t)}{p(\t)}  \nn
	&= \kB\frac{\lla \ddx|\t\rra - F(\t)}{D(\t)}\,\ddx(\t) - \kB\frac{\dot p(\t)}{p(\t)} \label{eq:EPR_s_traj} \;,
\end{align}
where we have written for short $p(\t)=p\big(x(\t),\t\big)$ (analogous for the other quantities).\\
The first term is of the form ``force times velocity'', in which, in the case of Brownian particles, the force becomes $\g\lla\ddx|\t\rra+V'(\t)$. In view of (\ref{eq:LE_en-bal_add}), we see that this force is analog to the thermal force in the equivalent LE, and we may interpret $\big(\g\lla\ddx|\t\rra+V'(\t)\big)\ddx(\t)$ as the average energy received along $x(\t)$ by the particle undergoing collisions with the fluid molecules.\remark{average weil $\lla\ddx|\t\rra$ obwohl $\dds(\t)$ before average. das average was noch kommen würde, wäre über alle schnitte $x(\t)$. so noch nicht ganz zufriedenstellend, siehe genLE p.16 für weitere ansätze. am besten mal für gezogene parabel die einzelnen terme anschauen zum überprüfen (analytisch+simulation).}\\
The second term of \ref{eq:EPR_s_traj} contributing to $\dds(\t)$ is the EP due to temporal change of $p(x,t)$, which is found in cases where the system is off equilibrium, that is for time dependent $D(x,t)$ or $F(x,t)$, or for an initial distribution $p_0(x)\neq\peq(x)$.\\
To obtain the EP rate in the steady state, we substitute $\pst(x,t)$ from (\ref{eq:FPE_td-interpretation}),
\begin{align} \label{eq:EPR_s_rev_traj}
	\ddsst(\t) =  - \kB\frac{F(\t)}{D(\t)}\,\ddx(\t) + \kB\big(\dot\p(\t) - \dot\GG(\t)\big)
\end{align}
which upon integration becomes the difference of entropy between the stationary initial and final states, 
\begin{align} 
	\D\Sst &= \kB\int_{t_0}^{t} \p'(\t)\,\ddx(\t) + \dot\p(\t) - \dot\GG(\t) \di\t \nn
	&= \kB\D\p - \kB\DG \label{eq:EPR_s_rev_traj_int} \;,
\end{align}
irrespective of particular trajectories $\xc$. For the Brownian particle we recover the equilibrium EP,
\begin{subequations}
\begin{align}
	\D\Sst &= \frac{-Q[\xc]+W[\xc]-\DF}{T} = \frac{\D U - \DF}{T} = \DSeq \label{eq:EP_EQ} \\
	&= \frac{-(\Qirr[\xc]+\Qrev)+\Wdiss[\xc]}{T} = -\frac{\Qrev}{T} = \DSeq \label{eq:EP_EQ_diss}  \;,
\end{align}
\end{subequations}
which is, again, valid for {\it arbitrary} individual trajectories $\xc$. We thus can determine the equilibrium entropy difference $\DSeq$ between two states at time $t_0$ and $t$ from non-equilibrium measurements of $Q$ and $W$ or $\Qirr$ and $\Wdiss$, provided, of course, that we know the free energy difference $\DF$.\\

Having identified the equilibrium EP from the FPE, we turn to the irreversible EP that obeys the second law which states that in an {\it isolated} and {\it macroscopic} system the entropy must increase. The specification 'macroscopic' is important in our considerations, since we deal with {\it individual} realisations. The bridge to macroscopic thermodynamics is established by considering the thermodynamic quantities in the {\it ensemble average} of in principle infinite many realisations. This macroscopic limit is equivalent to the thermodynamic limit in which the number of involved particles approaches infinity.\\
We introduced the Gibbs EP as the EP of the system. Being coupled to a heat bath, the system is not isolated, and accordingly, the system entropy does not need to obey the second law. Instead, as already indicated in the previous section, we consider the compound of system and heat bath as an isolated system.\\
We denote by $\ddsm(\t)$ the EP in the medium, and by 
\begin{align}
	\ddstot(\t) = \dds(\t) + \ddsm(\t)
\end{align}
the total EP of the compound of system and heat bath, which we expect to obey the second law. Availing ourselves of the picture of the Brownian particle and arguing that the EP in the medium is brought about by the heat $Q[\xc]$, we identify by comparison of (\ref{eq:EP_EQ}) with (\ref{eq:EPR_s_rev_traj}) the formal EP rate of the heat bath as
\begin{align}
	\ddsm(\t) \dfn \kB\frac{F(\t)}{D(\t)}\ddx(\t) \label{eq:def_ddsm} \;.
\end{align}
Substituting the system EP rate $\dds(\t)$ from (\ref{eq:EPR_s_traj}), we find for the total EP rate
\begin{align}
	\ddstot(\t) &= \dds(\t) + \ddsm(\t) \nn
	&= \kB\frac{\lla\ddx|\t\rra}{D(\t)}\,\ddx(\t) - \kB\frac{\dot p(\t)}{p(\t)} \label{eq:def_ddstot}\;,
\end{align}
or, in integrated form the total EP,
\begin{align}
	\Stot[\xc] = \D S + \Sm[\xc] \label{eq:EP_Stot} \;.
\end{align}
We will show in the next section, see (\ref{eq:FTs_secondlaw}), that $\Stot[\xc]$ is indeed non-negative in the ensemble average. The identification of the total EP from the FPE we discussed here has been achieved by Seifert in \cite{Seifert2005}.\\
In addition to defining the EP $\Sm[\xc]$ by comparing (\ref{eq:EP_EQ}) with (\ref{eq:EPR_s_rev_traj}) as done above, we may also define the entropic equivalent of work, $R[\xc]$,
\begin{subequations} \label{eq:def_Sm_R_Rdiss}
  \begin{align}
		\Sm[\xc] &= -\kB\int_{t_0}^{t} \p'(\t)\ddx(\t) \di\t \;, \label{eq:def_Sm} \\
		R[\xc] &\dfn \kB\int_{t_0}^{t} \dot\p(\t) \di\t \;,  \label{eq:def_R}
	\end{align}
\end{subequations}
With the definitions above and substituting the total differential\linebreak $\dd_\t\p(\t)=\p'(\t)\ddx(\t)+\dot\p(\t)$ into (\ref{eq:EP_Stot}), we can relate $\Stot[\xc]$ with $R(\xc]$\remark{mind that $\D\p$ is not the equivalent of the equilibrium internal energy difference, since here initial and final state are not steady states.},
\begin{align}
	\Stot[\xc] = R[\xc] - (\D\p\-\D S) \;. \label{eq:EP_Stot_R}
\end{align}
Note that in these considerations we did not demand the process to start and end in a steady state. If we do so, however, the equations for the total EP simplify to
\begin{subequations}
\begin{align}
	\Stot[\xc] &= \DSst + \Sm[\xc] \label{eq:EP_Stot_SS} \\
	&= R[\xc] - \DG \;. \label{eq:EP_Stot_R_SS}
\end{align}
\end{subequations}
In view of this simplification, we mention that a formulation in terms of the Kullback-Leibler divergence ($-\ln(p/\pst)$) instead of the Gibbs entropy ($-\ln p$) is advantageous.\remark{In addition to defining the EP $\Sm[\xc]$ by comparing (\ref{eq:EP_EQ}) with (\ref{eq:EPR_s_rev_traj}) as done above, we may also define the entropic equivalents of work $R[\xc]$ and dissipative work $\Rdiss[\xc]$,
\begin{subequations} \label{eq:def_Sm_R_Rdiss}
  \begin{align}
		\Sm[\xc] &= -\kB\int_{t_0}^{t} \p'(\t)\ddx(\t) \di\t \;, \label{eq:def_Sm_with_R_Rdiss} \\
		R[\xc] &\dfn \kB\int_{t_0}^{t} \dot\p(\t) \di\t \;,  \label{eq:def_R} \\[5pt]
		\Rdiss[\xc] &\dfn R[\xc] - \DG \;. \label{eq:def_Rdiss}
	\end{align}
\end{subequations}
With the definitions above and substituting the total differential\linebreak $\dd_\t\p(\t)=\p'(\t)\ddx(\t)+\dot\p(\t)$ into (\ref{eq:EP_Stot}), we can relate $\Stot[\xc]$ with $R(\xc]$,
\begin{align}
	\Stot[\xc] &= R[\xc] - (\D\p\-\D S)  \nn
	&= R[\xc] - \DG + (\D S-\DSst)  \;, \label{eq:EP_Stot_R}
\end{align}
and write the first law (\ref{eq:1st_law}) as
\begin{align}
	\D\p &= R[\xc] - \Sm[\xc] \nn
	&= \Rdiss[\xc] - \Stot[\xc] + \DG + \D S \;.
\end{align}
In equilibrium, i.e. $\Rdiss[\xc]\eq\Stot[\xc]$ and $\D S=\DSeq$, we recover the equivalent of (\ref{eq:EP_EQ}). Aufpassen, $\D\p$ ist nicht equivalenz der gleichgewichts-innere-energie differenz $\D U$, denn start und ende der trajektorien sind ja beliebig!
Note that $\Stot[\xc]$ is conceptually different from $\Sirr[\xc]=\Qirr[\xc]/T$, since, in our discussion after (\ref{eq:1st_law_irr}), we considered $\Qirr[\xc]$ to be the irreversibly dissipated heat for trajectories connecting the equilibrium initial and final states that defined the equilibrium state variables, whereas $\Ds$ was derived from the FPE with arbitrary initial distribution $p_0(x_0)$ and without requiring $p(x,t)=\pst(x,t)$ at the end of the process. Indeed, in the case $p(x,t)=\pst(x,t)$ we have
\begin{align}
	\D S = \DSst = \D\p - \DG
\end{align}
from (\ref{eq:FPE_td-interpretation_pst}), and by substitution into (\ref{eq:EP_Stot_R}) we find
\begin{align}
	\Stot[\xc] = R[\xc] - \DG = \Rdiss[\xc] \sep (\,p|_{t_0}^{t}=\pst|_{t_0}^{t}\,) \label{eq:EP_Stot}
\end{align}
being identical to $\Qirr[\xc]/T$ for $D(x,t)\equiv T$.\\
Note also, now again for the general case $\D S \neq \DSst$, that by combining (\ref{eq:EP_Stot}) with (\ref{eq:EPR_s_rev_traj_int}) and (\ref{eq:def_Sm_R_Rdiss}), we get
\begin{align}
	\D S - \DSst &= (\Stot[\uc] - \Rdiss[\uc]) \\
	\Rdiss[\xc] &= \DSst + \Sm[\xc]
\end{align}
in contrast to $\Stot[\xc] = \D S + \Sm[\xc]$. That demonstrates, that $\Stot$ considers also the EP for not starting in SS, but $\Rdiss$ not. The difference, 
\begin{align}
	\DI \dfn \Stot[\uc] - \Rdiss[\uc]
\end{align}
is therefore the EP for not starting in SS (penalty EP or entropic auslenkungs-arbeit (entropic deflection-work)).\\
From the above relation, we also get an alternative expression for the system EP,
\begin{align}
	\D S = \DSst + (\Stot[\uc] - \Rdiss[\uc]) \;,
\end{align}
which, in view of our interpretation of $\Stot[\uc] - \Rdiss[\uc]$, means that the system EP is the stationary EP plus the deflection entropy. Hence, by adding 
\begin{align}
	\Stot = \D S + \Sm = \D S - \DSst + \Rdiss[\xc] \;,
\end{align}
we subtract from $\D S$ the stationary EP which leaves us with the deflection entropy, which added to $\Rdiss$ results into $\Stot$. Siehe auch diss p.32-33. Insgesamt würd ich sagen, müsste man auf diesem formalen level anstelle von der Gibbs entropy $S=-\ln p$ die Kullback-Leibler divergenz $K=-\ln p/\pst$ zur stationären verteilung nehmen, siehe diss p.32-33.}\\

\paragraph{Spatial averaged EP}
The discussion of entropies presented so far was based on evaluating EP rates $\dds(x,t)$ along trajectories $x(\t)$ to get EP rates $\dds(\t)=\dds\big(x(\t),\t\big)$ and EPs $S[\xc]$ on the level of single $\xc$. This formulation in terms of functionals is of advantage when addressing EPs using WPIs, as will be discussed in the next section. In contrast to the {\it functional} formulation, it is also common to consider the {\it spatial average} of $\dds(x,t)$, as suggested by Esposito and van den Broeck \cite{Esposito2010,Esposito2010a,VandenBroeck2010}. For completeness, we briefly demonstrate the correspondence of both formulations.\\
Starting point is again the Gibbs EP
\begin{align}
	\ddS(t) &= -\kB\dd_t\int p(x,t)\,\ln p(x,t) \di x \nn
	&= -\kB\int \dot p(x,t)\,\ln p(x,t) \di x - \kB\int \dot p(x,t) \di x \;. \label{eq:EPR_Gibbs}
\end{align}
Substituting $\dot p(x,t) = -j'(x,t)$, integrating by parts and substituting $p'(x,t)$ from the definition of $j(x,t)$ in (\ref{eq:def_J_FD}), we find the splitting
\begin{align}
	\ddS(t) &= - \kB\int j(x,t)\,\frac{p'(x,t)}{p(x,t)} \di x \nn
	&= \kB\int \frac{j(x,t)^2}{D(x,t)p(x,t)} \di x - \kB\int \frac{F(x,t)j(x,t)}{D(x,t)} \di x \;. \label{eq:EPR_Gibbs_split}
\end{align}
The first term is the total irreversible EP and corresponds to $\Stot[\xc]$, and the second term is the EP in the heat bath and corresponds to $\Sm[\xc]$.\\
To clarify this correspondence, we rewrite 
\begin{align}
	\ddS(t) &= \int p(x,t)\,\big[\ddsi(x,t)-\ddse(x,t)\big] \di x
\end{align}
in terms of the EP fields
\begin{subequations} \label{eq:EPRFs_bruessel}
  \begin{align}
		\ddsi(x,t) &\dfn \kB\frac{\lla\ddx|x,t\rra^2}{D(x,t)}-\kB\frac{\ddp(x,t)}{p(x,t)} \label{eq:EPRFs_bruessel_si} \\
		\lolra\quad \ddstot(\t) &= \kB\frac{\lla\ddx|\t\rra}{D(\t)}\,\ddx(\t) - \kB\frac{\dot p(\t)}{p(\t)}\nn
		\ddse(x,t) &\dfn \kB\frac{F(x,t)}{D(x,t)}\lla\ddx|x,t\rra \label{eq:EPRFs_bruessel_se} \\
		\lolra\quad \ddsm(\t) &= \kB\frac{F(\t)}{D(\t)}\ddx(\t) \nonumber
	\end{align}
\end{subequations}
\remark{where we followed the notation of \cite{Esposito2010,Esposito2010a,VandenBroeck2010} (abgesehen vom vorzeichen in $\se$)}in correspondence to (\ref{eq:def_ddstot}) and (\ref{eq:def_ddsm}). Note that we included the second term in (\ref{eq:EPR_Gibbs}) into $\ddsi(x,t)$ which vanishes under spatial average due to probability conservation.\\
On spatial average, we arrive at the average EP rate discussed by Esposito and van den Broeck \cite{Esposito2010,Esposito2010a,VandenBroeck2010}
\begin{subequations} \label{eq:EPRs_bruessel}
  \begin{align}
		\ddSi(x,t) &= \kB\int \frac{j(x,t)^2}{D(x,t)p(x,t)} \di x \label{eq:EPRs_bruessel_Si} \\
		\ddSe(x,t) &= \kB\int \frac{F(x,t)}{D(x,t)}j(x,t) \di x \label{eq:EPRs_bruessel_Se} \\
		\ddS(x,t) &= \ddSi(x,t) - \ddSe(x,t)
	\end{align}
\end{subequations}
In view of (\ref{eq:EPRFs_bruessel}) and (\ref{eq:EPRs_bruessel}), we make three comments.\\
(i) The total EP is non-negative for all times $t$, $\ddSi(x,t)\geq0$, and is the irreversible EP that obeys the second law.  \\
(ii) On the trajectory level, we have besides the local average velocity along a trajectory, $\lla \ddx|\t\rra$, also the velocity of the trajectory itself, $\ddx(\t)$, a distinction that is lost in the spatially averaged formulation of the Gibbs EP.\remark{(udo's review auf seite 15-16)}\\
(iii) The term $\ddp/p$ in the irreversible EP rate only contributes on the trajectory level which suggests that $\ddp/p$ accounts for a local balance between EPs that cancel in the spatial average.\remark{It stands to reason to assume these contributions to be the EP from the dissipative work $\Wdiss$ and the irreversible heat dissipation $\Qirr$ as the two sources of irreversible EP that must cancel on average. This picture is further corroborated by noting that for $p=\pst$, where $\ddQirr\equiv0$, we find $\ddp/p=\ddWdiss$ which also vanishes in the quasi-steady limit. stimmt so wohl nicht ganz, besser mal anhand sim oder ana (gezpar) testen.)}\remark{splitting in boundary and driving term? superfluous}\\

\paragraph{Multidimensional generalisation}
So far, we limited ourselves to one-\linebreak dimensional continuous MPs. In higher dimensions, i.e. $\vek{X}(t)\in\mathbb{R}^n$ with $n\geq2$, the stationary solution of the FPE does not necessarily coincide with thermal equilibrium, the system is then referred to be in a {\it non-equilibrium steady state} (NESS). The defining property of a NESS is that the stationary probability current does no longer vanish, but acquires a constant value in time. The non-vanishing current in a steady state is possible because in higher dimensions a particle can run in closed loops, repeating each loop under the same conditions. Thus, in a NESS, the probability of observing a particle at time $t$ at a position $\vek{x}$ will be constant in time and the probability current $\vJst(\vek{x},t)$ will be divergence free.\\
We discussed this multidimensional setting already in terms of a FPE in section \ref{ss_FPE} after (\ref{eq:highdim_conti-equ}). The multidimensional drift velocity reads
(\ref{eq:highdim_Fcnc})
\begin{align}
	\vek{F}(\vek{x},t) =  \mat{D}(\vek{x},t)\,\big[\-\vek{\nabla}\p(\vek{x},t) + \vek{A}(\vek{x},t)\big] \;.
\end{align}
The forces that drive the particles in loops must be non-conservative, which we included into $\vek{F}(\vek{x},t)$ in terms of a divergence free vector field $\vek{A}(\vek{x},t)$. For a Brownian particle is 
\begin{subequations}
  \begin{align}
		\mat{D}_{ij}(\vek{x},t) &\equiv \G/\b\,\d_{ij} \\
		\vek{A}(\vek{x},t) &= \b \vek{F}_\mr{\!nc}(\vek{x},t) \\
		\p(\vek{x},t) &= \b V(\vek{x},t) \\
		\vek{F}(\vek{x},t) &= \G\,\big[\-\vek{\nabla} V(\vek{x},t) + \vek{F}_\mr{\!nc}(\vek{x},t) \big]
	\end{align}
\end{subequations}
and $\vek{F}_\mr{\!nc}$ the non-conservative force.\remark{In (\ref{eq:highdim_pst}) and (\ref{eq:highdim_J}), we defined stationary distribution and current as
\begin{subequations}
\begin{align}
	\pst(\vek{x},t)&=\exp\big[-\phi(\vek{x},t)\big] \;, \\
	\vJst(\vek{x},t)&=\lla\vek{\ddx}|\vek{x},t\rra_\mr{st} \pst(\vek{x}) \;,
\end{align}
\end{subequations}
where the non-equilibrium potential $\phi(\vek{x},t)$ has to be distinguished from the one-dimensional equivalent $\p(x,t)$.}\\
The non-zero current in a NESS has consequences for the EP. Although in a steady state, the persistent motion of the particle dissipates energy and the system needs a continuous intake of energy to compensate that heat loss. The continuous dissipated heat into the medium is called {\it housekeeping heat} $\Qhk$, in the sense that its compensation maintains the NESS. The heat transferred to the medium that comes in addition to $\Qhk$ is the so-called {\it excess heat} $\Qex$. Distinguishing the two EPs associated with these two heat fluxes into the medium, we split the EP in the medium
\begin{align} \label{eq:EPs_NESS_slitSm}\\
	\Sm[\vxc] = \Shk[\vxc] + \Sex[\vxc]
\end{align}
into the two contributions
\begin{subequations} \label{eq:EPs_NESS}
  \begin{align}
		\Shk[\vxc] &= \int_{t_0}^{t} \vek{\ddx}(\t)\,\mat{D}^{-\!1}(\t)\lla\,\vek{\ddx}|\t\rra_\mr{st} \di \t \label{eq:EPs_NESS_Shk}\\
		\Sex[\vxc] &= -\int_{t_0}^{t} \vek{\ddx}(\t)\,\vek{\nabla}\phi(\t) \di \t \label{eq:EPs_NESS_Sex}
	\end{align}
\end{subequations}
with the stationary mean local velocity $\lla\vek{\ddx}|\t\rra_\mr{st}$ from (\ref{eq:highdim_mean-ddx}) and the non-equilibrium potential $\phi(\vek{x},t)=-\ln\pst(\vek{x},t)$ from (\ref{eq:highdim_pst}). This splitting was introduced by Oono and Paniconi \cite{OonoPaniconi98PoTPS} and picked up by Hatano and Sasa \cite{HatanoSasa01PRL}, Speck and Seifert \cite{SpeckSeifert05JoPAMaG,SpeckSeifert06EL} and Chernyak, Chertkov and Jarzynski \cite{ChernyaChertkoJarzyn06JoSMTaE} and many others \cite{BokseMaesNetocPesek11EL,AbreuSeifert12PRL,Lee2012,Verley2012a,Spinney2012b}.\\
The housekeeping EP $\Shk[\vxc]$ is non-negative in the ensemble average, whereas $\Sex[\vxc]$, like $\Sm[\vxc]$, can remain positive on average, as it also includes reversible entropy transfer. To obtain the irreversible part of $\Sex[\vxc]$, we follow the same strategy as for $\Stot[\vxc]$ and find the non-negative EP\remark{Seiferts review p.14 and p.22 and p.30)} (\cite{Seifert2012} p.\,14)
\begin{align}
	\Sirrna[\vxc] &\dfn \Ds + \Sex[\vxc] \nn
	&= \Ds - \D\phi + \int_{t_0}^{t} \dot\phi(\t) \di \t	\nn
	&= -\ln\frac{p(\t)}{\pst(\t)}\Big|_{t_0}^{t} + \int_{t_0}^{t} \dot\phi(\t) \di \t \label{eq:EPs_NESS_Sirrna}
\end{align}
In the absence of non-conservative forces, $\vek{A}(\vek{x},t)\equiv0$, we have $\phi(\vek{x},t)=\p(\vek{x},t)-\GG(t)$ (and $\Shk[\vxc]\equiv0$) and are left with $\Sirrna[\vxc]=\Stot[\vxc]$. \\
Taking it all together, in the presence of NESSs we have three non-negative EPs, the total EP 
\begin{align} \label{eq:EPs_NESS_Stot}
	\Stot[\vxc] = \Shk[\vxc] + \Sirrna[\vxc] \geq0
\end{align}
which is composed of the entropy $\Shk[\vxc]\geq0$ produced for keeping up the stationary current $\Jst(x,t)$ in the ``instantaneous'' NESS $\pst(x,t)$, and the entropy $\Sirrna[\vxc]\geq0$ produced when not being in the NESS as in a relaxation process or a driven process.\\

The splitting into $\Shk$ and $\Sirrna$ can also be written in terms of spatial averaged EPs,
\begin{subequations} \label{eq:EPR_3faces}
  \begin{align}
		\ddSna(t) &\dfn \int p(\vek{x},t)\,\bigg[\frac{\vek{J}(\vek{x},t)}{p(\vek{x},t)}\-\frac{\vJst(\vek{x},t)}{\pst(\vek{x},t)}\bigg]\,\mat{D}^{-\!1}(\vek{x},t)\,\bigg[\frac{\vek{J}(\vek{x},t)}{p(\vek{x},t)}\-\frac{\vJst(\vek{x},t)}{\pst(\vek{x},t)}\bigg]\di^nx \nn
		&= -\int \dot p(\vek{x},t)\,\ln\frac{p(\vek{x},t)}{\pst(\vek{x},t)}\di^nx \label{eq:EPR_3faces_Sna} \\
		\ddSa(t) &\dfn \int p(\vek{x},t)\,\frac{\vJst(\vek{x},t)}{\pst(\vek{x},t)}\,\mat{D}^{-\!1}(\vek{x},t)\,\frac{\vJst(\vek{x},t)}{\pst(\vek{x},t)}\di^nx  \label{eq:EPR_3faces_Sa}\\
		\ddSi(t) &= \ddSna(t) + \ddSa(t) \label{eq:EPR_3faces_Si}
	\end{align}
\end{subequations}
where it can be shown using the chain rule and the FPE that the mixed term of the quadratic form in $\ddSna(t)$ vanishes,\remark{sumtab p.7 (alle ableitungen zu gradienten, und quadratische Form beibehalten. Geht dann durch Produktregel und FPE einsetzen.}
\begin{align}
	\int&p(\vek{x},t)\,\bigg[\frac{\vJst(\vek{x},t)}{\pst(\vek{x},t)}\bigg]\mat{D}^{-\!1}(\vek{x},t)\,\bigg[\frac{\vek{J}(\vek{x},t)}{p(\vek{x},t)}-\frac{\vJst(\vek{x},t)}{\pst(\vek{x},t)}\bigg]\di^nx \nn
	&= -\int\vek{\nabla}\bigg[p(\vek{x},t)\,\frac{\vJst(\vek{x},t)}{\pst(\vek{x},t)}\bigg]\di^nx = 0 \nonumber
\end{align}
and therefore the splitting (\ref{eq:EPR_3faces_Si}) holds.\\
Due to the quadratic forms it is apparent that $\ddSna(t)\geq0$ and $\ddSa(t)\geq0$ for all times $t$. The EP $\ddSna(t)$ is the analog of $\Sirrna[\vxc]$, and $\ddSa(t)$ the analog of $\Shk[\vxc]$. The index of $\Sna$ stands for ``non-adiabatic'', and in $\Sa$ for ``adiabatic'', where ``adiabatic'' refers to the quasi-steady limit of process control. In that sense, $\ddSna(t)$ is the irreversible entropy produced despite of adiabatic process control, which is non-zero if non-conservative forces are involved \cite{Esposito2010}.\\
Instead of non-conservative forces giving rise to NESSs, also in the case of multiple heat bathes at different temperature we encounter NESSs. The simplest example is the heat transport from one heat bath to a second at lower temperature. In general, we can think of concurrent processes driven by different heat baths, each of which trying to impose their temperature on the system. This setting was analysed by Esposito and van den Broeck for continuous MPs \cite{VandenBroeck2010} and discontinuous MPs \cite{Esposito2010,Esposito2010a}.\\
 
The analysis of EP can also be rather analogously extended to discontinuous MPs, which we consider beyond the scope of this thesis and direct the interested reader to the comprehensive article by Harris and Schütz \cite{Harris2007} and part 6 of \cite{Seifert2012} for a brief introduction.

\subsection{Fluctuation theorems and irreversibility} \label{ss_td-interpration_FTs}
In the previous section, we mentioned the relation between entropy and irreversibility: If a transition from state A to state B produces entropy in finite time, no transition from B to A will be able to entirely consume this entropy, resulting in a net increase of {\it irreversible} entropy after completing the cycle.\remark{isoliere einen kolben. komprimiere kräftig, erwärmung, entropie wird produziert. expandiere, abkühling, entropie wird konsumiert, aber nicht vollständig, es bleibt netto ne temperaturerhöhung übrig.}\todo{reicht hier die forderung 'in finite time', also nicht quasi-statisch?} This irreversibly produced amount of entropy is a measure of the loss of useful energy, typically in the form of dissipated heat.\remark{andere beispiele: deformierung, chemische reaktionen, mischungsentropie... Recall that irreversible entropy production, being a process function, has to be distinguished from the entropy as state variable.}\\

In this section, we aim at further exploring the connection of entropy and irreversibility. This connection will be stated in terms of fluctuation theorems (FTs), a rather new finding in the framework of stochastic thermodynamics, which were first observed 1993 by Evans et al. for entropy fluctuations in a shear driven fluid in contact with a heat bath \cite{Evans1993,Evans1994}, and two years later proved to hold for a subclass of dissipative dynamical systems \cite{Gallavotti1995,Gallavotti1995b}. In the sequel, FTs have been intensively studied and proved for many other settings. We will focus on FTs for continuous MPs, and give a short account on their applications in the next section.\\
Recent review articles on the relation between fluctuation theorems and irreversibility haven been written by van den Broeck \cite{VandenBroeck2010b} and Jarzynski \cite{Jarzynski2011b}, and is also covered by Seifert's review article \cite{Seifert2012}.

\paragraph{Entropy and irreversibility} As already mentioned, we ask the question at what entropic cost the consequences of an individual trajectory $\xc$ can be undone. This is best understood by parametrising the time dependency of the dynamics in terms of a protocol $\kc$, i.e. $F\big(x;\k(\t)\big)$, $D\big(x;\k(\t)\big)$. The protocol $\k(\t)$ takes different values when time evolves from $t_0$ to $t$. To assess the probability of individual trajectories $\xc$, we employ the WPIs introduced in section \ref{ss_WPI}. There, we defined the path probability
\begin{align}
	P[\xc|x_0;\kc] = \eee{-S[\xc;\kc]}
\end{align}
in terms of the action functional
\begin{align}
	S[\xc;\kc] &= \int_{t_0}^{t} \frac{\big(\ddx_\t - F(x_\t;\k_\t)\big)^2}{4D(x_\t;\k_\t)} + J(x_\t;\k_\t) \di \t \\
	J(x,t) &= \tfrac{1}{2}F'(x,t) + \tfrac{1}{4}D''(x,t) \;.
\end{align}
Here, we use the Stratonovich convention in order to use ordinary calculus, cf. (\ref{eq:WPI_action}) or (\ref{eq:A5_overviewFP-FDE_WPI}) for $\a=1/2$.\\
To measure irreversibility, consider the time-reversal of the dynamics,\linebreak $\bt = t-\t+t_0$, $\dd\bt=-\dd\t$. Imagine we fix a starting point $x_0$ and observe a trajectory $\xc$, to which we can assign the path-probability $P[\xc|x_0;\kc]\Di\xc$. As a conjugate process, we define the time-reversed process by $\t\mapsto\bt$, i.e. use the protocol $\bk(\t)=\k(t\-\t\+t_0)$ instead of $\k(\t)$. The path-probability of the time-reversed dynamics then reads $P[\xc|\x_0;\bkc]\Di\xc$.\\
In general, for a reversible trajectory $\xc$, the path probability of the time-reversed trajectory $\bxc$ under the time-reversed dynamics should be equal to the observed trajectory $\xc$ in the original dynamics, $P[\bxc|\bx_0;\bkc]=P[\xc|x_0;\kc]$. For arbitrary trajectories, but in a reversible process, we will still have $P[\lla\bxc\rra|\bx_0;\bkc]=P[\lla\xc\rra|x_0;\kc]$ for the average realisation $\lla\xc\rra$. For trajectories in an irreversible process, however, the path probabilities $P[\bxc|\bx_0;\bkc]$ and $P[\xc|x_0;\kc]$ will in general be distinct, since it is the very nature of an irreversible process that the original state can not be retrieved upon reversal of process control. This suggests, in analogy to the Gibbs entropy, to consider as a measure for irreversibility
\begin{align}
	\ln\,\frac{P[\xc|x_0;\kc]}{P[\bxc|\bx_0;\bkc]} &= \SS[\bxc;\bkc] - \SS[\xc;\kc] \;,
\end{align}
that is, the difference of the respective actions.\\
To calculate this difference explicitly, we make use of the substitution $\bt = t-\t+t_0$ and rewrite
\begin{align}
	\SS[\bxc;\bkc] &= \int_{t_0}^{t} \frac{\big(\bddx_\t - F(\bx_\t;\bk_\t)\big)^2}{4D(\bx_\t;\bk_\t)} + J(\bx_\t;\bk_\t) \di \t \nn
	&= -\int_{t}^{t_0} \frac{\big(-\bddx_\bt - F(\bx_\bt;\bk_\bt)\big)^2}{4D(\bx_\bt;\bk_\bt)} + J(\bx_\bt;\bk_\bt) \di \bt \nn	
	&= \int_{t_0}^{t} \frac{\big(\ddx_\t + F(x_\t;\k_\t)\big)^2}{4D(x_\t;\k_\t)} + J(x_\t;\k_\t) \di \t	\;.
\end{align}
The Jacobian cancels and only the mixed term of the square remains,
\begin{align}
	\SS[\bxc;\bkc] - \SS[\xc;\kc] &= \int_{t_0}^{t} \frac{\big(\ddx_\t \+ F(x_\t;\k_\t)\big)^2 - \big(\ddx_\t \- F(x_\t;\k_\t)\big)^2}{4D(x_\t;\k_\t)} \di \t \nn
	&= \int_{t_0}^{t} \frac{\ddx_\t F(x_\t;\k_\t)}{D(x_\t;\k_\t)} \di \t \nonumber \\[7pt]
	&= \Sm[\xc] \;,
\end{align}
and we recover as the measure of irreversibility, on the level of individual trajectories, the EP in the medium, 
\begin{align} \label{eq:FTs_WPI_Sm}
	\Sm[\xc] = \ln\,\frac{P[\xc|x_0;\kc]}{P[\bxc|\bx_0;\bkc]} \;.
\end{align}
Any trajectory $\xc$ connecting initial and final point with an EP\linebreak $\Sm[\xc]>0$ is not as likely to be observed in the time-reversed process.\\
There seems, however, to be a flaw in that argumentation, since $\Sm$ also includes the reversible heat transfer from system to heat bath, implying that a realisation $\xc$ can be reversible despite $\Sm[\xc]=-\DSst\neq0$ ($\DSst$ is the reversible EP in the system, and consequently $-\DSst$ the EP in the heat bath). This flaw gets unravelled by noting that the stationary EP $\Sst$ is not a process functional but a state variable and as such only depends on initial and final points $x_0$ and $x_t$, whereas $\Sm[\xc]$ only accounts for the EP by connecting a given initial and final state. In other words, the path probability $P[\xc|x_0;\kc]$ does not account for the probability $p_0(x_0)$ of the initial value $x_0$, and the associated entropy, $-\ln p_0(x_0)$, can not enter our considerations.\\
Augmenting the conditioned path-probabilities with the initial distribution for the forward process, $p_0(x)$, and the solution of the FPE as the initial distribution for the reverse process, $p_t(x)\dfn p(x,t)$, we indeed recover the total, irreversible EP
\begin{align} 
	\ln\,\frac{p_0(x_0)\,P[\xc|x_0;\kc]}{p_t(\bx_0)\,P[\bxc|\bx_0;\bkc]} &= -\ln\,\frac{p_t(x_t)}{p_0(x_0)} + \Sm[\xc] \nn
	&= \Ds + \Sm[\xc] = \Stot[\xc] \label{eq:FTs_WPI_Stot}
\end{align}
derived in the context of the corresponding FPE (\ref{eq:EP_Stot}).\\
We readily convince ourselves that the above augmentation with the initial distributions indeed resolves the flaw discussed in view of (\ref{eq:FTs_WPI_Sm})\remark{, that is for a reversible realisation $\xc$ it is $\Sm[\xc]=-\Dseq\neq0$}: For the case that the process starts in a steady state and for quasi-steady process control, we have $p_0(x)=\pst(x,t_0)$ and $p_t(x)=\pst(x,t)$ and therefore $\Ds=\Dseq$ and $\Stot[\xc]=\Dseq-\Dseq=0$\remark{(natürlich nur, wenn $\xc$ auch wirklich ne reversible realisierung ist, also aus GG-Lage startet und in GG-Lage endet)}. For any irreversible realisation $\xc$, we will have $\Stot[\xc]\neq0$ which can be due to start off equilibrium (relaxation process) or non-equilibrium process control (driven process) or both. In this way, the path integral formulation and the notion of time-reversal elucidates the intimate relation between entropy and irreversibility.\\

The EP $\Stot[\xc]$ features a time-reversal symmetry. This symmetry becomes evident when using (\ref{eq:FTs_WPI_Stot}) to write $\Stot[\xc]$ in the form
\begin{align}
	\Stot[\xc] &= \SS[\bxc;\bkc]\-\ln p_t(x_t) - \big(\SS[\xc;\kc]\-\ln p_0(x_0)\big)
\end{align}
and then express the EP of the reversed trajectory under the time-reversed dynamics, $\bStot[\bxc]$, in terms of the original EP,
\begin{align}		
	\bStot[\bxc] &= \SS[\bbxc;\bbkc]\-\ln \bar p_t(\bx_t) - \big(\SS[\bxc;\bkc]\-\ln \bar p_0(\bx_0)\big) \nn
	&= \SS[\xc;\kc]\-\ln p_0(x_0) - \big(\SS[\bxc;\bkc]\-\ln p_t(x_t)\big) \nn
	&= -\Stot[\xc] \label{eq:FTs_Stot_sym}
\end{align}
\remark{hab hier die notation für time-reversal auch einfach für die initial distributions angewandt, ist ja von definition $\bt = t-\t+t_0$ eigentlich auch klar.}\\
Building on (\ref{eq:FTs_Stot_sym}) and (\ref{eq:FTs_WPI_Stot}), we can express the forward path-probability in terms of the reversed path-probability
\begin{align} \label{eq:detbal_traj}
	P[\xc|x_0] = \frac{p_t(x_t)}{p_0(x_0)}\,\bP[\bxc|x_t]\;\eee{-\bStot[\bxc]}
\end{align}
where we have written for short $\bP[\bxc|x_t]=P[\bxc|x_t;\bkc]$. The above relation can be interpreted as the detailed balance condition on the level of single realisations, broken if the realisation irreversibly produces entropy, i.e. $\Stot[\xc]>0$.\\
This kind of observation is the starting point to the derivation of so-called {\it fluctuation theorems} (FTs). We use (\ref{eq:WPI_pdf_int-val}) to express the probability density of the integral value $\Stot=\Stot[\xc]$ in terms of a WPI and substitute (\ref{eq:detbal_traj}) to derive
\begin{align}
	p(\Stot) &= \int\< \dd x_t \int\< \dd x_0 p_0(x_0) \<\<\int\limits_{(x_0,t_0)}^{(x_t,t)}\<\<\< \Di\xc\,P[\xc|x_0]\,\d(\Stot\!-\Stot[\xc]) \nn
	&\<\<\<\<\<= \int\< \dd x_t \int\< \dd x_0 \<\<\int\limits_{(x_0,t_0)}^{(x_t,t)}\<\<\< \Di\xc\,\eee{-\bStot[\bxc]}\;p_t(x_t)\bP[\bxc|x_t]\,\d(\Stot\+\bStot[\bxc]) \nn
	&\<\<\<\<\<= \eee{\Stot}\int\< \dd x_0 \int\< \dd x_t p_t(x_t) \<\<\int\limits_{(x_t,t)}^{(x_0,t_0)}\<\<\< \Di\bxc\,\bP[\bxc|x_t]\,\d(\Stot\+\bStot[\bxc]) \nn
	&\<\<\<\<\<= \eee{\Stot}\,\bp(-\Stot) \label{eq:FTs_dFT_Stot_deriv}
\end{align}
For the third line we used the $\d$-function to write $\ee{\Stot}$ before the integral, and we performed the variable transformation $\xc\mapsto\bxc$ of which the Jacobian is the identity as the transformation merely corresponds to a reversal of integration order (cf. the discrete definition of a WPI in (\ref{eq:WPI_discr})).\\
The resulting relation,
\begin{align}
	\frac{p(\Stot)}{\bar p(-\Stot)} = \eee{\Stot} \;, \label{eq:FTs_dFT_Stot}
\end{align}
is known as a {\it detailed fluctuation theorem} (dFT), in this case for the irreversible EP $\Stot$. The dFT relates the probability of observing a {\it production} of entropy to the probability of observing the {\it consumption} of entropy in the time-reversed process. The statement is that the odds of a consumption of entropy $\Stot$ in the time-reversed process compared to the probability of a production of entropy $\Stot$ in the forward process goes down exponentially with the entropy $\Stot$ itself. In contrast to addressing the concept of irreversibility in terms of a WPI, as done in (\ref{eq:FTs_WPI_Stot}), the dFT is a probabilistic formulation of irreversibility in terms of the PDF $p(\Stot)$.\\
This probabilistic formulation can be further reduced by integration of (\ref{eq:FTs_dFT_Stot}),
\begin{align}
	\int \eee{-\Stot}\,p(\Stot)\di\Stot = \lla\eee{-\Stot}\rra = 1 \;, \label{eq:FTs_iFT_Stot}
\end{align} 
which is known as an {\it integral} fluctuation theorem (iFT). Along similar lines as for the derivation of the dFT in (\ref{eq:FTs_dFT_Stot}), this iFT can also be derived in terms of a WPI by using the normalisation condition of the WPI as introduced in (\ref{eq:WPI_norm}). Note that the iFT goes without the notion of time-reversal, implying that the mere existence of time-reversal rather than its explicit form is enough for the iFT to hold.\\
The iFT and dFT for the total EP $\Stot$ was derived from the path integral representation by Seifert \cite{Seifert2005}.\\

In the previous chapter, we postponed the proof that upon ensemble average it is $\lla\Stot[\xc]\rra\geq0$. This proof follows now immediately by applying the inequality $\lla\exp\,x\rra\geq\exp\lla x \rra$ to the iFT
\begin{align}
	1 &= \lla\eee{-\Stot[\xc]}\rra \geq \eee{-\lla\Stot[\xc]\rra} \nn
	&\Ra \lla\Stot\rra \geq 0 \;. \label{eq:FTs_secondlaw}
\end{align}
Remarkably, considering that the second law is an inequality and the iFT an equality, we reproduced the second law from the iFT. Or, in other words, in equilibrium thermodynamics we have $\lla\Stot[\xc]\rra=0$, whereas in non-equilibrium thermodynamics we find $\lla\exp\big[-\Stot[\xc]\big]\rra=1$. In that sense, the iFT refines the second law. Note that in order to observe the iFT to hold true for a limited number of realisations, the exponential average, $\lla\exp\big[-\Stot[\xc]\big]\rra$, requires a sufficient portion of realisations $\Stot[\xc]\!<\!0$.\\
Having retrieved the second law of classical thermodynamics from the iFT in the framework of stochastic thermodynamics, we discuss the scope of applicability of an iFT. To do so, we restate that stochastic thermodynamics differs from classical thermodynamics with respect to two defining properties: {\it Nanoscopic} systems, implying an inherent stochasticity, and {\it non-equilibrium}, brought about by driving the process or by preparing an initial state off equilibrium. Accordingly, we distinguish the following situations.\\
i) Nanoscopic, non-equilibrium thermodynamics, being the scope of stochastic thermodynamics. The realisations $\xc$ are stochastic and due to this stochasticity, the total EP $\Stot[\xc]$ is not restricted to non-negative values, and the exponential in the iFT, $\lla\exp\big[-\Stot[\xc]\big]\rra$, approaches for a reasonable number of realisations its theoretical value of one. Due to the non-equilibrium conditions, the total EP will on average be strictly positive, $\lla\Stot[\xc]\rra>0$.\\
ii) Nanoscopic, equilibrium thermodynamics. The realisations $\xc$ remain stochastic and the resulting realisations of $\Stot[\xc]$ still may be negative. In the ensemble average, the total EP will be zero, $\lla\Stot[\xc]\rra=0$, implying that negative and positive realisations of $\Stot[\xc]$ are in exact balance and the validity of the iFT is observed for very few realisations.\\
iii) Macroscopic, equilibrium thermodynamics. Instead of individual trajectories, the trajectories $\xc$ represent the evolution of collective degrees of freedom of macroscopic systems which do not display a notable stochasticity. In other words, all realisations $\xc$ are practically identical. Due to equilibrium conditions, the EP will be zero, $\Stot[\xc]\equiv0$, and the iFT is satisfied trivially.\\
iv) Macroscopic, non-equilibrium thermodynamics. The realisations $\xc$ remain practically identical, but due to the non-equilibrium condition, the total EP is strictly positive, $\Stot[\xc]>0$. The iFT still holds in principle, but needs an infinite number of realisations (thermodynamic limit) which also includes the practically impossible realisations with negative EP.\\
We will discuss in the next section that the trade-off between non-equilibrium and stochasticity is vital for the application of FTs.\remark{weit im NEQ-regime brauch ich auch starke stochasticity.}\\

\paragraph{General methodology} The procedure determining the dFT in (\ref{eq:FTs_dFT_Stot}) is a special case of a general methodology to determine a variety of FTs. To allow for a better insight into that procedure\remark{insbesondere auch die genaue instruktion um dFT zu testen}, and to derive two more important dFTs, we present a more abstract derivation of a dFT. In this derivation, the key point is to devise a conjugate dynamics from the original dynamics, and to define a transformation to be applied to trajectories $\xc$. The conjugate dynamics gives rise to a modified path-probability, which we denote by $\tP[\xc|x_0]$, and also includes an altered initial distribution, $\tp_0(x)$. For the transformed trajectories we will write $\sxc$, evolving in time $\st$.\\
In the derivation of the dFT for $\Stot$ in (\ref{eq:FTs_dFT_Stot}), the conjugate dynamics was the time-reversed dynamics which originated from reversing the protocol, $\tP[\xc|x_0]=P[\xc|x_0;\bkc]$, and the initial distribution, $\tp_0(x)=p_t(x)$, and likewise for the transformation of trajectories, $\sxc=\bxc$. \\
We now turn to the more abstract level and define the observable
\begin{align} \label{eq:FTs_def_Y}
	\YY[\xc] \dfn \ln\,\frac{P[\xc|x_0]\,p_0(x_0)}{\tP[\sxc|\sx_0]\tp_0(\sx_0)} \;.
\end{align}
In most cases, the transformation of the dynamics is an involution (i.e. its own inverse), and the transformation applied to the trajectories is either the identity or the time-reversal (which both are involutions). We therefore restrict ourselves to these cases and find, analogous to (\ref{eq:FTs_Stot_sym}), the symmetry
\begin{align} \label{eq:FTs_Y_sym}
	\tY[\sxc] = -\YY[\xc] \;,
\end{align}
which enables us to express $P[\xc|x_0]$ in terms of the transformed trajectories in the conjugate dynamics,
\begin{align} \label{eq:FTs_Y_tildeP}
	P[\xc|x_0] = \frac{\tp_0(\sx_0)}{p_0(x_0)}\tP[\sxc|\sx_t]\;\ee{-\tY[\sxc]}
\end{align}
and to obtain the dFT for $Y$ along similar lines as in (\ref{eq:FTs_dFT_Stot_deriv}),
\begin{align}
	p(Y=\YY[\xc]) &= \int\< \dd x_t \int\< \dd x_0\,p_0(x_0) \<\<\int\limits_{(x_0,t_0)}^{(x_t,t)}\<\<\< \Di\xc\,P[\xc|x_0]\,\d(Y\!-\YY[\xc]) \nn 
	&= \eee{Y}\!\!\int\< \dd \sx_t \int\< \dd \sx_0\,\tp_0(\sx_0) \<\<\int\limits_{(\sx_0,\st_0)}^{(\sx_t,\st)}\<\<\< \Di\sxc\,\tP[\sxc|\sx_0]\,\d(Y\+\tY[\sxc]) \nn
	&\<\<\<\<\<= \eee{Y}\,\tp\big(-Y=\tY[\sxc]\big) \;. \label{eq:FTs_dFT_Y_deriv}
\end{align}
Note that in the second line we did not only use (\ref{eq:FTs_Y_tildeP}) to substitute $P[\xc|x_0]$ but also used the symmetry (\ref{eq:FTs_Y_sym}) to substitute $\YY[\xc]$ in the $\d$-function, with the result that the complete integrand is written as a function of the integration variable $\sxc$. The discussion of FTs on this abstract level is pursued in greater detail by Verley and Lacoste in their article \cite{Verley2012a}, a unification of practically all FTs based on this abstract formulation was achieved by Seifert in his overview article \cite{Seifert2012}.\\ \renewcommand{\c}{_\mr{c}}
From the path integrals in (\ref{eq:FTs_dFT_Y_deriv}) we can also read off the general prescription to apply an iFT or dFT to a simulation or experiment:\\
To determine $p(Y)=p(Y=\YY[\xc])$, we need a set of realisations $\{\xc\}$ of the stochastic process fixed by the coefficients $F(x,t)$ and $D(x,t)$ and the initial distribution $p_0(x)$ which define $P[\xc]$. The set  $\{\xc\}$ can be obtained by simulating the corresponding LE using, e.g., the integration scheme (\ref{eq:A3_SDE_num_alpha}), or, the other way around, find $p_0(x)$, $F(x,t)$ and $D(x,t)$ that describes an already existing set of $\{\xc\}$ we got, e.g., from an experiment. To each realisation $\xc$, we determine $Y=\YY[\xc]$ from (\ref{eq:FTs_def_Y}). From the resulting set $\{Y\}$ we can readily verify the iFT
\begin{align} \label{eq:FTs_iFT_Y}
	\lla\eee{-Y}\rra = 1 \;.
\end{align}
If we have furthermore access to the set of realisations $\{\sxc\}$ that obey the conjugate dynamics as used in the second path integral in (\ref{eq:FTs_dFT_Y_deriv}), we can obtain from the functional $\tY[\sxc]$ the set of values $\{Y\c\}$, where the index $c$ declares that the $Y_c$ were determined from the conjugate dynamics, and are be distinguished from the set $\{Y\}$ obtained from the original dynamics. From the set $\{Y\}$ we produce a histogram to approximate $p(Y)$, and likewise, we build from $\{Y\c\}$ a histogram to approximate $\tp(Y)$. Making use of a suitable interpolation method to plot $\ln\big(p(Y)/\tp(-Y)\big)$ versus $Y$, we will, according to the dFT
\begin{align} \label{eq:FTs_dFT_Y}
	\frac{p(Y)}{\tp(-Y)} = \eee{Y}
\end{align}
get a straight line of slope one. \remark{die bezeichnung der rückwärtstrajektorien mit $\sx$ soll nicht heißen, dass diese transformiert sind. dadurch dass sie im zweiten pfad-integral zusammen mit den konjugierten funktionalen als integrationsvariable stehen, hab ich jetzt einfach die bezeichnung $\sx$ für die aus der konjugierten dynamik entsprungenen trajektorien übernommen.}\remark{diese detalierte auflistung der vorgehensweise um das iFT zu beobachten mag für zB andreas klar sein. aber wenn man auf diesem abstrakten level das mal machen möchte, insbesondere für andere konjugierte dynamiken wie für mein $\e$FT, kam ich zumindest ins schleudern.}\remark{Derive along similar lines as for $\Stot$ in (\ref{eq:FTs_dFT_Stot_deriv}) the corresponding dFT 
\begin{align}
	p(Y=\YY[\xc]) &= \int\< \dd x_t \int\< \dd x_0 p_0(x_0) \<\<\int\limits_{(x_0,t_0)}^{(x_t,t)}\<\<\< \Di\xc\,P[\xc|x_0]\,\d(Y\!-\YY[\xc]) \nn
	&\<\<\<\<\<= \int\< \dd x_t \int\< \dd x_0 \<\<\int\limits_{(x_0,t_0)}^{(x_t,t)}\<\<\< \Di\xc\,\eee{\YY[\xc]}\;\tp_0(\sx_0)\tP[\sxc|\sx_0]\,\d(Y\!-\YY[\xc]) \nn
	&\<\<\<\<\<= \eee{Y}\int\< \dd \sx_t \int\< \dd \sx_0 \tp_0(\sx_0) \<\<\int\limits_{(\sx_0,t_0)}^{(\sx_t,t)}\<\<\< \Di\sxc\,\tP[\sxc|\sx_0]\,\d(Y\!-\YY\big[x\big(\sxc\big)\big]) \nn
	&\<\<\<\<\<= \eee{Y}\,\tp(Y=\YY\big[x\big(\sxc\big)\big]) \label{eq:FTs_dFT_Stot_deriv}
\end{align}
We see, in opposition to $Y$ in the original dynamics, that we can not derive $Y$ in the conjugate dynamics from the conjugate functional $\tY$, therefore weak dFT.}\\

\paragraph{Non-equilibrium steady states} Having established the general formalism to set up observables that obey a dFT, we can readily discuss two more important FTs. To this end, recall that in the case of more than one degree of freedom, i.e. $\vek{X}\in\mathbb{R}^n$ with $n\geq2$, non-conservative forces that do not arise from a scalar potential entail non-equilibrium steady states (NESSs). The distinguishing property of a NESS is that its probability current $\Jst(\vek{x})$ does not vanish, instead it is divergence free, $\vek{\nabla}\Jst(\vek{x})=0$.\\
In (\ref{eq:highdim_Fcnc}), we divided the drift velocity into a conservative part deriving from the scalar potential $\p(\vek{x},t)$ and an additional non-conservative force arising from the vector-field $\vek{A}(\vek{x},t)$,
\begin{align} \label{eq:highdim_Fcnc}
	\vek{F}(\vek{x},t) =  \mat{D}(\vek{x},t)\,\big[-\vek{\nabla}\p(\vek{x},t) + \vek{A}(\vek{x},t)\big] \;.
\end{align}
In view of the Fokker-Planck operator in the form (\ref{eq:FPE_OP_st}),
\begin{align}
	\hLLFP(\vek{x},t) = -\vek{\nabla}\,\Big[\ee{\phi(\vek{x},t)}\vJst(\vek{x},t) - \mat{D}(\vek{x},t)\ee{-\phi(\vek{x},t)}\vek{\nabla}\ee{\phi(\vek{x},t)}\Big] \,,
\end{align} 
we found that the triple $\{\phi,\,\mat{D},\,\vJst\}$ fixes the dynamics of the process, and by varying $\vJst$ at fixed $\p$ and $\mat{D}$, we can define a family of dynamics with identical $\phi$ and $\mat{D}$ but distinct currents, where the non-equilibrium potential $\phi(x,t)$ fixes the stationary distribution $\pst(\vek{x},t)=\exp\big(-\phi(\vek{x},t)\big)$.\\
A particular meaningful modification of the original dynamics is found by reversing the current, $\{\phi,\mat{D},-\vJst \}$. In this dynamics, particles will run the loops in opposite direction, as is evident from equation (\ref{eq:highdim_FPEst}) since the velocity $\lla\dot{\vek{x}}|\vek{x},t\rra$ acquires the opposite sign. In the current reversed dynamics the Fokker-Planck operator now reads
\begin{align} \label{eq:FTs_FPO_dagger}
	\hLLFPd(\vek{x},t) = \vek{\nabla}\,\Big[\ee{\phi(\vek{x},t)}\vJst(\vek{x},t) + \mat{D}(\vek{x},t)\ee{-\phi(\vek{x},t)}\vek{\nabla}\ee{\phi(\vek{x},t)}\Big]
\end{align}
or, equivalently, instead of $\vek{F}(\vek{x},t)$ from (\ref{eq:highdim_Fcnc}), we may use the current reversed drift velocity
\begin{align} \label{eq:FTs_F_dagger}
	\vek{F}^\dagger(\vek{x},t) &= -\ee{\phi(\vek{x},t)}\vJst(\vek{x},t)-\mat{D}(\vek{x},t)\vek{\nabla}\phi(\vek{x},t) \nn
	&= - \big(\vek{F}(\vek{x},t) + 2\mat{D}(\vek{x},t)\vek{\nabla}\phi(\vek{x},t)\big) \;.
\end{align}
At this point, it suggests itself to take the current reversed dynamics as the conjugate dynamics and consider the observable 
\begin{align} \label{eq:FTs_Y_dagger}
	\YY[\vxc] = \ln\,\frac{P[\vxc|\vek{x}_0]\,p_0(\vek{x}_0)}{P^\dagger[\vxc|\vek{x}_0]p_0(\vek{x}_0)}
\end{align}
Note that current reversal does not affect the initial distributions. For the transformation of trajectories we chose the identity, as the system is in the same steady state for both dynamics (using the time reversed trajectories in the denominator would in fact result into $\YY[\vxc]\equiv0$ for all $\vxc$).\\
Verifying that current reversal is an involution,
\begin{align}
	\hLLFPdd(\vek{x},t) &= \vek{\nabla}\,\Big[\ee{\phi(\vek{x},t)}\big(-\vJst(\vek{x},t)\big) + \mat{D}(\vek{x},t)\ee{-\phi(\vek{x},t)}\vek{\nabla}\ee{\phi(\vek{x},t)}\Big] \nn
	&= -\vek{\nabla}\,\Big[\ee{\phi(\vek{x},t)}\vJst(\vek{x},t) - \mat{D}(\vek{x},t)\ee{-\phi(\vek{x},t)}\vek{\nabla}\ee{\phi(\vek{x},t)}\Big] \nn
	&= \hLLFP(\vek{x},t) \,
\end{align}
is the only requirement needed to state that $Y$ obeys the FTs (\ref{eq:FTs_iFT_Y}) and (\ref{eq:FTs_dFT_Y}). The interesting bit is to see whether $\YY[\vxc]$ as defined in (\ref{eq:FTs_Y_dagger}) has a physical meaning. Carrying out the necessary calculations\footnote{Details can be found in \cite{ChernyaChertkoJarzyn06JoSMTaE} along with the multidimensional formulation of WPIs.}, we indeed end up with the housekeeping heat from (\ref{eq:EPs_NESS_Shk}),
\begin{align}
	\YY[\vxc] &= \int_{t_0}^{t} \vek{\ddx}(\t)\,\mat{D}^{-\!1}(\t)\lla\,\vek{\ddx}|\t\rra_\mr{st} \di \t \nn
	&= \Shk[\vxc] \;,
\end{align}
which is nothing else than the entropy produced irreversibly while keeping the NESS up. This makes perfect physical sense, as the statement is that the odds to observe the same realisation in the current reversed dynamics as in the original dynamics goes down exponentially with the entropy $\Shk[\vxc]$ produced by this realisation. It further confirms that $\Shk[\vxc]$ indeed obeys the second law $\lla\Shk[\vxc]\rra\geq0$ in the ensemble average. The iFT for the housekeeping heat was first derived by Speck and Seifert \cite{SpeckSeifert05JoPAMaG}.\\
Having this in mind, it may not be surprising that taking the time reversal as conjugate dynamics and reversing also the trajectories, we recover the EPs given by (\ref{eq:EPs_NESS_slitSm}), (\ref{eq:EPs_NESS}) and (\ref{eq:EPs_NESS_Stot}),
\begin{subequations} \label{eq:FTs_Y_bar}
	\begin{align}
		\ln\,\frac{P[\vxc|\vek{x}_0]}{\bP[\vbxc|\vek{\bx}_0]} &= \Sm[\vxc] = \Sex[\vxc] + \Shk[\vxc] \label{eq:FTs_multidim_Sm} \\
		\ln\,\frac{P[\vxc|\vek{x}_0]\,p_0(\vek{x}_0)}{\bP[\vbxc|\vek{\bx}_0]\,p_t(\vek{x}_t)} &= \Stot[\vxc] = \Sirrna[\vxc] + \Shk[\vxc] \label{eq:FTs_multidim_Stot}
	\end{align}
\end{subequations}
where $\Sm[\vxc]$ is the EP in the medium, $\Stot[\vxc]$ is the total, irreversible EP, and $\Sex[\vxc]$ and $\Sirrna[\vxc]$ are the excess entropies, being the respective equivalents to $\Sm[\vxc]$ and $\Stot[\vxc]$ for dynamics without non-conservative forces.\\
To retrieve also the irreversible EP $\Sirrna[\vxc]$, associated with the entropy produced when not being in a steady state (e.g. in a relaxation process and/or driven process), we combine both time reversal and current reversal for the conjugate dynamics and obtain
\begin{align}
	\ln\,\frac{P[\vxc|\vek{x}_0]\,p_0(\vek{x}_0)}{\bP^{\,\dagger}[\vbxc|\vek{\bx}_0]\,p_t(\vek{x}_t)} &= \bar\SS^{\,\dagger}[\vbxc] \- \ln \bp_t(\vek{\bx}_t) - \big(\SS[\vxc] \- \ln p_0(\vek{x}_0)\big) \nn
	&= \bar\SS^{\,\dagger}[\vbxc] \- \ln \bp_t(\vek{\bx}_t)- \big(\SS[\vxc] \- \ln p_0(\vek{x}_0)\big) \nn
	&\quad - \bar\SS[\vbxc] + \bar\SS[\vbxc] \nonumber\\[10pt]
	&= \bar\SS[\vbxc]\-\ln \bp_t(\vek{\bx}_t) - \big(\SS[\vxc] \- \ln p_0(\vek{x}_0)\big) \nn
	&\quad - \big(\bar\SS[\vbxc] -  \bar\SS^{\,\dagger}[\vbxc]\big) \nonumber\\[10pt]
	&= \Stot[\vxc] - \Shk[\vxc] = \Sirrna[\vxc] \label{eq:FTs_multidim_Sex_noneg} \;.
\end{align}
We hence have, for systems with non-conservative forces, three non-negative entropies that obey a FT, the total EP $\Stot$, the housekeeping EP $\Shk$ and the non-negative excess EP $\Sirrna$. The dFT for $\Sirrna$ is also known as Hatano-Sasa relation \cite{HatanoSasa01PRL}. A comprehensible presentation of how $\Shk$ and $\Sirrna$ arise from the path integral formulation was achieved by Chernyak, Chertkov and Jarzynski \cite{ChernyaChertkoJarzyn06JoSMTaE}. A derivation of the three FTs and the corresponding second laws in the framework of a master equation was achieved by Esposito and van den Broeck \cite{Esposito2010,Esposito2010a}.

\subsection{Applications} \label{ss_td-interpration_FTs_applications}
In this chapter we give a brief account on various forms of FTs, their applications, and arising difficulties.\\

\paragraph{Steady-state FT}
The FT that was first observed by Evans et al. \cite{Evans1993} for entropy fluctuations in simulations of a two-dimensional shear-driven fluid belongs to the class of steady-state FTs which hold in a NESS for a fixed value of the protocol $\k$. Under certain chaotic assumptions, Gallavotti and Cohen proved that steady-state FTs hold in dissipative dynamical systems \cite{Gallavotti1995,Gallavotti1995b}.\\
In a NESS, reversal of time does not affect the dynamics, we therefore have the same path probability for the original and the time-reversed dynamics, and consequently, the same PDF for the observable that obeys the FT. For the total EP $\Stot=\Sirrna + \Shk$, the resulting dFT follows by combining (\ref{eq:FTs_dFT_Y}) and (\ref{eq:FTs_multidim_Stot}) as
\begin{align}
	\frac{p(\Stot)}{p(-\Stot)} = \eee{\Stot}
\end{align}
which is known as the Gallavotti-Cohen relation.\\
The Gallavotti-Cohen relation can conveniently be applied to experimental measurements as it only involves the knowledge of the PDF $p(\Stot)$.\\
In early works involving the Gallavotti-Cohen relation, the EP in the medium, $\Sm = \Sex + \Shk$ from (\ref{eq:FTs_multidim_Sm}), was considered instead of $\Stot$. For large time intervals $t-t_0$, the Gallavotti-Cohen relation holds approximately for $\Sm$, and asymptotically, that is in the limit $t-t_0\to\infty$, even exact. This asymptotic validity can be attributed to the divergent property of $\Shk$, as, for large times, the contribution of $\Shk$ to $\Sm$ dominates over the contribution from $\Sex$, such that the error made by using $\Sex$ instead of $\Sirrna$ becomes negligible. Taking it to extremes, it is sufficient to measure in an experiment a quantity $Y$ that is somehow related to $\Shk$, such that the influence of $\Shk$ prevails for large times, and then check whether $\ln\big(p(Y)/p(Y-)\big)$ takes a straight line for large measuring times. Examining the asymptotic validity of the Gallavotti-Cohen relation for entropy related quantities is a popular course of action to judge whether the system under consideration does in principle hold the symmetry of a dFT.\\

\begin{figure}[t] 
  \subfloat[][\figsubtxt{Work distribution}]{\label{sf:sun_W}
	\includegraphics[width=0.45\textwidth]{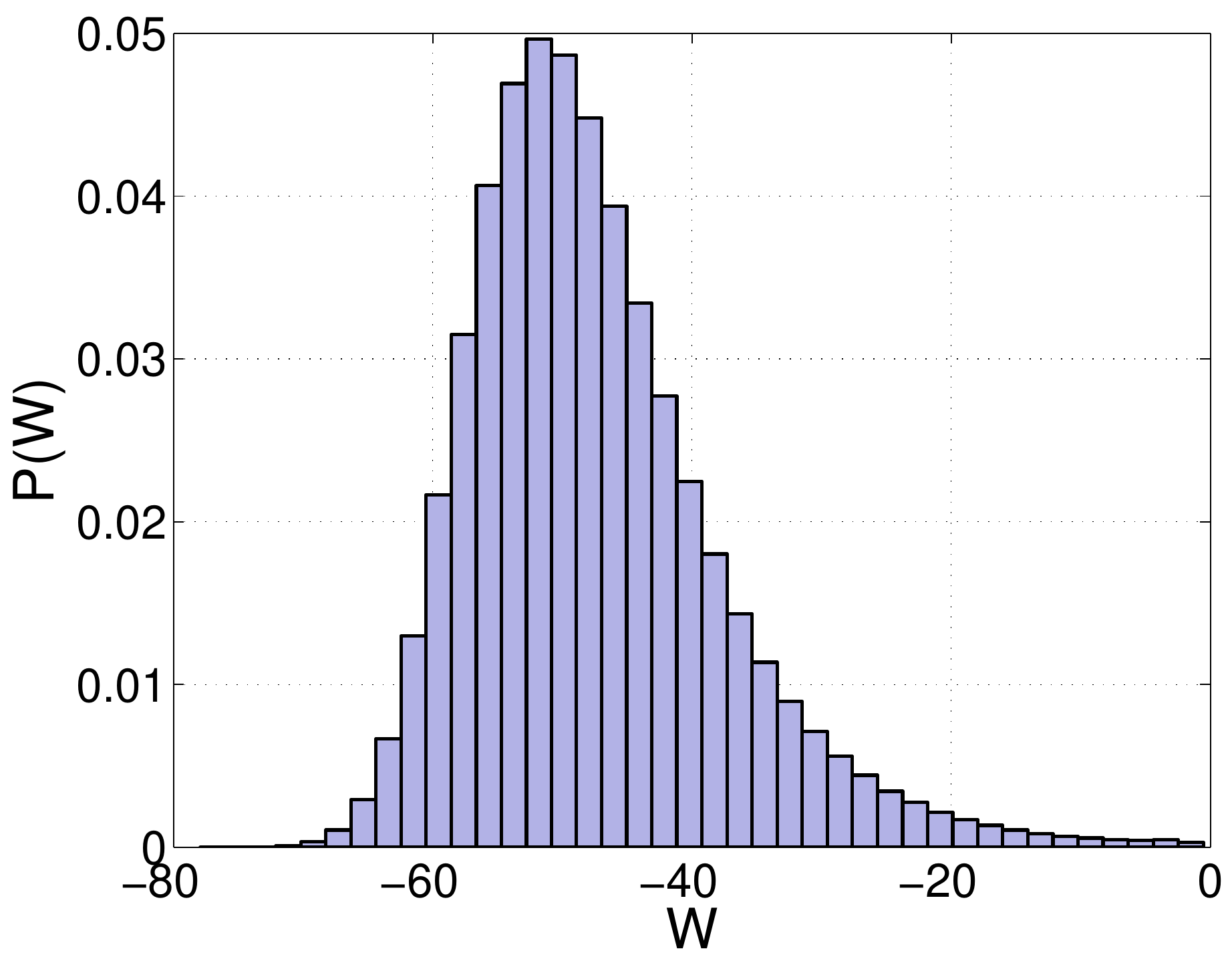}}
  \subfloat[][\figsubtxt{Integrand of exponential average}]{\label{sf:sun_W_exp}
	\includegraphics[trim=0mm 0mm 0mm 10mm,clip,width=0.472\textwidth]{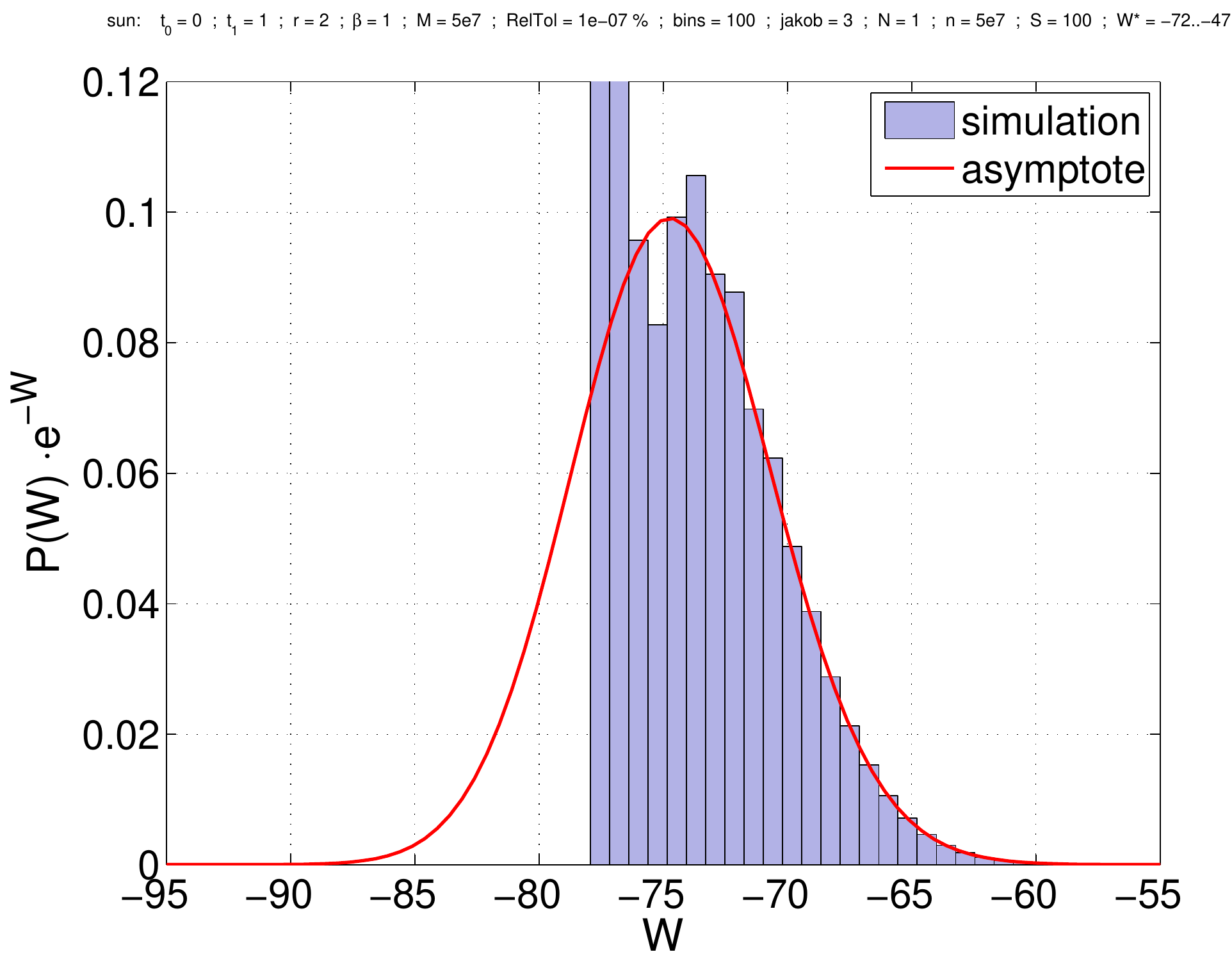}}
  \caption{\label{ff:sun}\figtxt{Work distribution of a driven bistable system. The work values were obtained by simulating the LE with a time-dependent potential $V(x,t)$ that has initially a single minimum and becomes a double-well potential as time evolves, cf. (4.17) of \cite{Nickelsen2011} reprinted in section \ref{ss_asymp_application_toy}. A number of $5\cdot10^7$ work values have been sampled in a time interval $t_0=0,\,t=1$ and for $\b=1$. The resulting histogram of work values is shown in (a), the biased distribution that is needed to be integrated to obtain the exponential average (\ref{eq:expavg}) in the Jarzynski relation (\ref{eq:JR}) is shown in (b). The red line is an asymptotic solution obtained in [1] which will be discussed in the next chapter, in particular in section \ref{ss_asymp_application_toy}.}}
\end{figure}

\begin{figure}[t] 
  \subfloat[][\figsubtxt{Original and exponential biased PDF}]{\label{sf:expavg_PDFs}
	\includegraphics[width=0.48\textwidth]{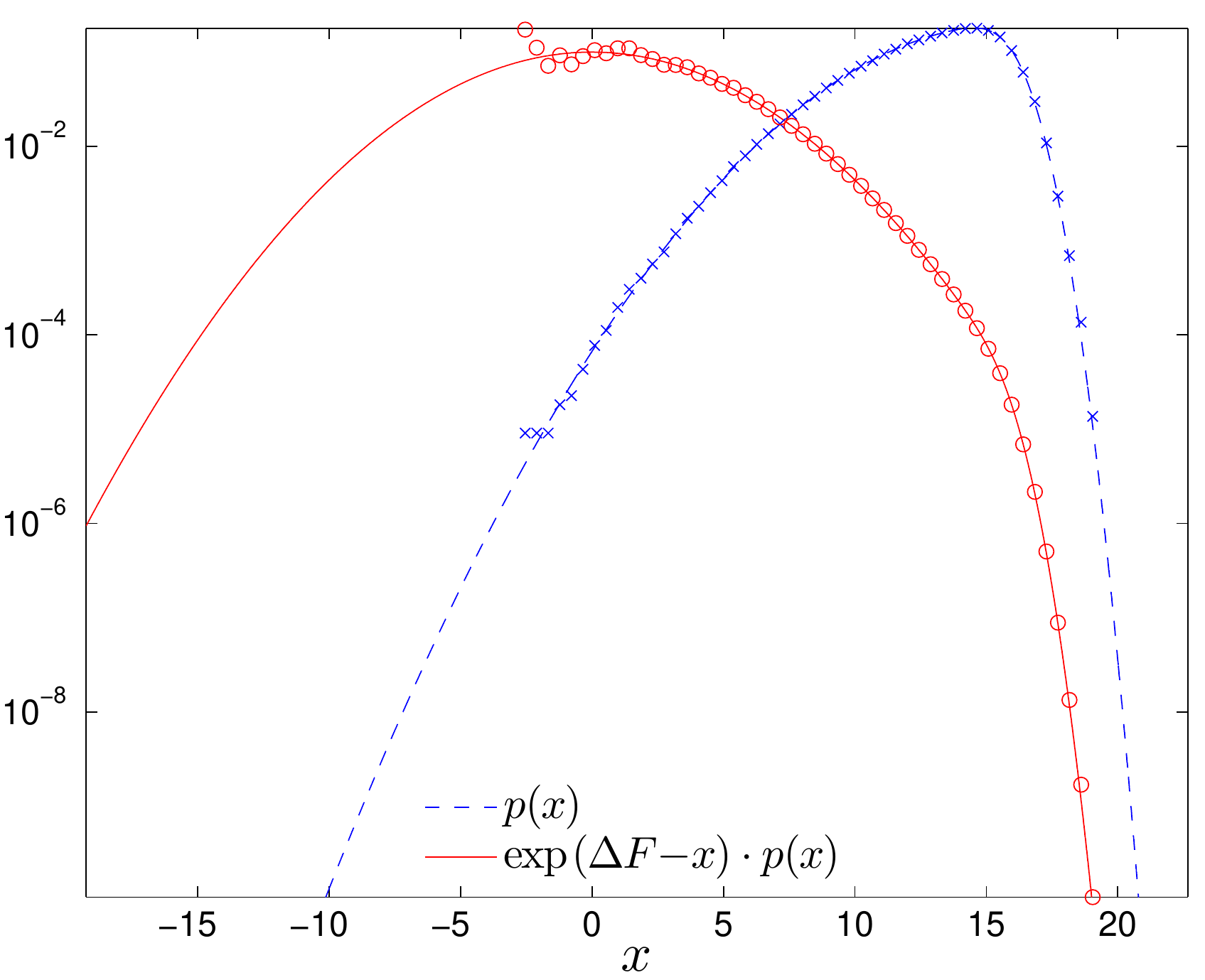}}
  \subfloat[][\figsubtxt{Running averages}]{\label{sf:expavg_AVGs}
	\includegraphics[width=0.46\textwidth]{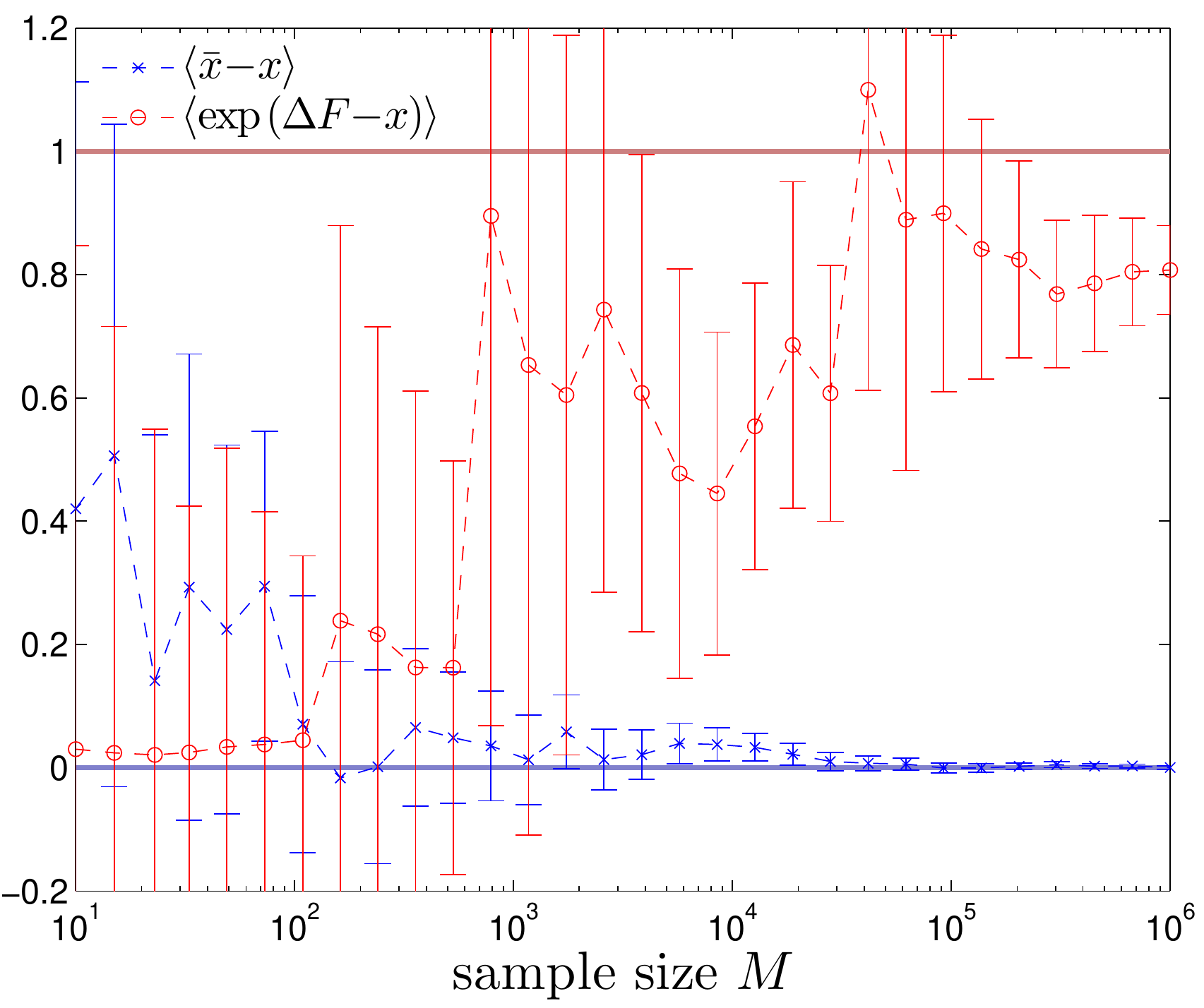}}
  \caption{\label{ff:expavg}\figtxt{Illustration of the difficulties of exponential averages compared to linear averages. A number of $10^6$ samples have been drawn from an artificial distribution $p(x)$, depicted in (a) as a blue line together with the histogram resulting from the sampled values as symbols. The quantity $\D F$ is analytically determined such that the exponentially weighted distribution shown in red is normalised. The running averages for the deviation of the arithmetic average $\lla x \rra$ to the exact mean $\bar x$ and the exponential average that should give one due to the choice of $\D F$ are shown in (b) for an increasing number $M$ of considered samples. The error bars are the corresponding standard errors. Note that for the linear average the true value is practically for all sample sizes within the error margin, whereas the convergence of the exponential average to its true value is hard to judge without knowing the true value.}}
\end{figure}

\paragraph{Estimation of $\bs{\DF}$}
A widely used variant of the FTs for $\Stot$ are so called work relations. The first work relation was proved by Jarzynski \cite{Jarzynski97PRL,Jarzynski97PRE}, and is known as the Jarzynski relation. In our formulation, Jarzynski used the alternative form $\Stot\eq R\-\DG$ of the total EP, which we obtained in (\ref{eq:EP_Stot_R_SS}) under the assumption that the initial state $x_0$ of the process is the steady state at initial time $t_0$.\footnote{In fact, we assumed that also the final state is the corresponding steady state, but the Jarzynski relation holds for arbitrary final states, since in the derivation (\ref{eq:FTs_iFT_Stot}) of the iFT for $\Stot$ only the initial distribution enters.}\remark{die annahme dass start und ende stationär sind brauch ich nur um $\DG$ zu definieren, die explizite wahl von $\st(x,t)$ am ende des prozesses geht dann aber in die berechnung des iFTs gar nicht mehr ein.} The quantities $R$ and $\DG$ are the entropic equivalents of work $W$ and free energy difference $\DF$. In the usual thermodynamic setting, the diffusion coefficient follows from the Einstein relation (\ref{eq:D_einstein}), $D\equiv \G/\b$, we therefore have $R=\b W$ and $\DG=\b\DF$ and the FT for $\Stot$ becomes the Jarzynski relation
\begin{align} \label{eq:JR}
	\lla\eee{-\b W}\rra = \eee{-\b\DF} \;.
\end{align}
A remarkable feature of the Jarzynski relation is that it relates work values $W$, measured in an arbitrary {\it non-equilibrium} transition from state A to state B, with the {\it equilibrium} free energy difference $\DF=\FF_\mr{B}-\FF_\mr{A}$ between these two states. Since the derivation of the iFT for $\Stot$, (\ref{eq:FTs_iFT_Stot}), holds for arbitrary final distributions $p(x,t)$, the process needs only to be prepared to start from equilibrium initially, whereas the sampled trajectories giving rise to the work values $W$ may be arbitrarily far from equilibrium. Practically, instead of {\it once} driving the transition {\it as slow as possible} in order to reach the final state in almost equilibrium and obtain the approximate free energy of the final state, $\FF_\mr{B}$, the Jarzynski relation enables the determination of $\DF$ from {\it many} realisations of {\it arbitrarily fast} driven transitions.\\
Determining $\DF$ from the Jarzynski relation (\ref{eq:JR}) also entails a difficulty: The exponential average is dependent on negative-large realisations of $W$. Typically, these realisations are the rare realisations where $W<\DF$, demonstrating that the second law, $\Wdiss=W-\DF\geq0$, can not directly be transferred to nanoscopic systems exhibiting pronounced stochasticity. In the literature, these relations are often referred to as transiently violating the second law, a misleading statement, since the second law is a result of macroscopic thermodynamics. Indeed, by applying again the inequality $\lla\exp\,x\rra\geq\exp\lla x \rra$ as we did to derive the second law from the iFT for $\Stot$ in (\ref{eq:FTs_secondlaw}), we retrieve that in the ensemble average
\begin{align}
	\lla W\rra \geq \DF \;.
\end{align}
The difficulty with the exponential average can be visualised by writing
\begin{align} \label{eq:expavg}
	\lla\eee{-\b W}\rra = \int \eee{-\b W}\,P(W) \di W \;,
\end{align}
where $P(W)$ is the work distribution to be obtained from an ensemble of work values $W$. It is evident, that the factor $\ee{-\b W}$ incorporates a weight on large-negative $W$, shifting the work values that dominate the exponential average to the left tail of $P(W)$. A typical example is depicted in figure \ref{ff:sun}. The exponential weight manifests itself in a slow convergence of the exponential average to the theoretical value $\exp(-\b\D\FF)$, cf. figure \ref{ff:expavg}. In contrast to linear averages, the standard error of the set of realisations is not suited to judge the deviation from the true value of the average for a finite number of samples.\\
It arises the question, when it is feasible to employ the Jarzynski relation to estimate the free energy difference $\DF$ between two states. The answer is, (i) in systems where the underlying potential $V(x,t)$ is not accessible, and (ii) in nanoscopic systems that exhibit work-fluctuations of the order of magnitude $W\sim\DF$. Condition (i) is rather obvious, since $\DF$ follows per definition (\ref{eq:def_FF}) by integrating the Boltzmann factor $\exp\big[\-\b V(x,t)]$. In systems with many degrees of freedom, however, a high dimensional integration is needed, and the application of the Jarzynski relation may become advantageous despite an explicit knowledge of $V(\vek{x},t)$. Condition (ii), $W\sim\DF$, implies that frequent fluctuations $W<\DF$ are realistic, and the convergence of the exponential average to $\exp(-\b\DF)$ is to be expected for a reasonable number of work values.
\\
The direct use of the Jarzynski relation to estimate $\DF$ was only the starting point of many refinements in the procedure of $\DF$ estimation. A major improvement can be made, if also realisations of the reverse dynamics are accessible and the corresponding dFT of the Jarzynski relation can be employed,
\begin{align}
	\frac{P(W)}{\bar P(-W)} = \eee{\b(W-\DF)}
\end{align}
which was first derived by Crooks \cite{Crooks1999a,Crooks00PRE}, along with a further generalisation to other observables, and is known as the Crooks FT. The improvement in $\DF$ estimation via the Crooks FT is due to $P(W)=\bar P(-W)$ for $W=\DF$, that is, the value of $\DF$ can be estimated by determining the intersection point of work distribution $P(W)$ for the forward process and work distribution $\bar P(W)$ of the reverse process.\\
A further improvement of $\DF$ estimation was achieved by Hummer and Szabo, who were able to derive in their article \cite{Hummer2001b} a variant of the Jarzynski relation that does not only involve the free energy {\it difference} $\DF$, but the free energy {\it landscape} $\FF(\vek{x}_\mr{r})$ along a reaction coordinate $\vek{x}_\mr{r}$.\\
A prominent example of using FTs for $\DF$ estimation is force-spectroscopy of single biomolecules. In these experiments, single molecules are pulled along a reaction coordinate which forces the molecule to unfold, and then folded back to the original state by reversing the process. Monitoring the work needed to unfold the molecule and the work retrieved by re-folding the molecule enables an estimation of $\DF$ from the Crooks FT after a sufficient number of repetitions of the experiment. In similar experiments, the Hummer-Szabo relation allows to determine the free energy landscape $\FF(\vek{x}_\mr{r})$ of single biomolecules. Experimental realisations include RNA and DNA molecules \cite{Liphardt2001,Collin2005,Mossa2009,DudkoGrahamBest11PRL,Alemany2012} and other molecules \cite{Gebhardt2010,Jarzynski11NP,Chung2012a}.\\
Being an iFT, the Hummer-Szabo relation involves the same difficulties regarding the convergence of the exponential average, and the bias generated by the fact that the exponential average equals $\exp(-\b\DF)$ and not $\DF$ itself. These issues are still keeping researchers busy, considerable progress was made by Ritort and co-workers \cite{Gore2003,Palassini2011}, Dellago and co-workers \cite{Lechner2007,Oberhofer2009} and Zuckerman and Woolf \cite{Zuckerman2000,Zuckerman2002,Zuckerman2004}. While the bias generated by taking the logarithm of the exponential average is tractable, the incomplete convergence of the exponential average persists to be a problem \cite{Pohorille2010}. A possible approach is to deliberately bias the sampling of work values in order to improve the convergence of exponential average, and correct for the introduced bias in the final result, a technique that is known as umbrella sampling or importance sampling \cite{Zuckerman2000,Oberhofer2005,Lechner2007,Williams2013}. Another approach is to augment $P(W)$ with an analytic expression for the left tail of $P(W)$ derived analytically from the specifics of the system under consideration \cite{Engel2009,Nickelsen2011,Nickelsen2012}. The asymptotic method to determine the tails of observables in systems subject to stochastic dynamics is discussed in the next chapter.\\
In general, FTs have been observed in a variety of experimental situations, prominent examples are colloidal particles \cite{Wang2002,Carberry2004,Blickle2006}, electrical circuits \cite{Altland2010,Saira2012,Ciliberto2013}, quantum dots \cite{Campisi2013} and turbulent flows as discussed in \cite{Nickelsen2013} (chapter \ref{s_turbulence_FT}) and references therein.\\

\paragraph{Dominant realisations}
In the context of the Crooks FT, Jarzynski discussed in his appealing article \cite{Jarzynski06PRE} {\it typical} and {\it dominant} realisations of the non-equilibrium process from which work values $W$ are obtained. The typical realisations refer to realisations giving rise to work values from the centre of $P(W)$, whereas the dominant realisations give rise to the rare work values $W<\DF$  that dominate the exponential average and hence are needed for the convergence of the Jarzynski relation. The statement made by Jarzynski is that the dominant realisations of the forward process are the typical realisations of the reverse process. In other words, the set of realisations of the forward process features a subset of realisations that undo the effects of the typical realisations and are responsible for the convergence of the exponential average. Or, vice versa, if for the dynamics of a certain system the validity of the Jarzynski relation is observable, then exists a subset of realisations that correspond to the time-reversed dynamics of the system.\\
The considerations regarding work $W$ and free energy $\FF$ can be directly transferred to the total EP $\Stot$. From equation (\ref{eq:EP_Stot_SS}) follows that the entropic equivalent to the condition $W\sim\DF$, which indicates that dominant realisations are to be expected, reads $-\Sm\sim\DSeq$, where $\DSeq$ is the equilibrium difference between the initial and the final states, and the minus sign is due to the convention that $\Sm$ is the EP in the medium. For entropy fluctuations of the order $\DSeq$, or equivalently but less significant, $\Stot\sim0$, we can hence expect to observe $\Stot<0$ which indicates that a subset of time-reversed realisations exists in the forward dynamics.\\
A complication arises for processes that are not initially in equilibrium. For these cases, we have resort to $\Stot=\Sm+\D S$ which we derived in (\ref{eq:EP_Stot}), where $\D S$ is the difference of Gibbs entropy between initial and final state which depends explicitly on the initial distribution of the process. We will come back to this kind of discussion of EPs in chapter \ref{s_turbulence_FT} in the context of cascade processes in fully developed turbulence. \\

In closing this chapter, we mention that the application of the above considerations to biological and nanoscopic machines is intriguing, since the microscopic fluctuations are not constrained by the second law, and mechanisms rectifying thermal noise are possible. On the macroscopic level, of course, the second law remains to hold true. It is a focus of current research to explore the possibilities of such nanoscopic devices \cite{Esposito2010b,Blickle2011,Seifert2011,Esposito2012a,VandenBroeck2012,Zimmermann2012,Barato2013,Hoppenau2013}. In the course of this research, it proved that apart from work and entropy, a third thermodynamic quantity enters the picture, {\it information}. In particular the relation between entropy and information in nanoscopic devices is keeping researchers busy \cite{Toyabe2010,Mandal2012a,Berut2012,Mandal2013,Petrosyan2013,Andrieux2013,Deffner2013,Barato2014,Hoppenau2014}.

\cleardoublepage
\section{Asymptotic analysis} \label{s_asymp}
The discussed applications of FTs to nanoscopic systems in the previous section pointed up that the characterisation of rare fluctuations in the dynamics of nanoscopic systems is indispensable. These rare fluctuations typically give rise to a consumption of entropy, crucial for the functioning of biological machines and in designing nanoscopic devices, and required to estimate free energy profiles using FTs. The importance of rare fluctuations calls for a method to asymptotically analyse the PDFs of thermodynamic quantities. In this chapter, we develop such a method.\\
We will begin in section \ref{ss_gauss_approx} with a pedagogical one-dimensional example to give an understanding of the asymptotic method which will be introduced in the two subsequent sections \ref{ss_asymp_method} and \ref{ss_asymp_ELE}. A major improvement of the method, along with applications to physically relevant toy-models including single molecule force-spectroscopy, is presented in section \ref{ss_asymp_application_toy}, being the first two publications \cite{Nickelsen2011,Nickelsen2012} included into this thesis.

\subsection{Gauß approximation (GA)} \label{ss_gauss_approx}
The asymptotic method to determine the tails of entropy or work distributions relies on the path integral representation discussed in chapter \ref{ss_WPI}. Before we turn to WPIs, let us consider a pedagogical example - the $\G$-function.\\
The $\G$-function is defined by
\begin{align} \label{eq:asymp_Gamma}
	\Gamma(n+1) = \dint_{0}^{\infty} \ee{-x}\,x^n\di x \;.
\end{align}
Successive integrations by parts show that $n! = \Gamma(n+1)$. Knowledge of the asymptotic behaviour of the factorial is vital in statistical physics. A {\it Gaussian approximation} (GA) of the integral in (\ref{eq:asymp_Gamma}) yields this asymptotic behaviour and is the one-dimensional equivalent of the method we will apply to WPIs in the next chapter. The GA is also known as {\it saddle-point approximation}, since the equivalent method in function theory, the method of steepest descent, deals with a saddle-point.\\
To perform the GA, we define the function 
\begin{align} \label{eq:asymp_Gamma_f-fct}
	f(x) = x-n\ln x
\end{align}
and rewrite the integrand of (\ref{eq:asymp_Gamma}) as $\ee{-f(x)}$. The Taylor expansion up to second order of this function at a point $\bx$ reads
\begin{align}
	f(x) \simeq f(\bx) + f'(\bx)\,(x\-\bx) + \mfrac{1}{2}f''(\bx)\,(x\-\bx)^2 \;.
\end{align}
The crux of the GA is to choose $\bx$ such that the linear term vanishes, and we are left with
\begin{align}
	f(x) &\simeq f(\bx) + \mfrac{1}{2}f''(\bx)\,(x\-\bx)^2 \sep f'(\bx)\meq0 \;, \\ 
	\Gamma(n+1) &= \dint_{0}^{\infty} \ee{-f(x)} \di x \;\simeq\; \ee{-f(\bx)} \dint_{0}^{\infty} \ee{-\frac{f''(\bx)}{2}\,(x\-\bx)^2}  \di x \;.
\end{align}
The remaining integral is a Gaussian integral, thus the name of the method, which can be readily carried out, and we end up with\footnote{This is, in fact, not entirely correct, since the integration covers not the whole $\mathbb{R}$. However, our approximation includes $\bx\gg0$, and the error we make is negligible considering the quality of the GA.}
\begin{align} \label{eq:asymp_Gamma_GA}
	\Gamma(n+1) \simeq \frac{\sqrt{2\pi}\,\ee{-f(\bx)}}{\sqrt{f''(\bx)}} \;.
\end{align}
By solving $f'(\bx)\meq0$ we find $\bx=n$, which substituted into the above GA formula yields Stirling's approximation of the factorial,
\begin{align} \label{eq:asymp_Stirling}
	n! \simeq \sqrt{2\pi n}\;\ee{n(\ln n-1)} \;.
\end{align}
Stirling's approximation is good for $n>10$ and becomes excellent for $n>100$, hence, it is indeed an asymptotic approximation for large $n$. But where did the asymptotic assumption enter?\\
To answer this question, we first note that $\bx=n$ is a minimum of $f(x)$, and accordingly, a maximum of the integrand $\ee{-f(x)}$. We then observe from the definition of $f(x)$, (\ref{eq:asymp_Gamma_f-fct}), that with increasing $n$ the value of this maximum, $\ee{-f(n)}\propto n^n$, increases exponentially. For large $n$, the integral in (\ref{eq:asymp_Gamma}) is therefore dominated by the maximum and its immediate vicinity.\remark{leider wird der integrand nicht immer schmaler, sonder die peaks steigen mit $n^n$ während die varianz nur mit $n$ geht. irgendwie skaliert ließe sich das bestimmt gut darstellen...} Accordingly, for large $n$, the Gaussian integration captures the predominant contribution to the integral in (\ref{eq:asymp_Gamma}). This is the desired situation of the GA.\\
If only the far asymptotic behaviour is of interest, which is in this case $\ln n! \sim n(\ln n - 1)$, it is sufficient to replace the integral with the maximum value of the integrand. This $0$-th order approximation spares us the Gaussian integration, which is in this simple example of course pointless, but represents a major simplification when applying the GA to more involved integrals like WPIs. In the following, we will refer to the factor $\ee{-f(\bx)}$ as the {\it exponential factor}, and we will call the result of the Gaussian integration, $\sqrt{2\pi/f''(\bx)}$, the {\it pre-exponential factor}.

\subsection{The asymptotic method} \label{ss_asymp_method}
Having set the conceptional basis in the previous section, we now discuss the spirit of the GA of constrained WPIs, and will perform the explicit calculations in the next section.\\
We have seen in chapter \ref{ss_WPI} that we can write the solution of a FPE as the WPI
\begin{align} \label{eq:asymp_WPI}
	p(x_t,t) = \dint \dd x_0\,p_0(x_0) \dint\limits_{(x_0,t_0)}^{(x_t,t)} \lDi\xc\,\eee{-\SS[\xc]}
\end{align}
with the action and Jacobian
\begin{align}
	S[\xc] &= \dint_{t_0}^{t} \mfrac{\big[\ddx_\t\-F\big(x_\t,\t\big)\big]^2}{4D\big(x_\t,\t\big)}-J\big(x_\t,\t\big)\di \t \;, \\
	J(x,t) &= \mfrac{1}{2}F'(x,t) + \mfrac{1}{4}D''(x,t) \;.
\end{align}
Note that we used Stratonovich convention and are therefore not bothered with modified calculus.\\
Suppose we are interested in an integral observable $\OmO=\OmI[\xc]$,
\begin{align}
	\OmI[\xc] \dfns \int_{t_0}^{t} \om\big(x_\t,\ddx_\t,\t\big) \di\t \;,
\end{align}
where $\om(x,\ddx,\t)$ is the kernel function that defines $\OmI[\xc]$.
To incorporate the constraint into the WPI, we use the Fourier representation of the $\d$-function,
\begin{align}
	\d(\OmO-\OmI[\xc]) = \frac{1}{2\pi}\int \eee{ik(\OmO-\OmI[\xc]} \di k \;.
\end{align}
Including $\d(\OmO-\OmI[\xc])$ and the initial distribution via $\L(x_0)\dfns-\ln p_0(x_0)$ into the new action $\widetilde\SS[\xc;\,k]$, and performing the integral with respect to $x_t$, we get from (\ref{eq:asymp_WPI}) the PDF of the observable in terms of a constrained WPI,
\begin{subequations} \label{eq:asymp_constrWPI}
  \begin{align}
		p(\OmO) &= \dint \di x_t\dint \di x_0 \dint\frac{\dd k}{2\pi} \dint\limits_{(x_0,t_0)}^{(x_t,t)} \lDi\xc\, \widetilde P[\xc;\,k,\OmO] \label{eq:asymp_constrWPI_pOmO} \\		
		\widetilde P[\xc;\,k,\OmO] &= \exp\big[-\widetilde\SS[\xc;\,k,\OmO]\big] \label{eq:asymp_constrWPI_POmO} \\[10pt]	
		\widetilde\SS[\xc;\,k,\OmO] &= \L(x_0) + \dint_{t_0}^{t} s(x_\t,\ddx_\t,\t;\,k) \di \t - ik\OmO \label{eq:asymp_constrWPI_SOmO} \\
		s(x_\t,\ddx_\t,\t;\,k) &= \mfrac{\big[\ddx_\t\-F(x_\t,\t)\big]^2}{4D(x_\t,\t)}-J(x_\t,\t) + ik\,\om(x_\t,\ddx_\t,\t) \label{eq:asymp_constrWPI_sOmO}
	\end{align}
\end{subequations}
In the spirit of our pedagogical example, the claim is that by applying a GA to the constrained WPI in the first line, we yield an approximation for $p(\OmO)$. We state that this approximation is asymptotically correct for rare $\OmO$. This statement is plausible since by choosing rare values for $\OmO$, the realisations $\xc$ giving rise to that value are rare by itself and dominate the integrand $P[\xc;\OmO]$. In other words, by putting in place an exceptionally hard constraint, the variation of $\xc$ gets substantially narrow, and the path integral (\ref{eq:asymp_constrWPI}) is dominated by the mode of $P[\xc;\OmO]$ and its immediate vicinity, analogous to the pedagogical example discussed in the previous section.\\
However, as in the example of the $\G$-function, this argumentation is of rather qualitative nature. The GA of the $\G$-function can be put quite straight forwardly into a systematic shape, for the GA of the WPI, however, the situation is intricate. In essence, our asymptotic method is equivalent to the weak noise limit, in which the GA gets asymptotically exact in the limit of $D(x,t)\to0$. We could therefore rewrite the diffusion coefficient $D(x,t)\mapsto\frac{1}{\din}\tilde D(x,t)$ such that $\tilde D(x,t)$ is dimensionless, and put in \ref{eq:asymp_constrWPI_POmO} the weak noise parameter $\din$ as a prefactor of the action $\widetilde\SS$.\footnote{The terms $\L$, $J$, $\om$ and $\OmO$ obtain a prefactor $1/\din$. By substituting $q\dfns k/\din$ and taking the stationary distribution as initial distribution, only the Jacobian $J$ is left with the prefactor. But since in the weak noise limit the realisations $\xc$ loose their stochasticity, $\SS$ is not a stochastic integral anymore, and $J$ does not contribute to the $0$-th order GA.} In the weak noise limit, $\din\!\to\!\infty$, the integrand of the WPI becomes a singularly narrow peak, just as choosing a practically impossible value for $\OmO$. The GA of the WPI (\ref{eq:asymp_constrWPI}) therefore becomes also exact in the weak noise limit. This formal correspondence can be used to give the asymptotic method a systematic order of approximation. In section \ref{ss_asymp_application_toy}, the weak noise parameter will be the inverse temperature $\b=1/T$.\remark{Gauß integral, $\quad \int\ee{-ax^2}\di x = \sqrt{\frac{\pi}{a}}$, also for arbitrary dimension, $\quad \int\ee{-\vek{x}\bs{A}\vek{x}}\di^n x = \sqrt{\frac{\pi^n}{\det\bs{A}}}$.}\\

\subsection{Euler-Lagrange equations} \label{ss_asymp_ELE}
In the previous section we have formulated in (\ref{eq:asymp_constrWPI_POmO}) the constrained path-probability $P[\xc;\OmO]$. We argued that the GA consists in substituting the integral of $P[x(\cdot,\OmO)]$ with the value of the path-probability at its mode $\bxc$.\remark{substitute the path integral with a Gaussian distribution with mode $\bxc$ (not average! in prl, e.g., figure 3 shows the average, which can be quite different) of $P[x(\cdot,\OmO)]$ as the mean and an appropriate standard deviation; note that the Gaussian is not normalised, the height of the peak is rather given by $P[\bxc]$ - the exponential factor, the remaining integration of the Gaussian with height of peak equals one is then the pre-exponential factor.} The mode of $P[\xc;\OmO]$ is given by the minimising trajectory $\bxc$ and parameter $\bar k$ of the action $\widetilde\SS[\xc;\,k]$ under the constraint $\OmO=\OmI[\bxc]$\,; a variational problem with constraint, where $k$ is the Lagrange parameter.\\ 
To obtain $\bxc$ and $\bar k$, we need to solve the associated Euler-Lagrange equation (ELE)
\begin{align} \label{eq:asymp_ELE_gen}
	0 \;=\; \dfrac{\pt s(x,\ddx,\t)}{\pt x}\Big|_{x=x(\t)} \;-\; \dfrac{\dd}{\dd\t}\dfrac{\pt s(x,\ddx,\t)}{\pt\ddx}\Big|_{x=x(\t)}	
\end{align}
with boundary conditions
\begin{align}
	\frac{\pt\L(x_0)}{\pt x_0} &\;=\; \dfrac{\pt s(x,\ddx,\t)}{\pt\ddx}\Big|_{x=x_0} \sep	0 \;=\; \dfrac{\pt s(x,\ddx,\t)}{\pt\ddx}\Big|_{x=x_t} \;.  \label{eq:asymp_ELE-BC_gen}
\end{align}
The solution of the ELE will depend on the Lagrange parameter $k$ and has to be adjusted such that the constraint
\begin{align}
	\OmI[\bxc] = \int_{t_0}^{t} \om\big(\bx_\t,\bddx_\t,\t\big) \di\t
\end{align}
is satisfied. Due to the analogy to optimisation problems, the minimising trajectory $\bxc$ is also called {\it optimal path}.\\
Taking the function $s(x,\ddx,\t)$ from (\ref{eq:asymp_constrWPI_sOmO}) and carrying out the differentiations yields the ELE
\begin{subequations} \label{eq:asymp_ELE}
  \begin{align}
  \begin{split}
    \dddx &= \mfrac{\ddx \- F}{2D}\big(D'(\ddx \+ F)+2\ddD\big) \\
     &\hspace{70pt}+FF'+\ddF+2DJ''+2Dik(\pt_x\-\dd_t\pt_{\ddx})w \;,
  \end{split}  \label{eq:asymp_ELE_ode}\\
		0&=\mfrac{\ddx_0-F_0}{2D_0}+ik\pt_{\ddx}w_0-\L'_0 \;, \label{eq:asymp_ELE_BC0}\\
		0&=\mfrac{\ddx_t-F_t}{2D_t}+ik\pt_{\ddx}w_t \;, \label{eq:asymp_ELE_BC1}
	\end{align}
\end{subequations}
where we dropped the arguments of $x_\t$, $F(x_\t,\t)$, $D(x_\t,\t)$, $w(x_\t,\ddx_\t,\t)$ and derivatives, and indices denote evaluation at time $\t$, $t_0$ and $t$ respectively.\\
The $0$-th order result of the GA is accordingly
\begin{align} \label{eq:asymp_finalGA_0th}
	p(\OmO) \sim \eee{-\SS[\bxc,\bk]} \;.
\end{align}
The pre-exponential factor can also be obtained, as will be demonstrated in the next chapter for additive noise. For multiplicative noise, the equations for the pre-exponential factor gets acutely involved, we therefore refrain from reporting preliminary results in that general case.

\subsection{The pre-exponential factor} \label{ss_asymp_application_toy}
The determination of the pre-exponential factor for the asymptotic approximation was achieved in \cite{Nickelsen2011}, which is the first publication included in this thesis. The second publication \cite{Nickelsen2012} of this thesis follows directly after \cite{Nickelsen2011} and is devoted to a particular application. In the provided reprints of the publications the force $F(x,t)$ will arise from a time dependent potential, $F(x,t)=-\pt_xV(x,t)$, the diffusion coefficient is simply proportional to temperature, $D=2/\b$, and the processes are taken to start from equilibrium, $p_0(x_0)=\exp[-\b V(x_0,t_0]/Z_0$. The observable of interest is work $W$, it therefore is $w(x,t)=\dot V(x,t)$, cf. (\ref{eq:def_W}). The application of the method will be demonstrated for physically relevant choices of $V(x,t)$, in particular, \cite{Nickelsen2012} addresses single molecule spectroscopy.
\includepdf[pages=1,trim=6mm 10mm 13mm 30mm,clip,scale=0.85,pagecommand={\pagestyle{headings}}]{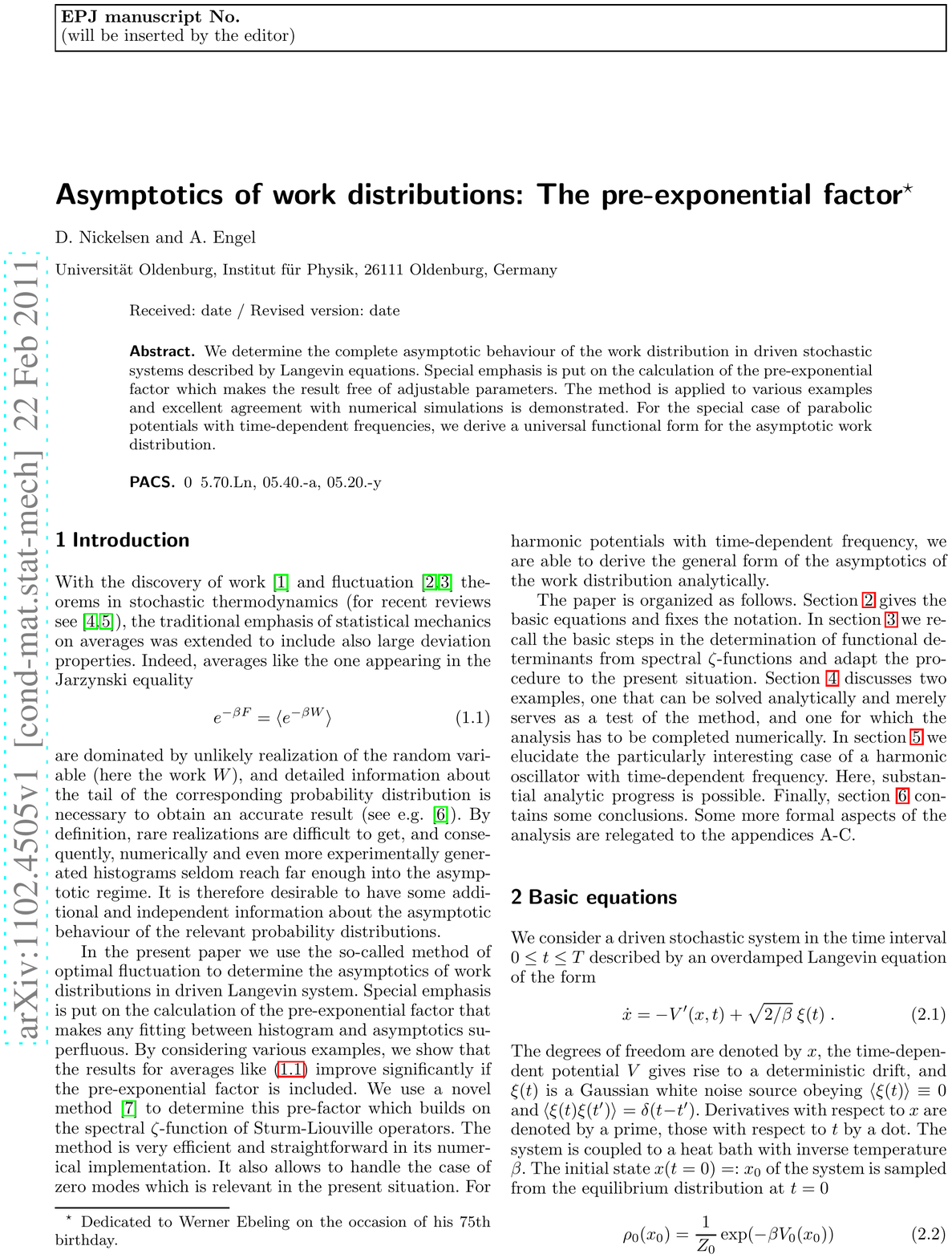} 
\includepdf[pages=2-12,trim=13mm 10mm 13mm 24mm,clip,scale=0.85,pagecommand={\pagestyle{headings}}]{EPJB_arxiv.pdf}
\includepdf[pages=1,trim=6mm -5mm 14mm 16mm,clip,scale=0.85,pagecommand={\pagestyle{headings}}]{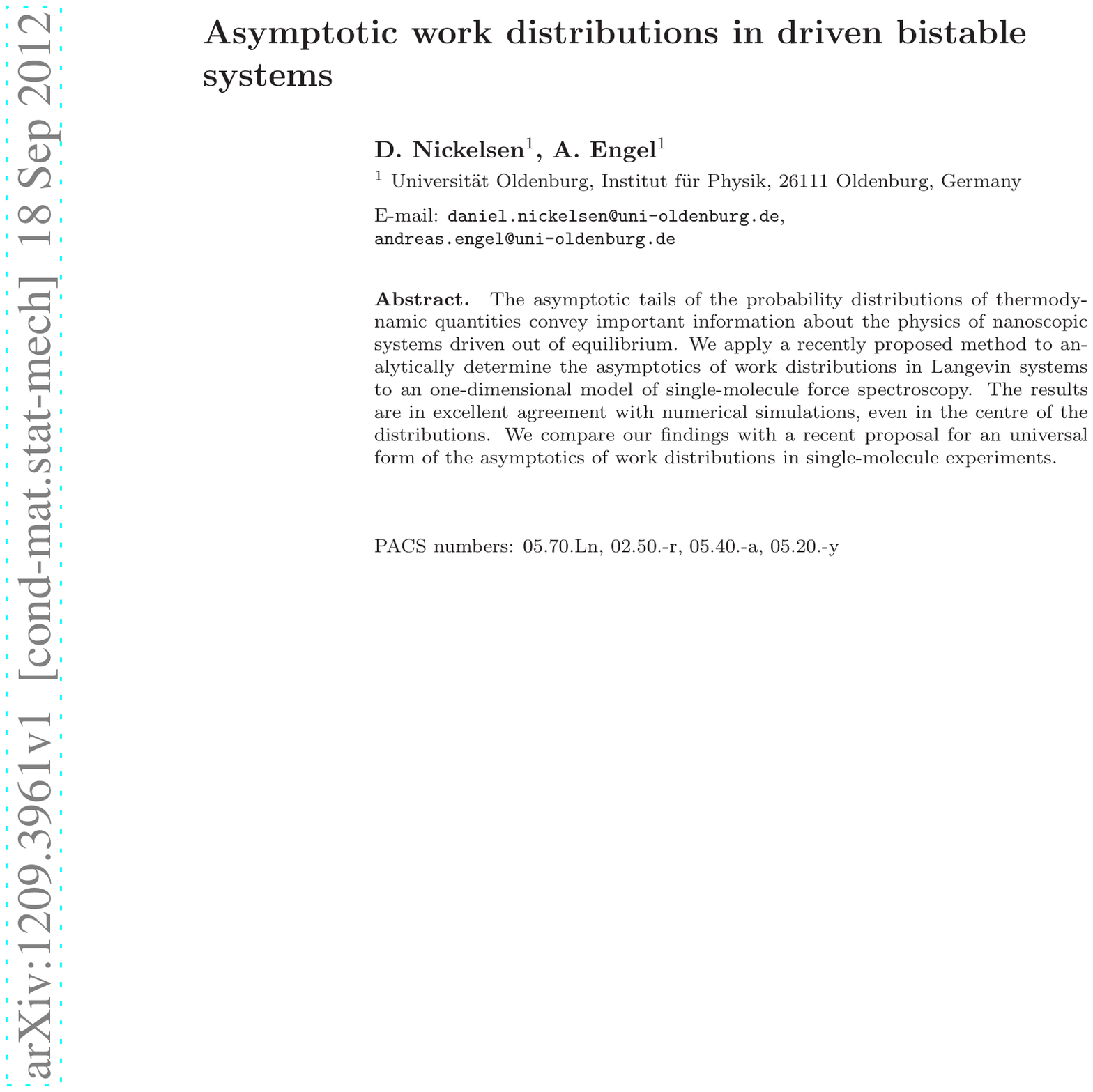}
\includepdf[pages=2-10,trim=16mm -5mm 14mm 16mm,clip,scale=0.85,pagecommand={\pagestyle{headings}}]{PS_arxiv.pdf}

\cleardoublepage
\section{Conclusions}
In the first of the two parts of this thesis, we have introduced the mathematical description of Markov processes, and discussed its relevance for stochastic thermodynamics, being the thermodynamics of nanoscopic systems in non-equilibrium.\\

We focused on continuous Markov processes, for which three equivalent descriptions were introduced, stochastic differential equations, the Fokker-Planck equation and Wiener path integrals. The embedding into discontinuous Markov process was carried out via the Kramers-Moyal expansion, with coefficients being directly related to the moments of the jump density which fixes the discontinuous component of the Markov process. Estimating the first two Kramers-Moyal coefficients from realisations of a Markov process yields a Fokker-Planck equation which approximately describes the Markov process, provided that the discontinuous component of the Markov process is negligible.\\

As a paradigm for thermodynamic interpretation of continuous Markov processes, we introduced the Langevin equation as a stochastic differential equation that models Brownian particles suspended in a fluid which experience molecular friction and thermal collisions while dragged through the fluid by an external force. The Brownian particles constitute a canonical ensemble, coupled to a heat bath at constant temperature being realised by the fluid. By examining the energy balance, we identified the work done by the external force and the heat dissipated into the fluid on the level of single trajectories of Brownian particles, including the first law relating work, heat and the internal energy difference between initial and final position of the particle.\\
Besides the first law, we also formulated the second law on the level of individual trajectories in terms of the total entropy production of particle and fluid. The total entropy production is the sum of the entropy difference between initial and final position of the particle plus the entropy produced in the fluid. Formulated as Wiener path integrals, the value of total entropy production proved to be a direct measure of irreversibility, in the sense that with increasing entropy production, the likelihood of concurrent trajectories that consume that entropy decreases exponentially with that entropy production.\\
This relation between entropy production and irreversibility was shown to be quantitatively captured by fluctuation theorems which involve the exponential average of thermodynamic variables. Due to their relation to irreversibility, observing a convergence of the exponential average to the theoretical value implies that the ensemble of trajectories used for the exponential average include realisations that correspond to a time-reversal of the underlying dynamics. In other words, fluctuation theorems can be used to assess whether the stochastic dynamics of a system generates a subset of realisations that are typical for the reverse of that dynamics.\\
The connection to macroscopic thermodynamics was established by noting that using fluctuation theorems to estimate the lower bound of entropy production reproduces the second law.\\

Fluctuation theorems relate thermodynamic non-equilibrium quantities with equilibrium state variables. The prominent application of fluctuation theorems is therefore the recovery of free energy profiles from non-equilibrium measurements. The occurrence of reversed realisations in the non-equilibrium dynamics are crucial for the performance of these applications. If the non-equilibrium dynamics is known explicitly, we demonstrated that a asymptotic method is capable of assessing the reversed realisations in order to quantify the probability of these rare realisations and improve the recovery of free energy profiles.\\

As a special case of non-equilibrium states, we introduced non-equilibrium steady states which are possible for multidimensional dynamics or multiple heat baths. A non-equilibrium steady state is maintained by a non-conservative force which gives rise to an extra entropy production, which we introduced as the housekeeping entropy production. The housekeeping entropy production, the total entropy production and the difference between these two obey a fluctuation theorem, which in turn imply three faces of the second law. The irreversibility related to the housekeeping entropy accounts for the reversal of a current of particles or heat generated by non-conservative forces or multiple heat baths, respectively.


\cleardoublepage
\pagestyle{plain}
\quad
\cleardoublepage
\pagestyle{headings}
\chapter{Universal features of turbulent flows}
\vspace*{30pt}
Turbulent flows are omnipresent, ranging from the atmospheric boundary layer on large scales to applications like turbulent drag optimisation or turbulent mixing on small scales. The characterisation of intermittent velocity fluctuations is central in the majority of situations involving turbulent flows. Examples include predictions of extreme weather conditions, wind loads on buildings, air-planes, wind energy converters, site assessment for wind energy, turbulent mixing and combustion.\\[30pt]

\noindent The introductory first chapter gives an account on basic aspects of fluid mechanics. In the second chapter, the idealised concept of fully developed turbulence is introduced, including established theories of fully developed turbulence which statistically approach the mentioned intermittent velocity fluctuations. The introduced approaches to turbulence can be recast as Markov processes, which is demonstrated in the third chapter. Section four and five exploit the consequences of fluctuation theorems and the asymptotic analysis developed in the first part. Finally, chapter six closes this part by classifying the discussed approaches to turbulence by their Markov representation and suggesting possible interpretations of the implications.

\renewcommand{\kB}{k_\mr{B}}

\cleardoublepage
\section{General theory} \label{s_hydro}
In this chapter we give a brief account on how the concepts of turbulence generation and energy transfer arise from fundamental considerations. The ruling equation for a viscid turbulent flow is the Navier-Stokes equation (NSE).\\
We briefly demonstrate in the first section \ref{ss_hydro_fluiddynamics} how the NSE arises from Newton's equation of motion by discussing the forces that act on an infinitesimal fluid volume. Using the NSE, we show that an expression for the rate of energy dissipation in a turbulent flow can be derived. The second section \ref{ss_hydro_turbulencegeneration} explicates the generation of the highly irregular motion in a turbulent flow due to energy injection into the fluid, and addresses the applicability of statistical physics due to emerging degrees of freedom for increasing energy injection. In section \ref{ss_hydro_energybudget} we identify the mechanisms of energy transfer from large to small structures in a turbulent flow.

\subsection{The Navier-Stokes equation (NSE)} \label{ss_hydro_fluiddynamics}
The material of this section is adapted from the book by Landau and Lifshitz on fluid mechanics \cite{landau1987fluid} and the book by Frisch on turbulence \cite{Frisch95}.\\

The motion of a fluid at position $\vek{x}$ and time $t$ is completely determined by the knowledge of fluid velocity $\vek{v}(\vek{x},t)$, the fluid density $\r(\vek{x},t)$\footnote{Quantities in this part of the thesis will be measured per unit mass.} and the pressure $p(\vek{x},t)$\remark{(es gehen auch andere td größen als druck und dichte, da durch zustandsgleichung des betreffenden fluids die jeweils fehlenden folgen)}. With the three components of velocity $\vek{v}$, the pressure $p$ and the density $\r$, we have five quantities which have to be fixed by five equations to describe the motion of the fluid.\\
The first equation is the well known continuity equation\remark{LL p.2}
\begin{align} \label{eq:Hydro_conti}
	\frac{\pt\r(\vek{x},t)}{\pt t} + \vek{\nabla}\big(\r(\vek{x},t)\vek{v}(\vek{x},t)\big) = 0
\end{align}
as the consequence of conservation of mass.\\
The quantities fixed by the continuity equation are fluid velocity and density, the pressure field $p(\vek{x})$ does not enter. To formulate an equation including also $p(\vek{x})$, we make us of the fact that pressure gives rise to a force acting on a fluid volume, which allows us to set up Newton's equation of motion for a certain fluid volume. In differential form, this equation reads 
\begin{align} \label{eq:Hydro_EuE}
	\frac{\pt\vek{v}(\vek{x},t)}{\pt t} + \big(\vek{v}(\vek{x},t)\cdot\vek{\nabla}\big)\,\vek{v}(\vek{x},t) = - \frac{\vek{\nabla}p(\vek{x})}{\r(\vek{x},t)}
\end{align}
and is known as the {\it Euler equation} (\cite{landau1987fluid} p.\,4,14]). The l.h.s. is the total time-derivative of $\vek{v}(\vek{x},t)$, the r.h.s. is the acting force due to $p(\vek{x})$.\remark{(the second term on the l.h.s. stems from writing the equation of motion in terms of positions and not in terms of trajectories (p.3))} The Euler equation includes three equations for each component of $\vek{v}$.\\
In the derivation of the Euler equation we did not include dissipation and transfer of energy. Therefore, the fluid motion takes place adiabatic, a requirement that constitutes the remaining fifth equation to fix all five quantities $\vek{v}$, $\r$ and $p$.\\
Finally, the equations need to be supplemented with boundary conditions such as vanishing tangential velocity at confining walls.\remark{(p.52-55) Can write Euler equations in tensor form $\r\pt_t v_i = - \pt \varPi_{ik}/\pt x_k$ in which $\varPi_{ik}$ is the tensor of momentum flux density. Augmenting $\varPi_{ik}$ with a term accounting for viscous momentum transfer, we obtain the stress tensor $\sigma_{ik}$. From detailed considerations it follows that the general form of $\sigma_{ik}$ only depends on two viscosity coefficients, $\eta_1$ and $\eta_2$. siehe auch entropy-bilanz-gleichung s. 214 im vergleich zu (15,6) auf seite 54 um einfluss der koeffizienten zu deuten.)} \\

The Euler equation describes inviscid fluids, that is, no molecular friction is taken into consideration. Accordingly, the kinetic energy of the fluid is conserved, and all processes in the fluid take place reversibly.\\
In a viscous fluid, molecular friction entails a dissipation of kinetic energy in form of heat, thus an irreversible process is added to the otherwise reversible processes of an inviscid fluid.\\
We now explicate the effect of taking viscosity into account. In a viscous fluid it can be shown that two more forces contribute to the equation of motion (\cite{landau1987fluid} p.\,53f), namely
\begin{subequations} \label{eq:Hydro_viscous-forces}
	\begin{align}
		\vek{f}_1(\vek{x},t) &= \eta_1 \Big(\laplace\vek{v}(\vek{x},t) + \mfrac{1}{3}\,\vek{\nabla}\big(\vek{\nabla}\,\vek{v}(\vek{x},t)\big)\Big) \\
		\vek{f}_2(\vek{x},t) &= \eta_2 \vek{\nabla}\big(\vek{\nabla}\,\vek{v}(\vek{x},t)\big)
	\end{align}
\end{subequations}
These forces simplify considerably for an incompressible fluid since then the velocity field $\vek{v}(\vek{x},t)$ is divergence free. When dealing with liquid fluids, incompressibility can generally be assumed. Also a gaseous fluid can be assumed to be incompressible, as long as flow velocities stay well below the speed of sound.\\ 
Setting therefore $\vek{\nabla}\vek{v}(\vek{x},t)\equiv0$ in (\ref{eq:Hydro_viscous-forces}) and including $\vek{f}_1(\vek{x},t)$ into the Euler equation (\ref{eq:Hydro_EuE}), we arrive at the so-called {\it Navier-Stokes equation} (NSE)
\begin{align} \label{eq:Hydro_NSE}
	\frac{\pt\vek{v}(\vek{x},t)}{\pt t} + \big(\vek{v}(\vek{x},t)\cdot\vek{\nabla}\big)\,\vek{v}(\vek{x},t) = -\frac{1}{\r}\vek{\nabla}p(\vek{x}) + \nu\laplace\vek{v}(\vek{x},t) \;,
\end{align}
governing the dynamics of a viscous, incompressible fluid. Here we defined the {\it kinematic} viscosity $\nu\dfn\eta_1/\r$ ($\eta_1$ is called {\it dynamic} viscosity).\remark{The continuity equation is now simply the statement that we have a divergence free velocity field $\vek{\nabla}\vek{v}(\vek{x},t)\equiv0$. Schließungsproblem nur bei Reynolds-gemittelte NSE}\\
The convection term $(\vek{v}\cdot\vek{\nabla})\vek{v}$ couples the vector components of the NSE, as it couples the gradient of $\vek{v}$ with all components of $\vek{v}$. It is due to this non-linear coupling of the vector components of (\ref{eq:Hydro_NSE}) that makes an analytical solution of the NSE practically impossible. Furthermore, this non-linearity is responsible for a highly irregular and chaotic motion of the fluid, the very phenomenon which we call {\it turbulence}.\remark{convection, weil es beschleunigung des fluids durch aneinander vorbeiströmen hervorruft.} On the other hand, we have the frictional force $\nu\laplace\vek{v}(\vek{x},t)$ which damps this irregular motion and causes eventually a decay of turbulence. The higher the viscosity $\nu$, the stronger the smoothing effect of friction. The interplay between the effects of the convection term and the frictional term governs the phenomenology of decaying turbulence.\\

To explore the effects of the frictional term, we examine the energy dissipation it provokes. To this end, we consider the total kinetic energy content of an incompressible fluid,
\begin{align}\label{eq:Hydro_Ekin}
	\Ekin = \frac{\r}{2}\int v(\vek{x},t)^2 \di^3x \;.
\end{align}
The time derivative of $\Ekin$ will include a term corresponding to energy flux, and another term accounting for dissipation of energy. To identify these contributions, we differentiate the integrand of $\Ekin$, make use of the NSE (\ref{eq:Hydro_NSE}) and obtain after some manipulations (\cite{landau1987fluid} p.58f)
\begin{align} \label{eq:Hydro_Eflux}
	\frac{\pt}{\pt t}\frac{\r v(\vek{x},t)^2}{2} &= -\vek{\nabla}\bigg[\r\vek{v}(\vek{x},t)E(\vek{x},t) - \vek{j}(\vek{x},t)\bigg] - \eps(\vek{x},t)
\end{align}
with the terms
\begin{subequations} \label{eq:Hydro_Eflux_terms}
	\begin{align}
		E(\vek{x},t) &= \frac{\vek{v}(\vek{x},t)^2}{2} + \frac{p(\vek{x})}{\r} \;, \label{eq:Hydro_Eflux_terms_E}\\
		j_k(\vek{x},t) &= \nu\,\sum\limits_{i} \, v_i\,\bigg(\frac{\pt v_i(\vek{x},t)}{\pt x_k}+\frac{\pt v_k(\vek{x},t)}{\pt x_i}\bigg)  \label{eq:Hydro_Eflux_terms_j} \;, \\
		\eps(\vek{x},t) &= \frac{\nu}{2}\,\sum\limits_{ik}\, \bigg(\frac{\pt v_i(\vek{x},t)}{\pt x_k}+\frac{\pt v_k(\vek{x},t)}{\pt x_i}\bigg)^{\!2} \;. \label{eq:Hydro_Eflux_terms_eps}
	\end{align}
\end{subequations}
The term $E(\vek{x},t)$ is the local energy content which, multiplied with $\r\vek{v}(\vek{x},t)$ in (\ref{eq:Hydro_Eflux}), is the flux of energy due to motion of the fluid. As indicated by the dependence on viscosity $\nu$, the terms $\vek{j}(\vek{x},t)$ and $\eps(\vek{x},t)$ result from molecular friction. To tell the difference between $\vek{j}(\vek{x},t)$ and $\eps(\vek{x},t)$, consider the integration of (\ref{eq:Hydro_Eflux}) over a certain volume. Applying Gauss's theorem, we see that the first term in (\ref{eq:Hydro_Eflux}) turns into a surface integral, accounting for the flux into and out of the volume, whereas the positive term $\eps(\vek{x},t)$ appears to represent a sink of energy in that volume. We therefore attribute $\eps(\vek{x},t)$ to energy dissipation due to molecular friction. \\

\subsection{Turbulence generation and decay} \label{ss_hydro_turbulencegeneration}
Energy dissipation due to molecular friction is responsible for the decay of turbulence, as discussed in the previous chapter. In this chapter, we take into account the generation of turbulence, and relate it to the dissipation of kinetic energy.\\

Basically, turbulence is generated by imposing a certain initial condition on $\vek{v}(\vek{x},t)$ which entails the highly irregular motion of a turbulent flow. Due to the frictional term in the NSE, the irregular motion will be smoothed out until the flow becomes laminar.\\
Practically, turbulence is forced by blocking a laminar flow with an obstacle, the wake of the obstacle is then a turbulent flow. This kind of turbulence generation is characterised by the incoming velocity $\vchar$ of the laminar flow and a typical length scale $\lchar$ of the obstacle. Examples for $\lchar$ of obstacles are the diameter of a cylinder or the mesh size of a grid. A similar kind of turbulence generation are {\it free jets}, in which a flow is accelerated by directing it through a narrow nozzle, after which it hits a resting fluid. In this case, $\vchar$ is typically taken to be the flow velocity at the nozzle and $\lchar$ the diameter of the nozzle.\\
In all cases, a stirring force is acting on the fluid implying an injection of turbulent energy into the fluid. The product of $\vchar$ and $\lchar$ characterise the magnitude of this energy injection.\remark{$\vchar\lchar$ is fläche pro zeit: $\lchar$ legt die breite und $\vchar$ legt die pro zeiteinheit erzeugte länge einer erzeugten turbulenzstruktur fest.}\\

The form of the NSE (\ref{eq:Hydro_NSE}) introduced in the previous chapter does not account for energy injection. We therefore augment the NSE with a stirring force $\vek{f}(\vek{x},t)$ and obtain the {\it forced} NSE 
\begin{align} \label{eq:Hydro_NSE_forced}
	\frac{\pt\vek{v}(\vek{x},t)}{\pt t} + \big(\vek{v}(\vek{x},t)\cdot\vek{\nabla}\big)\,\vek{v}(\vek{x},t) =  \nu\laplace\vek{v}(\vek{x},t)-\frac{1}{\r}\vek{\nabla}p(\vek{x}) + \vek{f}(\vek{x},t) \;.
\end{align}
The force $\vek{f}(\vek{x},t)$ is usually assumed to be a random force with homogeneous, isotropic and stationary statistical properties. We do not need any more specifics on $\vek{f}(\vek{x},t)$ as we are only interested in the mean energy injection caused by $\vek{f}(\vek{x},t)$ (\cite{Frisch95} p.77).\\
\newcommand{\vv}{\tilde v}
\newcommand{\xx}{\tilde x}
\newcommand{\ttt}{\tilde t}
\newcommand{\pp}{\tilde p}
\newcommand{\ff}{\tilde f}
It is instructive to recast the forced NSE in dimensionless form, which we achieve by multiplying (\ref{eq:Hydro_NSE_forced}) with $\lchar/\vchar^{\,2}$ and obtain\footnote{in terms of dimensionless quantities  $\vek{\vv}=\vek{v}/\vchar$, $\vek{\xx}=\vek{x}/\lchar$, $\ttt=t\cdot \vchar/\lchar$, $\pp=p\cdot\r/\vchar^{\,2}$, $\vek{\ff}=\vek{f}\cdot \lchar/\vchar^{\,2}$}
\begin{align} \label{eq:Hydro_NSE_dimless}
	\<\<\frac{\pt\vek{\vv}(\vek{\xx},\ttt)}{\pt \ttt} + \big(\vek{\vv}(\vek{\xx},\ttt)\!\cdot\!\vek{\nabla}\big)\vek{\vv}(\vek{\xx},\ttt) = \frac{1}{\Reych}\laplace\vek{\vv}(\vek{\xx},\ttt) -\vek{\nabla}\pp(\vek{\xx}) + \vek{\ff}(\vek{\xx},\ttt)
\end{align}
where $\vek{\nabla}$ and $\laplace$ apply to $\xx$, and we have defined the dimensionless number
\begin{align} \label{eq:Hydro_Rey}
	\Reych = \frac{\vchar\,\lchar}{\nu}
\end{align}
which is known as {\it Reynolds number}.\\
Taking the viscosity as characteristic for the magnitude of energy dissipation, we may say that the Reynolds number relates the magnitude of energy injection with the magnitude of energy dissipation. It is furthermore evident from the dimensionless NSE (\ref{eq:Hydro_NSE_dimless}) that $\Reych$ determines the impact of dissipation on the dynamics; for $\Reych\!\to\!0$ we get the forced version of the Euler equation (\ref{eq:Hydro_EuE}). We will denote $\Reych$ also by $\Rey$.\\
The Reynolds number also has practical implications: If we aim at investigating the properties or effects of a turbulent flow on a too large scale to be done in a laboratory, we can instead set up a smaller version of the situation and rescale the results of the experiment according to the ratio of Reynolds numbers. The underlying concept is referred to as similarity principle (\cite{landau1987fluid} p.\,67ff, \cite{Frisch95} p.2).\\

There is no theorem that guarantees the existence and uniqueness of a solution of the NSE (\cite{Frisch95} p.38). But a {\it stationary} solution of the NSE in principle does exist for any reasonable flow situation (\cite{landau1987fluid} p.114,131f). The stationary solution, however, will only be observable in reality if the solution is stable. A linear stability analysis shows that below a critical Reynolds number $\Reycr$, which is found to be of order 100, the stationary solution is insusceptible to small perturbations. This is reasonable, since small Reynolds numbers imply that molecular friction is predominant in the dynamics of the flow, and hence small perturbations are absorbed sufficiently fast. For $\Reych\,$\mscale[0.8]{\gtrsim}$\,\Reycr$, however, the flow becomes unstable when subjected to arbitrary small perturbations; ruling out the realisation of a stationary flow. The transition to an unstable flow for increasing Reynolds number can therefore be attributed to a magnitude of energy injection that is sufficiently larger than the magnitude of energy dissipation.\\
In the course of the stability analysis it turns out that with $\Reych>\Reycr$, periodic modulations superimpose with the stationary solution of the flow. The frequencies and amplitude of the modulations are fixed by the perturbations, but the attendant phases enter the solution as additional degrees of freedoms which have to be fixed by extra initial conditions. Raising $\Reych$ well beyond $\Reycr$, more frequencies of broader range prevail in the flow involving an increasing number of degrees of freedoms.\footnote{A dimensional analysis shows that the number of degrees of freedom in a turbulence flow scales with $(\Reych/\Reycr)^{9/4}$ (\cite{landau1987fluid} p.139)}. In addition to that, the flow becomes chaotic and therefore depends highly sensitively on the initial conditions.\remark{which is a consequence of the fact that only phases between $0$ and $2\pi$ create new physics, and therefore any possible state of the flow will be approached arbitrary closely in finite time}\\ 
The generation of additional degrees of freedom for Reynolds numbers $\Reych>\Reycr$ in the flow allows for a close analogy to statistical physics: On the microscopic level, a macroscopic system has an immense number of degrees of freedom. If we were to predict the behaviour of the macroscopic object, it would be hopeless to do so by solving the equations of motion for each individual molecule. Instead it is feasible to consider the statistics of the degrees of freedom and extract integral values to describe the macroscopic system. The same applies for a flow at high Reynolds numbers. The attempt to solve the NSE would involve the knowledge of an immeasurable number of initial conditions, in extreme, the knowledge of position and velocity of all fluid molecules. But as the motion of flow is highly irregular and chaotic\footnote{it is indeed possible to recast the NSE as a dynamical system featuring high sensitivity to initial conditions}, after a finite time any reference to the initial conditions would be lost. This renewal property provokes a random behaviour of flow properties which enables us to treat the flow velocity by statistical means, which will be the strategy in the following chapters.\remark{frisch p.58f: Taylor hypothesis of frozen turbulence for sufficient low turbulence intensities (threshold is 0.1, for higher turbulence intensities resampling techniques exist). frisch p.74f [LL wohl nur korrelationen, auch gut]: velocity increments between two points of distance $r$ and direction $\vek{e}$ to measure spatial structures.}

\subsection{Energy budget} \label{ss_hydro_energybudget}
We have argued that the Reynolds number relates energy injection with energy dissipation. It also seems likely that the immediate effect of energy injection is large scale motion, whereas molecular friction acts only on small scale motion. Taking $\lchar$ as the scale of the large scale motion, it stands to reason that the Reynolds number also relates the respective scales on which energy injection and molecular friction are acting, being in accord with the increasing range of frequencies prevailing in the flow for increasing Reynolds numbers as discussed in the previous section.\\
The impact of $\Rey$ on the frequencies in the flow conveys an intuitive picture of energy injection and dissipation in the flow (\cite{landau1987fluid} p.\,134): Energy injection acts on small wavenumbers, heat dissipation takes place at large wavenumber. If we denote by $\Kchar$ the order of magnitude of wavenumbers influenced by energy injection and by $\Knu$ the wavenumbers where energy is dissipated, we can expect that for sufficient high Reynolds numbers a range of wavenumbers $\Kchar\ll k\ll\Knu$ emerges in which neither energy injection nor dissipation has an influence. Consequently, in this range pure energy transfer from small to large wavenumbers must prevail.\\
The absence of forcing and dissipation reduces the forced NSE (\ref{eq:Hydro_NSE_forced}) to the Euler equation (\ref{eq:Hydro_EuE}) in which we only have inertia terms, which is why the range of wavenumbers $\Kchar\ll k\ll\Knu$ is referred to as the {\it inertial range} (\cite{Frisch95} p.\,86). The transfer of energy to higher wavenumbers in the inertial range is known as {\it energy cascade}.\\

To further formalise the picture of subsequent injection, transfer and dissipation of energy, we define the mean energy injection rate
\begin{align} \label{eq:Hydro_WW_injection}
	\WW(t) = \r\int \vek{f}(\vek{x},t)\cdot\vek{v}(\vek{x},t) \di^3x \;,
\end{align}
and the mean energy dissipation rate as
\begin{align} \label{eq:Hydro_EE_dissipation}
	\EE(t) = \r\int \eps(\vek{x},t) \di^3x
\end{align}
with $\eps(\vek{x},t)$ given by (\ref{eq:Hydro_Eflux_terms_eps}), as suggested by Frisch (\cite{Frisch95} p.\,18ff,\,76ff). We then restrict our considerations to wavenumbers $k\leq K$ which we denote symbolically by an index $K$. For the energy transfer rate through wavenumber $K$ we write $\varPi_K(t)$, we will come back to the explicit form of $\varPi_K(t)$ in a simplified setting in the next chapter.\\
Based on the forced NSE (\ref{eq:Hydro_NSE_forced}) and conservation laws for energy, momentum and helicity, and with $\Ekin$ given by (\ref{eq:Hydro_Ekin}), it is possible to formulate a cumulative energy flux equation in the form (\cite{Frisch95} p.\,25)
\begin{align} \label{eq:Hydro_Ebudget}
	\pt_t {\Ekin}_{,K}(t) = \WW_K(t) - \EE_K(t) - \varPi_K(t) \;.
\end{align}
\remark{$\EE_K(t)$ ist mittlere energydissipation für wellenzahlen $k<K$, also identisch mit $\lla\eps(\vek{x},t)_K\rra$, siehe definition $\lla\rra$ in (2.16) p.18, und (2.23) p.19 und (2.50) p.\,25}The statement of the equation is as follows: In the range of wavenumbers $k\leq K$, the kinetic energy carried by fluid elements is increased by injection of energy and decreased by dissipation and by transfer of energy to wavenumbers $k>K$.\\

We now successively simplify the cumulative energy flux equation by reasonable assumption. First, we restrict the energy injection $\WW_K$ to be time-independent. Consequently, after a sufficient amount of time, the flux of energy through wavenumbers will reach a steady state.\remark{(nicht zu verwechseln mit instabilen steady state für $\vek{v}$)} In the steady state, the kinetic energy content of the flow will be constant for all $K$ and the cumulative energy flux equation becomes
\begin{align} \label{eq:Hydro_Ebudget_SS}
	\WW_K = \varPi_K + \EE_K \;.
\end{align}
Extending the wavenumber $K$ to infinity, no transfer of energy to higher wavenumbers $k>K$ is possible, and we find 
\begin{align} \label{eq:Hydro_Ebudget_SS_inf_beps}
	\WW_\infty = \EE_\infty \dfnrv \beps \;,
\end{align}
where we defined the mean energy dissipation rate $\beps$, which due to energy conservation must equal the mean energy injection rate $\WW_\infty$.\\
For wavenumbers in the inertial range, $\Kchar\ll k\ll\Knu$, we have $\EE_k\approx0$ and $\WW_k=\WW_\infty=\beps$, such that we find for the energy transfer rate
\begin{align} \label{eq:Hydro_Ebudget_SS_Pi-beps}
	 \varPi_k = \beps \sep \text{for } \Kchar\ll k\ll\Knu \;.
\end{align}
This equivalence between average energy transfer and energy dissipation will be useful in the next chapter. However, it should be kept in mind that $\varPi_k$ and $\beps$ are quantities average over the fluid volume of consideration, the {\it local} versions $\varPi_k(\vek{x},t)$ and $\eps(\vek{x},t)$ have to be distinguished.

\cleardoublepage
\section{Fully developed turbulence} \label{s_FDT}
So far, we kept our description on a rather general level. Imposing certain restrictions, an idealised picture of turbulence generation and decay emerges, which is referred to as fully developed turbulence, for which universal features of a statistical description may be expected. It are these universal features this chapter is devoted to.\\
We will begin in the first section with defining the assumptions that lead to fully developed turbulence, together with the conception of a statistical ensemble building the basis to address the phenomenology of fully developed turbulence by statistical means. In section \ref{ss_FDT_energycascade} we come back to the energy flux through wavenumbers discussed in the previous section. The last two sections will give an account on attempts to capture the universal features of fully developed turbulence that have been brought forward in the last 70 years.

\subsection{Homogeneity, Isotropy, Stationarity} \label{ss_FDT_HIS}
Shortly after the generation of turbulence, that is in closed distance to the obstacle, the flow field resembles the symmetries of the obstacle. As the turbulent flow develops further downstream, in a comoving frame the statistics of $\vek{v}(\vek{x},t)$ restores the symmetries of the free NSE. These symmetries are homogeneity and isotropy. The freely decaying turbulence of such a homogeneous and isotropic flow far from boundaries is termed {\it fully developed turbulence}.\\
In addition to homogeneity and isotropy, we also assume that the generation and decay of turbulence has reached a steady state, as discussed in the context of the cumulative energy flux equation (\ref{eq:Hydro_Ebudget}). The additional assumption of stationarity is crucial in constructing the statistical ensemble.\remark{Fully developed turbulence constitutes a clear distinction to general case discussed in the previous chapter. Separate turbulence generation and turbulence decay, only possible for large $\Rey$. Now freely decaying turbulence, no interaction with boundaries. Assume that turbulence generation (energy injection) is completed, like resting fluid that has been stirred. Assume that this turbulent flow field is proceeding downstream, assume time scale separation between inner dynamics and mean velocity (Taylor's hypothesis, like clouds). From now on in moving frame ($w=v-U$). The defining features of fully developed turbulence (FDT) are then homogeneity, isotropy and stationarity (HIS) regarding the statistical properties of flow quantities (in particular velocity increments). Most frequent experimental realisations of turbulence generation by: (i) vicinity of boundaries (wall flows), (ii) nozzles (jets), (iii) bodies (wakes).}

\paragraph{Statistical ensemble} 
An intuitive understanding of fully developed turbulence is conveyed by an idealised experiment: Consider a closed sphere containing a fluid which we violently shake to stir that fluid. After we let the fluid rest shortly for a fixed time, we instantly measure the flow velocity far enough from the wall to preclude boundary effects, and thus yield the velocity field $\vek{v}(\vek{x})$. We repeat the experiment sufficiently often to obtain a good statistics of the random field $\vek{v}(\vek{x})$ in the freely decaying turbulence, each sample at the same stage of the decay. Due to the symmetry of the set-up, the statistics of $\vek{v}(\vek{x})$ will be equal at each position $\vek{x}$ and for each component of $\vek{v}$, it is hence sufficient to measure the magnitude of the velocity along an arbitrary symmetry axis to obtain $v(x)$. This is precisely what is meant by a homogeneous and isotropic flow field.\\
We now transport this picture to the experimental realisations of turbulence generation explained in the previous chapter. The shaking of the sphere is realised by directing the laminar flow at an obstacle. The few seconds rest of the stirred fluid before the measurement is taken corresponds to let the flow evolve downstream and then measure the streamwise flow velocity $v(t)$ as it passes a measuring probe at fixed position. To convert the velocity signal $v(t)$ into a velocity field $v(x)$, we must avail ourselves of an approximation in which we use $x=\lla v(t)\rra (t_\mr{f}-t)$ with $t_\mr{f}$ being the length of the time signal\footnote{The coordinate systems points downstream. Accordingly, the ``end'' of the flow sample passes the probe first.}. The assumption underlying this approximation is that the time scale at which the ``real'' flow field $\vek{v}(\vek{x})$ evolves is sufficiently smaller than the timescale at which the flow field proceeds downstream. In other words, the changes in the flow field are so slow that it practically passes the measuring probe as an entity. This assumption is known as the {\it Taylor hypothesis of frozen turbulence} after G. Taylor \cite{Taylor1935} and has been well tested experimentally (\cite{Frisch95} p.\,58f). The Taylor hypothesis is assumed to hold true for $\lla (v-\lla v\rra)^2 \rra < 0.1\lla v \rra^2$.\\
Finally, the repetition of shaking and measuring to obtain an ensemble of flow field realisations $\{v(x)\}$ is simply achieved by recording one sufficient long velocity signal $v(t)$, rewrite it as a single velocity field $v(x)$ and then chop it into segments of a certain length $L$. In that sense, this experimental realisation continuously generates a turbulence flow, which we group into spheres of size $L$ that proceed downstream. In the comoving frame and at sufficient distance to the position of turbulence generation, each sphere contains a sample of freely decaying, homogeneous and isotropic turbulence. By fixing the position at which we probe the flow velocity and by ensuring that the experiment is in a steady state, we always catch the same stage of decay of the generated turbulence.\\

The above procedure is established practice to obtain $\{v(x)\}$ from a turbulent flow by experimental means. The remaining issue that needs clarification lies in choosing an appropriate value for the chopping length $L$. In the idealised experiment the largest coherent structure in each individual realisation $v(x)$ is of the order of magnitude of the diameter of the sphere. As a measure for the size of the largest coherent structures in the flow, we make use of the normalised autocorrelation function $R(r)=\lla v(x)v(x+r)\rra/\lla v(x)^2\rra$ and choose for $L$ the correlation length
\begin{align}
	L = \int\limits_{0}^{\infty} R(r) \di r
\end{align}
which is known as the {\it integral length scale} \cite{Frisch95,Taylor1935,Renner2001}.\remark{eigentlich müsste ich also immer ne wartezeit von $\lla v(t)\rra\l$ zwischen den segmenten einschieben, damit die einzelnen segmente wirklich unkorreliert sind. Aber da ich ja nur $u(r)$ betrachte, und da tatsächlich genau immer $\l$ zwischen den segmenten übrig bleibt $\l<r$, sind meine trajectorien $\uc$ tatsächlich als uncorreliert anzusehen (theoretisch könnte ich ja auch immer Stücke der Länge $L-\l$ rausschneiden). Anders ausgedrückt, mehr information ist aus den daten nicht rauszuholen. Das $\l\approx\rME$ rechtfertigt im Nachhinein, also nach Jahrzehnten gängiger experimenteller Praxis, diese Vorgehensweise.} Indeed, experiments show that $L$ is in good agreement with the characteristic length $\lchar$ of the external forcing, being the counterpart of the sphere's diameter in the idealised experiment.\\

\begin{figure}[t] 
	\begin{center}
		\includegraphics[width=0.5\textwidth]{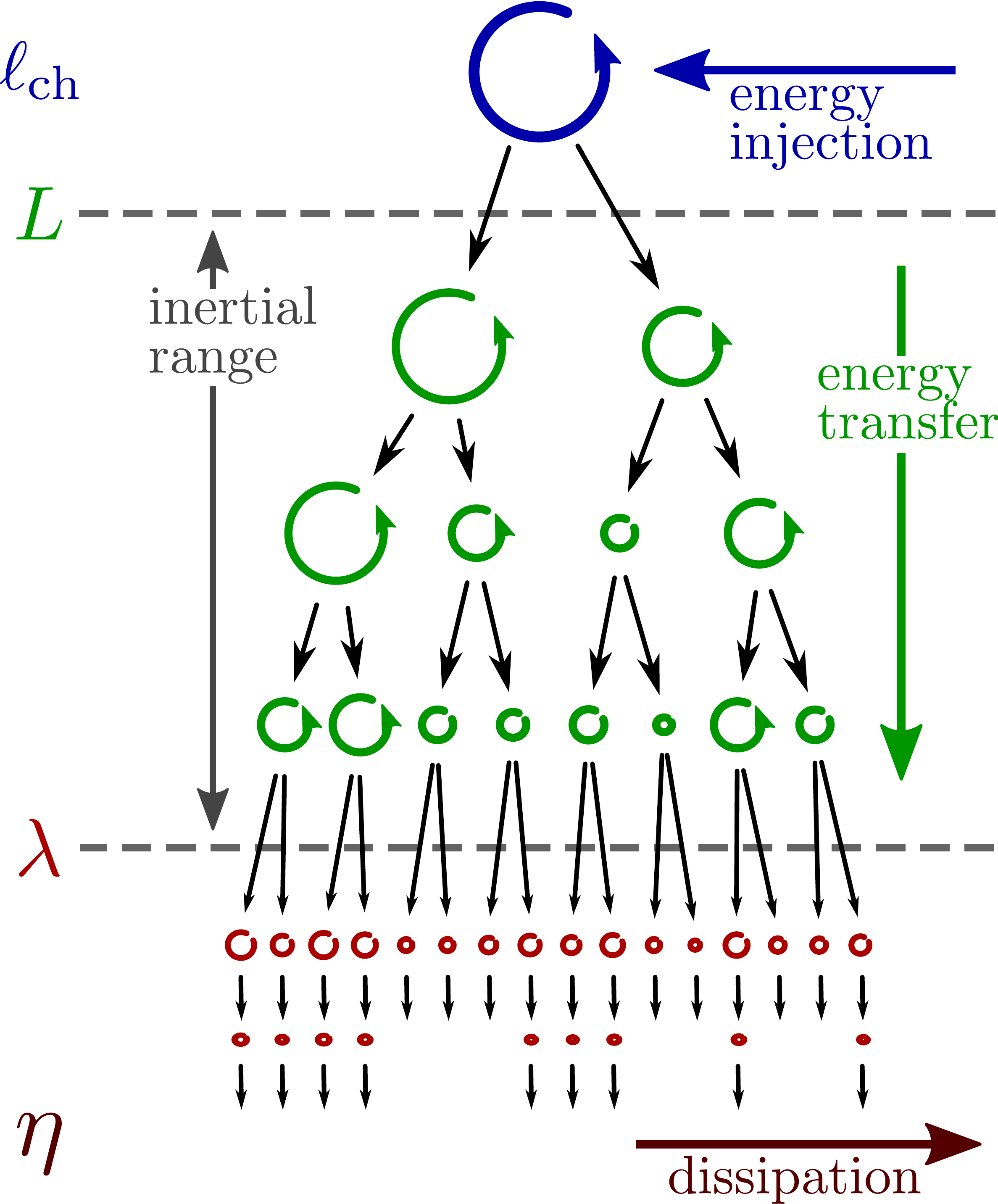}
		\caption{\label{ff:ecascade}\figtxt{Illustration of the Richardson cascade. Energy is injected by creation of eddies of size $\lchar$, by repeated break-up of eddies this energy is transported in the inertial range, until it is eventually dissipated in the dissipative range. The integral length scale $L$ and the Taylor microscale define the range of scale for which effects of energy injection and energy dissipation are negligible. The cascade ceases at the Kolmogorov dissipation scale $\eta$.}}
	\end{center}
\end{figure}

The notion of length scales leads us to revisit the picture of the energy cascade, which we discussed in terms of wavenumbers $k$ in the previous chapter. The different values of the wavenumbers $k$ are linked to eddies of various sizes $r\sim1/k$, and the corresponding amplitudes account for the persistence of the respective eddies in the flow.\\
The energy cascade can then be perceived as follows: The external force acting on the flow results into coherent structures of dimension $L\sim1/\Kchar$ which may be taken as eddies of diameter $L$. Due to the inertia terms in the NSE (\ref{eq:Hydro_NSE}), non-linear interaction causes the break-up of these eddies. As long as molecular friction is negligible, the ensemble of the resulting smaller eddies carry the same amount of energy as the ensemble of larger eddies they arose from. In the inertial range, this break-up of eddies repeats itself and thus transfers energy towards smaller scales. At some threshold $r=\l$, the molecular friction is no longer negligible, and for scales $r<\l$ dissipative effects emerge in addition to the non-linear interaction.\remark{It is a misconception that for scales $r<\l$ the non-linear interactions are negligible (\cite{Frisch95} p.\,92).} While the break-up of eddies continues, molecular friction strengthens until eventually the input of energy from larger scales equals the dissipation of energy at a scale $r=\eta\sim1/\Knu$ (\cite{Frisch95} p.\,91f). For scales $r<\eta$ practically no eddies remain.\\
This phenomenological picture of cascading eddies is known as {\it Richardson cascade} \cite{Richardson22,Kolmogorov62JoFM,Frisch95}\remark{Frisch p. 103}. However, we should keep in mind that the picture of eddies has its origin in the Fourier decomposition of the flow field and therefore is a merely conceptual picture.\\
The threshold $\l$ is referred to as {\it Taylor microscale}, and the scale $\eta$ is known as the {\it Kolmogorov dissipation scale}. The inertial range then is $L>r>\l$,\footnote{By separation of turbulence generation and turbulence decay in experimental realisations, we can mitigate $k\gg\Kchar$} and $\l\geq r\geq\eta$ is known as the {\it dissipative range}. See figure \ref{ff:ecascade} for an illustration of the Richardson cascade.
\\
To define the Kolmogorov dissipation scale $\eta$, we redefine the {\it internal} Reynolds number as function of the scale $r$,
\begin{align} \label{eq:FDT_Rey_r}
	\Rey_r \dfn \frac{r\,v_r}{\nu} \;,
\end{align}
where $v_r$ denotes the typical velocity within eddies of size $r$. The {\it integral} scale Reynolds number $\Rey_L=Lv_L/\nu$ with $v_L=\lla v(t)\rra$ is then of the same order of magnitude as the Reynolds number $\Reych=\lchar\vchar/\nu$ defined in terms of characteristic quantities of the external forcing in (\ref{eq:Hydro_Rey}).\remark{In fact, $\Rey_L$ is expected to be slightly smaller than $\Reych$, as in general $\lla v(t)\rra<\vchar$ due to blockage of the incoming flow. In that sense, $\Reych$ is associated with the forcing and $\Rey_L$ with the largest coherent structures of the flow.} The scale where the dissipation of energy is in balance with input of energy from higher scales has an internal Reynolds number of one, we therefore define $\nu$ via $\Rey_\eta=\eta v_\eta/\nu=1$ (\cite{landau1987fluid} p.\,134).\\
The Taylor microscale $\l$, which separates the inertial range $L>r>\l$ from the dissipative range $\l\geq r\geq\eta$, is defined by expanding the normalised autocorrelation function for small $r$ \cite{Taylor1935},
\begin{align}
	R(r) &= R(0) + \frac{r^2}{2}\frac{\pt^2 R(r)}{\pt r^2}\Big|_{r=0} + \dots \;,
\end{align}
where the first order term vanishes because $R(r)$ is even, and then set $\l$ such that $R(\l)=1/2$,\remark{teilt also in näherung kleiner $r$ den bereich $R=1$ ($r=0$) und $R=0$ ($r=L$) bei der $R$ auf die Hälfte angestiegen ist (von $r=L$ zu kleinen Skalen ausgehend)}\remark{There are various formulae to determine $\l$ from measured $v(t)$ which build on this definition, e.g. [Aronson,Löfdahl], from which we refrain to report here.}
\begin{align} \label{eq:FDT_def_lambda}
	\frac{1}{\l^2} \dfn - \frac{\pt^2 R(r)}{\pt r^2}\Big|_{r=0} \;.
\end{align}
The definition of the scales $L$ and $\l$ fixing the inertial range allows to formulate the inertial range condition in terms of the scale dependent Reynolds number $\Rey_r$ as
\begin{align} \label{eq:FDT_in-range_Rey}
	\Reych\gg\Rey_L>\Rey_r>\Rey_\l\gg1 \;,
\end{align}
where $\Rey_r$ is defined in (\ref{eq:FDT_Rey_r}).

Before we proceed with the explicit statistical analysis of $\{v(x)\}$, we corroborate the assumption of homogeneity, isotropy and stationarity by the symmetries of free NSE without an external force term\footnote{If the forcing respects these symmetries, then of course also the forced NSE (\ref{eq:Hydro_NSE}) exhibits the symmetries.} (\ref{eq:Hydro_NSE}). We express the symmetries in terms of invariances with respect to 
\begin{subequations} \label{eq:FDT_NSEinvar}
	\begin{alignat}{3}
		&\text{space-translations}\quad &&\vek{x} &&\;\mapsto\; \vek{x}+\vek{x}' \;, \label{eq:FDT_NSEinvar_x}\\
		&\text{time-translations}\quad &&t &&\;\mapsto\; t+t' \;,  \label{eq:FDT_NSEinvar_t}\\
		&\text{rotations}\quad &&(\vek{x},\,\vek{v}) &&\;\mapsto\; (\mat{A}\vek{x},\,\mat{A}\vek{v})  \;, \label{eq:FDT_NSEinvar_vec}\\
		&\text{Galilean transformation}\quad &&(t,\,\vek{x},\,\vek{v}) &&\;\mapsto\; (t,\,\vek{x}+\vek{v}'t,\,\vek{v}+\vek{v}') \;, \label{eq:FDT_NSEinvar_Gal}\\
		&\text{scaling (for $\Reych\to\infty$)}\quad &&(t,\,\vek{x},\,\vek{v}) &&\;\mapsto\; (c^{1-h}t,\,c\vek{x},\,c^h\vek{v}) \;, \label{eq:FDT_NSEinvar_scal}
	\end{alignat}
\end{subequations}
where $\mat{A}$ is a rotation matrix, $h$ a positive and $c$ an arbitrary real number (\cite{Frisch95} p.\,17).\remark{(bei scaling invariance muss nicht $3h$ an $\vek{r}$, damit $v\sim r^{1/3-(D-3)/3}$ gilt (siehe $\b$-model und notebook p.64).)}\\
In general, a forcing of the flow breaks some or all of these symmetries. But due to the chaotic nature of the turbulent flow, the forcing looses quickly its influence and the above symmetries are restored as the flow proceeds downstream. In fact, the flow continues to be forced by integral scale motion of the fluid, but this internal forcing adopts the symmetries of the unforced NSE. In that sense, at the instance of turbulence generation, the integral scale Reynolds number equals the characteristic Reynolds number, $\Rey_L=\Reych$, and decreases downstream due to turbulence decay. Hence, at sufficient distance to the turbulence generation and far from boundaries, it is justified to assume homogeneity and isotropy for the statistics of $\{\vek{v}(\vek{x})\}$, as long as $\Reych$ is large enough to guarantee a distinct regime to turbulence generation in which $\Rey_L\gg1$.\remark{es gilt auch noch parity: vorzeichenwechsel von $r$ und $v$, aber ist ja eigentlich schon in space- und rotation-invariance enthalten.}

\paragraph{Velocity increments} With the above discussion we have established the conceptual basis for the statistical analysis of $\{v(x)\}$. In most of the cases, the statistical analysis is carried out either in terms of
the spatial correlation tensor (\cite{landau1987fluid} p.\,140ff)
\begin{align} \label{eq:FDT_corr-fct}
	R_{ik}(\vek{r};\,\vek{x},t) = \lla\, v_i(\vek{x},t)\,v_k(\vek{x}+\vek{r},t) \,\rra
\end{align}
or by inspecting the moments of velocity increments projected on a unit vector $\vek{e}$ (\cite{Frisch95} p.\,57ff)
\begin{equation} \label{eq:FDT_moments}
 \Str^n(\vek{r};\,\vek{x},\vek{e},t) = \lla \big(\vek{e}\vek{v}(\vek{x} \+ \vek{r},t) - \vek{e}\vek{v}(\vek{x},t)\big)^n \rra \;,
\end{equation}
where the angular brackets denote an ensemble average over the samples $\{\vek{v}(\vek{x},t)\}$.\\
We begin with a short discussion of the correlation tensor and demonstrate that both descriptions are in fact equivalent, and then focus on the analysis of velocity increments throughout the remaining part of this thesis.\\
The tensor $R_{ik}(\vek{r};\,\vek{x},t)$ can be diagonalised by choosing the coordinate system such that $\vek{r}$ and the first component of $\vek{x}$ point in the same direction. The diagonal entry $R_{11}$ is then the {\it longitudinal} autocorrelation function in the direction of $\vek{r}$, and $R_{22}$ and $R_{33}$ are the {\it transversal} autocorrelation functions in the two perpendicular directions to $\vek{r}$. Owing to the assumed stationarity, homogeneity and isotropy of the statistics of the velocity field $\vek{v}(\vek{x},t)$, we can restrict the arguments of $R_{ik}(\vek{r};\,\vek{x},t)$ to the absolute value of $\vek{r}$ and drop the dependency on $\vek{x}$ and $t$ and simply write $R_{ik}(r)$.\\
Denoting the longitudinal component by $R_{\ell\ell}(r)$ and the two transversal components by $R_{tt}(r)$ and making use of the continuity equation, it can be shown that the longitudinal and transversal components are not independent but related by (\cite{landau1987fluid} p.\,142)
\begin{align} \label{eq:FDT_corr-fct_tt-rr}
	R_{tt}(r) = \frac{1}{2r}\frac{\dd\big(r^2R_{\ell\ell}(r)\big)}{\dd r} \;.
\end{align}
It is therefore sufficient to analyse either the longitudinal or the transversal component. In view of the experimental procedure described above, the longitudinal component is the most common choice.\\
The same argumentation applies to the moments of the velocity increments, we therefore restrict ourselves to the longitudinal moments and write
\begin{equation} \label{eq:FDT_struc-fct}
	\Str^n(r) = \lla \big(\vek{e}\vek{v}(\vek{x} \+ r\vek{e},t) -  \vek{e}\vek{v}(\vek{x},t)\big)^n \rra \;.
\end{equation}
The underlying longitudinal velocity increments
\begin{equation} \label{eq:FDT_vel_incr}
	u(r;\,\vek{x},\vek{e},t) \dfn \vek{e}\,\big(\vek{v}(\vek{x} \+ r\vek{e},t) -  \vek{v}(\vek{x},t)\big)
\end{equation}
of course still depend on $\vek{x}$, $\vek{e}$ and $t$ since stationarity, homogeneity and isotropy is only valid in the statistical sense, but for the sake of clarity we simply write $u(r)=u(r;\,\vek{x},\vek{e},t)$.\\
Denoting by $v_\ell(\vek{x})=\vek{e}\vek{v}(\vek{x},t)$ the longitudinal velocity component, we exemplify the equivalence of moments and correlation functions by means of the structure function of second order
\begin{align}
	\mfrac{1}{2}\Str^2(r) &= \mfrac{1}{2}\lla \big(v_\ell(\vek{x} \+ r\vek{e}) -  v_\ell(\vek{x})\big)^2 \rra \nn
	&= \mfrac{1}{2}\lla v_\ell(\vek{x} + r\vek{e})^2 - 2v_\ell(\vek{x} \+ r\vek{e})v_\ell(\vek{x}) + v_\ell(\vek{x})^2\rra \nn
	&= \lla v_\ell^2\rra - \lla v_\ell(\vek{x} \+ r\vek{e})v_\ell(\vek{x})\rra \nn
	&= \lla v_\ell^2\rra - R_{\ell\ell}(r) \;. \label{eq:FDT_S2-to-R}
\end{align}
The relation between moments and correlation functions of higher order can be derived along similar lines (\cite{Rennerdiss} p.\,20).\\

The $n$-th moment of velocity increments is usually referred to as the {\it structure functions of order $n$}\footnote{We will also say the $n$-th (order) structure function.}, indicating that the velocity increments address the spatial structure of the velocity field.\\
To obtain the structure functions from the ensemble $\{v(x)\}$, we simply apply the one-dimensional equivalent of (\ref{eq:FDT_vel_incr}),
\begin{equation} \label{eq:FDT_vel_incr_1D}
	u(r) = v(x \+ r) - v(x) \;,
\end{equation}
to $\{v(x)\}$ for scales in the inertial range, $r=L..\l$, and end up with an ensemble $\{u(r)\}$. Evaluating the ensemble average in the one-dimensional version of (\ref{eq:FDT_struc-fct}) finally yields the structure functions,
\begin{equation} \label{eq:FDT_struc-fct_1D}
	\Str^n(r) = \lla \big(u(r)\big)^n \rra \;.
\end{equation}
Note that by keeping $r>\l$, we can approximately assume that the samples $\{u(r)\}$ are sufficiently uncorrelated, as $\l$ was defined in (\ref{eq:FDT_def_lambda}) as the distance in $v(x)$ for which correlations have considerably subsided. By taking only every second sample of $\{v(x)\}$ to obtain the ensemble $\{u(r)\}$, we would practically rule out any correlations, but it would also reduce the sample size by a factor two. In that sense, keeping $r>\l$ is a trade-off between uncorrelated samples and sample size.\remark{Stationarity follows automatically when stirring force is time independent (incoming laminar flow, resting obstacle), and waiting until steady state has established.}\remark{The consequence of isotropy in that way is not that all odd structure functions are zero but that $\Str^n(r)$ is odd in $r$ for odd $n$, and even in $r$ for even $n$. In fact, Kolmogorov's four-fifths law states that $\Str^3(r)$ is negative and not zero. details in [castaing] (chapter 3.2).}

\subsection{Energy cascade} \label{ss_FDT_energycascade}
The energy cascade, that is the transfer of energy towards smaller scales due to repeated break-up of turbulent structures, is the central mechanism in fully developed turbulence. Having discussed the energy flux for general turbulent flows in the first chapter, we will now address the energy cascade in terms of structure functions.\\

\paragraph{The four-fifths law} Recall the cumulative energy flux equation in (\ref{eq:Hydro_Ebudget}). In the subsequent discussion we found for a stationary energy flux a constant energy transfer rate $\varPi_k\equiv\beps$ in the inertial range, where $\beps$ is the mean energy dissipation rate. In the case of isotropic and homogeneous turbulence, it can be shown that $\varPi_k$ can also be expressed in terms of the third order structure function (\cite{Frisch95} p.\,81),
\begin{align} \label{eq:FDT_PiK_S3}
	\varPi_k = -\frac{1}{6\pi} \int\limits_{0}^{\infty} \frac{\sin(kr)}{r}\Big[(1+r\pt_r)(3+r\pt_r)(5+r\pt_r)\Big]\frac{\Str^3(r)}{r} \di r \;.
\end{align}
Hence, we expect a relation between $\Str^3(r)$ and $\beps$, which we are going to derive now.\\
To this end, we define the auxillary function
\begin{align}
	f(r) \dfn \Big[(1+r\pt_r)(3+r\pt_r)(5+r\pt_r)\Big]\frac{\Str^3(r)}{6\pi r}
\end{align}
and use the dimensionless integration variable $z\dfn kr$ to rewrite (\ref{eq:FDT_PiK_S3}) as
\begin{align}
	\varPi_k = -\int\limits_{0}^{\infty} \frac{\sin(z)}{z}\,f(z/k)\di z \;.
\end{align}
On the one hand, the amplitude of the oscillating function $\sin(z)/z$ rapidly decreases for increasing $z$, on the other hand, the argument $r=z/k$ of $f(z/k)$ changes only slowly for large $k$. In the limit of infinite Reynolds numbers, we can extent $k$ to arbitrary large values without leaving the inertial range. We therefore take $f(z/k)$ as constant for the $z$ values that contribute to the integral and demand (\cite{Frisch95} p.\,84ff)\remark{frisch ist ja eigentlich im $\nu\to0$ grenzfall, also $\Reych\to\infty$, und braucht deswegen die annahme, dass $\beps(\nu)$ endlich für $\nu\to0$ bleibt.}
\begin{align} \label{eq:FDT_4/5_ODE}
	f(r) \approx -\frac{2}{\pi}\,\beps
\end{align}
to satisfy $\varPi_K=\beps$. The above condition constitutes an ODE for $\Str^3(r)$ and its solution reads
\begin{align}
	\Str^3(r) = -\mfrac{4}{5}\,\beps\,r + c_1r^{-a_1} + r^{-a_2}\big[c_2\cos(a_3\ln r)+c_3\sin(a_3\ln r)\big]\,,
\end{align}
where $a_1\approx0.75$, $a_2\approx1.1$, $a_3\approx2.0$ and $c_1$, $c_2$, $c_3$ are integration constants. In the limit $r\to0$, the energy transfer rate must vanish, we therefore set all integration constants to zero and are left with
\begin{align} \label{eq:FDT_four-fifths}
	\Str^3(r) = -\frac{4}{5}\,\beps\,r \sep \text{for } \l<r<L \;.
\end{align}
This relation is known as Kolmogorov's {\it four-fifths law} \cite{Kolmogorov1941c} and is one of the most important results for fully developed turbulence. For homogeneous and isotropic turbulence and in the limit of infinite Reynolds numbers, it is an exact\footnote{The approximation of the integral (\ref{eq:FDT_PiK_S3}) becomes exact for $k\to\infty$} implication of the NSE (\ref{eq:Hydro_NSE_forced}) without any further assumptions (\cite{Frisch95} p.\,76ff). For sufficient high Reynolds numbers, the four-fifths law still holds in the inertial range which is often used to identify the inertial range in experimental data.\\
The fact that (\ref{eq:FDT_four-fifths}) is an exact result of the NSE awards the four-fifths law a special role: Every model of fully developed turbulence must obey (\ref{eq:FDT_four-fifths}) in the limit $\Reych\to\infty$, or violate the assumptions of homogeneity and isotropy.\\
In his original derivation Kolmogorov actually found \cite{Kolmogorov1941c}
\begin{align} \label{eq:FDT_KHK}
	\Str^3(r) + \frac{4}{5}\,\beps\,r = 6\nu\frac{\pt \Str^2(r)}{\pt r}
\end{align}
which is known as K\`{a}rm\`{a}n-Howarth-Kolmogorov equation, as Kolmogorov built on results by von K\`{a}rm\`{a}n and Howarth (\cite{landau1987fluid} p.\,140). This equation also holds for $r<\l$. For $r$ in the inertial range, the second term is negligibly small  (\cite{landau1987fluid} p.\,145), and we reproduce the four-fifths law.\\
Solved for $\beps$, (\ref{eq:FDT_KHK}) is a scale-resolved balance equation for the total energy dissipation, in which the term with $\Str^3(r)/r$ accounts for local energy transfer and $1/r\pt_r\Str^2(r)$ for local energy dissipation. Augmented with a term accounting also for energy injection, it is possible to examine the main contributions to the energy cascade for all scales $r$, see for instance the comprehensive article \cite{Mouri2006}. For completeness, we mention that (\ref{eq:FDT_KHK}) can be generalised to anisotropic turbulence (\cite{Frisch95} p.\,77).\\
The proportionality between $\Str^3(r)$ and $\beps$ has another implication: The signature of energy dissipation in the dissipative range is a negative skewness of the velocity increments. This connection might at first seem surprising, but it can qualitatively explained as follows. A negative skewness implies that the relative flow velocities at positions $x+r$ and $x$ point more frequently in the opposed direction than in the same direction. This unbalance is assumed to be a signature of deforming turbulent structures; due to molecular friction, deformation goes with irreversible heat dissipation at small scales, slowing down the fluid motion. It is this irreversibility that makes the odds of observing the reversal of a compressing deformation smaller than the compression itself, which causes ultimately the excess of negative velocity increments in the statistics of $\{\vek{v}\}$. In the literature, the keyword that goes with this picture is {\it vortex stretching} (see, e.g., \cite{Frisch95} p.\,156 or \cite{Castaing1990} chapter 3.2).\remark{vortex stretching heißt dann konzeptionell, dass ein kreisförmiger wirbel zusammengedrückt wird, und dadurch in die länge gezogen wird (stretched). dabei wird ``im innern'' des wirbels durch reibung wärme dissipiert, und das ``rückwärts-stretchen'' wird unwahrscheinlicher. auch wenn die wärme nur auf kleinen skalen dissipiert, schlägt sich das eben doch auf die dynamik der größeren strukturen nieder.}\remark{As a signature of irreversibility, the negative skewness allows a trained eye to decide whether a freely decaying, homogeneous and isotropic velocity field $\vek{v}(\vek{x},t)$ at an instance of time $t$ is a snapshot of the forward or backward evolution of $\vek{v}(\vek{x},t)$ [G. Falkovich and K. Sreenivasan, Phys. Today 59, 43 (2006)].}\remark{die skewness steht nicht im widerspruch zur homogenität und isotropie des {\it geschwindigkeits}feldes, denn das heißt ja nur $\lla v(x)\rra_x=0$ und $\lla\vek{v}\rra_{\vek{e}}=0$, aber die höheren momente der geschwindigkeits{\it inkremente} verschwinden eben nicht $\Str^3(r) = 
\lla v(x)^2 v(x+r) \rra - \lla v(x) v(x+r)^2 \rra < 0$. das zeigt auch, dass two-point closure fatal ist wenn $\Str^3(r) = \lla v^2\rra \lla v\rra  - \lla v\rra\lla v^2 \rra = 0$}\\
To verify the four-fifths law by experimental means, we need to determine also $\beps$ from experimental data. Combining (\ref{eq:Hydro_Eflux_terms_eps}) with (\ref{eq:Hydro_EE_dissipation}) and (\ref{eq:Hydro_Ebudget_SS_inf_beps}), we find as an exact result of the NSE
\begin{align}
	\beps = \r \int \frac{\nu}{2}\sum\limits_{ik} \bigg(\frac{\pt v_i(\vek{x})}{\pt x_k}+\frac{\pt v_k(\vek{x})}{\pt x_i}\bigg)^{\!2} \di^3x \;.
\end{align}
Owing to homogeneity and isotropy this formula simplifies to the one-dimensional surrogate (\cite{landau1987fluid} p.\,142f)
\begin{align}
	\beps = \r\nu \int \sum\limits_{i} \bigg(\frac{\pt v_i(\vek{x})}{\pt x_\ell}\bigg) \di^3x
	\;\approx\; 15\nu \,\bigg\langle\,\bigg(\frac{\pt v(x)}{\pt x}\bigg)^{\!\!2}\,\bigg\rangle \;,\label{eq:FDT_eps_surrogate}
\end{align}
where the last expression corresponds to an one-dimensional cut through the three-dimensional flow field (\cite{Rennerdiss} p.\,22). We mention that the reliability of the one-dimensional surrogate is under scrutiny \cite{Stolovitzky1998}.\\

\paragraph{Inverse energy cascade} The energy dissipation rate $\eps(\vek{x})$ has to be distinguished from the fluctuating rate of energy transfer throughout the cascade. Whereas $\eps(\vek{x})$, as defined in (\ref{eq:Hydro_Eflux_terms_eps}), is always positive\remark{no transformation of heat into kinetic energy}, negative fluctuations of the energy transfer rate are known to be possible \cite{Kraichnan1974,Mouri2006}. On average, of course, we retain a positive energy transfer that must equal the energy dissipation. A negative energy transfer rate corresponds to an inverse energy cascade, that is to say, turbulent structures combine to larger structures and hence transfer energy to larger scales.\\
In two dimensional turbulence \cite{Boffetta2012}, the inverse energy cascade is known to be the dominant mechanism of energy transfer\footnote{in addition to transfer of energy from forced scales to large scales comes a vorticity cascade from forced scales to small scales.} which is attributed to the natural alignment of eddies in the same plane, as vortices of equal orientation will coalesce \cite{Kraichnan1967,Friedrich2012}; however, the precise mechanism responsible for the inverse energy cascade is subject to current research \cite{Chertkov2007,Xiao2008,Cencini2011}. Under flow conditions that are reminiscent of two-dimensional turbulence, an inverse energy cascade is also observed in three-dimensional turbulence \cite{Francois2013,Marino2013} and is of great importance for geophysical fluid dynamics \cite{Xiao2008}\remark{oceanic and atmospheric flows, popular theory to explain strong jets in oceans and atmospheres (large scale zonal flow)}. Recent results demonstrate that even full three-dimensional, homogeneous and isotropic turbulence always features a subset of non-linear evolution responsible for an inverse energy cascade \cite{Biferale2012,Lopez-Caballero2013a}.\remark{p.83 frisch: Note that the spectral energy transfer rate at wavenumber $k$ follows by differentiation of $\varPi_K$ with respect to $K$,
\begin{align} \label{eq:FDT_Pik_S3}
	T(k) = \pt_k\varPi_k = -\frac{1}{6\pi} \int\limits_{0}^{\infty} \cos(kr)\Big[(1+r\pt_r)(3+r\pt_r)(5+r\pt_r)\Big]\frac{\Str^3(r)}{r} \di r
\end{align}
Could test $T(k)$ for the extreme trajectories w.r.t. $\Sm$ by using instances of $\uc^3$ instead of $\Str^3$, but numerically difficult due to high derivatives.
Note that energy spectrum $E(k)$ also follows by $E(k)=\pt_k {\Ekin}_{,k}$ (for which Kolmogorov's famous $-5/3$ law applies - also doch hier einfach volle gleichung nochmal hinschreiben?).}\\

\subsection{Scaling laws} \label{ss_FDT_scalinglaws}
We now turn to the universal features of turbulence, which manifest in scaling laws of the structure functions. The theory of fully developed turbulence has benefited a lot from dimensional analysis initiated by the work of Kolmogorov \cite{Kolmogorov1941a,Kolmogorov1941b,Kolmogorov1941c} and Obhukov \cite{Obukhov1941a,Obukhov1941b} in 1941. At this point, we will give a short survey of resulting scaling relations which will lead us to scaling laws proposed by Kolmogorov and others.

\paragraph{Dimensional analysis} Starting point of the dimensional analysis is to educe from the phenomenological picture of the Richardson cascade the physical quantities that affect the statistics of velocity increments $u(r)$ in the inertial range. We can rule out viscosity $\nu$, and characteristic length scales and velocities associated with turbulence generation, as per definition effects of energy injection and dissipation are negligible in the inertial range. The only relevant energetic quantity in the inertial range is the energy transfer rate, which on average equals the mean energy dissipation rate $\beps$. Hence, we are left with $\beps$, $r$ and $u$ as the relevant quantities in the inertial range.\remark{ich kann ich ja auch nur $\beps$ nehmen, weil ich an der overall mittleren energie-transfer-rate interessiert bin, und die ist ja im SS gleich injection gleich dissipation. Und eigentlich könnte ich dann auch sagen, ich nehm $\nu$ statt $\beps$, nur dass dann alles komplizierter wird.} As $u$ itself is zero on average, we inspect the second moment of $u$ and find that, to respect the dimensions of the involved quantities, it can only be expressed in terms of $\beps$ and $r$ by
\begin{align} \label{eq:FDT_scals_two-thirds}
	\lla u^2 \rra \sim \beps^{2/3}r^{2/3} \;.
\end{align}
The above relation is known as Kolmogorov's {\it two-thirds law} \cite{Kolmogorov1941a}.\remark{, which he derived in his first paper from the year 1941. In his third paper he was even able to determine the constant of proportionality from the NSE and found $\Str^2=1/15\beps/\nu r^2$ [41c,LL p.143]. Aber wahrscheinlich stimmt da was nicht mit, sonst hätte four-fifths nicht mehr son alleinstellungsmerkmal, und ich seh grad nicht, wie das mit obigem two-thirds zusammenpasst.}\\
Using (\ref{eq:FDT_S2-to-R}), we can extract the $r$-dependency of the autocorrelation function, $R(r) \sim \const-r^{2/3}$, from which we obtain by means of the Fourier transformation for the energy spectrum
\begin{align} \label{eq:FDT_scals_five-thirds}
	E(k) = \frac{1}{2\pi}\int \ee{ikr} R(r) \di r \sim \beps^{2/3} k^{-5/3} \frac{1}{2\pi}\int \ee{iy} \di y \sim \beps^{2/3} k^{-5/3} \;,
\end{align}
a prediction which proved to be particular well verified in experiments. In addition, Kolmogorov postulated that the constant of proportionality in (\ref{eq:FDT_scals_two-thirds}) is universal and in his third paper from 1941 determined it to be $3/2$ using experimental measurements \cite{Kolmogorov1941c}. The constant of proportionality is now known as {\it Kolmogorov constant} $C_2$ and experimental results show that $C_2\approx1.9\pm0.2$ \cite{Hosokawa1996, Ishihara2009, Chien2013}.\footnote{The Kolmogorov constant is often also defined as the constant of proportionality in the five-thirds law and then takes the value $0.53\pm0.01$ \cite{Sreenivasan1995}.}\\
Applying the same dimensional arguments to the mean energy dissipation rate we find
\begin{align} \label{eq:FDT_scals_four-fifths}
	\beps \sim \frac{u^3}{r} = \frac{u^2}{\t}
\end{align}
which is in agreement with the four-fifths law. The second relation follows by defining the {\it turn-over time} $\t\dfn r/u(r)$ which is the characteristic time scale within eddies of size $r$, suggesting that $\beps$ is the fraction of the kinetic energy $u^2$ available for energy transfer on scale $r$.\\
Finally, from analogous argumentation and combinations of above findings, we find for the Kolmogorov dissipation scale (\cite{Frisch95} p.\,91, \cite{landau1987fluid} p.\,138)
\begin{align} \label{eq:FDT_scals_eta}
	\eta\sim\Big(\frac{\nu^3}{\beps}\Big)^{1/4} \;,
\end{align}
and that the relative size of the inertial range scales with the square root of the Reynolds number\remark{aus meiner scaling-matrix in maple (einfach auch LL p.138 citieren)},
\begin{align} \label{eq:FDT_scals_Rey}
	\Reych \sim \bigg(\frac{L}{\l}\bigg)^2 \qquad \text{or} \qquad \Reych \sim \bigg(\frac{L}{\eta}\bigg)^{4/3} \;.
\end{align}

\paragraph{Kolmogorov scaling}
Observing a turbulent flow conveys an impression of {\it self-similarity}, that is to say, structures observed on a large scale $L$ recur on smaller scales $r$ in a similar fashion. Kolmogorov was led by the perception of cascading eddies proposed by Richardson \cite{Richardson22}, to consider, in the statistical sense, self-similarity for turbulent structures like eddies \cite{Kolmogorov1941a}. As the structure functions $\Str^n(r)$ are a statistical measure for structures in the flow, it stands to reason to postulate in the limit of infinite Reynolds numbers\remark{damit wir unten das four-fifths nehmen können um $h=1/3$ zu finden.} a {\it scaling law} of the form
\begin{align} \label{eq:FDT_self-sim}
	\Str^n(r) = \big(\mfrac{r}{L}\Big)^{\z_n} \Str^n(L) \propto r^{\z_n}
\end{align}
with {\it scaling exponent} $\z_n$ to hold for all $r<L$. In words, the $n$-th moment of velocity increments on scale $L$ has only to be multiplied with the scaling factor $(r/L)^{\z_n}$ to obtain the $n$-th moment on scale $r$.\\
In the simplest case, the scaling exponents account for the spatial dimensionality of the structures to be rescaled: The length of a line on scale $L$ is simply multiplied with $r/L$ to be reproduced on the smaller scale $r$, a surface needs to be multiplied with $(r/L)^2$, a volume with $(r/L)^3$. To obtain $\z_n$ for the structure functions in (\ref{eq:FDT_self-sim}), we take from (\ref{eq:FDT_scals_four-fifths}) the scaling of velocity increments as $u\sim(\beps r)^{1/3}$. Consequently, the scaling law (\ref{eq:FDT_self-sim}) becomes
\begin{align}\label{eq:FDT_K41}
	\Str^n(r) \sim \lla(\beps r)^{n/3}\rra = \beps^{\,\z_n}\,r^{\,\z_n} \sep \z_n = \mfrac{n}{3} \qquad \text{(K41)}
\end{align}
in accordance with Kolmogorov's works from the year 1941 (\cite{Frisch95} p.\,89ff]; we will therefore refer to this scaling as {\it K41 scaling}.\footnote{Kolmogorov discussed in \cite{Kolmogorov1941c} only the structure functions of second and third order but seemed to be aware of the possible extension to arbitrary orders (\cite{Frisch95} p.\,97).} Note that being a result of the dimensional analysis above, the K41 scaling $\Str^n(r)\propto r^{n/3}$ reproduces the two-thirds law (\ref{eq:FDT_scals_two-thirds}) and the four-fifths law (\ref{eq:FDT_scals_four-fifths}).\\
Since the scaling $u\sim(\beps r)^{1/3}$ coincides with the scale-invariance property of the NSE (\ref{eq:FDT_NSEinvar_scal}) for choosing $h=1/3$, the scaling (\ref{eq:FDT_K41}) should hold universally in the limit of infinite Reynolds numbers. Kolmogorov claimed furthermore that in this limit also the constant of proportionality in his two-thirds and four-fifths law are universal. Concerning this claim, Landau objected that $\eps(\vek{x})$ is a strongly fluctuating quantity with considerable effects in the inertial range, the use of its average $\beps$ in (\ref{eq:FDT_K41}) and the claim of its universality is therefore not justified (\cite{landau1987fluid} p.\,143, \cite{Frisch95} p.\,93ff, \cite{Mouri2006}).\remark{Kolmogorov hat gesagt dass im four-fifths und im two-thirds jeweils eine universelle konstante vor dem $r$ steht ($\beps$ einschließend) [41a,41c], was landau dann objected hat ist das $\beps$ nicht universal sein kann, weil die statistik von $\eps(\vek{x})$ nicht für alle beliebige strömungen anwendbar ist, also auch $\beps$ nicht universal sein kann.[LL p143, footmark] das mit den fluktuationen hat kolmogorov dann eingesehen und K62 erfunden, und die universalität hat er des Vorfaktors hat er wohl aufgegeben (ist in [62] jetzt $C(\vek{x})$ und hängt von der makrostruktur der strömung ab). frisch meint, er hat wohl auch nicht K62 aufgrund von Landau gemacht, sondern kam wegen seinen arbeiten an pulverisierung von eisenerz selbst drauf, die er mit kaskaden in verbindung gebracht hat. nachdem obukhov 1962 das bemerkte und log-normale fluktuationen von $\beps$ vorschlug, griff Kolmogorov das auf und machte K62 (frisch 178f). Die etwas uneindeutige referenz zu Landau in K62-paper entstand wohl erst, nachdem das paper quasi fertig war, nachdem Yaglom ihn auf Landaus bemerkung aufmerksam machte (frisch 178). in meinem k62 paper ist es alles andere als uneindeutig, aber das ist wohl auch ne übersetzung die aus nem vortrag hervorging, also nicht original... oh, und experimente von Mouri2006 und Sreenivasan1995 zeigen tatsächlich dass Kolmogorovs konstante universal zu sein scheint, aber $\beps$ trotzdem in größenordnung seines wertes fluktuiert (was sich dann aber nur in den höheren wirklich niederschlägt), also beide irgendwie recht haben.} However, as already mentioned above, the constants of proportionality appear indeed to be universal \cite{Sreenivasan1995,Mouri2006}.\\

It took 21 years until Kolmogorov resumed the question of an universal scaling law and incorporated fluctuations of $\eps(\vek{x})$ into the scaling ansatz (\ref{eq:FDT_self-sim}) \cite{Kolmogorov62JoFM}. He based his derivation on the suggestion by Obukhov \cite{Oboukhov1962} to use a local average of the energy dissipation rate,
\begin{align} \label{eq:FDT_K62_epsr}
	\eps_r(\vek{x}) \dfn \frac{3}{4\pi r^3} \int\limits_{|\vek{x}-\vek{x'}|\leq r} \eps(\vek{x'}) \di^3x' \;,
\end{align}
instead of the overall mean $\beps$. The idea is again that the local energy dissipation $\eps_r(\vek{x})$ on scale $r$ also fixes the energy transfer rate on that scale. To include the fluctuations of $\eps_r(\vek{x})$ into the analysis, Obukhov argued further that $\eps_r(\vek{x})$ should follow a log-normal distribution due to the following phenomenological picture (\cite{Frisch95} p.\,171ff, \cite{Rennerdiss} p.\,24ff):\\
In view of the Richardson cascade, the energy transfer towards smaller scales is realised by the break-up of eddies. Let us single out one of the smaller eddies on scale $r$. Its energy is a fraction $h_1$ of the energy of the larger eddy it emerged from. This larger eddy itself received a fraction $h_2$ of energy from an even larger eddy, and so on. We can therefore trace back that the energy of our small eddy on scale $r$ has derived from the energy $\eps_0$ of a mother eddy. Thus, the energy that is to be dumped on scale $r$ can be written as $\eps_r(\vek{x})=\eps_L(\vek{x})\mscale[0.85]{\prod}_j h_j$.\remark{das ist ganz schön subtle. ich will die energie-transfer-rate, die ist aber schwer zu kriegen, also nehm ich die dissipations-rate. je nach weg den die kaskade nahm, fluktuiert diese aber stark. um also auch die fluktuationen der transfer-rate zu berücksichtigen, adoptieren wir auch die fluktuation der dissipations-rate, und kommen aufgrund dieser multiplikativen kaskade auf ne log-normal verteilung. beim nutzen der momente der dissipations-rate $\eps_r$ hoffen wir dann dass diese den momenten der transfer-rate entspricht.} Considering the logarithm of the dissipation rate, $\ln\eps_r(\vek{x})=\ln\eps_L(\vek{x})+\mscale[0.85]{\sum}_j \ln h_j$, and owing to the central limit theorem, the random variable $\ln\eps_r(\vek{x})$ should be approximately normal distributed, if the number of cascade steps is large enough. Obhukov therefore claimed that $\eps_r(\vek{x})$ is distributed according to the log-normal distribution
\begin{align} \label{eq:FDT_K62_lognorm}
	p(\eps_r) = \frac{1}{\sqrt{2\pi\s_r^{\,2}\eps_r^2}}\,\exp\Bigg[\frac{\big(\ln(\eps_r/\beps)+\s_r^{\,2}/2\,\big)^2}{2\s_r^{\,2}}\Bigg]
\end{align}
such that the mean of $\eps_r$ is $\beps$. The remaining freedom in this model is the standard deviation $\s_r$, for which Kolmogorov and Obukhov assumed
\begin{align} \label{eq:FDT_K62_sigr}
	\s_r^2(\vek{x}) = A(\vek{x}) + \mu\ln(L/r) \;,
\end{align}
accounting for the increasing variance of energy transfer fluctuations towards the end of the cascade. The increase of energy transfer fluctuations can indeed be observed in experiments \cite{Mouri2006}. Furthermore, the logarithm in (\ref{eq:FDT_K62_sigr}) ensures that the moments of $p(\eps_r)$ obey a scaling law,
\begin{align} \label{eq:FDT_K62_moms}
	\lla \eps_r(\vek{x})^n \rra &= \Big(\frac{r}{L}\Big)^{\displaystyle\tfrac{\mu n}{2}(1-n)} \, \eee{\tfrac{A(\vek{x})}{2}n(n-1)+n\beps} \nonumber \\[7pt]
	&\propto r^{\displaystyle\tfrac{\mu n}{2}(1-n)} \;.
\end{align}
The constant $\mu$ is called {\it intermittency factor}, for reasons we will come to shortly, and is, due to the scale-invariance (\ref{eq:FDT_NSEinvar_scal}), assumed to be universal. The universality of $\mu$ has indeed been proved well by experiments and it is found to be $\mu\approx0.26$ \cite{Arneodo1996}. The function $A(\vek{x})$ was meant to weaken the previous claim of universal constants of proportionality in the K41 theory, but has no influence on the scaling exponents.\\
Replacing the mean value $\beps$ by $\eps_r$ to take into account the scale-dependent fluctuations of the energy transfer and making use of the scaling (\ref{eq:FDT_K62_moms}) of the moments of $\eps_r$, we arrive at the {\it K62 scaling}\remark{die bedingung dass $\beps$ mittelwert von $\eps_r$ ist, macht das four-fifths enthalten ist ($n/3$ in momente oben einsetzen!)}
\begin{align}\label{eq:FDT_K62}
	\Str^n(r) \sim \lla(\eps_r r)^{n/3}\rra \sim r^{\,\z_n} \sep \z_n = \frac{n}{3} + \frac{\mu}{18}(3n-n^2) \qquad \text{(K62)} \;.
\end{align}
The K62 model is also often referred to as the {\it log-normal model}.\\

Compared to the K41 scaling (\ref{eq:FDT_K41}), the K62 scaling exponents are most notably not linear but have a term quadratic in $n$. The quadratic term provokes that $\z_n$ becomes a decreasing function in $n$ for \linebreak $n>(6+3\mu)/(2\mu)\approx11$ and eventually turns negative. This behaviour of $\z_n$ is commonly considered as unphysical due to two reasons. First, it can be shown that a decreasing $\z_n$ can not be reconciled with the incompressibility assumption in the derivation of the NSE\remark{supersonic velocities become too likely}, and second, the form of $\z_n$ violates the so-called Novikov inequality \cite{Novikov1971} valid for fully developed turbulence in the limit $\Rey\to\infty$ (\cite{Frisch95} p.\,172f). By substituting $n=3$, we see that the K62 scaling respects the four-fifths law (\ref{eq:FDT_four-fifths}), whereas the two-thirds law (\ref{eq:FDT_scals_two-thirds}) receives a minor correction, $\Str^2(r)\sim r^{2/3-\mu/9}$ where $\mu/9\approx0.03$. This correction applies analogously to the five-thirds law (\ref{eq:FDT_scals_five-thirds}).\\
We mention that scaling exponents that depart from the K41 scaling are also called {\it anomalous} scaling exponents, and the deviation $\mu/9\approx0.03$ is often referred to as the anomaly parameter.\\
The K41 scaling only holds for the structure function of second and third order (\cite{Frisch95} p.\,91), whereas the K62 scaling agrees well with experimental results up to order of about eight (\cite{Frisch95} p.\,132). The incorporation of violent small-scale fluctuations of energy transfer hence improved the probabilistic description of extreme velocity fluctuations considerably. Note that in the case of K62 scaling, the limit $\Rey\to\infty$ is essential for the central limit theorem to be applicable, since we see from (\ref{eq:FDT_scals_Rey}) that large Reynolds numbers imply a large inertial range, which in turn ensures a large number of steps in the cascade of eddies.

\subsection{Small-scale intermittency} \label{ss_FDT_sm-sc-intm}
\begin{figure}[t] 
	\begin{center}
		\includegraphics[width=0.9\textwidth]{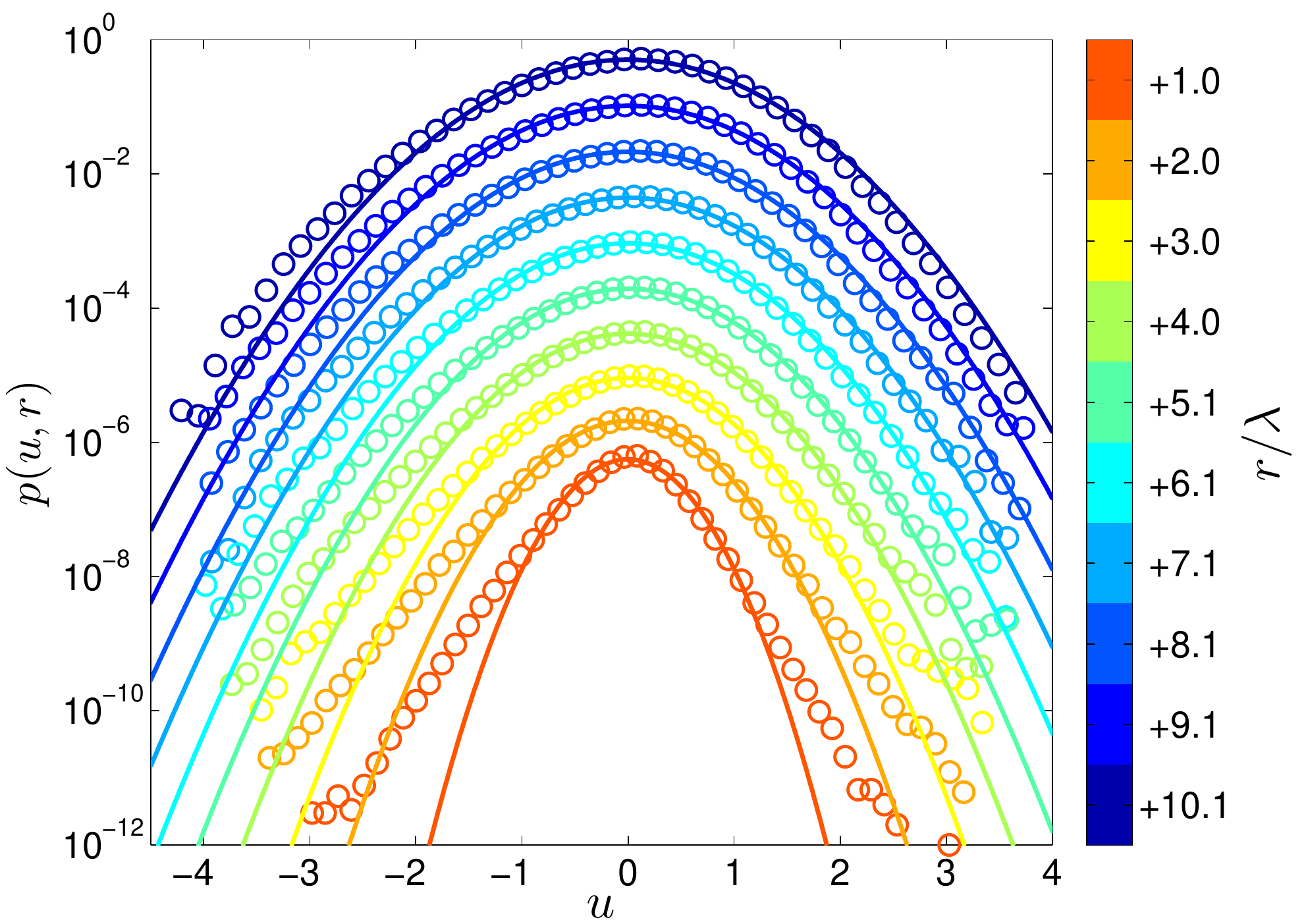}
		\caption{\label{ff:sm-sc-intm}\figtxt{Small-scale intermittency. The symbols are histograms for velocity increments $u(r)$ computed from the data used in \cite{Renner2001} for various scales $r$ covering the whole inertial range. Gaussian fits are included to emphasise the heavy tails towards smaller scales. Note also the negative skewness apparent on all scales. The PDFs $p(u,r)$ are shifted vertically for the sake of clarity.}}
	\end{center}
\end{figure}

The K62 scaling belongs to the class of random cascade models. These models share the idea of Kolmogorov and Obhukov that due to repeated break-up of coarse structures into finer structures, fluctuations of the energy transfer at large scales entail rampant fluctuations of energy transfer and dissipation at small scales (\cite{Frisch95} p.\,165ff). These strong energetic fluctuations at small scales cause similar strong velocity fluctuations at small scales, a phenomenon known in turbulence research as {\it small-scale intermittency}. Random cascade models are just one of many attempts to incorporate small-scale intermittency into the description of small-scale turbulence.\\
Small-scale intermittency is widely considered to be one of the most challenging and intriguing phenomena in turbulence \cite{Sreenivasan1997}\remark{einfach mal volltextsuche in mendeley, steht überall.}. Small-scale intermittency is particularly pronounced in atmospheric turbulence \cite{Bottcher2003,Bottcher2006,Pouquet2013}, and is also of importance in technical application, examples include turbulent mixing and combustion \cite{Warhaft2000,Dimotakis2005}, but also for loads on wind energy converters \cite{Peinke2004,Mucke2011,Wachter2012,Rinn2012a,Milan2013} and other constructions exposed to extreme weather conditions.\remark{the short (small scale) but violent wind gusts cause the most damages (rütteleffekt). Main statement from small scale intermittency: Small eddies are not always the slow ones, particular high energetic small eddies are more likely than comparable large eddies, in addition to the stronger rütteleffekt (the windturbine experiences small eddies as a strike, as it can not react to fast windspeed changes). This effect of course is also valid for other constructions than wind energy converters, as long as they are subject to wind.}\\
The term intermittency refers to short bursts of high activity in an otherwise moderately fluctuating signal. There are many ways to formally define intermittency. One is to take the fluctuations of a signal $f(t)$ and inspect the kurtosis of these fluctuations. A signature of intermittency is that the kurtosis exceeds considerably a value of three, which is the kurtosis of a Gaussian distribution. Or more specific, to accommodate in particular that intermittent bursts are short, after cutting out frequencies higher than a certain cut-off frequency, the kurtosis of the resulting signal should grow unbounded with this cut-off frequency (\cite{Frisch95} p.\,120). Regarding velocity fluctuations in a turbulent signal $v(t)$, it is common to change to the spatial domain by use of the Taylor hypothesis and inspect the distribution of velocity increments $u$ on various scales $r$. The signature of small-scale intermittency are then {\it heavy tails} of the distribution $p(u,r)$, i.e. towards small scales the frequency of large velocity increments exceed any Gaussian prediction. According to the above discussion of the strong fluctuations of energy dissipation, there is no doubt that $p(u,r))$ will develop heavy tails when the scale $r$ tends to zero. On the other hand, the frictional forces at very small scales will eventually damp down all intermittent fluctuations. The more intriguing question therefore is, how far does the intermittent behaviour extend into the inertial range. See figure \ref{ff:sm-sc-intm} for a typical signature of small-scale intermittency in $p(u,r)$.\\
\\
The reason to use the rather qualitative criterion of heavy tails is that it accounts for all high order structure functions $\Str^n(r)$, which in turn are characterised by scaling laws (\ref{eq:FDT_self-sim}) and in particular by the presumably universal scaling exponents $\z_n$. Hence, a physical model for small-scale turbulence which predicts $\z_n$ that are consistent with all available theoretical and experimental results would be a great step towards understanding the mechanism responsible for small-scale intermittency in any turbulent flow. The K41 scaling, for instance, predicts a Gaussian form of $p(u,r)$ in the inertial range which clearly conflicts with the features of intermittency. Corrections to the K41 scaling law are therefore called {\it intermittency corrections}, of which the K62 scaling is just one example. Other approaches apart from K62 include the $\b$-model, multifractal models, non-linear scaling exponents, and, going beyond the scaling approach, extended self-similarity, random cascade models and Burger's turbulence. In this section, we will briefly introduce these approaches.\\
We note that, however, our discussion is by no means complete. For a broader overview we recommend chapter 8 and 9 of the book by Frisch \cite{Frisch95}, which includes a discussion of the mentioned fractal models and random cascade models, large deviation theory, shell models, dynamical systems, field theoretic approaches, functional and diagrammatic approaches, the closure problem of the averaged NSE, multiscale methods and renormalisation groups. In addition, the overview article by Sreenivasan and Antonia \cite{Sreenivasan1997} primarily addresses scaling phenomenology, multifractality and kinematics of small-scale structures, and the review by Biferale \cite{Biferale2003} is devoted to shell models.\remark{Small-scale intermittency wird auch bei skalaren der strömung beobachtet, we zB temperatur, siehe Castating 1990.}

\paragraph{Modifications of scaling laws}
The $\b$-model takes into account that turbulent structures can not be as space-filling as the structures they evolved from \cite{Frisch1978}\remark{frisch p.135ff}\remark{dh. ich kann einen zerbrochenen wirbel nicht wieder so zusammensetzen, dass er das gleiche volumen wie vorher ausfüllt.}, that is, the ensemble of non space-filling structures does not occupy the full space with dimension $d=3$, but instead a subspace of fractal dimension $d=\dfr<3$. The consequence is that the scaling of extensive quantities (such as energy) is corrected by the volume-fraction $(r/L)^{3-\dfr}$\remark{, where $L$ in the denominator ensure that at the integral length scale all structures remain space-filling, siehe auch notebook p. 65}. In terms of the scale-invariance (\ref{eq:FDT_NSEinvar_scal}), we then have $h=1/3-(\dfr-3)/3$ which results into the linear scaling
\begin{align} \label{eq:FDT_beta-model}
	\z_n=\mfrac{n}{3}+(3-\dfr)\big(1-\mfrac{n}{3}\big) \qquad (\b\text{-model}) \;,
\end{align}
consistent with experimental results up to an order of about six (\cite{Frisch95} p.\,135ff).\\
The notion of fractals in the $\b$-model has initiated {\it multifractal models} in which the cascade of eddies takes place in fractal subspaces of $\mathbb{R}^3$ with fractal dimensions $\dfr<3$ \cite{Meneveau1987,Benzi1991,Arneodo1995,Yakhot2004,Aurell2006,ChakrabSahaBhatta09PRE,Zybin2012}, see also p.\,143ff in \cite{Frisch95}.\remark{This assumption is corroborated by the {\it local} scale-invariance of the NSE which allows an infinite number of scaling exponents.} The multifractality is then expressed by a continuous superposition of scaling laws, each of which featuring a scaling exponent $h_\mr{min}<h<h_\mr{max}$ and valid in a subspace of fractal dimension $\dfr(h)$. In contrast to pure scaling in the form (\ref{eq:FDT_self-sim}), the multifractal model puts forward a {\it spectrum} of scaling laws. In the limit $r\to0$, however, the method of steepest decent relates scaling exponent $\z_n$ and fractal dimension $\dfr$ via a Legendre transformation\remark{es ist dann $\z_n=\inf\limits_{h}\big[nh+3-\dfr(3)\big]$}, where a special choice of $\dfr(h)$ reproduces the K62 model. Therefore, the K62 model is also referred to as a multifractal model. In that sense, the $\b$-model generalises the K41 model and the multifractal model the K62 model by inclusion of non space-filling turbulent structures at small scales.\\
\\
As an illustrative example for fractal subspaces, {\it vortex filaments} may be named \cite{Kraichnan1974,JIMENEZ1998}, see also p.\,184ff in \cite{Frisch95}. Vortex filaments are basically long and thin swirls\remark{swirl im sinne von strudel}, their diameter is of the order of the Kolmogorov dissipation scale $\eta$ and their extension can reach the integral length scale $L$. As filaments, their fractal dimension is close to one.\\
The role that vortex filaments play for the statistics of velocity increments, in particular for small-scale intermittency, is an objective of current research \cite{Pirozzoli2012,Zybin2012a,Zybin2013}. In the course of these efforts, She and Leveque proposed in \cite{SheLeveque94PRL} a phenomenological characterisation of the energy dissipation field, itemised into an hierarchy of fluctuating structures ranging from the mean dissipation rate $\beps$ on the largest scale to the intermittent impact of vortex filaments on the smallest scale. Upon coarse-graining the dissipative scales to inertial range scales, they predict the universal scaling law,
\begin{align} \label{eq:FDT_zn_SL_gen}
	\z_n = \Big(1-\mfrac{C_0}{3}\Big)\mfrac{n}{3} + C_0\Big(1 - \b^{\,\frac{n}{3}}\Big) \;.
\end{align}
Here, $C_0=3-\dfr$ is the codimension of the dominant intermittent structures, and the intermittency parameter $\b$ accounts for intermittency strength as it various from one to zero. She and Leveque determined from their theory that $\b=2/3$. Note that K41 scaling is recovered if we set $\b=1$ and codimension $C_0=0$. Taking $\dfr=1$ for vortex filaments and $\b=2/3$, the SL scaling exponents become
\begin{align} \label{eq:FDT_zn_SL_parfree}
	\z_n = \frac{n}{9} + 2\Big(1 - \big(2/3\big)^{\,\frac{n}{3}}\Big) \qquad \text{(SL)} \;,
\end{align}
which She and Leveque claim to hold universally. Indeed, the agreement with experimental data is excellent, which is remarkable considering the phenomenological nature of this proposal and that it goes without any adjustable parameters.\remark{in einer Taylor reihe wird K62 (K41) in sehr guter näherung aber nicht exakt reproduziert} The phenomenological model giving rise to (\ref{eq:FDT_zn_SL_parfree}) can also be formulated as a random cascade model with Poissonian distributed multipliers and is therefore also called log-Poisson model \cite{Dubrulle1994,She1995}.\remark{der erste term $n/9$ weicht hier sehr stark von $n/3$ ab, weil die dominierenden wirbel als extrem fraktal angesehen werden ($\dfr=1$!).}\\

Another approach to model small-scale intermittency was proposes by L'vov and Procaccia \cite{L'vov1996,L'vov1996a}, in which they address the $n$-point correlation function
\begin{align}
	R_n(r_1,r_2,\dots,r_n) = \lla\,u(r_1)u(r_2)\cdots u(r_n)\,\rra
\end{align}
where all $L>r_i>\l$. For the special case that all $r_1\eq r_2\eq\dots\eq r_n\equiv r$, the $n$-point correlation function becomes the $n$-th order structure function $\Str^n(r)=\lla u(r)^n\rra$.\remark{doppelt: For the intermediate case in which the scales $r_1,\dots,r_n$ coalesce into groups of similar values, the $n$-point correlation function factorises into correlation functions associated with these groups and a homogeneous function.}\\
Two hypotheses of Kolmogorov type form the basis of the analysis of $R_n(r_1,\dots,r_n)$. The first hypothesis states that the correlation functions are homogeneous functions with scaling exponents $\z_n$,
\begin{align}
	R_n(cr_1,cr_2,\dots,cr_n) = c^{\z_n} R_n(r_1,r_2,\dots,r_n) \;,
\end{align}
which is the analogue of proposing scaling laws for structure functions. The second hypothesis addresses universality, in the sense that velocity increments in the inertial range are not correlated with the velocity increments on the scale of turbulence generation. A precise formulation of this hypothesis is given by the discussion of equation (3) in \cite{L'vov1996}.\\
The proposal of L'vov and Procaccia involves multipoint correlation functions that are intermediate between the extremes $R_n(r_1,\dots,r_n)$ and $\Str^n(r)$. More precisely, they consider the case in which the scales $(r_1,r_2,\dots,r_n)$ can be grouped into $p$ small scales $(r_1,\dots,r_p)\sim r_0$ and $n\-p$ larger scales $L>(r_{p+1},\dots,r_n)\gg r_0$ and then propose that the full correlation function factorises into the intermediate $p$-point correlation function
\begin{align}
	R_p(r_1,\dots,r_p)=\lla u(r_1)\cdots u(r_p)\rra
\end{align}
with scaling exponent $\z_p$ and a homogeneous function $\Psi_{n,p}(r_{p+1},\dots,r_n)$ with scaling exponent $\z_n\-\z_p$,
\begin{align}
	R_n(r_1,r_2,\dots,r_n) = R_p(r_1,r_2,\dots,r_p)\,\Psi_{n,p}(r_{p+1},r_{p+2},\dots,r_n) \;.
\end{align}
where the function $\Psi_{n,p}(r_{p+1},\dots,r_n)$ derives from the underlying equations of motions, in this case the NSE or adequate models of turbulence. In the limit that $r_1\!\simeq\! r_2\!\simeq\! \dots\! \simeq\! r_p$, the intermediate correlation function becomes the structure function $\Str^p(r)$. Instead of two groups of scales, an arbitrary number of groups can be constructed, where each group is associated with an intermediate correlation function, only the group of the largest scales enters in terms of a homogeneous function $\Psi$. The relations that govern this kind of coalescence of scales into groups are known as {\it fusion rules} from which various scaling relations can be determined. A simple example of such a scaling relation has been verified experimentally \cite{Fairhall1997,Benzi1998} and reads
\begin{align}
	R_{pq}(r_1,r_2) = \lla u(r_1)^p u(r_2)^q \rra \sim \frac{S_p(r_1)}{S_p(r_2)}S_{p+q}(r_2)
\end{align}
where $L>r_2>r_1>\l$.\\
In \cite{LvovProcaccia00PRE}, L'vov and Procaccia use fusion rules in a diagrammatic perturbation theory that builds on the NSE in the limit of infinite Reynolds numbers. They use as a small parameter the anomaly parameter\linebreak $\d_2\dfns \z_2-2/3\approx0.03$,\remark{The deviation from K41 scaling is also called anomaly of scaling exponents.} and find in second order of $\d_2$,
\begin{align} \label{eq:FDT_zn_LP}
	\z_n = \frac{n}{3} - \frac{n(n-3)}{2}\,\big(\d_2 + 2\d_2^{\,2}(n-2)b\big) \qquad \text{(LP)}
\end{align}
with parameter $b\approx-1$.\footnote{Renner and Peinke found that $b=-3/4$ is consistent with the scaling exponents implied by Yakhot's model which we introduce below. \cite{Renner2011}} Choosing $\d_2=0$, we obtain the K41 scaling as it should be. The first order correction in $\d_2$ recovers the K62 scaling, and the second order introduces to $\z_n$ a cubic term in $n$. The main advantage over K62 is that $\z_n$ does not become negative for large $n$. Compared to the K62 scaling exponents, the agreement between the LP $\z_n$ above and experimental results improves for $n\geq8$.

\paragraph{Extended self-similarity}
\remark{even if a model predicts $\z_n$ consistent with existing theory and experimental results, the explicit form of $\Str^n(r)$ does not necessarily be consistent, i.e. have the form of a scaling law.\\
In the previous sections we discussed the two basic scaling laws K41 and K62, followed by an overview of approaches that aim at incorporating the phenomenon of small-scale intermittency based on pure scaling in the form $\Str^n(r)\propto r^\z_n$. The simplest appreciable intermittency correction is the K62 scaling.
Other scaling laws are under scrutiny, ESS seems promising.\\
scaling relation $u(r)\sim r^h$ in the inertial range is a direct consequence of the NSE in the limit $\Rey\to\infty$.
according to K41, scaling exponent is $h=\z_1=1/3$. models that assume that eddies towards the bottom of the cascade are not space-filling introduce a small correction to $\z_1$. In the inertial range and for $\Rey\to\infty$, the four-fifths law $\lla u^3\rra=-4/5\beps r$ is an exact result of the NSE.\\
The scaling relation for $u$ and the four-fifths law for $\lla u^3\rra$ are the origin of a plethora of scaling laws $\lla u^n\rra\sim r^\z_n$, which all have $\z_1\approx1/3$ and $\z_3=1$. These scaling laws aim at modelling the statistics of extreme velocity fluctuations at small scales, known as small-scale intermittency.\\
There is little doubt that for small $n$ scaling laws of the structure functions correctly describe the statistics of velocity increments $u$. For large $n$, however, the applicability of scaling laws is not as clear. In this section we introduce three approaches that do not rest on the assumption that also for large $n$ the structure functions obey scaling laws. (auch wenn wenn yakhot scaling exponents aus seinen gleichungen für $pu,r$ macht, aber eben eher als vereinfachung und zum vergleich mit bestehenden theorien. und vor allem nicht auf annahme eines scaling laws aufbaut.)}
There is no doubt that Kolmogorov's four-fifths law must hold within the bounds of the assumption made in the derivation, as it has been derived from the full three dimensional NSE. Still, experimental data often satisfies the four-fifths law rather approximately. Deviation from the four-fifths law can accordingly only be attributed to non-compliance of the assumptions made in the derivation and experimental imperfections (\cite{Frisch95}, p.\,129ff), including i) remaining inhomogeneities and anisotropy, ii) not negligible molecular friction (i.e. no clear-cut crossover from inertial to dissipative range), iii) uncertainties in the determination of $\beps$, iv) unjustified use of the Taylor hypothesis.\remark{auch in numerischen simulationen kann man nicht so einfach daten ideal voll entwickelte turbulenz erstellen, denn mit steigender $\Reych$ wird der numerische Aufwand größer, und natürlich haben wir hier mit numerischen ungenauigkeiten zu kämpfen (finite size)}\\
The fit of scaling exponents in the inertial range suffers from the same impairments as the verification of the four-fifths law. In addition to that, the structure functions of order six and higher exhibit undulations, of which the origin remains rather unclear.\remark{p.130 in frisch sagt auch, dass vor allem in höheren momenten periodische schwankungen auftreten (undulations). die stammen wohl daher, dass in der wirbelkaskade bestimmte skalenverhältnisse beim bevorzugt werden, zB immer ein zerfallender wirbel immer ein drittel des mutterwibels. das führt zu einer faint lacunarity (löchrichkeit) in $u$, die in höheren momenten zu diesen undulations führen.} An improvement in the determination of the scaling exponents from experimental data can be achieved by a method known as {\it extended self-similarity} (ESS) by Benzi et al. \cite{Benzi1993}. In this method, instead of the structure functions itself, the ratio to the third structure function, $\Str^n(r)/\Str^3(r)$, are used to fit the scaling exponents.\footnote{Instead of $\Str^3(r)$ any other structure function could be used here. But the fact that $\Str^3(r)\sim r$\remark{as a result of the four-fifths law goes without the self-similarity assumption} makes $\Str^3(r)$ a reasonable choice.} The crucial assumption is that the above mentioned impairments affect all moments in a similar manner. The ESS method accounts for these imperfections as it introduces a correction to pure scaling in the form
\begin{align} \label{eq:FDT_ESS_Sn}
	\Str^n(r) \propto \big[\Str^3(r)\big]^{\z_n} \;. \qquad \text{(ESS)}
\end{align}
For the case of perfect measurements of ideally developed turbulence in the limit $\Reych\to\infty$, we would recover the pure scaling behaviour discussed in the previous section. In that sense, ESS allows a determination of the scaling exponents $\z_n$ minus the effects induced by the mentioned impairments. In their investigations, Benzi et al. even presented convincing experimental evidence that the validity of ESS extents considerably far into the dissipative range.

\paragraph{Random cascade models} \label{rcms}
The two main assumptions in setting up the K62 model were (i) the log-normal distribution for the local averaged energy dissipation $\eps_r$ and (ii) a standard deviation $\s_r$ of the log-normal distribution that scales with $\ln r$. The log-normal distribution resulted naturally from the perception of a multiplicative cascade of eddies, whereas $\s_r\sim\ln(L/r)$ was a mere pragmatic assumption to retain the scaling law for the structure functions. This has led Castaing and co-workers to explore the $r$-dependency of $\s_r$ experimentally \cite{Castaing1990}, and they found that $\eps_r$ is indeed log-normal distributed, but that a scaling law for the variance of the log-normal distribution,
\begin{align} \label{eq:FDT_castaing_sigr}
	\s_r^2 = c_\s \Big(\frac{r}{L}\Big)^{-\b} \;,
\end{align}
is in better agreement with their measurements than the logarithmic dependency used in the K62 model. In addition, they were able to show that this power law results from applying an extremum principle to the probability of $\eps_r$. The parameter $\b$ is found to be dependent on $\Rey$ and not universal; for the measurements discussed in \cite{Castaing1990}, $\b\approx0.3$.\\
The moments of $\eps_r$ become
\begin{align}
	\lla \eps_r^{\,n} \rra = \exp\Big[\,c_\s\frac{n(n-1)}{2}\Big(\frac{r}{L}\Big)^{-\beta}+n\beps\,\Big]
\end{align}
and the K62 scaling (\ref{eq:FDT_K62}) changes into
\begin{align}\label{eq:FDT_castaing_Sn}
	\Str^n(r) \sim \lla(\eps_r r)^{\,\frac{n}{3}}\rra \sim r^{\,\frac{n}{3}}\,\ee{\frac{c_\s\,n(n-3)}{18}\,\left(\!\frac{r}{L}\!\right)^{-\beta}} \;.
\end{align}
The authors of \cite{Castaing1990} also explored the consequence for the PDF of velocity increments, $p(u,r)$, on the basis of two assumptions. Their first assumption is that for a fixed value of $\eps$, the velocity increments are normal distributed,
\begin{align} \label{eq:FDT_castaing_p1}
	p(u|\eps) = \frac{1}{\sqrt{2\pi\s^2}}\exp\bigg[-\frac{u^2}{2\s_\eps^2}\bigg] \;.
\end{align}
Motivated by the scaling relation $\s\sim(\eps r)^{1/3}$ and the log-normal model, the second assumption is that the standard deviation $\s_\eps$ is log-normal distributed
\begin{align} \label{eq:FDT_castaing_p2}
	p(\s_\eps,r) = \frac{1}{2\pi\L_r^2}\exp\bigg[-\frac{\big(\ln(\s_\eps/\s_0)\big)^2}{2\L_r^2}\bigg] \;,
\end{align}
where the variance $\L_r^2$ is typically of the form (\ref{eq:FDT_castaing_sigr}). Both assumptions are well corroborated by their measurements.\\
Combining (\ref{eq:FDT_castaing_p1}) and (\ref{eq:FDT_castaing_p2}) yields the distribution
\begin{align} \label{eq:FDT_castaing_p12}
	p(u,r) = \frac{1}{2\pi\L_r}\int\limits_{0}^{\infty} \,\exp\bigg[-\frac{u^2}{2\s_\eps^2}\bigg] \exp\bigg[-\frac{\big(\ln(\s_\eps/\s_0)\,\big)^2}{2\L_r^{\,2}}\,\bigg] \,\frac{\dd \s_\eps}{\s_\eps^{\,2}}
\end{align}
which is a superposition of normal distributions and involves the free parameters $\s_0$ and $\L_r$.\\
The normal distribution in (\ref{eq:FDT_castaing_p12_skew}) causes $p(u,r)$ to be even in $u$ which contradicts the four-fifths law (\ref{eq:FDT_four-fifths}). To overcome this shortcoming, the normal distribution is augmented with a phenomenologically motivated skewness correction, and (\ref{eq:FDT_castaing_p12_skew}) becomes
\begin{align} \label{eq:FDT_castaing_p12_skew}
	p(u,r) = \frac{1}{2\pi\L_r}\int\limits_{0}^{\infty} \,\exp\bigg[-&\frac{u^2}{2\s_\eps^2}\bigg(1+a_\mr{s}\frac{u/\s_\eps}{\sqrt{1\+u^2/\s_\eps^{\,2}}}\bigg)\bigg] \nn
	&\times\,\exp\bigg[-\frac{\big(\ln(\s_\eps/\s_0)\,\big)^2}{2\L_r^{\,2}}\,\bigg] \,\frac{\dd \s_\eps}{\s_\eps^{\,2}}
\end{align}
with the skewness parameter $a_\mr{s}\approx-0.18$ claimed to be universal.\\
The parameter $\L_r$ accounts for the fluctuations of energy transfer on scale $r$ and is therefore a quantity of interest in itself. Accordingly, the important result of \cite{Castaing1990} is not so much the explicit form of $p(u,r)$ in (\ref{eq:FDT_castaing_p12_skew}), but rather to obtain $\L_r$ by fitting (\ref{eq:FDT_castaing_p12_skew}) to experimentally determined $p(u,r)$. As already mentioned, they found that $\L_r$ is well described by the power law in (\ref{eq:FDT_castaing_sigr}), for which $p(u,r)$ from (\ref{eq:FDT_castaing_p12_skew}) is in excellent agreement with measurements.\\
In the sequel of this article by Castaing and co-workers, their model has been cast in a more general formalism. Models using this formalism are known as {\it random cascade models} or {\it processes} (\cite{Frisch95} p.\,165) and have been widely used \cite{Dubrulle1994,Castaing1995,Castaing1996,Chilla1996,Arneodo1996,Amblard1999,Dubrulle2000,Chanal2000,Chevillard2005}. In random cascade models, multipliers $h<1$ are used to express the statistics of velocity increments on scales $r<L$ as $u=hu_L$. The multipliers are assumed to be a random variable and follow a scale dependent PDF $G_{rL}(\ln h)$, also referred to as a propagator. The propagator $G_{rL}(\ln h)$ then defines the PDF of velocity increments via a superposition of scaled PDFs $p_L(u/h)$ on the integral scale $L$ \cite{Castaing1996},
\begin{align} \label{eq:FDT_random-cascade}
	p(u,r) = \int G_{rL}(\ln h)\, p_L\Big(\mfrac{u}{h}\Big) \,\mfrac{\dd\ln h}{h} \;. \qquad \text{(RCM)} \;.
\end{align}
The above prescription of how random multipliers determine the distribution of velocity increments at scales $r<L$ is the essence of random cascade models. The {\it randomness} enters through the random variable $h$, and the association with a {\it cascade} originates from writing the propagator as a convolution \cite{Castaing1996}
\begin{align}
	G_{rL}(\ln h) = \int G_{rr_1}(\ln h_1)\,G_{r_1L}(\ln h-\ln h_1) \di\ln h_1
\end{align}
with the intermediate scale $L>r_1>r$ and multiplier $h_1$ to express the statistics of velocity increments on scale $r_1$ as $u_1=h_1u_L$. Instead of only one intermediate scale $r_1$, it is of course possible to set up $G_{rL}(\ln h)$ as a convolution of a series of intermediate scales $L>r_1>r_2>\dots>r$, and the connection to a cascade becomes obvious.\\
In view of (\ref{eq:FDT_random-cascade}), various special cases are included in the formalism of random cascade models.\\
(i) For the simplest choice, $G_{rL}(\ln h)=\d(\ln h-\ln h_{rL})$, and assuming a normal distribution for $p_L(u)$, the PDF $p(u,r)$ will be Gaussian for all scales $r$ which is the signature of the K41 model. In other words, for multipliers that are not random, the K41 model is recovered.\\
(ii) By taking for $G_{rL}(\ln h)$ a log-normal distribution with a variance $\L_r\sim\ln(r/L)$, the PDF $p(u,r)$ develops the non-Gaussian tails known from small-scale intermittency. The structure functions exhibit a scaling law with scaling exponents of K62 form (\ref{eq:FDT_K62}). This not surprising, as in the similar fashioned derivation of the K62 scaling we used the log-normal distribution (\ref{eq:FDT_K62_lognorm}) with variance (\ref{eq:FDT_K62_sigr}).\\
(iii) Sticking to the log-normal distribution but taking a variance that is not proportional to $\ln(L/r)$, we abandon a scaling law and the $r$-dependence of the variance determines the function of $r$ that supersedes the $r$ in $r^{\z_n}$.\remark{A variance that goes with $\pt_r \ln \Str^3(r)$, for instance, leads to the ESS scaling (\ref{eq:FDT_ESS_Sn}), and for the variance (\ref{eq:FDT_castaing_sigr}) obeying a scaling law the structure functions scale as in (\ref{eq:FDT_castaing_Sn}).} The PDF $p(u,r)$ will take the form (\ref{eq:FDT_castaing_p12_skew}) and the link to the afore introduced model by Castaing is established.\\
(iv) In \cite{Dubrulle1994,She1995} it has been shown that by using for $G_{rL}(\ln h)$ a log-Poisson distribution instead of the log-normal distribution, the statistics of velocity increments exhibit the SL scaling law (\ref{eq:FDT_zn_SL_parfree}) proposed by She and Leveque. The expectation of the log-Poisson distribution is related to the logarithm of the local energy transfer. The connection between log-Poisson statistics of energy transfer and the scaling exponents proposed by She and Leveque is intriguing, since Possion distributions are the natural distributions in the context of rare events \cite{Dubrulle1994}. Consider, for instance, the binomial distribution which gives the probability of success for $n$ independent trials, where each trial has probability $p_\mr{s}$ for success and probability $1-p_\mr{s}$ for failure. For finite $p_\mr{s}$ but $n\to\infty$, the binomial distribution approaches the normal distribution. Whereas $p_\mr{s}\to0$ (indicating rare events) and  $n\to\infty$ such that $np_\mr{s}$ stays finite is the limit of the Poisson distribution. In that sense, the log-Poisson model accounts for rare realisations of the cascade process, whereas the log-normal model concerns typical realisations. The experimentally demonstrated supremacy of the scaling exponents arising from the log-Poisson model over the exponents from the log-normal model also for high orders of structure functions hence suggests that the cascade process is dictated by rare events.\remark{das das bei seltenen $u$ der fall ist, ist nicht so verwunderlich, aber log-Poisson ist ja auch für moderate $u$ sehr gut, die werden als auch durch seltene realisierungen dominiert. Anders sehe es aus, wenn K62 für moderate $n$ und SL nur für große $n$ gut ist. Außerdem wäre an dieser stelle doch mal interessant zu untersuchen, ob es die formulierung eines binomial, bzw multinomial processes möglich ist, schließlich sind wir ja wegen Markov-Eigenschaft eigentlich diskret. Geht dann in Richtung Quantenstatistik...?}\\
(v) In general, cascade models are closely related to the notion of fractals. Therefore, random cascade models are often combined with multifractal models and vice versa \cite{Meneveau1987,Benzi1991,ChakrabSahaBhatta09PRE}. 

\paragraph{Burger's turbulence and Galilean invariance} \label{Yakhotsmodel}
Polyakov and Yakhot were able to derive a partial differential equation for the distribution of velocity increments,  $p(u,r)$, in a field theoretic approach to turbulence \cite{Polyakov95PRE,Yakhot98PRE}. They explicitly take into account the effect of the large scale parameters $\vrms^2=\big\langle(v-\lla v\rra)^2\big\rangle$ and $\lchar$.\remark{anstelle von $\lchar$ nehmen sie $L$ und bezeichnen $L$ sowohl als system size als auch als integral length scale. Polyakov hat es als system size eingeführt, das entspräche meinem $\lchar$. außerdem ist four-fifths ja im ganzen inertial-bereich gültig, und nicht erst $r\ll L$, und für $r\ll L$ nutzt Yakhot four-fifths. und bei mir ist inertial bereich $\lchar\gg L>r>\l\gg\eta$.} In the limit of small scales $r\ll \lchar$ and increments $u\ll\vchar$, they recover a scaling law.\\ 
Starting point of their considerations is the Burgers' equation, an one-dimensional NSE without the pressure term,
\begin{align} \label{eq:FDT_Burger}
	\dot v(x,t) + v(x,t)\,v'(x,t) = \nu v''(x,t) \;.
\end{align}
The Burger's equation is a\remark{, under moderate conditions solvable,} fundamental partial differential equation in mathematics and well studied\remark{It is known, for instance, that the Burgers' equation is bifractal in the sense that it gives rise to a superposition to two scaling laws. Each scaling law dominates for different $n$, i.e. there is a critical $n_c$ that separates the two scaling regions. The scaling of the $\b$-model is therefore exact, and machinery (e.g. diagrammatic methods, field theoretic stuff) can be tested first on Burgers' turbulence and than on NS turbulence. And for appropriate initial conditions, the Burger equation develops shockwaves.} and therefore often serves as a basis of models for turbulence.\\
To introduce the external force responsible for turbulence generation, the Burgers' equation is typically augmented with a white noise random force
\begin{align} \label{eq:FDT_Burger_forced}
	\dot v(x,t) + v(x,t)\,v'(x,t) = \nu v''(x,t) + f(x,t) \nn
	\lla f(x,t)f(x',t') \rra = \k(x-x')\d(t-t') \;,
\end{align}
where the spatial correlation function $\k(x\-x')$ depends on the details of turbulence generation and is generally assumed to act only on large scales.\\
Polyakov used in \cite{Polyakov95PRE} the notion of a generating functional $\ZZ[\omega(\cdot)]$ where $\omega(x,t)$ is conjugate to velocity $v(x,t)$. By substituting the above forced Burgers' equation and applying field theoretic techniques (operator product expansion\remark{Laurent expansion of operator (see wiki) - fusion rules determine the exact decomposition of the product}), he derives a partial differential equation for $\ZZ[\omega(\cdot)]$ for a steady energy flux and in the limit $\Rey\to\infty$. His result also explicitly includes large scale properties in form of the root-mean-square fluctuations of velocity, $\vrms$, which allows to incorporate different flow conditions and leads to intermittency corrections to K41 scaling. But this comes at the price of the breakdown of the Galilean invariance (\ref{eq:FDT_NSEinvar_Gal}) for the equations for $\ZZ[\omega(\cdot)]$, all other symmetries of the NSE (\ref{eq:FDT_NSEinvar}) are respected.\\
Permitting the breakdown of Galilean symmetry is the marking distinction of this approach compared to other field-theoretical attempts to account for intermittency corrections of scaling exponents. Preserving the Galilean invariance and recovering the K41 model is particular delicate in functional approaches to a field-theoretic description of turbulence (\cite{Frisch95} p.\,215).\\

\begin{figure}[!t] 
  \subfloat[][\figsubtxt{Structure functions.}]{\label{sf:zetan_ESS_S}
	\includegraphics[width=0.48\textwidth]{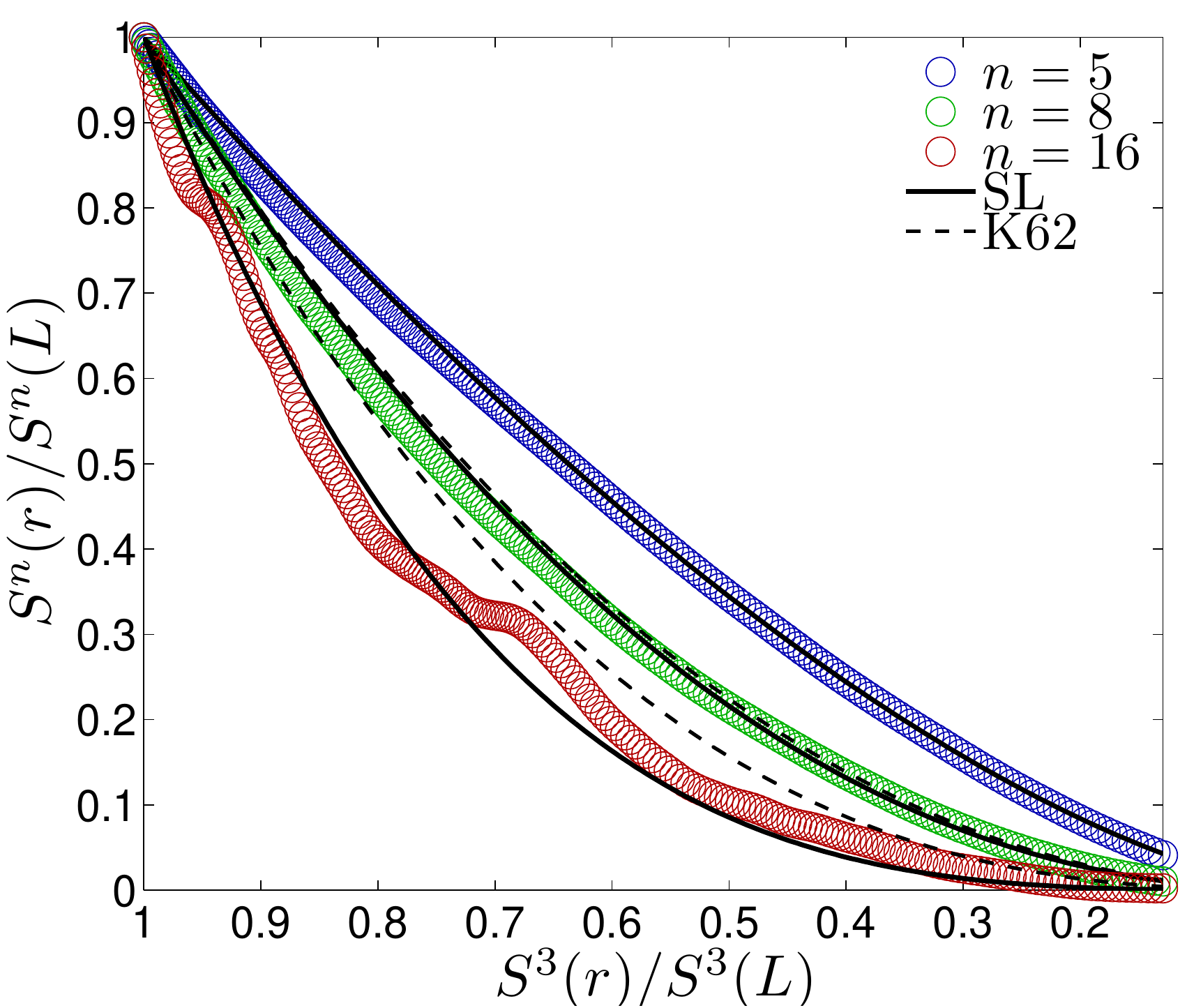}}
  \subfloat[][\figsubtxt{Scaling exponents.}]{\label{sf:zetan_ESS_zetan}
	\includegraphics[width=0.51\textwidth]{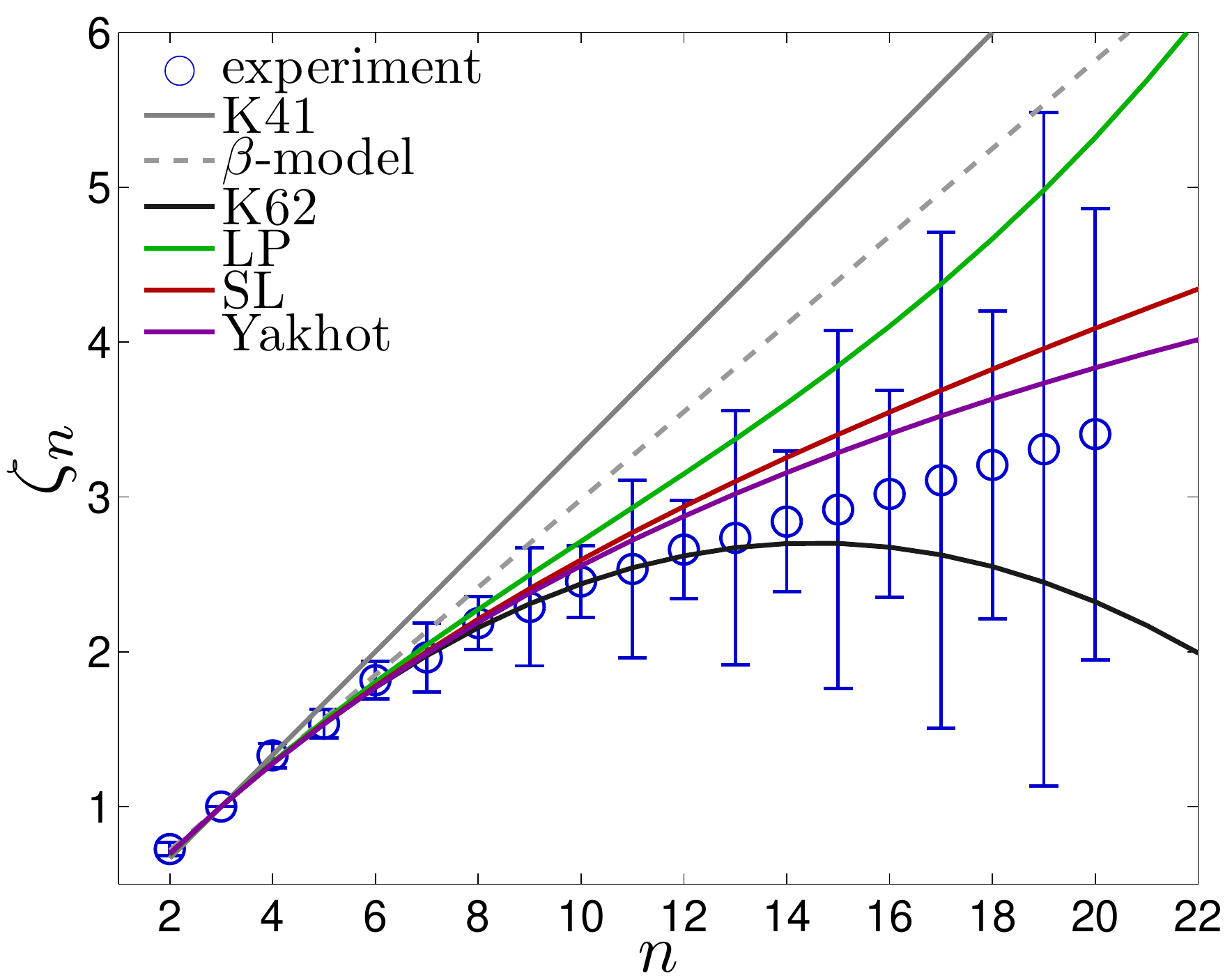}}
  \caption{\label{ff:zetan_ESS}\figtxt{Comparison of various theoretical predictions of scaling exponents with an experimental result using ESS. The experimental data consists of about $10^7$ realisations $u(r)$\remark{not as trajectories, but as $u$ values for fixed $r$, therefore we do not have to bother with cutting out non-overlapping trajectories.} and is the same as used in \cite{Renner2001} and figure \ref{ff:sm-sc-intm}. The scaling exponents were obtained from fitting the structure functions shown in (a) to the ESS formula (\ref{eq:FDT_ESS_Sn}), the error bars are the standard deviations of the $\z_n$ determined from $10^3$ blocks of $10^4$ realisations $u(r)$. The scaling laws for the K62 model, (\ref{eq:FDT_K62}), and suggested by She and Leveque (SL), (\ref{eq:FDT_zn_SL_parfree}), are included in (a). The resulting scaling exponents are shown in (b), which also includes the scaling exponents for the K41 model from (\ref{eq:FDT_K41}), the $\b$-model from (\ref{eq:FDT_beta-model}), the scaling exponents derived by L'vov and Procaccia in (\ref{eq:FDT_zn_LP}) and the scaling exponents in (\ref{eq:FDT_yakhot_zn}) implied by Yakhots model. For the $\b$-model the fractal dimension $\dfr=2.85$ is used, for the K62 model the intermittency factor $\mu=0.23$ and for the LP scaling exponents the parameter $b=-0.75$ as suggested in \cite{Renner2011}. The K41 scaling, the SL scaling and the scaling suggested by Yakhot are parameter-free.}}
\end{figure}

However, the model by Polyakov does not account for the influence of pressure. Resuming the work by Polyakov, Yakhot was able to include the influence of pressure to the model by making use of the three-dimensional Gaussian forced NSE \cite{Yakhot98PRE}.\remark{das macht er, indem er die transverale geschwindigkeitskomponente als kopplung zum druckterm der NSE missbraucht, oder so.} He further derived from the augmented equations for $\ZZ[\omega(\cdot)]$ the following partial differential equation for the PDF of velocity increments
\begin{align} \label{eq:FDT_yakhot_pde}
	-\frac{\pt\big(u\pt_r\,p(u,r)\big)}{\pt u} + B&\,\frac{\pt p(u,r)}{\pt r} \nn
	&= - \frac{A}{r} \frac{\pt\big(u\,p(u,r)\big)}{\pt u} + \frac{\vrms}{\lchar}\frac{\pt^2\big(u\,p(u,r)\big)}{\pt u^2}\;.
\end{align}
Note that the flow parameters $\vrms$ and $\lchar$ explicitly enter the equation.\\
To explore the consequence of his theory for scaling laws, Yakhot derived from the above equation the corresponding equation for the structure functions,
\begin{align} \label{eq:FDT_yakhot_moms}
	\frac{\pt \Str^n(r)}{\pt r} = \frac{An}{n+B}\frac{\Str^n(r)}{r} + \frac{\vrms}{\lchar}\frac{n(n-1)}{n+B}\Str^{n-1}(r) \;,
\end{align}
and substituted $\Str^n(r)=c_nr^{\z_n}$ to get
\begin{align} \label{eq:FDT_yakhot_zn_gen}
	\z_n = \frac{An}{B+n} + \frac{r}{\lchar}\,\frac{\vrms}{c_n/c_{n\-1}}\,\frac{n\,(n-1)}{B+n}\,r^{\z_{n\-1}-\z_n} \qquad \text{(YAK)} \;.
\end{align}
The last term with prefactor $r/\lchar$ takes effects of energy injection explicitly into account. To resort to scales $r$ where energy injection has no effect, that is assuming $r\ll \lchar$, Yakhot obtained
\begin{align} \label{eq:FDT_yakhot_zn}
	\z_n = \frac{An}{B+n} \sep r\ll \lchar \;.
\end{align}
In this limit the four-fifths law must hold and it follows for $n=3$,
\begin{align} \label{eq:FDT_yakhot_A}
	A = \frac{B+3}{3} \;.
\end{align}
Demanding that the moment equation (\ref{eq:FDT_yakhot_moms}) implies $\Str^3(\lchar)\equiv0$ and using $c_2=C_2\beps^{2/3}$ with Kolmogorov constant $C_2\approx2$ predicts a value of $B\approx20$. In the limit $B\to\infty$, the K41 scaling is recovered, and maximum intermittency is obtained for $B=0$.\remark{tja, $B\approx20$ konnte ich nicht reproduzieren, seine allgemeine Lösung (26) scheint mir nicht die allgemeine von (22) zu sein. siehe auch yakhot\_paper.mw. Achso, da geht noch ne Annahme für $\Str^2$ ein, siehe Renner, Peinke, J.Stat.Phys.2011}\\
With the numerical value $B\approx20$ and $A$ given by (\ref{eq:FDT_yakhot_A}), the scaling exponents (\ref{eq:FDT_yakhot_zn}) resulting from Yakhot's model for the scaling in the inertial range agree remarkably well with experimental data. See figure \ref{ff:zetan_ESS} for a comparison of the scaling laws discussed in this section together with an experimental result using ESS.
\\
In \cite{Yakhot2001a}, Yakhot was able to extend his model such that also transversal velocity increments are taken into account, and he generalised the formulation to two-dimensional turbulence. In the extended form, the formalism of Yakhot's model can be used to demonstrate that ESS is a consequence of the NSE \cite{Yakhot2001}, and to predict that for large $n$ the scaling exponents $\z_n$ are asymptotically a linear function of $n$ \cite{Yakhot03JoFM} as is the case for fractal models of turbulence and the SL scaling exponents (\ref{eq:FDT_zn_SL_parfree}).\remark{\cite{Yakhot03JoFM} also expresses $\z_n$ in terms of the prefactors of scaling laws} Indeed, in \cite{Yakhot2006}, the connection between Yakhot's model and multifractal models of turbulence is demonstrated.\remark{für mich zu merken: Polyakov und Yakhot nehmen im wesentlichen die turbulenz-erzeugung mit in ihre betrachtung, weil sie so auf intermittenz korrekturen kommen. da die erzeugung aber nicht GI ist ($\vrms=\lla\sqrt{u^2}\rra$ is not invariant under $v=u+U$), sind es ihre gleichungen auch nicht. Aber für $r\ll\lchar$ und $u\ll\vrms$ bleibt GI erhalten (also rechetrick?)}\\
In \cite{Renner2011}, Renner and Peinke pointed out that expansion of the scaling exponent (\ref{eq:FDT_yakhot_zn}) reproduces the scaling exponents (\ref{eq:FDT_zn_LP}) determined by L'vov and Procaccia, where the small parameter of the expansion was chosen to be $\d_2=\z_2-2/3$ as in (\ref{eq:FDT_zn_LP}) but with $\z_2$ taken from (\ref{eq:FDT_yakhot_zn}),
\begin{align}
	\z_n = \frac{n}{3}\frac{2+3\d}{2+3(n-2)\d} = \frac{n}{3} - \frac{n(n-3)}{2}\,\big(\d_2 + 2\d_2^{\,2}(n-2)b\big)
\end{align}
with $b=-3/4$ in agreement with $b\approx-1$ as predicted by L'vov and Procaccia.\remark{Whether the scaling law (\ref{eq:FDT_yakhot_zn}) from Yakhot's model constitutes the continuation of the systematic expansion by L'vov and Procaccia is speculative. The accordance of Yakhot's model with the scaling exponent derived by L'vov and Procaccia, however, valorises both results.}

\remark{im limes $\Rey\to\infty$ geht ja auch $\l\to0$, wo findet dann eigentlich dissipation statt? drei antworten. (i) $\Rey\to\infty$ is nicht realistisch. (ii) die turbulenz leute nehmen $\Rey\to\infty$ nicht wörtlich, es soll eher heißen 'sehr sehr, sehr groß'. (iii) es ist eine annahme in Kolmnogorov theorie dass $\beps$ remains finite in the limit of infinite Reynolds numbers.}

\cleardoublepage
\section{Markov analysis} \label{s_MAR}
Having introduced the relevant results of fully developed turbulence for what follows in the remaining part of the thesis, let us take one step back and relate this part with the first part of the thesis.\\
We started the first part with Newton's equation of motion for Brownian particles and ended up with an intimate relation between entropy and irreversibility at the nanoscale. This part, we also started with Newton's equation of motion, but for fluid volumes. We worked us through various assumptions and concepts to arrive at a statistical description of velocity fluctuations in fully developed turbulence. Overall, the analysis for both cases involved exploring the balance between external energy injection and internal heat dissipation. The microscopic mechanism for heat dissipation traces back to non-zero viscosity both for Brownian particles and for decaying turbulence. By including the dynamics of Brownian particles on a time scale smaller than the Markov-Einstein time scale $\tME$ into an ideal heat bath, and likewise, by limiting ourselves in the case of decaying turbulence to the inertial range where dissipative effects are negligible, we coarse grained in both cases the microscopic dynamics to obtain a mesoscopic description of the physics.\remark{mesoskopisch ist auch auf der nanoskala, und liegt irgendwo zwischen mikroskopisch (volle auflösung) und makroskopisch (anfassbar). in STD haben wir die stöße der suspendierten Brownschen teilchen mit den fluid-molekülen als weißes rauschen behandelt, im FDT beschreiben wir anstelle $\vek{v}(\vek{x},t)$ nur die statistik eindimensionaler $u(r)$ oberhalb dissipativer skalen.}\\
In this part we attempt to push this analogy further and explore the implications of describing the statistics of velocity increments $u(r)$ as realisations of a MP.

\subsection{Interpretation as stochastic process} \label{ss_MAR_stochasticprocess}
As for the scaling law in the previous section, we can again say that the following approach uses the phenomenology of the Richardson cascade. The idea is that the eddy which evolved from a larger eddy bears no reference to that larger eddy after sufficient time. The concept of self-similarity conveys the same picture, as each eddy, regardless on its stage in the cascade, initiates its own cascade. In other words, by observing a cascade, it is not possible to decide whether it is the whole cascade or only a part of a larger cascade.\\
This perception -- an eddy determines the next smaller eddy directly and the subsequent eddies only indirectly -- suggests that the cascade of eddies is a MP. In terms of velocity increments $u$ on scales $r$ we can state the Markov assumption as follows:  During the cascade process, an eddy of size $L$ breaks into an eddy of size $g_1L$ with fraction $0\!<\!g_1\!<\!1$. The break-up of the smaller eddy results into an eddy of size $g_1g_2L$ where $g_2$ is independent from $g_1$. Repeating this operation, we write the scale as $r=g_1g_2\dots g_{s(r)}L$, where $s(r)$ specifies the number of stages the cascade has to go through until it arrives at the scale $r$. Expressed on a logarithmic scale, we get $\ln r = \ln L \+ \ln g_1 \+ \ln g_2 \+ \dots \+ \ln g_{s(r)}$\remark{where all $\ln(g_i)<0$}, or by assuming for the sake of clarity that all fractions are equal, $g_i\equiv g_0$, 
\begin{align} \label{eq:MAR_cascade_r}
	\ln(L/r) = s(r)\ln(1/g_0) > 0 \;.
\end{align}
Solving for the cascade stage yields
\begin{align} \label{eq:MAR_cascade_s}
	s(r) = \frac{\ln (r/L)}{\ln g_0} \;.
\end{align}
We apply the same argument to the velocity increments as they evolve down the cascade, that is $u_r=h_1h_2\dots h_{s(r)}u_L$, and obtain
\begin{align} \label{eq:MAR_cascade_u}
	\ln(u_L/u_r) = s(r)\ln(1/h_0) \;,
\end{align}
where $h_i\equiv h_0$ determines the fraction of velocity that an eddy receives from the eddy it evolved from. Solving for $w_r\dfn\ln(u_r/u_L)$ yields
\begin{align} \label{eq:MAR_cascade_w}
	w_r = s(r)\,\ln h_0
\end{align}
which relates the logarithmic velocity increment $w_r$ to the cascade stage $s(r)$ on scale $r$. Note that $\ln u_r\!\sim\!\ln r$ from (\ref{eq:MAR_cascade_w}) and (\ref{eq:MAR_cascade_s}) is in agreement with the scaling-invariance (\ref{eq:FDT_NSEinvar_scal}). By considering $w(r)=\ln(u(r)/u_L)$ as a function of $r$ and differentiating (\ref{eq:MAR_cascade_w}) with respect to $r$, we finally arrive at the ODE
\begin{align}
	-\frac{\pt_r u(r)}{u(r)} &= -\pt_r s(r) \, \ln h_0 = -\frac{\ln h_0}{\ln g_0}\frac{1}{r} \sep a\dfn\frac{\ln h_0}{\ln g_0} \nn
	\Ra\quad  \pt_r u(r) &= a\,\frac{u(r)}{r} \sep u(L) = u_L \;. \nn \label{eq:MAR_cascade_K41}
\end{align}
The solution of this initial value problem is a scaling law
\begin{align} \label{eq:MAR_cascade_K41_scal}
	u(r) = u_L\,(r/L)^a \;.
\end{align}
\newcommand{\xxi}{q}%
In these considerations, we only considered the velocity increments at single cascade stages and ignored the dynamics between these stages. In principle, the dynamics between the stages is deterministic and dominated by the non-linear terms of the NSE, but of course we can not take this dynamics explicitly into account. Instead, we make use of the chaotic property of a turbulent flow which implies that the outcome of each turbulent structure evolving to the next stage of the cascade exhibits a certain randomness. In the spirit of the log-normal model, we add the random variable $Z\big(s(r)\big)$ with amplitude $w_0$ to the logarithmic velocity increments in (\ref{eq:MAR_cascade_w}),
\begin{align} \label{eq:MAR_cascade_random_w}
	w(r) = s(r)\,\ln h_0 + w_0Z\big(s(r)\big) \;.
\end{align}
We define $Z\big(s(r)\big)$ as the outcome of the cumulated randomness in the cascade up to stage $s$, 
where random variables $\xxi_i$ account for the (logarithmic) velocity fluctuations at each stage. As in the log-normal model, we refer to the central limit theorem and infer that the cumulative random variable $Z\big(s(r)\big)$ is normal distributed. The statistical properties of $Z\big(s(r)\big)$ can readily be determined to be
\begin{subequations} \label{eq:MAR_cascade_random_Z_properties}
	\begin{align}
		\lla Z\big(s(r)\big)\rra &= \sum\limits_{i=1}^{s(r)}\lla \xxi_i \rra = 0 \;, \label{eq:MAR_cascade_random_Z_properties_1} \\
		\lla Z\big(s(r)\big)^2\rra &= \sum\limits_{i,j=1}^{s(r)}\lla \xxi_i\xxi_j \rra = \sum\limits_{i,j=1}^{s(r)} \d_{ij} = s(r) \;. \label{eq:MAR_cascade_random_Z_properties_2}
	\end{align}
\end{subequations}
Here, we assumed in the first line that the fluctuations $\xxi_i$ are zero on average, and in the second line we used the Markov assumption that there are no correlations between the stages of the cascade.\\
To proceed, we write (\ref{eq:MAR_cascade_random_Z_properties}) as the continuous approximation
\begin{align} \label{eq:MAR_cascade_random_Z_cont}
	Z\big(s(r)\big) = \int\limits_{0}^{s(r)}\xxi(x)\di x
\end{align}
and require $\lla\xxi(x)\xxi(x')\rra=\d(x-x')$ to retain $\big\langle Z\big(s(r)\big)^2\big\rangle=s(r)$ as in (\ref{eq:MAR_cascade_random_Z_properties_2}). The differentiation of (\ref{eq:MAR_cascade_random_w}) with respect to $r$ now yields
\begin{align}
	-\frac{\pt_r u(r)}{u(r)} &= -\frac{a}{r} - \frac{w_0}{r\ln g_0}\xxi\big(s(r)\big) \sep b\dfn-\frac{w_0^{\,2}}{2\ln g_0}>0 \nn
	\Ra\quad  -\pt_r u(r) &= -a\,\frac{u(r)}{r} + \sqrt{2b\frac{\,u(r)^2}{r}}\,\xi(r)  \sep u(L) = u_L \;, \label{eq:MAR_cascade_K62}
\end{align}
where the minus sign before $\pt_r$ accounts for $\dd r<0$ and we defined the new random variable $\xi(r) \dfns {\xxi\big(s(r)\big)}/\sqrt{-r\ln g_0}$ such that the stochastic integral of $\xi(r)$ has the usual properties as in (\ref{eq:A1_Z_properties}), which will allow us in the next section to identify drift and diffusion for an equivalent formulation with the corresponding FPE (cf. the transformation rule for the independent variable in (\ref{eq:A2_SDE_trans2})).\\
We stress, however, that the continuous approximation (\ref{eq:MAR_cascade_random_Z_cont}) involving $\xi(r)$ implies an infinitely divisible cascade process \cite{She1995}, otherwise $\xi(r)$ is not $\d$-correlated. Only for differences in scales that exceed the typical scale-interval covered by one cascade step, $\ln(r/r')>\ln(1/g_0)$, we can assume that the correlation $\lla\xi(r)\xi(r')\rra\to0$ vanishes. Solving (\ref{eq:MAR_cascade_K62}) yet for arbitrary scales hence implies infinite many cascade stages, an assumption that is justified in the limit $\Rey\to\infty$.\todo{dann darf ich ja eigentlich in die FTs auch nur die auf skalenschritte $\rME$ diskretisierte $\uc$ einsetzen! Mal für K62 probieren! Für jfm $\Dfg$ darf ich wohl die voll-aufgelösten $\uc$ nehmen, da $\Dfg$ auch aus voll aufgelösten $v(x)$ gewonnen wurde. Und darf ich eigentlich die dimlosen $u(r)$ nehmen? wahrscheinlich schon, weil kürzt sich in $u_L/u_r$ und wegen wk-trafo.}\\ From the mathematics presented in \ref{ss_LE} it is clear that (\ref{eq:MAR_cascade_K62}) is a stochastic differential equation (SDE), and as such needs completion by specifying the rule of discretisation. In the derivation of (\ref{eq:MAR_cascade_K62}) we clearly distinguished the deterministic and stochastic origin. Following van Kampen (cf. p.\,\pageref{a-dilemma}), we therefore identify $\sqrt{bu^2/r}\xi(r)$ as external noise for which the mid-point rule should be taken \cite{Kampen1981,vanKampen2007,Gardiner2009}\remark{van Kampen p.232ff, Gardiner p.88 p.99}, which comes with the convenience that we do not have to bother with modified calculus.\footnote{In principle, any other discretisation is possible, but then we have to make sure to modify the rules of calculus accordingly. In the case of the pre-point rule, for instance, we would need to replace ordinary calculus by \Ito~calculus and use the \Ito~lemma for the variable transformation $s(r)$. The resulting SDE would then have to be interpreted in \Ito,~and manipulations of the SDE have to comply with \Ito~calculus. Furthermore, the coefficients of the SDE would loose their physical meaning.}\\
We observe that on the logarithmic scale $s(r)$ the stochastic process defined by (\ref{eq:MAR_cascade_K62}) is in fact nothing else then geometric Brownian motion (GBM) discussed as example in \ref{s_Markov_processes} and \ref{AA_Itocalc}. We will discuss the solutions of the above SDE in the following section.\\

The phenomenology of the calculation presented in this section is the essence of random cascade model \cite{Mandelbrot1974,Benzi1984,Meneveau1987}, see also p.\,165ff in \cite{Frisch95}. The connection of random cascade models to a SDE has been first\todo{(wirklich?)} discussed by She in \cite{She1995} for the log-Poisson model (\ref{eq:FDT_zn_SL_parfree}) and picked up by others \cite{Arneodo1997,Naert1997,Chanal2000,Dubrulle2000,Siefert2007}. We will come back to the relation between random cascade models and SDEs in more detail in the next section.\todo{Can I trace back a geometrical meaning for $\Dg$ in the sense of an entropic barrier?}\\

We explicate the analogy to Brownian motion:\\
The exact process of collisions between Brownian particles and the fluid molecules is coarse grained by considering only the random outcome of the collisions. The energy injection by the random force is provided by an ideal heat bath which models the kicks received by the Brownian particle from the fluid molecules. The interpretation of the ideal heat bath is the essential ingredient in the thermodynamic interpretation of the MP accomplished in chapter \ref{s_td_interpretation}.\\
In the case of a turbulent cascade, the velocity fluctuations throughout the cascade is a result of the energy injection on large scales and the subsequent transfer of energy to smaller scales due to the non-linear interaction of fluid elements. In the SDE (\ref{eq:MAR_cascade_K62}), the energy injection is realised by randomising $u_L$, and the stochastic term $b\sqrt{u^2/r}\xi(r)$ accounts for the fluctuations in velocity that result from the transfer of energy to scale $r$. The energy dissipation at small scales does not enter this model which hinders a thermodynamic interpretation. But the exclusion of dissipation effects in the Markovian description of the cascade suggests that the Markov condition can only be met at scales where dissipative effects are negligible. Indeed, in chapter \ref{ss_MAR_estimateD1D2} we will see that the Markov assumption holds for $r>\rME\approx\l$, where $\rME$ is the spatial analog of the Markov-Einstein time scale $\tME$ discussed after (\ref{eq:Dk_esti}).\\
The SDE (\ref{eq:MAR_cascade_K62}) further constitutes a prescription to artificially generate realisations $\uc$. In the limit $\Rey\to\infty$, in which $\l\to0$, we could compose a flow field $v(x)$ by connecting the realisations $\uc$ in series. We hence obtain an ideal flow field $v(x)$ that respects the characteristics of the cascade process introduced above, ideal in the sense that $\Rey\to\infty$.\\
We mention that it should be kept in mind that a realisation $\uc$ is not one specific cascade of one eddy. Instead, a large ensemble of realisation $\uc$ reflects the statistics caused by the cascade. Considering $u(r)$, we just probe the spatial structures of the flow field being composed of cascades, without being able to pick out a certain cascade.

\subsection{Drift and diffusion} \label{ss_MAR_D1D2}
We have seen in the previous section that a log-normal random cascade model can be represented by a SDE. Formulating the process as a SDE implies the definition of drift and diffusion coefficients $\Df(u,r)$ and $\Dg(u,r)$. In this section, we will explore the representation of the approaches to turbulence, which we introduced in \ref{s_FDT}, in terms of $\Dfg(u,r)$.

\paragraph{Kolmogorov scaling} 
For the cascade model (\ref{eq:MAR_cascade_w}) without stochasticity, we already found the scaling (\ref{eq:MAR_cascade_K41_scal}) implying the following scaling law for the structure functions
\begin{align}
	\Str^n(r) = \lla u_L^{\,n}\rra\,\Big(\mfrac{r}{L}\Big)^{\z_n} \sep \z_n=n\,\mfrac{\ln h_0}{\ln g_0} \;.
\end{align}
It is apparent that $g_0=h_0^{\,3}$ recovers the K41 scaling (\ref{eq:FDT_K41}). If we want to formulate this model in terms of a stochastic process, we can set
\begin{align}
	\Df(u,r) = -\frac{1}{3}\,\frac{u}{r} \sep
	\Dg(u,r) \equiv 0 \qquad \text{(K41)} \;, \label{eq:D1D2_K41}
\end{align}
where the only randomness arises from drawing the initial value $u_L$ from a distribution $p_L(u)$.\\
By including stochasticity in the cascade model, we found the Stratonovich SDE (\ref{eq:MAR_cascade_K62}). Integration of this SDE yields (see appendix \ref{AA_Itocalc}, applying ordinary calculus\remark{diss p.19-21})
\begin{align} \label{eq:D1D2_K62_SDE_sol}
	\frac{u(r)}{u_L} = \Big(\frac{r}{L}\Big)^{a}\,\exp\big[\sqrt{2b\ln(L/r)}\,Z(r)\big] \;,
\end{align}
where $Z(r)$ is a Gaussian random variable with zero mean and variance one. Hence, $u(r)/u_L$ is a log-normal distributed random variable with mean $a\ln(r/L)$ and variance $2b\ln(L/r)$.\remark{diss p.19} \\
The equivalent FPE that governs the PDF $p(u,r)$ reads according to (\ref{eq:FPE_fg}) and (\ref{eq:def_D12})\remark{$\a=1/2$}
\begin{subequations}
	\begin{align}
		-&\pt_r p(u,r) = \big[-\pt_u\Df(u,r) + \pt_u^2\Dg(u,r)\big]p(u,r) \sep p(u,L)=p_L(u) \;, \nonumber \\[5pt]
		&\Df(u,r) = -(a\-b)\,\frac{u}{r} \sep
		\Dg(u,r) = b\,\frac{u^2}{r} \label{eq:D1D2_cascade_K62}
	\end{align}
\end{subequations}
with initial distribution $p_L(u)$, and the minus sign before $\pt_r$ again accounts for $\dd r\!<\!0$. Using the substitutions $s=\ln(L/r)$ and $w=\ln(u/u_L)$, the solution of the FPE can be determined to be\remark{diss p.15a, but there Ito, therefore $a$ instead $a-b$ in mean.}
\begin{align}
	p(u,r) = \frac{1}{u\,\sqrt{4\pi b\ln\!\frac{L}{r}}}\int p_L(u_L)\,\exp\!\Bigg[\-\frac{\big(\ln\!\frac{u}{u_L}-a\ln\!\frac{r}{L}\big)^2}{4b\ln\!\frac{L}{r}}\Bigg]\di u_L  \label{eq:D1D2_K62_FPE_sol}
\end{align}
which is a log-normal distribution for $u/u_L>0$ with mean $a\ln(r/L)$ and variance $2b\ln(L/r)$ in agreement with the solution of the SDE above.
\begin{figure}[!t] 
  \subfloat[][\figsubtxt{Vertically shifted PDF $p(u,r)$ for $\mu=0.26$}]{\label{sf:pur_K62_nu27_shiftzoom}
	\includegraphics[width=0.48\textwidth]{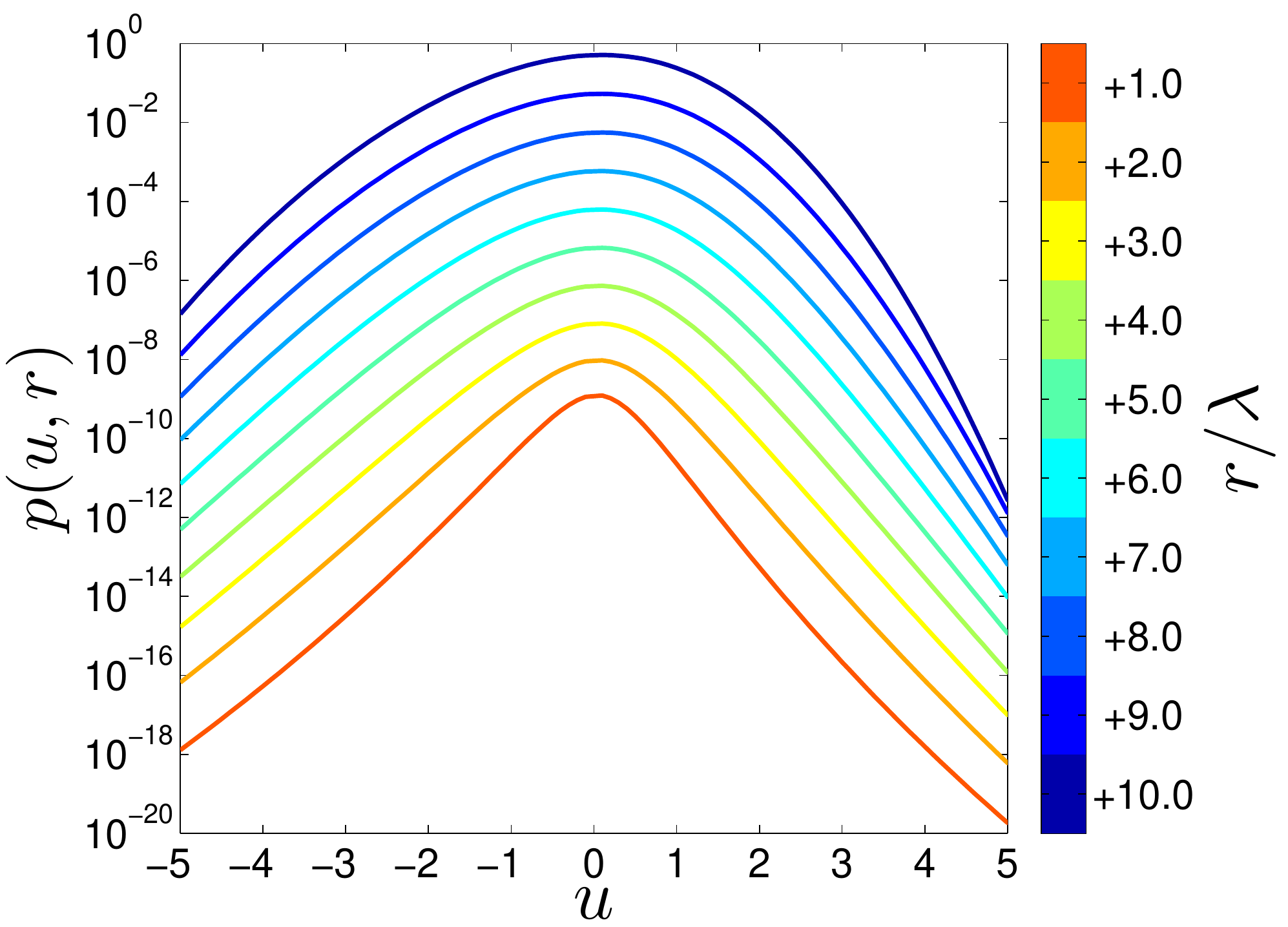}} \hfil
  \subfloat[][\figsubtxt{PDF $p(u,r)$ for $\mu=0.26$}]{\label{sf:pur_K62_nu27}
	\includegraphics[width=0.48\textwidth]{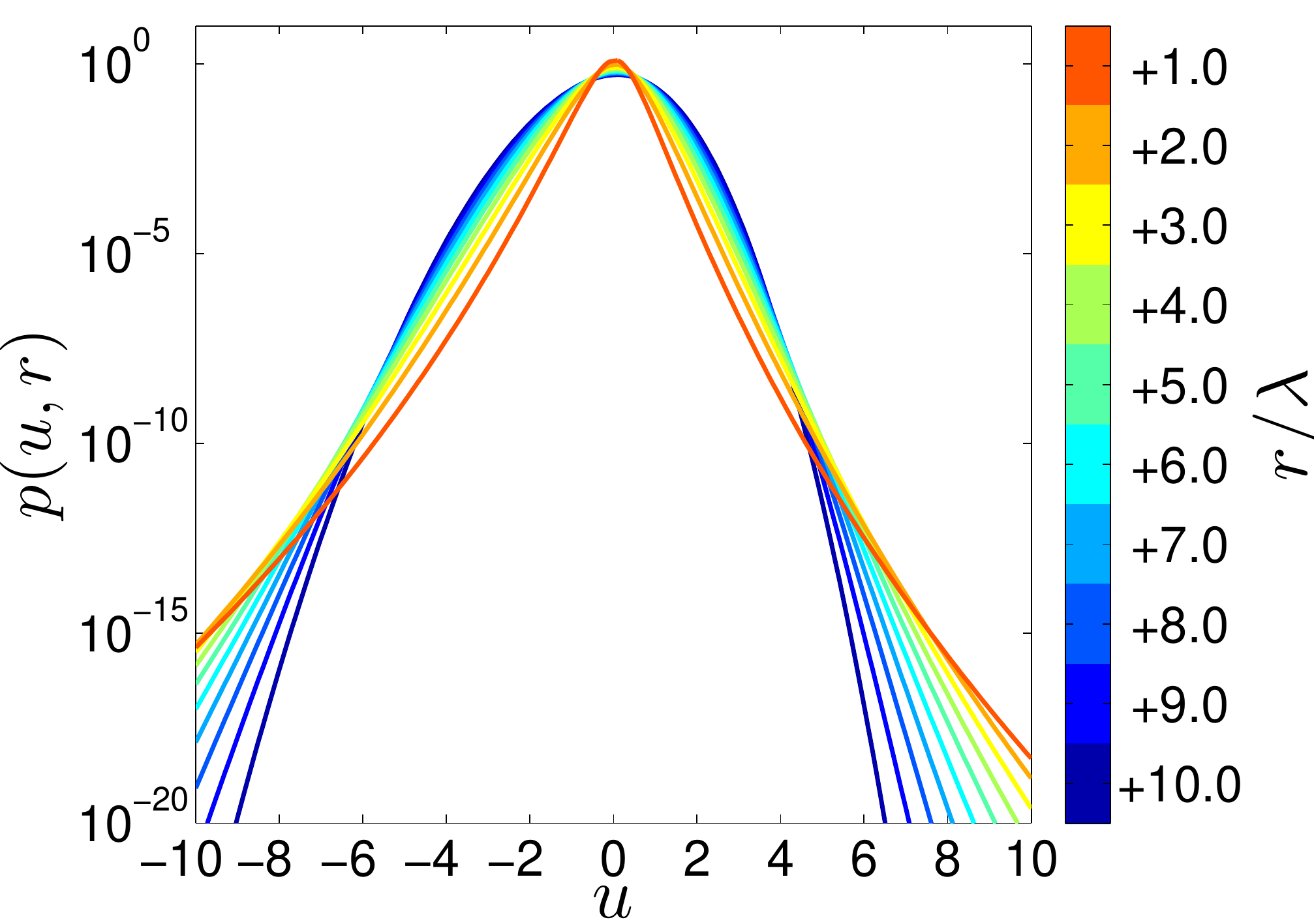}} \\
  \subfloat[][\figsubtxt{PDF $p(u,r)$ for $\mu=0.02$}]{\label{sf:pur_K62_nu304}
	\includegraphics[width=0.48\textwidth]{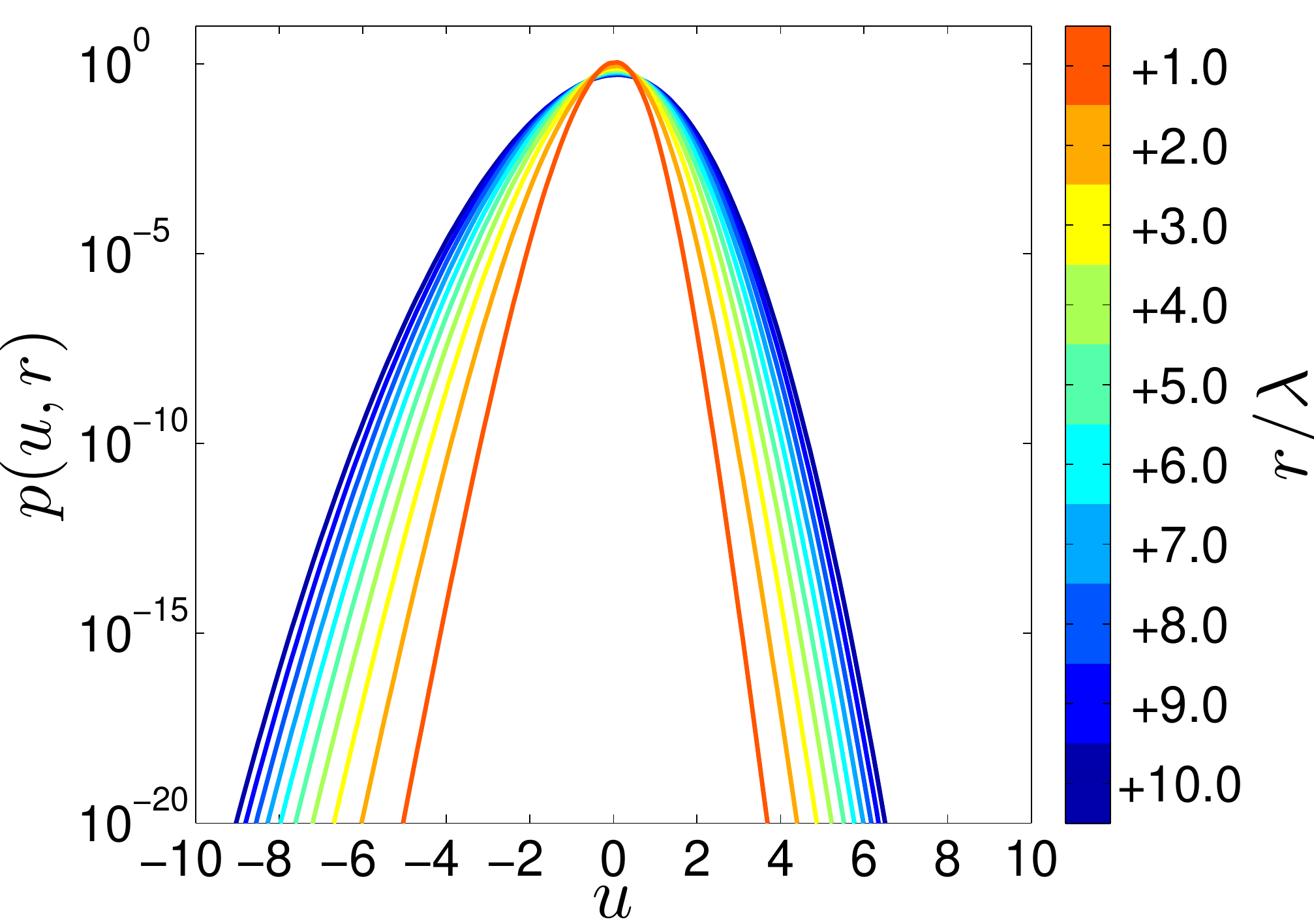}} \hfil
  \subfloat[][\figsubtxt{PDF $p(u,r)$ for $\mu=2$}]{\label{sf:pur_K62_nu7}
	\includegraphics[width=0.48\textwidth]{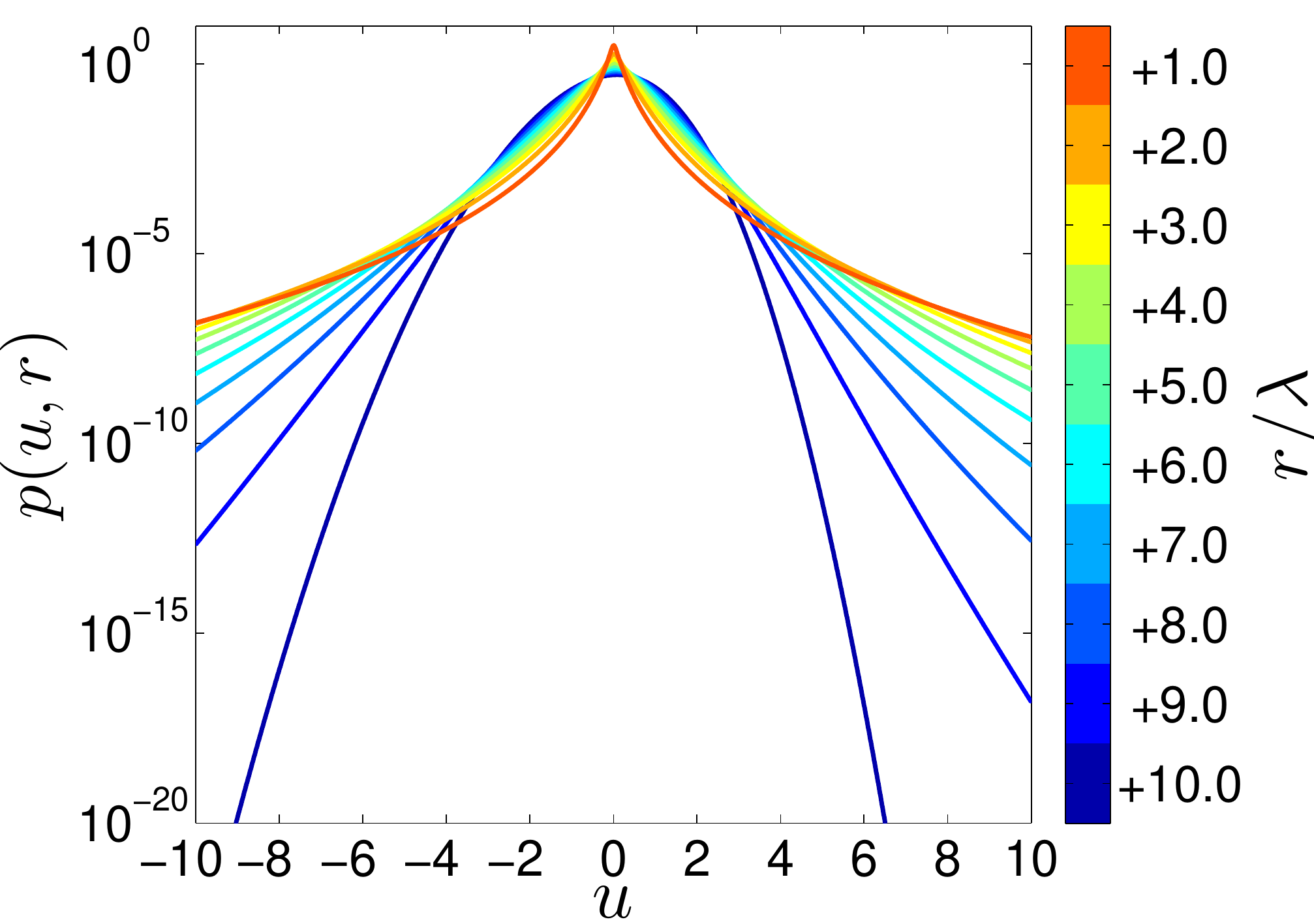}}
  \caption{\label{ff:pur_K62}\figtxt{Analytic solutions $p(u,r)$ of the FPE for the K62 process for various scales $r$ and intermittency factors $\mu$. The analytic solution is given by (\ref{eq:D1D2_K62_FPE_sol}), for the initial distribution $p_L(u_L)$ a skew-normal distribution was taken that was fitted to experimental data from \cite{Renner2001}.}}
\end{figure}\\
From the FPE (\ref{eq:D1D2_cascade_K62}) above and the moment equation (\ref{eq:FPE_moments}), we can conveniently obtain the scaling of the structure functions for the stochastic cascade,
\begin{align}
	&-r\pt_r \Str^n(r) = \bigg[-(a\-b)\,n + b\,n(n-1)\bigg]\, \Str^n(r) \nn
	\Ra\quad& \frac{\Str^n(r)}{\Str^n(L)}  = \bigg(\frac{r}{L}\bigg)^{\displaystyle an -bn^2} =  \bigg(\frac{r}{L}\bigg)^{\displaystyle (a-3b)n -bn(n-3)}\;.
\end{align}
In comparison with the moments in (\ref{eq:FDT_K62_moms}) for the K62 model, we indeed find that the phenomenological introduced SDE (\ref{eq:MAR_cascade_K62}) is equivalent to the K62 model.\\
We recover the K62 scaling for $b=\mu/18$ and $a=(2\+\mu)/6\approx1/3+0.04$\remark{notebook p.75 und diss p.19, $F(u,r)=(3+2\mu)/9$}. The solution for $p(u,r)$ above is therefore the PDF for velocity increments $u$ in an ideal K62 process. See figure \ref{ff:pur_K62} for a plot of $p(u,r)$ for various scales $r$ and intermittency factors $\mu$.\\
The resulting drift and diffusion for the K62 process read
\begin{align}
	\Df(u,r) = -\frac{3+\mu}{9}\,\frac{u}{r} \sep 
	\Dg(u,r) = \frac{\mu}{18}\,\frac{u^2}{r} \qquad \text{(K62)} \;. \label{eq:D1D2_K62}
\end{align}
The equivalence between the K62 model and the FPE with $\Dfg(u,r)$ as above have been first found by Friedrich and Peinke \cite{FriedrichPeinke97PRL,FriedrichPeinke97PDNP}.\\
In the picture of the multiplicative cascade, the K62 model implies a scaling $g_0=h_0^{\,3-\dfr}$ with fractal dimension $\dfr=3-1/a=3\mu/(2+\mu)\approx0.35$ and an amplitude of velocity fluctuations of $w_0^2=\mu/9\,\ln(1/g_0)$.\\
The equivalence between the K62 scaling on a logarithmic scaling, i.e. the velocity increment as a function of the cascade stage, and the geometric Brownian motion, i.e. the Black-Scholes model for stock prices \cite{Black1973,PaulBaschnagel1999}, establishes a curious connection between the classical models of turbulent flows and stock markets.

\renewcommand{\l}{h}
\paragraph{Multifractal model} We have shown that the Kolmogorov model K62 is equivalent to a FPE, if we use a drift coefficient linear in $u$ and a diffusion coefficient quadratic in $u$, as specified in (\ref{eq:D1D2_K62}). We now aim at exploring the general connection between drift and diffusion coefficients $\Dfg(u,r)$ and scaling exponents $\z_n$.\\
To this end, we assume a general power series for $\Dfg(u,r)$,
\begin{align} \label{eq:D1D2_powser}
	\Df(u,r) = -\sum_{k=0}^\infty \df_k(r)\,u^k \sep \Dg(u,r) = \sum_{k=0}^\infty \dg_k(r)\,u^k \;.
\end{align}
Substitution of (\ref{eq:D1D2_powser}) into the moment equation (\ref{eq:FPE_moments}) yields
\begin{align}
	-\pt_r \Str^n(r) &= \sum_{k=0}^\infty\left[ -\!n\df_k(r)\,\Str^{n+k-1}(r) + n(n\-1)\dg_k(r)\,\Str^{n\+k\-2}(r)\right] \nn
  &= \sum_{k=0}^\infty\left[ -\!n\,\df_{k\-1}(r) + n(n\-1)\,\dg_k(r) \right]\Str^{n\+k\-2}(r) \;, \label{eq:D1D2_powser_S_sum}
\end{align}
where we have set $\df_{-1}(r)\equiv0$.\\
By defining the matrix
\begin{align} \label{eq:D1D2_powser_Amat}
  \mat{A} =
  \begin{pmatrix} 
	 a_{02} &  a_{03} &  a_{04} &  a_{05} & \ldots \\ 
	 a_{11} &  a_{12} &  a_{13} &  a_{14} & \ddots \\ 
	 a_{20} &  a_{21} &  a_{22} &  a_{23} & \ddots \\ 
	 0      &  a_{30} &  a_{31} &  a_{32} & \ddots \\
	 0      &  0      &  \ddots &  \ddots & \ddots \\
  \end{pmatrix}
\end{align}
with diagonals
\begin{align} \label{eq:D1D2_powser_Adiag}
	a_{nk}(r) &= n\df_{k-1}(r) - n(n-1)\dg_k(r) \nn
    &= \big(\df_{k-1}(r)+\dg_k(r)\big)\,n - \dg_k(r)\,n^2 \;,
\end{align}
we can rewrite (\ref{eq:D1D2_powser_S_sum}) as a linear system of ordinary differential equations
\begin{align} \label{eq:D1D2_powser_S_SLODE_gen} 
	\pt_r \Str^n(r) = A^n_j(r)\,\Str^j(r) \;.
\end{align}
For the case that $\df(r)$ and $\dg(r)$ depend reciprocally on $r$, we can rewrite (\ref{eq:D1D2_powser_S_SLODE_gen}),
\begin{align} \label{eq:D1D2_powser_S_SLODE} 
	r\pt_r \Str^n(r) = A^n_j\,\Str^j(r) \;,
\end{align}
with now constant $\df$, $\dg$, $a_{nk}$ and $A^n_j$.\\
Substitution of the scaling ansatz $\Str^n(r)=r^\l$ in (\ref{eq:D1D2_powser_S_SLODE}) results into the condition $\det(A^n_j-\l\d^n_j)=0$ in order to obtain non-trivial solutions for $\Str^n(r)$. The general form of this solution yields the sought connection between a scaling law and drift and diffusion coefficients,
\begin{align}
	\Str^n(r) &= \sum_i v^n_i \, r^{\l_i} \nn 
	&= v^n_1\,r^{\l_1} + v^n_2\,r^{\l_2} + v^n_3\,r^{\l_3} + \dots \;\meq\; c_n\,r^{\z_n} \;, \label{eq:D1D2_powser_spec}
\end{align}
with $\l_i$ being the $i$-th eigenvalue to the $i$-th eigenvector $\vek{v}_i$ of matrix $\mat{A}$.\remark{Application of the Gershgorin disc theorem shows that the eigenvalues $\l_k$ are unbounded from above which can not be clear a-priori, considering the specific form of $A^i_j$.}\\
The task might be to find for given $\z_n$ a matrix $\mat{A}$ of form (\ref{eq:D1D2_powser_Amat}) that has eigenvalues $\l_i$ und eigenvectors $\vek{v}_i$ such that (\ref{eq:D1D2_powser_spec}) is satisfied. The entries $a_{nk}$ of that matrix $\mat{A}$ then define $\df_k$, $\dg_k$ and thus $r\Df$, $r\Dg$ using (\ref{eq:D1D2_powser}). But as we are equating a power series in $r$ on the l.h.s. with {\em one} power of $r$ on the r.h.s. of (\ref{eq:D1D2_powser_spec}), we are forced to set $v^n_i=c_n\,\d_{h_i\z_i}$ such that $\mat{A}$ is diagonal. The entries of the diagonal are then the desired scaling exponents, that is, $a_{n2}=\z_n$.\remark{In other words, there is only one $v^n_{i(n)}\!\neq\!0$ such that $\Str^n(r)\!=\!s^n_{i(n)}\,r^{\l_{i(n)}}\!=\!c_n r^{\z_n}$. Hence, by sorting the eigenvalues according to $i(n)=n$, the matrix $\mat{A}$ must be diagonal to allow a scaling law for $\Str^n(r)$. The entries of the diagonal are then the scaling exponents, $a_{n2}\eq=\z_n$.} Using (\ref{eq:D1D2_powser_Adiag}) and (\ref{eq:D1D2_powser}), we get
\begin{subequations}
	\begin{align}
		&\z_n = a_{n2} = \big(\df_1+\dg_2\big)n - \dg_2n^2 \sep \df_{k\neq1}=\dg_{k\neq2}=0 \;, \\
    \Ra\quad &\Df(u,r) = - \df_1\,\frac{u}{r} \sep \Dg(u,r) = \dg_2\,\frac{u^2}{r} \;. \label{eq:D1D2_powser_K62}
	\end{align}
\end{subequations}
\remark{Die allgemeine Summe der skalengesetze hat eine immer nur eine zahl als exponent, je nach fraktaler dimension des unterraums für den dieses skalengesetz gilt. Die ordnung $m$ der strukturfunktion gibt dann vor, wie sich die zusammensetzung dieser skalengesetze ändert, zum beispiel ein verschieben der schwerpunke zu skalengesetze mit größeren exponenten. Für K62 wird immer genau ein skalengesetz rausgesammelt, und der skalenexponent dieses skalengesetzes ist dann identisch mit dem K62 skalenexponent. Dass heißt, der index $k$ an den exponenten $\l_k$ hängt dann von der ordnunge $m$ ab, $k(m)$. Fordere ich also das multifraktale integral wie frisch p.\,144, dann hab ich auch $k(m)$. durch ne substitution oder so müsste ich beide formen übereinander kriegen.}%
The form of $\Dfg(u,r)$ above is the most general form that allows a scaling law. Hence, by comparing the $\Dfg(u,r)$ above with the K62 form of $\Dfg(u,r)$ in (\ref{eq:D1D2_K62}), we see that the K62 scaling is already the most general scaling law covered by a FPE. This limitation of a FPE to K62 scaling has also been found along other lines by Hosokawa \cite{Hosokawa2002}.\\
Writing the scaling exponents in (\ref{eq:D1D2_powser_spec}) as \mbox{$\l_i=\frac{1}{3}-(1-\frac{1}{3}\dfr(i))$} with fractal dimension $\dfr(i)$, we may interpret the sum in (\ref{eq:D1D2_powser_spec}) as the superposition of scaling laws in subspaces of fractal dimension $\dfr(i)$, very much like the multifractal model we mentioned in (\ref{ss_FDT_scalinglaws}) (cf. discussion of equation (8.40) in \cite{Frisch95}). For a full correspondence to the multifractal model, however, a continuous superposition of powers in the form
\begin{align} \label{eq:D1D2_power_multifrac}
	\Dfg(u,r) = \frac{1}{r}\int_{k_\mr{min}}^{k_\mr{max}} \dfg(k) \,u^k \di k
\end{align}
seems promising, which we leave for further studies.

\paragraph{Scaling beyond K62}
In the previous paragraph, we found that the restriction to a FPE to describe the stochastic process in $u(r)$ also limits the resulting scaling law of $\Str^n(r)$ to the form of K41 or K62. In particular, this restriction rules out the promising SL scaling law (\ref{eq:FDT_zn_SL_parfree}). By extending our analysis to discontinuous MPs, however, we will now demonstrate that the SL scaling does have a representation as a MP.\\
In chapter \ref{ss_ME} we learnt that a MP is in general a jump process with a continuous component. The continuous component can be fixed by a drift coefficient $A(u,r)$ and a diffusion coefficient $B(u,r)$, whereas the occurrence of jumps of width $w$ is given by a jump density $\jd(w|u,r)$ with moments $\Ak(u,r)$. Realistic MPs always have a discontinuous component, but in many applications this component is small such that\linebreak $\Df=A+\Af\simeq A$ and $\Dg=B+\Ag/2\simeq B$, and the FPE with $\Dfg$ is a reasonable approximation of the process.\\
To bridge to scaling laws of structure functions $\Str^n(r)$, recall that for the choice $A(u,r)=a(r)\,u$, $B(u,r)=b(r)\,u^2$ and $\Ak(u,r)=\dk\!(r)\,u^k$, the moments follow from the Kramers-Moyal expansion (KME) as 
\begin{align} \label{eq:D1D2_bK62_moms-gen}
	\Str^{\,n}(r) = \Str^{\,n}(L)\,\exp\!\bigg[-\!\int_{L}^{r} na(r)+n(n\-1)b(r) + \sum\limits_{k=1}^{n}\mbinom{n}{k}\dk\!(r)\di r\bigg] \;.
\end{align}
(The minus in the exponent is due to $\dd r<0$, see (\ref{eq:KME_contin}) and (\ref{eq:KME_moms_special_sol}).)\\
Let us temporarily restrict ourselves to a pure jump process, i.e. $a(r)\equiv0$ and $b(r)\equiv0$. For the special case $\dk\!(r)=\dk/r$ with constant $\dk$, we obtain the scaling law
\begin{align} \label{eq:D1D2_bK62_moms-jump}
	\Str^{\,n}(r) = \Str^{\,n}(L)\,r^{\,-\!\tsum_{k=1}^{n}\!\tbinom{n}{k}\dk} \;.
\end{align}
Hence, to obtain a scaling law with exponents $\z_n$, we require that the $\dk$ satisfy
\begin{align} \label{eq:D1D2_bK62_condition_dk}
	\sum\limits_{k=1}^{n}\!\mbinom{n}{k}\dk\meq-\z_n \;.
\end{align}
In some sense, the $\dk$ are the coefficients of a binomial expansion of $-\z_n$. For the SL scaling, it is indeed possible to solve (\ref{eq:D1D2_bK62_condition_dk}) for the $\dk$, if we also take an appropriate drift term into account,
\begin{subequations} \label{eq:D1D2_bK62_SL_expo}
  \begin{alignat}{2}
		A(u,r) &= -\mfrac{1}{3}\Big(1-\mfrac{C_0}{3}\Big)\,\frac{u}{r} &&\sep B(u,r)\equiv0 \label{eq:D1D2_bK62_SL_expo_A} \;,\\
		\Ak(u,r) &= C_0\,\Big(\b^{1/3}-1\Big)^k\,\frac{u^k}{r} && \qquad \text{(SL)}\;. \label{eq:D1D2_bK62_SL_expo_Psik}
	\end{alignat}
\end{subequations}
The corresponding KMCs are $\Df(u,r)=A(u,r)+\Af(u,r)$ and\linebreak $\Dk(u,r)=\Ak(u,r)/k!$ for $k\geq2$.\footnote{The KME with this KMCs should be solvable.}\\
As expected, the K41 process in (\ref{eq:D1D2_K41}) is recovered if we set the intermittency parameter to $\b=1$, which corresponds to zero intermittency, and insert the codimension $C_0=0$.\\
Due to $\b^{1/3}\-1\!\approx\!-0.13$, the odd moments $\Ak(u,r)$ turn out to be negative which incorporates a negative skewness of $\jd(w|u,r)$ for positive $u$. From the $k$-dependence we see also that the moments $\Ak(u,r)$ decrease with increasing $k$ only if $|u|<|1/(\b^{1/3}\-1)|$, implying that intensive intermittent bursts are to be expected if $|u|$ is already quite large. The KMCs, however, are still decreasing rather fast for increasing $k$ and reasonable values for $u$, suggesting that an approximation of the jump process as a continuous process is reasonable.\\
However, it is desirable to set up the master equation of the form (\ref{eq:ME}) or (\ref{eq:ME_discr}) that governs the dynamics of $u$. To do so, we need to solve the moment problem
\begin{align}
	\int w^k \jd(w|u,r) \di w = \frac{C_0}{r}\,(b\,u)^k \sep b\dfn\b^{1/3}-1
\end{align}
in order to get hold of the jump density $\jd(w|u,r)$. This is a challenging problem. Exponential distributions seem to be a promising candidate, although a (compound) Poisson distribution would be the more natural choice (cf. \cite{vanKampen2007} p.\,237ff). Also promising is to formulate the characteristic function of $\jd(w|u,r)$ in terms of its moments $\Ak(u,r)$. At this point, however, further analysis has to be left for future study.\\
We note that similar considerations were pursued by Hosokawa in which he considered the moments of $\ln\eps_r$ instead of $u(r)$ \cite{Hosokawa2002}.\remark{Davoudi und Tabar haben in ihrem 2000 paper nach gleich (10) in einem nebensatz fallen lassen, dass für den adjungierten KM-OP die $\z_n$ die EW zu EF $u^n$ sind.}\\

\renewcommand{\l}{\lambda}
\paragraph{Random cascade models}
Regarding scaling laws, only the two Kolmogorov scalings can be reproduced by a FPE. We will now discuss the representation of turbulence models that do not rest on scaling laws of the form $r^{\z_n}$. We begin with random cascade models.\\
In random cascade models, the PDF of velocity increments, $p(u,r)$, is obtained by propagating the PDF at integral scale, $p_L(u_L$), to smaller scales making use of random multipliers $h$, see (\ref{eq:FDT_random-cascade}).\\
Amblard and Brossier discuss in \cite{Amblard1999} the connection of random cascade models to stochastic processes. They considered the \Ito~SDE
\begin{align} \label{eq:D1D2_castaing_SDE}
	-\pt_r u(r) = -a(r)u(r) + \sqrt{2b(r)}u(r)\xi(r) \sep u(L)=u_L
\end{align}
with positive coefficients $a(r)$ and $b(r)$. The general solution reads (see appendix \ref{AA_Itocalc})
\begin{align} \label{eq:D1D2_castaing_SDE_sol}
	u(r) &= \Phi_{rL}\,u_L \\
	\Phi_{rL} &= \exp\bigg[ \int\limits_{L}^{r} a(r')+b(r') \di r' + \int\limits_{L}^{r} \sqrt{2b(r')}\xi(r') \di r' \bigg] \;, 
\end{align}
where $\ln\Phi_{rL}$ is a normal distributed random variable with mean $\mu_\Phi$ and variance $\s_\Phi^{\,2}$ given by
\begin{subequations} \label{eq:D1D2_castaing_SDE_sol_musig}
  \begin{align}
		\mu_\Phi(r,L) &= \int_{L}^{r} a(r')+b(r') \di r' \;, \label{eq:D1D2_castaing_SDE_sol_mu} \\
		\s_\Phi^{\,2}(r,L) & = -2\int_{L}^{r} b(r') \di r' \;. \label{eq:D1D2_castaing_SDE_sol_sig}
	\end{align}
\end{subequations}
\remark{sollte mit vorzeichen so jetzt stimmen, siehe diss p.22 and p.19-21. das minus für $\s_\Phi^2$ in (\ref{eq:D1D2_castaing_SDE_sol_sig}) kommt aus dem integral in (\ref{eq:D1D2_castaing_SDE_sol}) durch das minus in $\lla\xi(r)\xi(r')\rra=-\d(r-r')$, siehe diss p.22 unten.}%
The form of the solution (\ref{eq:D1D2_castaing_SDE_sol}) is the same construction that led to the propagator formulation (\ref{eq:FDT_random-cascade}) in random cascade models: The statistics of velocity increments on scale $r$ is expressed by the statistics of a random multiplier $\Phi_{rL}$. This equivalence becomes explicit by writing down the PDF for $u(r)$\remark{see diss p.15ab},
\begin{align}
	p(u,r) = \frac{1}{\sqrt{4\pi\s_\Phi(r,L)}}\int\limits_{0}^{\infty} \exp\bigg[-\frac{\big(\ln h-\mu_\Phi(r,L)\big)^2}{2\s_\Phi^{\,2}(r,L)}\bigg]\,p_L\bigg(\frac{u}{h}\bigg)\frac{\dd \ln h}{h} \;, \label{eq:D1D2_castaing_SDE_sol_PDF}
\end{align}
where we defined $h\dfns u/u_L$, and $p_L(u_L)$ is the PDF of the velocity increments on integral scale $L$. Clearly, the propagator $G_{rL}$ from (\ref{eq:FDT_random-cascade}) reads in this case
\begin{align}
	G_{rL}(\ln h) = \frac{1}{\sqrt{4\pi\s_\Phi(r,L)}}\exp\bigg[-\frac{\big(\ln h-\mu_\Phi(r,L)\big)^2}{2\s_\Phi^{\,2}(r,L)}\bigg] \;. \label{eq:D1D2_castaing_SDE_sol_G}
\end{align}
This propagator is equivalent to Green's function of the FPE\remark{
\begin{align} \label{eq:D1D2_castaing_FPE}
	\pt_r p(u,r) &= \Big[-a(r)\,\pt_uu - b(r)\,\pt_u^{\,2}u^2\,\Big]\,p(u,r) \sep p(u,L)=p_L(u)
\end{align}
or}
\begin{align}
	-\pt_r p(u,r) &= \Big[-\pt_u\Df(u,r) + \pt_u^{\,2}\Dg(u,r)\,\Big]\,p(u,r)  \sep p(u,L)=p_L(u) \nn
	\Df(u,r) &= -a(r)\,u \sep \Dg(u,r) = b(r)u^2 \qquad \text{(RCM)} \;. \label{eq:D1D2_castaing_FPE_D1D2}
\end{align}
\remark{Solution according to [Donkov98] reads
\begin{align}
	p(u,r) &= \frac{\ee{\,\a(r)+2\b(r)}}{\sqrt{4\pi\b(r)}} \int\limits_{0}^{\infty} \exp\bigg[\-\frac{\big(\ln h+\a(r)+3\b(r)\big)^2}{4\b(r)}\bigg]\,p_L\bigg(\frac{u}{h}\bigg)\,\frac{\dd h}{h} \nn
	\a(r) &= -\int_{L}^{r} a(r') \di r' \sep \b(r) = -\int_{L}^{r} b(r') \di r' \label{eq:D1D2_castaing_FPE_sol}
\end{align}
which is indeed of the form (\ref{eq:FDT_random-cascade}).}According to (\ref{eq:A2_FPE}), the FPE corresponds to the SDE (\ref{eq:D1D2_castaing_SDE}). From the FPE also follows by use of the moment equation (\ref{eq:FPE_moments}) an ODE for the structure function,
\begin{align}
	-\pt_r \Str^n(r) = \big[-na(r) + n(n-1)b(r)\big]\,\Str^n(r) \;, \label{eq:D1D2_castaing_moms}
\end{align}
where substitution of the scaling ansatz $\Str^n(r)=\Str^n(L)\,r^{\z_n(r)}$ yields
\begin{align}
	\z_n(r) = \frac{1}{\ln r}\int\limits_{L}^{r}n\,\big(\,a(r')-(n\-1)b(r')\,\big)\di r' \;. \label{eq:D1D2_castaing_zn}
\end{align}
Note that solving the moment equation (\ref{eq:D1D2_castaing_moms}) results into
\begin{align}
	\Str^n(r)=\Str^n(L)\exp\Big[n\mu_\Phi(r,L) + \tfrac{1}{2}n^2\s_\Phi^{\,2}(r,L)\Big]
\end{align}
in agreement with the moments of the log-normal distribution in (\ref{eq:D1D2_castaing_SDE_sol_PDF}).\\
On the basis of the drift and diffusion coefficients $\Dfg(u,r)$ from (\ref{eq:D1D2_castaing_FPE_D1D2}) and the resulting scaling exponents $\z_n(r)$ in (\ref{eq:D1D2_castaing_zn}), we can discuss the same special cases as we did in view of the propagator formulation (\ref{eq:FDT_random-cascade}) for random cascade processes.\\
(i) In accordance with (\ref{eq:D1D2_K41}), setting $a(r)=1/(3r)$ and $b(r)\equiv0$ recovers the K41 scaling exponents (\ref{eq:FDT_K41}). We also see from (\ref{eq:D1D2_castaing_SDE_sol_sig}) that in this case the multiplier $\Phi_{rL}$ in (\ref{eq:D1D2_castaing_SDE_sol}) is not a random variable.\\
(ii) Only if $a(r)$ and $b(r)$ depend reciprocally on $r$, the scaling exponents become constant and take the K62 form (\ref{eq:FDT_K62}) which is in compliance with our discussion of (\ref{eq:D1D2_powser_K62}). In this case $\Phi_{rL}$ is a log-normal distributed random variable with variance $\L_r\sim\ln(L/r)$.\\
(iii) By leaving the K62 form of $\Dfg(u,r)$ but keeping the same scale dependency, say $a(r)=a_0\pt_r\ln f(r)$ and $b(r)=b_0\pt_r\ln f(r)$, we obtain a modified scaling law of the form $\Str^n(r)\propto f(r)^{\z_n}$ with now $\z_n=a_0n-b_0n(n-1)\big)$. The random cascade process defined by such a choice of $\Dfg(u,r)$ was experimentally examined by Arneodo \cite{Arneodo1997}, which he termed {\it continuous self-similar} cascade. Note that in this case we can always transform to the new scale $s=\ln f(r)$ such that drift and diffusion become scale-independent. The MP is then stationary in $s$. Note also that for the special case $f(r)=\Str^3(r)$, we recover the ESS scaling (\ref{eq:FDT_ESS_Sn})\remark{, as we have already seen from (\ref{eq:D1D2_ESS})}. In that sense, ESS corresponds to a special random cascade model.\\
(iv) For arbitrary $a(r)$ and $b(r)$, we still have a log-normal random cascade model where $a(r)$ and $b(r)$ determine the mean and variance of the associated log-normal distribution for the multipliers $\Phi_{rL}$ according to (\ref{eq:D1D2_castaing_SDE_sol_musig}). In view of the introductory derivation of the K62 model in terms of a random cascade model, we see that this restriction is due to the fact that a FPE is only equivalent to a SDE with {\it Gaussian} white noise $\xi(r)$ (\cite{Risken89} chapter\,3.3). A SDE with {\it Poissonian} white noise, for instance, would lead to the log-Poisson model by She and Leveque, for which a master equation is needed to describe how $p(u,r)$ evolves in scale. Note that the white noise property of $\xi(r)$ is equivalent to $\lla \xi(r)\xi(s)\rra=\d(r-s)$\remark{FT (ie. the spectrum) of $\d(r-s)$ is constant, all colours}, that is the Markov assumption of the cascade.\\
As a last remark regarding random cascade models, we mention that drift and diffusion of the form (\ref{eq:D1D2_castaing_FPE_D1D2}) exclude a skewness in the solution $p(u,r|u_L,L)$ of the corresponding FPE. A skewness in $p(u,r)$ can only enter through a skewness in the initial PDF $p_L(u)$. This is a known shortcoming of standard random cascade models of the form (\ref{eq:D1D2_castaing_SDE_sol_PDF}) \cite{Dubrulle2000}.
\remark{Straight forward is ESS, since we obtain from
\begin{align}
	\Df(u,r) &= -au\,\pt_r\ln \Str^3(r) \sep \nn
	\Dg(u,r) &= bu^2\,\pt_r\ln \Str^3(r) \sep \label{eq:D1D2_ESS}
\end{align}
immediately
\begin{align}
	&-\pt_r \Str^n(r) = \big(-\!a\,n + b\,n(n-1)\big)\,\Str^n(r)\,\pt_r\ln \Str^3(r) \nonumber\\[5pt]
	\Ra\quad& \frac{\Str^n(r)}{\Str^n(L)}  = \bigg(\frac{\Str^3(r)}{\Str^3(L)}\bigg)^{\displaystyle (a+b)\,n \,-\, b\,n^2}
\end{align}
using the moment equation (\ref{eq:FPE_moments}).}

\paragraph{Yakhot's model}
Recall Yakhot's model of turbulence which we introduced in section \ref{ss_FDT_sm-sc-intm} p.\,\pageref{Yakhotsmodel}. In this model two parameters, $B$ and $A$, enter, together with two characteristics of the specific turbulent flow under consideration, the root-mean-square fluctuations $\vrms$ and the characteristic length scale of turbulence generation $\lchar$. The result of Yakhot's efforts is the partial differential equation for $p(u,r)$ in \ref{eq:FDT_yakhot_pde}.\\
Davoudi und Tabar \cite{Davoudi1999,Davoudi2000} showed that solutions of the PDE (\ref{eq:FDT_yakhot_pde}) satisfy the Kramers-Moyal expansion (KME)\footnote{In \cite{Davoudi2000} the term $\b_k\,u^{k-1}$ enters $\Dk(u,r)$ with the opposite sign than in (\ref{eq:D1D2_Yak_KME_Dk}) which is not correct.}
\begin{subequations} \label{eq:D1D2_Yak_KME}
	\begin{align}
		\hspace*{-50pt}-\pt_r p(u,r) &= \sum\limits_{k=1}^{\infty} (-1)^k \pt_u^{\,k} \big[\Dk(u,r)p(u,r)\big] \;, \label{eq:D1D2_Yak_KME_pde}\\
		\Dk(u,r) &= \mfrac{\a_k}{r}\,u^k - \b_k\,u^{k-1} \;, \label{eq:D1D2_Yak_KME_Dk}\\
		\a_k &= (\-1)^kA\,\mfrac{\G(B\+1)}{\G(B\+k\+1)} = \mfrac{A\,(-1)^k}{(B\+1)(B\+2)\cdots(B\+k)} \;, \label{eq:D1D2_Yak_KME_ak}\\
		\b_k &= (\-1)^k\,\mfrac{\vrms}{\lchar}\,\mfrac{\G(B\+2)}{\G(B\+k\+1)} = \mfrac{\vrms}{\lchar}\,\mfrac{(-1)^k}{(B\+2)(B\+3)\cdots(B\+k)} \;, \label{eq:D1D2_Yak_KME_bk}
	\end{align}
\end{subequations}
where $\G(x)$ is the gamma function and we set $\b_1\dfns0$.\remark{In terms of the coefficients in the power series (\ref{eq:D1D2_powser}), $\dk_k(r)=\a_k/r$, $\dk_{k-1}(r)\equiv\b_k$.}\\
For a further discussion we substitute $A=(B+3)/3$ from (\ref{eq:FDT_yakhot_A}) and rewrite the coefficient $\b_2$ to get for the Kramers-Moyal coefficients (KMCs) the suggestive form
\begin{align} 
	\Dk(u,r) &= \mfrac{\a_k-\tilde\b_k(u,r)}{r}\,u^k \;, \label{eq:D1D2_Yak_KME_Dk_tilde} \\
	\a_k &= \mfrac{(-1)^k\,(B+3)}{3(B\+1)\cdots(B\+k)}\sep \tilde\b_k(u,r) = \mfrac{r/u}{\lchar/\vrms}\,\mfrac{(-1)^k}{(B\+2)\cdots(B\+k)} \;, \nonumber
\end{align}
where $\tilde\b_2(u,r)$ now is dimensionless. Four special cases are apparent.\remark{hier sieht man auch, dass die brechung der GI zu dem term $\b$ term führt der wegen seiner $r$-abhängigkeit das skalengesetz bricht.}\\
(i) It is clear that due to $\G(B+k+1)$ in the denominator of (\ref{eq:D1D2_Yak_KME_ak}) and (\ref{eq:D1D2_Yak_KME_bk}), the coefficients $\a_k$ and $\b_k$ will rapidly decrease with increasing $k$. It is hence a reasonable approximation to truncate the KME after the second term and obtain a FPE with drift and diffusion given by
\begin{align}
\Df(u,r) &= \mfrac{\a_1}{r}\,u \sep \Dg(u,r) = \mfrac{\a_2-\tilde\b_2(u,r)}{r}\,u^2 \;, \qquad \text{(YAK)} \label{eq:D1D2_Yak_FPE_tilde_D1D2} \\
\a_1 &= -\mfrac{B\+3}{3(B\+1)} \sep \a_2 = \mfrac{B\+3}{3(B\+1)(B\+2)} \sep \tilde\b_2(u,r) = \mfrac{r/u}{\lchar/\vrms}\,\mfrac{1}{B\+2} \;. \nonumber
\end{align}
This approximation is in opposition to a ``blind'' application of Pawula's theorem, as we not use $\Dfr(u,r)\approx0$ to argue that all coefficients $\Dgg(u,r)$ are negligible. If $\Dfr(u,r)$ is not exactly zero, the higher coefficients might as well diverge despite $\Dfr(u,r)\approx0$, but here we know that $\a_k\!\to\!0$ and $\b_k\!\to\!0$ for $k\!\to\!\infty$ without using Pawula's theorem. However, it is difficult to say whether the influence of the entirety of $\Dgg(u,r)$, despite the negligibility of the individual terms for $k\to\infty$, remains significant for the dynamics.\\
(ii) For $r/u\ll\lchar/\vrms$, expressing that typical time scales of the internal dynamics of turbulent structures are small compared to the characteristic time scale associated with turbulence generation, we may neglect $\tilde\b_2(u,r)$ in the KME and arrive at
\begin{align} \label{eq:D1D2_Yak_KME_scal}
	-r\,\pt_r p(u,r) &= \sum\limits_{k=1}^{\infty} (-1)^k \pt_u^{\,k} \big[\a_ku^k\,p(u,r)\big] \;.
\end{align}
The moments equation of this KME is of the same form as (\ref{eq:KME_moms_special}), for which we found the solution (\ref{eq:KME_moms_special_sol}) implying a scaling law with exponents\remark{siehe diss p.9}
\begin{align} \label{eq:D1D2_Yak_KME_Dk_scal_zn}
	\z_n = \frac{n}{3}\,\frac{B+3}{B+n}
\end{align}
in agreement with (\ref{eq:FDT_yakhot_zn}).\\
(iii) Neglecting also $\tilde\b_2(u,r)$ in the FPE, that is in the limit $\tilde\b_2(u,r)\to0$ for all $u$ and $r$, drift and diffusion coefficients of the FPE (\ref{eq:D1D2_Yak_FPE_tilde_D1D2}) take the K62 form in (\ref{eq:D1D2_K62}). Therefore, instead of resorting to inertial range scales $L\!>\!r\!>\!\l$, the K62 scaling is included into Yakhot's model in the limit of a clear-cut time scale separation between internal dynamics of turbulent structures and turbulence generation.\footnote{This is of course hard to achieve, since even for $L/\!\lla u_L\rra\ll\lchar/\vrms$ we can not rule out fluctuations of $u_L$ that bring $L/u_L$ close to $\lchar/\vrms$, which, to lower extent, also applies for time scales $r/u$ at smaller scales.}\\
(iv) In the limit $B\to\infty$, all $\Dk(u,r)$ vanish for $k\geq2$ and we recover the K41 form of $\Df(u,r)$ in (\ref{eq:D1D2_K41}) which excludes any intermittency corrections from the model. This role of $B$ is in accord with the discussion of scaling exponents after (\ref{eq:FDT_yakhot_zn_gen}), but here we do not need to resort to comparing scaling laws.\\
In total, Yakhot's model constitutes a generalisation of K62 in two aspects. First, by including all $\Dk(u,r)$ instead of only $\Dfg(u,r)$, the scaling exponents (\ref{eq:D1D2_Yak_KME_Dk_scal_zn}) also account for a jump process underlying the continuous K62 process. And second, direct effects of turbulence generation on the dynamics in the inertial range are accounted for by the coefficient $\tilde\b_k(u,r)$ in (\ref{eq:D1D2_Yak_KME_Dk_tilde}) and imply a departure from a scaling law.\\\label{yakhots_problems}%
Note, however, that for certain values of $u$ and $r$ all even $\Dk(u,r)$ become negative, which is contradictory to the general form of $\Dk(x,t)$ given in (\ref{eq:Dk_esti}). In terms of the dimensionless turn-over time $\t_r\dfns|(r/u)/(\lchar/\vrms)|$ and the critical value $\t_\mr{cr}=3(B\+1)/(B\+3)=1/\z_1$, the $\Dk(u,r)$ become negative for $\t_r<\t_\mr{cr}$.\\
In the Markov formalism, Yakhot's model is hence only valid as long as the normalised time scale of eddy dynamics is larger than the critical value $\t_\mr{cr}$\remark{$\;\approx1/2.7=1/\ee{}\,!$}. Instead of the inertial range condition $\Rey_L\!>\!\Rey_r\!>\!\Rey_\l$ from (\ref{eq:FDT_in-range_Rey}), the Markov representation of Yakhot's model suggests that the K62 scaling law can be expected if\remark{für maximale intermittenz ($B=0$) gibt es gar kein K62 scaling law mehr}
\begin{align}
	1 \gg \t_r > \t_\mr{cr} \approx 0.37 \;.
\end{align}
This condition is hardly fulfilled and gets more likely to be violated for large fluctuations of $u$ which are described by high order structure functions. This observation might explain the failure of the K62 scaling at high order.\\
It is further reasonable to assume that the restriction $\t_r>\t_\mr{cr}$ amounts to the principle of time scale separation underlying the Markov assumption. Accepting $\Rey_r>\Rey_\l$ as the analog of $\t_r>\t_\mr{cr}$ then implies that the Markov assumption is linked to $r>\l$, a guess which will be confirmed in the next section.\remark{Interessant wäre, sich ein model für kleine zeit-skalen zu überlegen, die eine gleichung liefert die ich für das kritische interval anflanschen kann. vielleicht liefert das sogar ne art gekoppelte Markov prozesse, wo die hier gegebene form energie-injektion und -transfer übernimmt, und die angeflaschte gleich die energie-dissipation. hätte dann zwei reserviore und könnte sogar NESS untersuchen! Vielleicht ist das sogar ein hinweis yakhots model zu erweitern? aber link dieses kritischen intervals zu yakhot ist unklar. Negative diffusion tritt auch für $\D r>0$ auf, irgendein bezug zu inverse kaskade??}\remark{der GI breaking term verursacht auch die negative diffusion}\\

In closing this chapter, we mention that it should also be interesting to tackle the moment problem for Yakhots scaling exponents (\ref{eq:D1D2_Yak_KME_Dk_scal_zn}),
\begin{align}
	\int w^k \,\jd(w|u,r) \di w = \frac{B!\,(B\+3)}{3r}\,\frac{(-\,u)^k}{(B+k)!} \;,
\end{align}
which would yield the jump density $\jd(w|u,r)$ and thus a master equation governing the evolution of $p(u,r)$.

\subsection{Experimental investigations} \label{ss_MAR_estimateD1D2}
In the hitherto discussion on MPs representing the eddy cascade in fully developed turbulence, we demonstrated how many major achievements in turbulence research find their counterpart in the theory of MPs. The point of origin was to assume that once an eddy has evolved into smaller eddies, these smaller eddies start their own cascade without reference to the eddy they came from, which we interpreted as being equivalent to the memoryless property of a MP. The equivalence of the Markov description to established theories of turbulence legitimates this interpretation, however, experimental evidence that the Markov assumption holds is called for. In this section we will give a survey of experimental investigations regarding the Markov assumption and also the applicability of a FPE.\\

At the beginning of the first part, we formally defined the Markov condition in (\ref{eq:MarCond}) which states that in a time-ordered series of events, the current event is only influenced by the most recent one. A direct consequence of the Markov condition is the Chapman-Kolmogorov relation (CKR) (\ref{eq:CKR_3conds}) which typifies the kind of Markov chain we associate with the eddy cascade. Building on the CKR, we found that continuous MPs are fixed by SDEs and the corresponding FPEs, and discontinuous MPs can be described by a KME. Hence, if the CKR holds, all results of the first part hold.\\
The first work that pursued the Markov representation of the eddy cascade was in fact experimental. In \cite{FriedrichPeinke97PRL,FriedrichPeinke97PDNP}, Friedrich and Peinke demonstrated that the PDF of velocity increments, $p(u,r)$, indeed satisfies the CKR. They determined conditional probabilities $p(u_1,r_1|u_2,r_2)$ from a free jet experiment and substituted these into the CKR for a number of triples $r_1<r_2<r_3$. In the spirit of a Markov chain, they used the experimentally determined $p(u_1,r_1|u_2,r_2)$ to propagate $p_L(u_L)$ successively downwards in scale in comparison to the directly measured $p(u,r)$. Both investigations demonstrated the validity of the CKR for all accessible $u$ and $r$\remark{even for scales in the dissipative range} which constitutes convincing evidence that the evolution of velocity increments $u$ in scale $r$ is a MP.\\
Friedrich and Peinke also addressed the question whether the MP is continuous or discontinuous by estimation of the first four KMCs $\Dff(u,r)$ according to the procedure described in the context of the definition of $\Dk$ in (\ref{eq:Dk_esti}). They found that, within the error margin, $\Dtf(u,r)$ vanish for all accessible $u$ and $r$, whereas $\Dfg(u,r)$ have well-defined limits. Referring to the theorem of Pawula as discussed after (\ref{eq:tAk-vs-Ak}), they argued that the MP should be continuous, that is to say, the jump density (\ref{eq:def_jump-distr}) is a $\d$-function for which all moments (including $\Dfr$) vanish. This is a bold statement, since it is hardly possible to prove $\Dfr(u,r)\equiv0$ for all $u$ and $r$ by experimental means, and $\Dfr(u,r)\approx0$ does not imply that the FPE is a good approximation, as the KME is not a systematic expansion in the sense that a small $\Dfr$ implies smaller $\Dggg$. In other words, a jump density that has a vanishing fourth moment needs only to be vanishing in the part that contributes to the forth moment and might as well involve a significant probability for extreme jumps, a scenario not unthinkable in the context of turbulence.\\
However, such a scenario remains to be peculiar, and $\Dfr(u,r)\approx0$ is a respectable indication that a FPE is a good approximation of the MP. Furthermore, the finding (\ref{eq:D1D2_Yak_KME}) by Davoudi and Tabar which implies that the higher $\Dk(u,r)$ rapidly tend to zero, a finding that builds on the PDE (\ref{eq:FDT_yakhot_pde}) for $p(u,r)$ which Yakhot derived by using the full NSE, corroborates the assumption that $\Dfr(u,r)\approx0$ indicates $\Dgg(u,r)\to0$ for all $u$ and $r$ in the limit $k\to\infty$.\\

The pioneering work by Friedrich and Peinke, now primarily referred to as {\it Markov analysis}, entailed more activity than we can present here. However, we review a selection of developments, for further information we refer the reader to the review article by Friedrich et al. \cite{Friedrich2011a}.\\
A major improvement of the analysis was achieved by Renner et al. in \cite{Renner2001,Rennerdiss}. The authors confirmed in a careful free jet experiment the validity of the Markov assumption\footnote{The next chapter will analyse the data taken in this experiment and also give some characteristics of this data. Histograms of $u$ approximating $p(u,r)$ were already presented in figure \ref{ff:sm-sc-intm}, the scaling behaviour was analysed in figure \ref{ff:zetan_ESS}.}, but identified a scale below which the Markov assumption breaks down. In a sophisticated statistical procedure, the so-called Wilcoxon test (for details see appendix A in \cite{Renner2001}), they pinpointed this scale to be about the Taylor length scale $\l$. The Wilcoxon test thus confirmed the Markov assumption for all scales $r\,\mscale[0.7]{\gtrsim}\,\l$ (and $r\mscale[0.85]{<}L$). As below the scale $\l$ the dissipative range succeeds the inertial range, the role of $\l$ in the Markov analysis suggests that molecular friction causes the break-down of the Markov assumption. In analogy to Einsteins theory of Brownian motion, the exact scale above which the Markov assumption holds is termed {\it Markov-Einstein length scale} $\rME$, the spatial analogue of the Markov-Einstein time scale $\tME$ discussed after (\ref{eq:Dk_esti}). Various experimental investigations confirmed that $\rME\simeq\l$ holds for different flow conditions and a wide range of Reynolds numbers \cite{Luck2006,Siefert2007,Stresing2011}.\\
Renner and co-workers also addressed the question whether a FPE is a suitable approximation of the MP. The estimated $\Dfr(u,r)$ is of similar shape as $\Dg(u,r)$ but by three orders of magnitude smaller and within the error margin not distinguishable from being zero. Using the KME truncated after the fourth term with $\Dg(u,r)$ and $\Dfr(u,r)$ estimated form their data, they further demonstrated that the influence of $\Dfr(u,r)$ on the eighth-order structure function $\Str^8(r)$ is negligible compared to the influence of $\Dg(u,r)$\remark{faktor 10 bis 100, wobei der einfluss für $r<2\l$ nicht mehr vernachlässigbar erscheint}. Truncating therefore the KME after the second term, they obtained the FPE\footnote{Velocity increment $u$ and scale $r$ are given in units of $\sigi=0.54\,\mr{m/s}$ at infinite scales and the Taylor length scale $\l=6.6\,\mr{mm}$, respectively.}
\begin{align} \label{eq:D1D2_JFM_FPE}
	-\pt_r p(u,r|u_L,L) = \big[-\pt_u \Df(u,r) + \pt_u^2 \Dg(u,r)\big] p(u,r|u_L,L)
\end{align}
with the estimated drift and diffusion coefficients
\begin{subequations} \label{eq:D1D2_JFM_D1D2}
  \begin{alignat}{9}
		&\Df(u,r) =\;& -&a_0r^{0.6} \;&-\;& a_1r^{-0.67}\,u \;&+\;& a_2\,u^2 \;&-\;& a_3r^{0.3}\,u^3 \;&  \label{eq:D1D2_JFM_D1} \\
		&\Dg(u,r) =\;&  &b_0r^{0.25} \;&-\;& b_1r^{0.2}\,u \;&+\;& b_2r^{-0.73}\,u^2 & & & \label{eq:D1D2_JFM_D2}
	\end{alignat}
	\vspace{-30pt}
	\begin{alignat}{7}
		&a_0 = 0.0015 &\sep& a_1 = 0.61 &\sep& a_2 = 0.0096 &\sep& a_3 = 0.0023& \; , \\
		&b_0 = 0.033 &\sep& b_1 = 0.009 &\sep& b_2 = 0.043 \;. & & & 
	\end{alignat}
\end{subequations}
Solving this FPE numerically, Renner et al. found for various scales that $p(u,r|u_L,L)$ fixed by the FPE above is in perfect agreement with the experimental result, confirming again the permissibility to approximate the MP with a FPE. Note that the estimation of $\Dfg(u,r)$ involves the conditional PDFs $p(u,r|u_L,L)$, in order to describe the measured $p(u,r)$ the FPE (\ref{eq:D1D2_JFM_FPE}) needs to be complemented by the measured initial distribution $p_L(u)$.\\
The form of the drift $\Df(u,r)$ in (\ref{eq:D1D2_JFM_D1}) is approximately the same as for the K62 model in (\ref{eq:D1D2_K62}), if we acknowledge that the term linear in $u$ is significant larger then the other terms. The diffusion coefficient $\Dg(u,r)$, however, is clearly not purely quadratic in $u$ as the K62 counterpart in (\ref{eq:D1D2_K62}). The constant term is responsible for background fluctuations that persist $u=0$ and ensures that $\Dg(u,r)>0$ for all $u$ and $r$. The linear term, though small, proves to be crucial in order to account for the skewness in $p(u,r)$.\remark{hab ich meinen spielerein mit $\eps$-FT gefunden.}\\
As pointed out by Renner et al. in \cite{Renner2001}, the corresponding SDE to the FPE (\ref{eq:D1D2_JFM_FPE}) can also be formulated involving an additive and a multiplicative noise term 
\begin{align} \label{eq:D1D2_JFM_SDE_two-xi}
	-\pt_r u(r) = \Df(u,r) + \xi_0(r) + u(r)\xi_2(r) \;,
\end{align}
where $\xi_0(r)$ is determined by the constant term of $\Dg(u,r)$ and $\xi_2(r)$ by the quadratic term of $\Dg(u,r)$, and the linear term of $\Dg(u,r)$ gives rise to a correlation between $\xi_0(r)$ and $\xi_2(r)$. It is this correlation between additive and multiplicative noise sources that is responsible for the skewness in a generalisation of random cascade models \cite{Dubrulle2000}.\todo{in hinblick auf 3faces auswerten! es ist tatsächlich $\xi_0$ groß für große $r$ und $\xi_2$ groß für kleine $r$!}\\

By the same authors, data from a cryogenic helium jet experiment \cite{Chanal2000} for Reynolds numbers ranging from $\Reych\approx10^4$ to $\Reych\approx10^6$ was used to characterise how the terms in $\Dfg$ depend on the Reynolds number. On the basis of ten sets of $\{v(x)\,|\,\Rey\}$, they determined drift and diffusion to be\remark{hier scheint der lineare term in $\Dg$ nicht verantwortlich für skewness zu sein, eher der konstante in $\Df$. widerspricht meiner erfahrung und der deutung der SDE mit zwei rauschtermen.}
\begin{subequations} \label{eq:D1D2_FPE_PRL}
  \begin{alignat}{2}
		&\Df(u,r) =\;& -&a_1(r)\,u  \label{eq:D1D2_PRL_D1} \;, \\
		&\Dg(u,r) =\;&  &b_0(r;\,\Rey) \,-\, b_1(r;\,\Rey)\,u \,+\, b_2(r;\,\Rey)\,u^2 \;, \label{eq:D1D2_PRL_D2}
	\end{alignat}
	\vspace{-25.06pt}
	\begin{align} 
		a_1(r) &= 0.67 + 0.2\sqrt{r/\l} \;, \\
		b_0(r;\,\Rey) &= 2.8\Rey^{-3/8}\,r/\l \sep b_1(r;\,\Rey) = 0.68\Rey^{-3/8}\,r/\l  \;.
	\end{align}
\end{subequations}
Accordingly, the drift $\Df(u,r)$ proved to be independent from $\Rey$, whereas the diffusion is in all terms Reynolds dependent. Specifically, the constant and linear terms of $\Dg(u,r)$ could explicitly be formulated as functions of $\Rey$, but the quadratic term evades such a simple dependency. In the limit of infinite Reynolds numbers, the quadratic term should acquire the asymptotic form $b_2(r;\infty)=a_1(r)/2-1/6$ in order to respect the four-fifths law (\ref{eq:FDT_four-fifths}). Indeed, the estimated form of $b_2(r;\Rey)$ exhibits a tendency towards $b_2(r;\infty)$ for increasing $\Rey$, but still remains distinct from $b_2(r;\infty)$ to a considerable extent.\\

Instead of velocity increments, Naert et al. used the one-dimensional surrogate (\ref{eq:FDT_eps_surrogate}) and (\ref{eq:FDT_K62_epsr}) to compute the locally averaged energy dissipation rate $\eps_r$ from a set $\{v(x)\}$ measured in a free jet experiment \cite{Naert1997}. Instead of the physical scale $r$, they used $l=\ln(L/r)$ which now varies from $l=0$ to $l\simeq\ln(L/\rME)$.\\
To describe the evolution of the stochastic variable $x(l)\dfns\ln\eps(l)$ from large to small scales, they estimated drift and diffusion coefficients in $x$ and $l$. It turned out that a drift $\Df(x,l)$ linear in $x$ and a constant diffusion $\Dg$, i.e. an Ornstein-Uhlenbeck process, is well suited to describe their data. The fourth order coefficient is found to be constant and $\sqrt{\Dfr}\approx0.05\Dg$, indicating that also for $\ln\eps(l)$ the underlying jump process is negligible. The Markov-Einstein length turned out to be $\rME\simeq\eta$, which means, in the case of the stochastic variable $x(l)$, the Markov assumption appears to hold even in the dissipative range.\\
The solution $p(x,l)$ of the FPE of an Ornstein-Uhlenbeck process is Gaussian which confirms the widely used assumption that $\eps(r)$ is log-normal distributed. The variance of $p(x,l)$ increases exponentially with $l$ implying that in physical scale $r$ the variance obeys a power law in agreement with the result (\ref{eq:FDT_castaing_sigr}) of Castaing and co-workers and in contradiction to K62.\\
However, in a subsequent investigation, Marcq and Naert examined the statistical properties of the noise by extracting $\xi(l)$ from the Ornstein-Uhlenbeck process and found that $\xi(l)$ is indeed uncorrelated for $r>\rME$ but exhibits a considerable skewness \cite{MarcqNaert98PDNPa}. They argue that this skewness is the consequence of neglecting the third order coefficient $\Dth(ur)$. Ultimately, a non-Gaussian $\xi(l)$ indicates that the approximation of the MP by a FPE does not hold, whereas the $\d$-correlation of $\xi(l)$ above the elementary step-size $\rME$ proves the Markov assumption to be justified.\\
In the later article, Renner and Peinke also investigated $\eps_r$, but coupled with the stochastic process for $u(r)$, that is to say, a two-dimensional coupled stochastic process in the variables $u(r)$ and $x(r)=\ln(\eps_r/\beps)$ \cite{Renner2002}.\remark{In this two dimensional case, the drift is a vector $\vDf(u,x,r)=\big(\,\Df_u(u,x,r),\,\Df_x(u,x,r) \,\big)$, and the diffusion coefficient $\mDg(u,x,r)$ a matrix with diagonal entries $\Dg_{uu}(u,x,r)$, $\Dg_{xx}(u,x,r)$ and off-diagonal entries $\Dg_{ux}(u,x,r)$, $\Dg_{xu}(u,x,r)$.} Performing a two-dimensional Markov analysis, they found a diagonal diffusion matrix and that the stochastic process in $u(r)$ is a scale-dependent Ornstein-Uhlenbeck process where the diffusion coefficient depends on $x$. The drift of the process in $x(r)$ is also linear in $x$ and scale-dependent, but the diffusion coefficient is of rather complicated form and depends on $r$ and both $u$ and $x$. Hence, the stochastic processes for $u(r)$ and $x(r)$ couple only via their stochastic term, a finding that supports the perception that the randomness of $u$ originates from fluctuations of energy dissipation. Indeed, if $\eps_r$ was not a stochastic variable, $u(r)$ would reduce to a Gaussian. Consequently, the intermittent fluctuations of $u(r)$ on small scales trace back to the stochastic and scarcely accessible\remark{weil $\Dg_{xx}(u,x,r)$ von unklarer form} nature of energy dissipation.\\
Note that the considerable simplification of the stochastic process in $u$ was achieved by extending the analysis to two dimensions, instead of mapping the high-dimensional problem of turbulence on the analysis of longitudinal velocity increments. A further simplification can be expected by also including transversal velocity increments into the stochastic analysis.\\

We discussed phenomenological achievements of the Markov analysis that were initiated by the work of Friedrich and Peinke.\\
On the methodological side, significant progress has been made on the procedure to extract drift and diffusion coefficients from experimental data involving maximum-likelihood estimators \cite{NawroPeinkKleinFried07PRE,Kleinhans2007,Kleinhans2012}, finite-size corrections in poor sampled statistics \cite{Ragwitz2001,Honisch2011,Honisch2012} and corrections in the presence of strong measurement noise \cite{Bottcher2006,Carvalho2011}. A comparison of different estimation procedures is provided in \cite{GottschallPeinke08NJoP}.\\

Finally, we mention that the Markov analysis is also applied to atmospheric turbulence \cite{Lind2005} and wind farms \cite{Milan2013,Raischel2013}, and even found its way to fields of research that are not related to turbulence wherever complexity is involved, see page 106 of \cite{Friedrich2011} for an overview.

\cleardoublepage
\section{Fluctuation theorems and irreversibility} \label{s_turbulence_FT}
In the first chapters of this part of the thesis, we have introduced various theories and models that address universal properties of velocity fluctuations in the idealised concept of fully developed turbulence. In the preceding chapter, we demonstrated that these 'traditional' approaches to turbulence all have a Markovian counterpart, implying drift and diffusion coefficients $\Dfg(u,r)$, and in principle also a jump density $\jd(w|u,r)$.\\
In chapter \ref{s_td_interpretation} of the first part, we have in particular discussed the total entropy production (EP) $\Stot$ of a thermodynamic MP as defined in (\ref{eq:EP_Stot}) in terms of the potential $\p$ defined in (\ref{eq:defs_pst_phi_FD}). The potential is basically the integral of $\Df/\Dg$ and implies by (\ref{eq:defs_pst_phi_FD}) an instantaneous stationary distribution $\pst$. The entropy $\Stot$ is directly related to the irreversibility of the process, as we learnt from (\ref{eq:FTs_WPI_Stot}), and as such obeys the fluctuation theorems (FTs) in (\ref{eq:FTs_iFT_Stot}) and (\ref{eq:FTs_dFT_Stot}). The FTs include the second law $\Stot>0$ as shown in (\ref{eq:FTs_secondlaw})\\
In view of this interpretation of MPs, the aim of this chapter is to explore the implications of the various Markov approaches to turbulence discussed in the previous chapter. In the following we will always use ordinary calculus, therefore, discretisations of stochastic integrals have to be made in mid-point, and equivalent SDEs need to be formulated in the Stratonovich convention, but we do not have to bother with modified calculus.
\subsection{Experimentally estimated versus K62} \label{ss_turbulence_PRL}
The ultimate aim would be to find a universal FT for velocity increments, in which the EP $\Stot[\uc]$ qualifies the irreversibility of the energy cascade. This EP would at the same time allow to distinguish realisations $\uc$ of the direct energy cascade ($\Stot[\uc]>0$), from the realisations of the inverse energy cascade ($\Stot[\uc]<0$). We will see that this is a particularly difficult endeavour.\\
On the one hand, we can estimate $\Dfg(u,r)$ directly from experimental data and expect that the resulting $\Stot[\uc]$ defines a FT that holds for this very data. The convergence of the FT for the available amount of data, however, is only guaranteed for sufficient number of realisations $\uc$ with $\Stot[\uc]<0$, a requirement that is typically not met in macroscopic systems such as turbulent flows, cf. the discussion of the scope of stochastic thermodynamics after (\ref{eq:FTs_secondlaw}). Furthermore, with $\Dfg(u,r)$ being tailored to match the statistics of experimental data without any physical input, the entropy $\Stot[\uc]$ evades a direct physical interpretation.\\
On the other hand, getting hold of proper $\Dfg(u,r)$ from first principles involving physical quantities, such as the energy transfer rate $\eps_r$, length scales $L$ and $\r$ or viscosity $\nu$, is an unsolved problem.\footnote{An exception are $\Dfg$ from Yakhot's model involving $L$ and $\vrms$, but these $\Dfg$ are not without problems as we have discussed in \ref{ss_turbulence_beyondscaling} p.\,\pageref{yakhots_problems} .} In addition, the extreme sensitivity of the FT to the correct modelling of extreme events is delicate, but at the same time constitutes a valuable tool to benchmark possible $\Dfg$.\remark{zwischenweg wäre also eine feste form von $\Dfg(u,r)$ mit parametern, die in irgendner weise nen physikalischen bezug haben, und diese dann mit FTs zu fitten. schließlich ist bekannt, dass viele deutlich verschiedene $\Dfg$ kaum unterscheidbare MPs definieren, die frage ist eine form von $\Dfg(u,r)$ zu haben, die physikalisch interpretierbar ist und nicht einfach nur funktioniert.}\\
At this point, we present our third publication \cite{Nickelsen2013}, in which we contrast the FT from experimentally estimated $\Dfg(u,r)$ (\ref{eq:D1D2_JFM_D1D2}) with the FT resulting from the K62 theory (\ref{eq:D1D2_K62}). By doing so, also the basic concepts of stochastic thermodynamics will be recapped. Note that the experimental data referred to in the publication is the data discussed in \cite{Renner2001} which we already used in the previous chapters. Note also that the parameter $\nu=(6+4\mu)/\mu\approx27$ introduced in the publication bears no reference to viscosity so far denoted by $\nu$. In the remaining part of this thesis it will always be $\nu=(6+4\mu)/\mu$.
\includepdf[pages=1-5,trim=6mm 0mm 17mm 7mm,clip,scale=0.85,pagecommand={\pagestyle{headings}}]{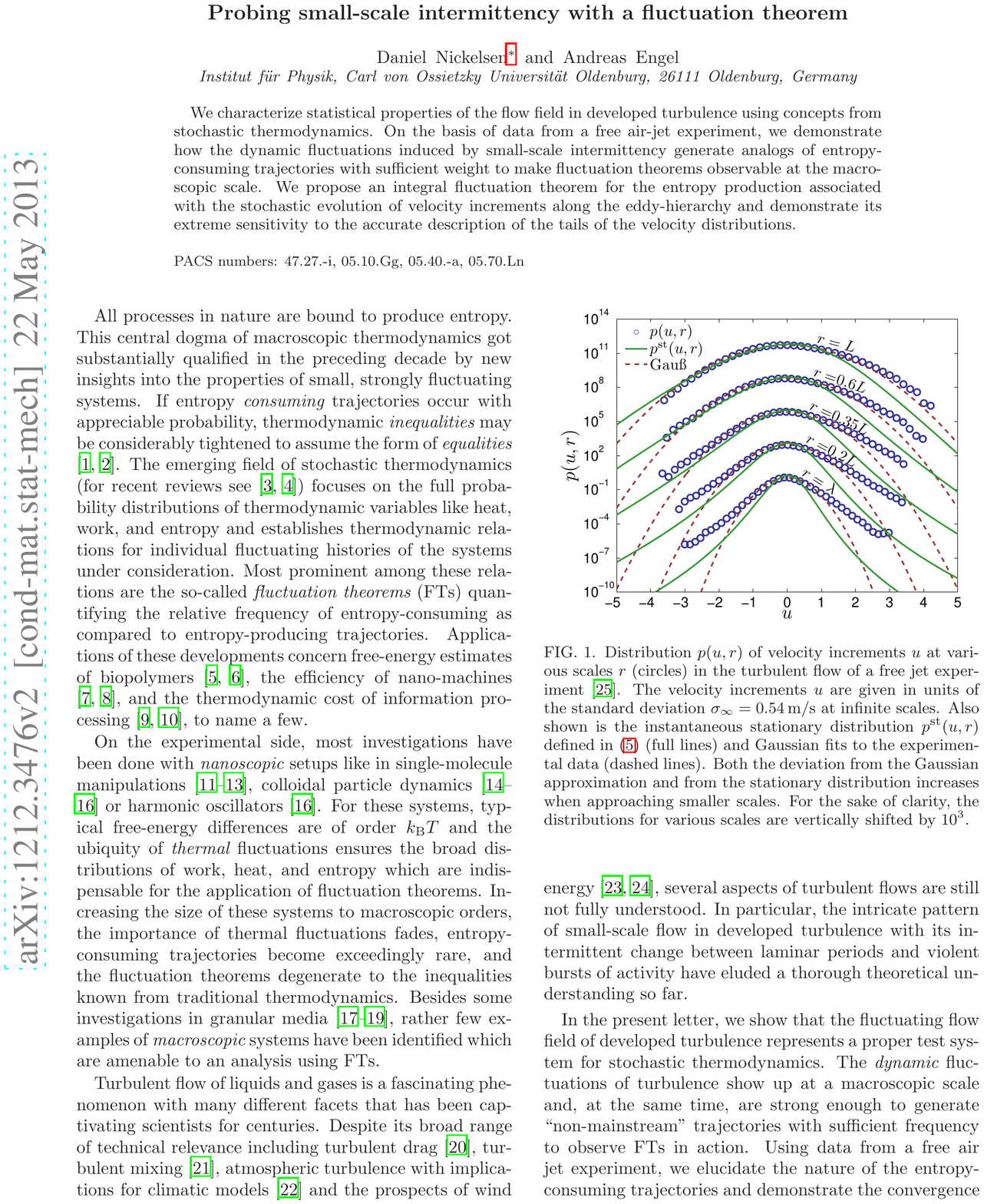} 
\paragraph{The K62 second law}
\remark{The connection to reversal of energy transfer rate is suggestive but unclear, as no physical background on $\Dfg$. FT observed although turbulent flow field is rather macroscopic (with $\l$~mm, L~cm). Dominant trajectories needed for convergence of exponential average in FT.}\remark{diss p.34}We add a discussion concerning the K62 iFT from equation (13),
\begin{align} \label{eq:turbFTs_K62_FT}
	\lla \frac{u_r^{\,\nu} \, p_r(u_r)}{u_L^{\,\nu} \, p_L(u_L)} \rra = 1 \;.
\end{align}
Note that via $g=2-\nu$, the parameter $\nu\approx27$ is directly related to the exponential growth rate $g$ of geometric Brownian motion, where\linebreak $g>1$ signifies divergence and $g<1$ signifies decay in the evolution of the stochastic variable, see the discussion after (\ref{eq:A1_GBM_Ito_momsol}). As $\nu\geq4$, we always have a negative growth rate $g\leq-2$ and therefore decaying turbulence, as it should be.\remark{dabei hab ich $\dd r<0$ schon berücksichtigt, hab ja schon ein minus vor $a$ in der SDE, siehe diss. p.23 (sonst wäre $g=\nu-2\geq2$ und ich hätte divergence für $r=t\to\infty$ und decay für $t=r\to\infty$)}\\
Analogous to recovering the second law from an iFT, cf. (\ref{eq:FTs_secondlaw}), we find from (\ref{eq:turbFTs_K62_FT}) by exponentiation and application of Jensen's inequality the second law like inequality
\begin{align} \label{eq:turbFTs_K62_secondlaw_pre}
	\lla \ln\frac{u_L}{u_r} \rra \geq \frac{1}{\nu}\lla\ln\frac{p_r(u_r)}{p_L(u_L)}\rra \;.
\end{align}
We recognise on the r.h.s. the Shannon entropy $S(r)=-\lla\ln p_r\big(u(r)\big)\rra$, and, in view of the K62 random cascade model, cf. (\ref{eq:MAR_cascade_u}), we identify the l.h.s. as the sum of the averaged logarithm of the multipliers $h_i=u_i/u_{i+1}$, and write
\begin{align} \label{eq:turbFTs_K62_secondlaw}
	\lla\ln H_r\rra = \sum\limits_{i=1}^{s(r)} \lla \ln h_i \rra \leq \frac{1}{\nu}\D S(r) \;.
\end{align}
Here, $s(r)$ is the number of stages the cascade took until it reached a scale $r$, cf. (\ref{eq:MAR_cascade_s}), $H_r=h_1h_2\cdots h_{s(r)}$ is the overall multiplier such that $u_r=H_ru_L$, and $\D S(r)=S(r)-S(L)$ is the difference in Shannon entropy between final and initial stage of the cascade.\\
In section \ref{ss_td-interpration_FTs_applications}, we discussed the condition $\Sm\sim-\D S$ under which the validity of a FT can be expected to be observable in a system, where $\Sm$ was defined as the entropy transferred into a heat bath. Here, we have $\Sm=-\ln H_r$, the formally introduced entropy $\Sm$ therefore accounts for the overall multiplier statistics.
\begin{figure}[!t] 
  \subfloat[][\figsubtxt{Shannon entropy $S(r)$}]{\label{sf:2ndlaw_K62_Shannon_nu27}
	\includegraphics[width=0.47\textwidth]{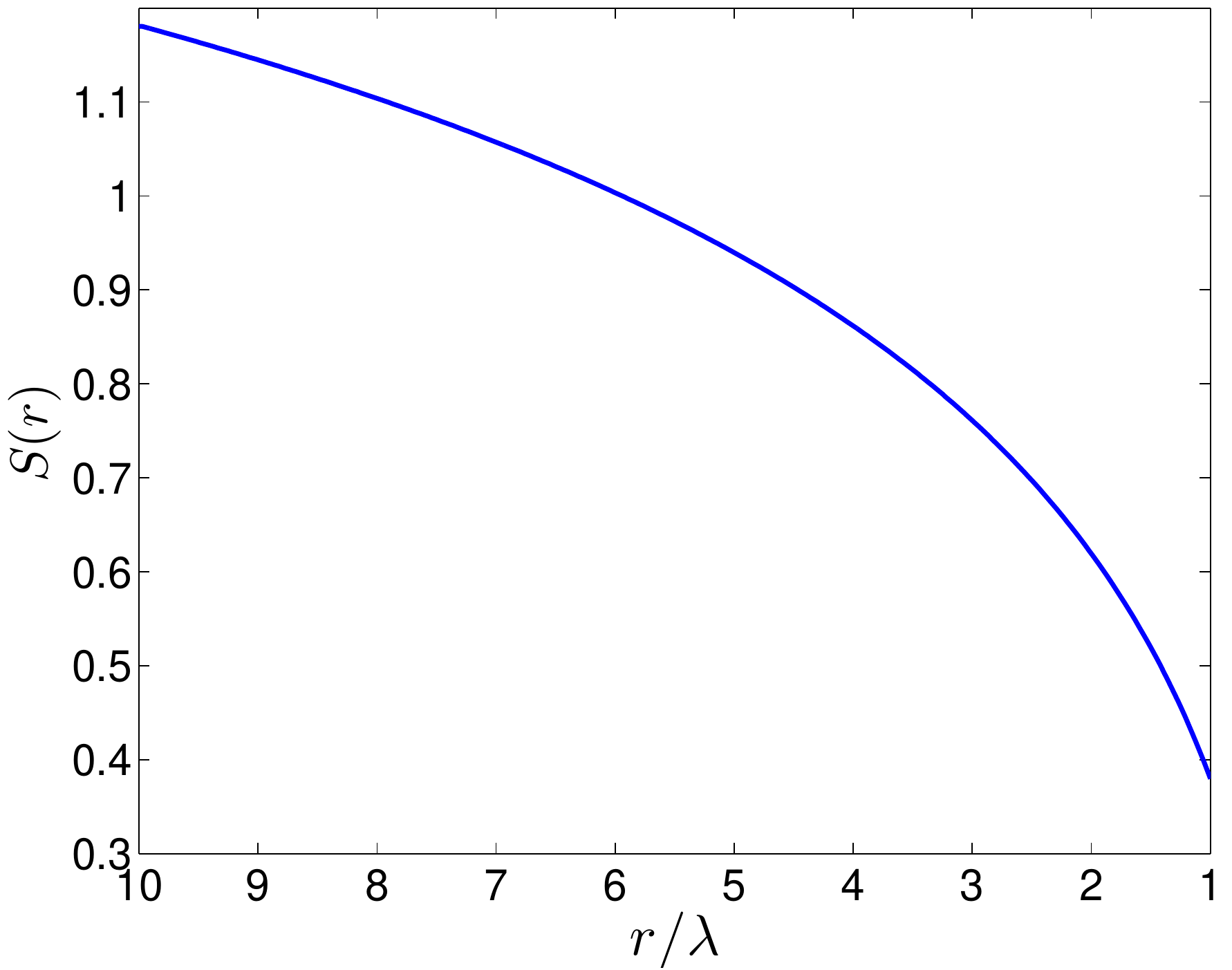}} \hfil
  \subfloat[][\figsubtxt{Multiplier statistics for $\mu=0.26$ ($\nu\approx27$)}]{\label{sf:2ndlaw_K62_2ndlaw_nu27}
	\includegraphics[width=0.45\textwidth]{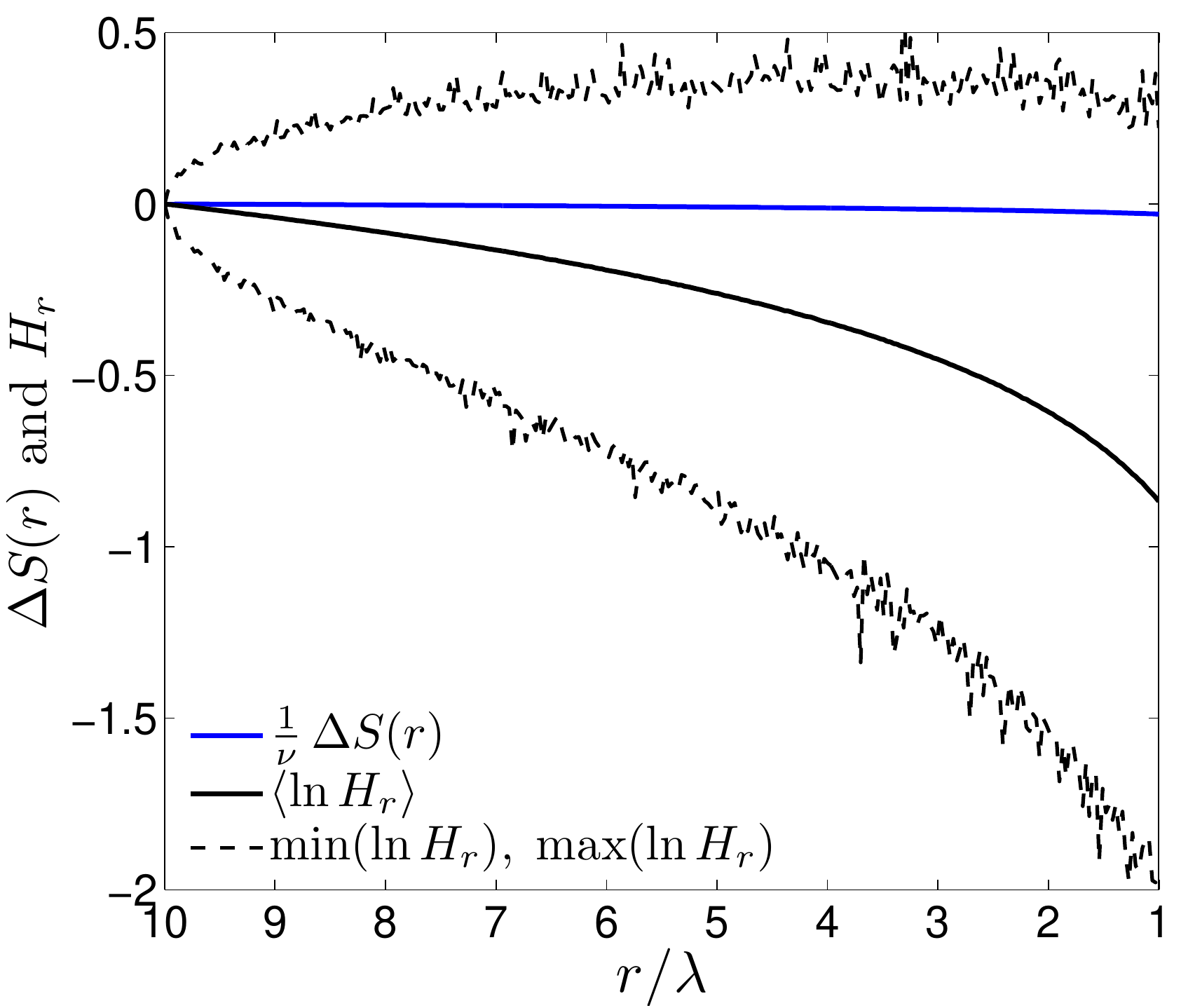}} \\
  \subfloat[][\figsubtxt{Multiplier statistics for $\mu=0.02$ ($\nu\approx300$)}]{\label{sf:2ndlaw_K62_2ndlaw_nu304}
	\includegraphics[width=0.45\textwidth]{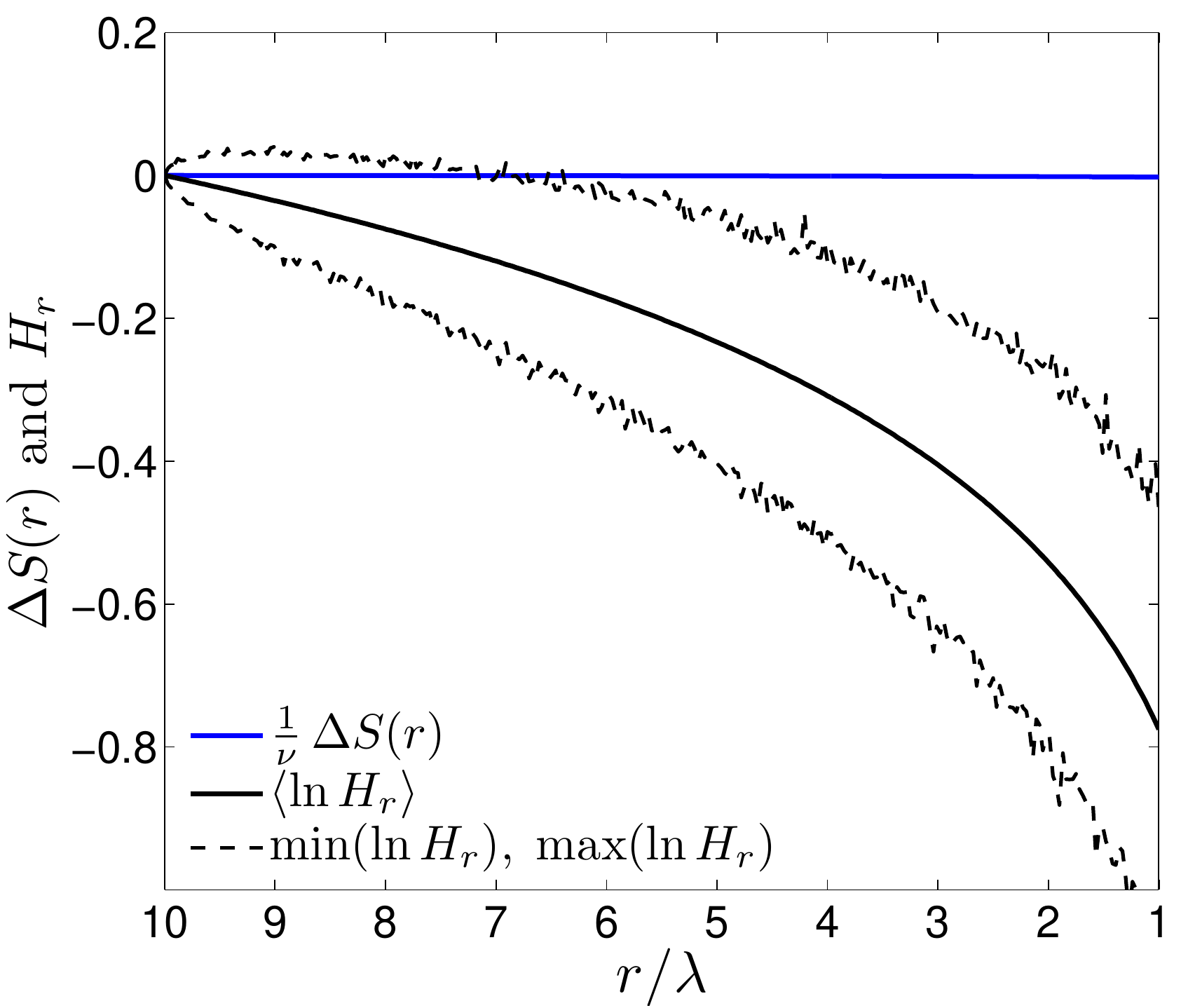}} \hfil
  \subfloat[][\figsubtxt{Multiplier statistics for $\mu=2$ ($\nu=7$)}]{\label{sf:2ndlaw_K62_2ndlaw_nu7}
	\includegraphics[width=0.45\textwidth]{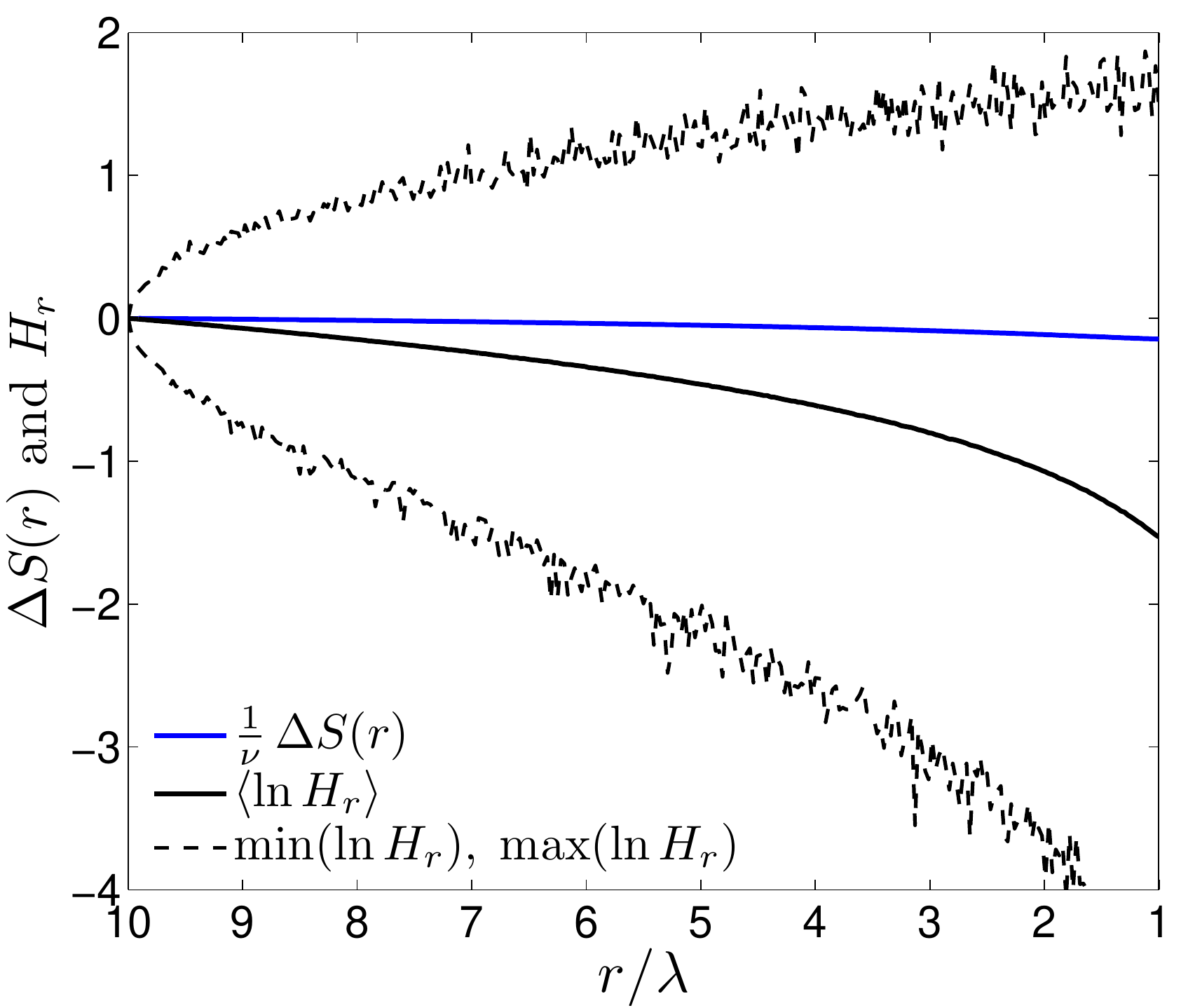}}
  \caption{\label{ff:2ndlaw_K62}\figtxt{Illustration of the second law for the K62 model for $L=10$ and $\l=1$. The Shannon entropy $S(r)=-\lla\ln p(u,r)\rra_u$, shown in (a), determines via \mbox{$\D S(r)=S(L)-S(r)<0$} the upper bound for the average of the overall log-multiplier $\ln H_r=\ln u_L/u_r$ in the second law like inequality (\ref{eq:turbFTs_K62_secondlaw}). The average log-multiplier $\lla\ln H_r\rra$, and its minimum and maximum value is depicted in (b), (c) and (d) for various values of intermittency factor $\mu$, together with the upper bound $\frac{1}{\nu}\D S(r)<0$ for $\lla\ln H_r\rra$. To determine $S(r)$, the analytic solutions for $p(u,r)$ from (\ref{eq:D1D2_K62_FPE_sol}) was used, see also figure \ref{ff:pur_K62}. The multiplier statistics was obtained from generating $10^6$ samples of $u(r)$ from the solution (\ref{eq:D1D2_K62_SDE_sol}) of the corresponding SDE.}}
  
\end{figure}\\
The Shannon entropy $S(r)$ can be computed up to quadrature from the solution (\ref{eq:D1D2_K62_FPE_sol}) of the K62 process (\ref{eq:D1D2_cascade_K62}) and, as shown in figure \ref{sf:2ndlaw_K62_Shannon_nu27}, is found to be positive for all scales and decreasing for increasing intervals $L-r$. Due to the latter property, the difference in Shannon entropy is negative, and therefore also the upper bound for the sum of multipliers. Hence, the multiplier statistics must be such that strictly $\lla H_r\rra<1$, that is, multipliers $h_i<1$ outbalance $h_i>1$ on average and velocity increments $u(r)$ predominantly decrease in the cascade process. This is a reasonable statement for a second\;law\,-\,like inequality with regard to a turbulent cascade.\\
Single realisations of $H_r$, however, may still be positive, corresponding to instances where $h_i>1$ outbalance $h_i<1$ and velocity increments $u(r)$ predominantly increase in the cascade process. Taking the commonly accepted value of $\mu=0.26$ for the intermittency factor, Figure \ref{sf:2ndlaw_K62_2ndlaw_nu27} shows the upper bound $\frac{1}{\nu}\D S$, together with the average multiplier $\lla \ln H_r\rra$ and maximum and minimum values of $\ln H_r$, where the set of multipliers $H_r$ was determined from generating $10^6$ samples of $u(r)$ from the solution (\ref{eq:D1D2_K62_SDE_sol}) of the K62 process (\ref{eq:D1D2_K62}). It is evident that the upper bound $\frac{1}{\nu}\D S$ only plays a minor role, compared to $\lla \ln H_r\rra$ it can be taken as zero. For increasing intervals $L-r$, the average multiplier $\lla \ln H_r\rra$ decreases, indicating that instances of increasing velocity increments over short ranges of scales are more likely than those over large ranges of scale, a result being in agreement with intuition. On the other hand, the maximal realisations $\ln H_r$ remain more or less constantly at a positive level, indicating that instances of increasing velocity increments over the complete range of scales remain possible. These instances, as discussed in \cite{Nickelsen2013}, are both indispensable for the convergence of the FT and responsible for small-scale intermittency.\\
We included into figure \ref{ff:2ndlaw_K62} the same analysis for two other values of the intermittency factor, $\mu=0.02$ corresponding to vanishing intermittency, and $\mu=2$ corresponding to extreme intermittency. For $\mu=0.02$, practically no instances $\ln H_r>0$ occur, in particular for large scale intervals $L-r$. For $\mu=2$, despite decreasing $\lla \ln H_r\rra$ for increasing $L-r$, the value of maximal multipliers $H_r$ even increases for large scale intervals $L-r$.\\
The last point to make concerns the relation to an inverse energy cascade which was, due to its intricacy, not discussed in \cite{Nickelsen2013}. If we accept that in an inverse cascade velocity increments increase while evolving downwards in scale, we can argue that the realisations $\uc$ giving rise to extreme negative values of $\Stot[\uc]$, shown in figure 3 in \cite{Nickelsen2013}, signify an inverse energy cascade.\footnote{Measurements show that the fluctuations of energy transfer are of the same order of magnitude as the average energy dissipation rate $\beps$ \cite{Mouri2006}, being reminiscent of $\DF\sim\kB T$, which is the condition in order to expect the convergence of FTs.} However, as mentioned in chapter \ref{ss_MAR_stochasticprocess}, it is disputable to interpret a single realisation $\uc$ as a single cascade. A realisation $\uc$ is merely a spatial snapshot of the outcome of a superordinate cascade process, and rather not the velocity of a certain eddy as it evolves down-scale.\\

\subsection{Beyond K62 scaling} \label{ss_turbulence_beyondscaling}
We now turn to discuss the expressions for entropy production arising from the various Markov representations of traditional approaches to turbulence discussed in the previous chapter. When appropriate, we also discuss the resulting integral fluctuation theorems and the implied second law. Starting point will always be the drift and diffusion coefficients.

\paragraph{Multifractal model}
We have seen in \ref{ss_MAR_D1D2} that by choosing
\begin{align} \label{eq:FTs_multifrac_powser}
	\Df(u,r) = -\sum_{k=0}^N \mfrac{\df_k}{r}\,u^k \sep \Dg(u,r) = \sum_{k=0}^N \mfrac{\dg_k}{r}\,u^k \;,
\end{align}
the structure functions obey a spectrum of scaling laws, which is the essence of the multifractal model discussed in \ref{ss_FDT_scalinglaws}. For the discussion of FTs it is more convenient to consider instead of the $\Df(u,r)$ the coefficient
\begin{align}
	F(u,r) = \Df(u,r) - \Dgx(u,r) = -\sum_{k=0}^N \mfrac{f_k}{r}\,u^k
\end{align}
where $f_k = (\df_k\+k\dg_{k+1})/\dg_0$, and we will use $d_k=\dg_k/\dg_0$.\\
The EP $\Sm[\uc]$ that is associated with the entropy transferred into a heat bath follows according to (\ref{eq:def_Sm}) from the integral
\begin{subequations}
\begin{align}
	\Sm[\uc] &= \int_{L}^{r}\pt_{r'} u(r')\,\big[-\pt_u \p\big(u(r'),r'\big)\big] \di r' \;,\\
	\pt_u \p(u,r) &= -\frac{F(u,r)}{D(u,r)} \;.
\end{align}
\end{subequations}
For this class of MPs, the integral in the EP $\Sm[\uc]$ can be explicitly calculated. To accomplish the integration, we consider the power series expansion\remark{beachte dass $\p'=-F/D$, aber wir haben $F$ bzw. $\Df$ ja schon ein minuszeichen verpasst, damit die $f_k$ tendenziell positiv sind.}
\begin{align} \label{eq:FTs_multifrac_phi_powser}
	\pt_u\p(u,r) = \frac{f_0+f_1u+f_2u^2+\dots+f_Nu^N}{1+d_1u+d_2u^2+\dots+d_Nu^N} = \sum_{n=0}^{\infty} a_n u^n \;,
\end{align}
where the coefficients $a_n$ are to be determined. Indeed, successive differentiation of the above fraction and an observant eye leads to the finding that the $a_n$ are the solution of the linear set of equation
\begin{align}
	D^{j}_{n}\,a^n = f_j \qquad\Ra\quad a^n = \big(D^{-1}\big)^{j}_{n}\,f_j
\end{align}
with the coefficient matrix and its inverse,
\begin{align} \label{eq:FTs_multifrac_Dmat}
  \mat{D} =
  \begin{pmatrix} 
	 1      & 0      & 0      & \ldots \\ 
	 d_1    & 1      & 0      & \ldots \\ 
	 d_2    & d_1    & 1      & \ldots \\ 
	 \vdots & \ddots & \ddots & \ddots \\
  \end{pmatrix}
  \quad \text{and} \quad 
  \mat{D}^{-1} =
  \begin{pmatrix} 
	 \bd_0    & 0        & 0        & \ldots \\ 
	 \bd_1    & \bd_0    & 0        & \ldots \\ 
	 \bd_2    & \bd_1    & \bd_0    & \ldots \\ 
	 \vdots   & \ddots   & \ddots   & \ddots \\
  \end{pmatrix}
  \; .
\end{align}
Owing to the diagonal structure of the matrix and the fact that the $\bd_k$ are the solution of the homogeneous equation, the entries of the inverse matrix can be determined from the recursion
\begin{align} \label{eq:FTs_multifrac_bdn}
	\bd_n = -\sum\limits_{k=0}^{n-1}\frac{d_{n-k}}{\bd_k} \;.
\end{align}
For the coefficients of the power series follows finally
\begin{align} \label{eq:FTs_multifrac_an}
	a_n = \sum\limits_{k=0}^{n}\bd_{n-k}f_k \;.
\end{align}
Having this set, the EP becomes
\begin{align} \label{eq:FTs_multifrac_Sm}
	\Sm[\uc] &= \int\limits_{L}^{r}\pt_{r'} u(r')\,\big[-\pt_u \p\big(u(r'),r'\big)\big] \di r' \nn
	&= \sum\limits_{n=0}^{\infty}\mfrac{a_n}{n\+1}\Big(u_L^{n+1} - u_r^{n+1}\Big) \;.
\end{align}
The associated second law constitutes a constraint for the structure functions,
\begin{align} \label{eq:FTs_multifrac_Sm}
	\sum\limits_{n=1}^{\infty}\mfrac{a_{n-1}}{n}\,\Big[\Str^{\,n}(L)-\Str^{\,n}(r)\Big] \;\geq\; \lla \ln\frac{p_r(u_r)}{p_L(u_L)}\rra = -\DS \;.
\end{align}
At this point, however, an explicit form of $f_k$ and $d_k$ is needed, in order to compute the coefficients $a_n$.\\
As already mentioned, the full correspondence to the multifractal model is to be expected if the power series in (\ref{eq:FTs_multifrac_powser}) is written as an integral, cf. (\ref{eq:D1D2_power_multifrac}). In this case, the recursive equation for $\bd(n)$ should turn into a differential equation, and the equation for the coefficient $a(n)$ becomes a convolution integral. However, we leave the continuous formulation for further study.

\paragraph{Random cascade models}
In the Markovian description, the class of log-normal random cascade models are defined by drift and diffusion of the form $\Df(u,r)=-\tilde a(r)\,u$ and $\Dg(u,r)=b(r)\,u^2$. Again, we rewrite the drift $\Df$ as the coefficient $F=\Dg-\Dgx$, that is
\begin{align}
	F(u,r) = -a(r)\,u \sep D(u,r) = b(r)\,u^2 \;,
\end{align}
where $a(r)=\tilde a(r)+2b(r)$.\\
We define the scale function $\L(r)\dfns a(r)/b(r)$ and the log-multiplier $H(r)\dfns \ln \frac{u(r)}{u_L}=\sum_{i=1}^{s(r)}\ln h_i$, cf. (\ref{eq:turbFTs_K62_secondlaw}), such that we can write for $\Sm$\remark{diss p.25}
\begin{align}
	\Sm[\uc] &= -\int\limits_{L}^{r} \frac{a(r')}{b(r')}\frac{\pt_{r'}u(r')}{u(r')} \di r' \nn
	&= -\int\limits_{L}^{r} \L(r') \frac{\dd H\big(r')}{\dd r'}\,\di r' \nn
	&= -\int\limits_{0}^{H(r)} \L\big(r(H')\big)\,\di H' \;.
\end{align}
Here, $r(H)$ is the scale reached after the cascade has performed $s(r)$ cascade stages. The scale function $\L(r)$ may be interpreted as the EP rate along the scale-path $r(H)$ as a function of the log-multiplier $H$. By choosing $\L(r)$ we can hence influence the form of the EP $\Sm[\uc]$, which raises hope that a suitable $f(r)$, defining a meaningful EP $\Sm[\uc]$ in the context of a turbulence cascades, also defines a meaningful MP in terms of the resulting $\Dfg(u,r)$. Attempts along such lines, however, remained unsatisfactory.\\
For K62 scaling it is $\L(r)\equiv\nu$ and we retrieve the K62 FT (\ref{eq:turbFTs_K62_FT}). Due to cancellation, the form of the K62 FT is also obtained for an arbitrary choice of $a(r)=b(r)$, i.e. continuous self-similarity. Accordingly, the FT for scaling laws of the form $\Str^n(r)=[f(r)]^{\,\z_n}$ is insensitive to the function $f(r)$, including the ESS case $f(r)=\Str^3(r)$.\\
To inspect a FT that deviates from the K62 form, we employ the experimental results by Castaing and co-workers which we discussed in chapter \ref{ss_FDT_sm-sc-intm}. Their object of investigation was the variance of energy transfer fluctuations, for which they found $\s_r\sim r^{-\d}$ with $\d\approx0.3$, cf. (\ref{eq:FDT_castaing_sigr}). This variance also determines the variance of the propagator in the associated log-normal random cascade model, for which we found in (\ref{eq:D1D2_castaing_SDE_sol_sig}) that $b(r)\sim\pt_r\s_r\sim r^{-\b-1}$. Accordingly, we keep the K41 coefficient $a(r)$ and modify the K62 coefficient $b(r)$,
\begin{align}
	a(r) &= \frac{3+2\mu}{9}\,\frac{1}{r} \sep b(r) = \frac{\mu}{18}\,\frac{1}{r^{1+\d}}  \nonumber \\[5pt]
	\Ra\quad \L(r)&=\nu\,r^\d
\end{align}
such that $\d=0$ recovers the K62 result.\remark{$\d=0$ ist komisch in $\s_r$ weil für K62 $\s_r\sim\ln r$.} The explicit formula for the associated EP $\Sm$ reads
\begin{align}
	\Sm[\uc] &= \int_{\L_0}^{\L(r)} \ln u\big(r(\L')\big) \di \L' - \Big[\L'\ln u\big(r(\L')\big)\Big]_{\L_0}^{\L(r)} \nn
	&= \d\,\int_{L}^{r} r'^{\d-1}\,\ln u(r') \di r' - \Big[r'^\d\,\ln u(r')\Big]_{L}^{r} \;.
\end{align}
To assess the validity of the resulting FT, we used the same data as in chapter \ref{ss_turbulence_PRL} to compute the ensemble $\{\Sm[\uc]\}$ from the above formula and plugged the result into the FT
\begin{align}
	1=\lla\frac{p_r(u_r)}{p_L(u_L)}\,\eee{-\Sm[\uc]}\rra \;.
\end{align}
Unfortunately, for the predicted value of $\d=0.3$ the FT exhibits a similar divergence as for K62, and even a variation of $\d$ did not improve the situation.\\
We offer an explanation for this failure of applying the FT for random cascade models to experimental data. Standard random cascade models, as considered by drift and diffusion coefficients above, do not account for the skewness of the PDFs $p(u,r)$ observed in experiments and demanded by the four-fifth law. A vanishing skewness of $p(u,r)$, however, implies a vanishing energy dissipation rate $\beps$ ruling out irreversible processes in the flow. But it is the very balance between irreversible process and their reversals the derivation of FTs rest on, suggesting that the skewness in a model of turbulence is crucial for the validity of the resulting FT for real data.\\

More promising might be the log-Poisson random cascade model by She and Leveque, for which we found a jump process underlying a Liouville process, cf. (\ref{eq:D1D2_bK62_SL_expo}). The Liouville process is defined by the deterministic drift $\Df(u,r)\eq-1/9\,u/r$, and the jump process is characterised by a transition probability $\j(u^+|u,r)$ for a jump velocity increment from $u$ to $u^+$ at scale $r$, for which we found the moments $\Ak(u,r)=C_0\,(bu)^k/r$ with $b=\b^{1/3}\-1$. The drift implies the scaling $u(r)=u_L(r/L)^{1/9}$.\\
In order to extract the pure jump process, it is convenient to consider the scaled velocity increments $\tilde u(r)\dfns (\frac{L}{r})^{\frac{1}{9}}u(r)$ and scaled transition probability $\tilde\j(\tilde u^+|\tilde u,r)\dfns(\frac{L}{r})^{\frac{1}{9}}\,\j\big((\frac{L}{r})^{\frac{1}{9}}\tilde u^+\,\big|\,(\frac{L}{r})^{\frac{1}{9}}\tilde u,\,r\big)$.\remark{kann man nicht theoretisch versuchen, auch $\tilde \j(\tilde u^+|\tilde u,r)$ mithilfe seiner definition (\ref{eq:CKR_3conds_drift}) zu schätzen?}\\
From considerations similar to those in chapter (\ref{ss_td-interpration_FTs}), it follows that the EP $\Sm[\tuc]$ is determined by \cite{Seifert2012}\remark{hier kürzt sich der pdf-trafo-term sowieso}
\begin{align}
	\Sm[\tuc] = \sum\limits_{j=1}^{n} \ln\,\frac{\tilde\j(\tilde u^+_j|\tilde u_j,r_j)}{\tilde\j(\tilde u_j|\tilde u^+_j,r_j)} \;,
\end{align}
where the sum is over all jumps. The associated FT remains unchanged,
\begin{align}
	1=\lla\,\frac{\tilde p_r(\tilde u_r)}{\tilde p_L(\tilde u_L)}\,\eee{\Sm[\tuc]}\,\rra \;.
\end{align}
A test of this FT with experimental data is still pending, since the explicit form of $\j(u^+|u,r)$ remains to be determined from the moments $\Ak(u,r)$.

\paragraph{Yakhot's model}
Recall drift and diffusion coefficients from Yakhot's model (\ref{eq:D1D2_Yak_FPE_tilde_D1D2})
\begin{align}
\Df(u,r) &= -\mfrac{a_1}{r}\,u \sep \Dg(u,r) = \mfrac{a_2}{r}\,u^2-b_2u  \label{eq:FTs_Yak_D1D2} \\[5pt]
a_1 &= \mfrac{B\+3}{3(B\+1)} \sep a_2 = \mfrac{B\+3}{3(B\+1)(B\+2)} \sep b_2 = \mfrac{\vrms\,\l}{\sigi\,L}\,\mfrac{1}{B\+2} \;, \nonumber
\end{align}
where $u$, $r$ and $b_2$ are dimensionless by using the same normalisation as in \cite{Nickelsen2013} reprinted in chapter \ref{ss_turbulence_PRL}.
We first simplify
\begin{align}
	\pt_u\p(u,r) &= \frac{(a_1+2a_2)\,u - b_2r}{a_2u^2-b_2ur} \nn
	&= \frac{1}{u}\,\frac{\nu\,u - \d_1\,r}{u - \d_1\,r} \sep \d_1\dfn\frac{b_2}{a_2}=\mfrac{3\,\vrms\,\l}{\sigi\,L}\,\mfrac{B\+1}{B\+3}
\end{align}
with the K62 parameter $\nu=(2a_2\-a_1)/a_2=B\+4\approx24$. In the following we assume that $|u(r)|\!>\!\d_1r$ in order to ensure $\Dg(u,r)\!>\!0$, which is the known complication with Yakhot's model in the Markov representation (cf. the last paragraph in \ref{ss_MAR_D1D2}).\\
We subtract the K62 contribution
\begin{align}
	u\cdot\pt_u \p(u,r) - \nu &= (\nu-1)\,\frac{\d_1\,r/u}{1-\d_1\,r/u} \nn
	&= \frac{\nu}{u} + \frac{\nu-1}{u}\,\sum_{k=1}^{\infty}\Big(\frac{\d_1\,r}{u}\Big)^k
\end{align}
to get the two equivalent expressions
\begin{subequations} \label{eq:FTs_Yak_phi-K62}
  \begin{align}
		\pt_u\p(u,r) &= \frac{\nu}{u} + \frac{\nu-1}{u}\,\frac{\d_1\,r}{u-\d_1\,r} \label{eq:FTs_Yak_phi-K62_1} \\[5pt]
		&= \frac{\nu}{u} + \frac{\nu-1}{u}\,\sum_{k=1}^{\infty}\Big(\frac{\d_1\,r}{u}\Big)^k \;. \label{eq:FTs_Yak_phi-K62_2}
	\end{align}
\end{subequations}
The first expression can be readily integrated to get
\begin{align}
	\p(u,r) = \nu\ln u + (\nu-1)\ln\Big(1-\mfrac{\d_1\,r}{u}\Big) 
\end{align}
which implies the non-normalised stationary distribution
\begin{align}
	\pstnn(u,r) \dfn \exp\big[-\p(u,r)\big] = u^{-\nu}\,\Big(1-\mfrac{\d_1\,r}{u}\Big)^{1-\nu} \;.
\end{align}
The K62 model is recovered in the limit $\d_1\to0$. The term involving $\d_1$ can thus be viewed as the Yakhot extension of the K62 model. Note that $\pstnn(u,r)$ in the above form is not normalisable\remark{kommt nur blödsinn bei raus, siehe nebenrechnung}. \\
The second expression for $\pt_u\p(u,r)$ in (\ref{eq:FTs_Yak_phi-K62}) is convenient to partially perform the integral in $\Sm[\uc]$,
\begin{align}
	\Sm[\uc] &= \int\limits_{L}^{r}\pt_{r'} u(r')\,\big[-\pt_u \p\big(u(r'),r'\big)\big] \di r' \nn
	&= -\nu\int\limits_{L}^{r}\frac{\pt_{r'} u(r')}{u(r')}\di r' - (\nu\-1)\int\limits_{L}^{r} \sum_{k=1}^{\infty}\frac{\pt_{r'} u(r')}{u(r')^{k\+1}}\,\big(\d_1\,r'\big)^k \di r' \nn
	&= -\nu\ln\frac{u(r)}{u(L)} + (\nu\-1)\bigg[\sum_{k=1}^{\infty}\frac{1}{k}\bigg(\frac{\d_1\,r'}{u(r')}\bigg)^{\!k}\,\bigg]_{L}^{r} + R[\uc]\nn
	&= R[\uc]-\Big[\nu\ln u(r') + (\nu\-1)\ln\Big(1\-\mfrac{\d_1\,r'}{u(r')}\Big)\Big]_{L}^{r} \nn
	&= R[\uc]-\D\p  \sep \label{eq:eq:FTs_Yak_SmRDp}
\end{align}
with the work-like functional
\begin{align}
	R[\uc] &= - (\nu\-1)\int\limits_{L}^{r}\frac{1}{r'}\sum_{k=1}^{\infty}\bigg(\frac{\d_1\,r'}{u(r')}\bigg)^{\!k} \di r' \nn
	&= -(\nu\-1)\int\limits_{L}^{r}\frac{1}{r'} \frac{\d_1\,r'/u(r')}{1-\d_1\,r'/u(r')} \di r' \nn
	&= \d_1(\nu\-1)\int\limits_{L}^{r} \frac{1}{u(r')-\d_1\,r'} \di r' \label{eq:eq:FTs_Yak_R}
\end{align}
The resulting Yakhot FT then reads
\begin{align}
	1 &= \lla\frac{p_r(u_r)/\pstnn(u_r,r)}{p_L(u_L)/\pstnn(u_L,L)}\,\eee{-R[\uc]}\rra \label{eq:eq:FTs_Yak_FTpst} \\
	&= \lla\frac{u_r}{u_L}\frac{(u_r-\d_1\,r)^{\,\nu-1}}{(u_L-\d_1\,L)^{\,\nu-1}}\,\frac{p_r(u_r)}{p_L(u_L)}\,\eee{-R[\uc]}\rra \label{eq:eq:FTs_Yak_FTnice}
\end{align}
\remark{The resulting Yakhot FT then reads
\begin{align}
	1 &= \lla\frac{p_r(u_r)}{p_L(u_L)}\,\exp\!\Big[\D\p -R[\uc]\Big]\rra \nn
	&= \lla\frac{p_r(u_r)/\pstnn(u_r,r)}{p_L(u_L)/\pstnn(u_L,L)}\,\eee{-R[\uc]}\rra \label{eq:eq:FTs_Yak_FTpst} \\
	&= \lla\frac{u_r^{\,\nu}}{u_L^{\,\nu}}\frac{(1-\d_1\,r/u_r)^{\,\nu-1}}{(1-\d_1\,L/u_L)^{\,\nu-1}}\,\frac{p_r(u_r)}{p_L(u_L)}\,\eee{-R[\uc]}\rra \nn
	&= \lla\frac{u_r}{u_L}\frac{(u_r-\d_1\,r)^{\,\nu-1}}{(u_L-\d_1\,L)^{\,\nu-1}}\,\frac{p_r(u_r)}{p_L(u_L)}\,\eee{-R[\uc]}\rra \label{eq:eq:FTs_Yak_FTnice}
\end{align}}In the limit $\d_1\to0$, we have $R[\uc]\equiv0$ and recover the K62 FT. Note that by equating $\nu=B+4$ with the K62 result $\nu=(6+4\mu)/\mu$, we obtain the reasonable prediction $\mu=6/B\approx0.3$ for $\d_1\to0$.\\
For the data from \cite{Renner2001} it is $\d_1\simeq0.19$.\footnote{taking $\vrms=0.3818\,\mr{m/s}$, $\l=6.6\,\mr{mm}$, $L=67\,\mr{mm}$, $\sigi=0.54\,\mr{m/s}$ and $B=20$ in $\d_1=\frac{3\vrms\l(B+1)}{L\sigi(B+3)}$} If we cut out all realisations violating $|u(r)|\!>\!\d_1r$, being about one half of all $\{\uc\}$, and substituting the remaining $\uc$ into the r.h.s. of the FT above, we get approximately zero instead of one. Being well below the theoretical value of one implies that negative EPs are too rare, indicating that Yakhot's model underestimates the intermittency of realisations $\uc$\remark{Yakhot's EP sagt, das ist keine intermittente realisierung, obwohl es (laut estimated $\Dfg$) eine ist.}, it therefore stands to reason to expect that incorporating the underlying jump process would correct this discrepancy.\remark{ich fürchte unsere analoge K62 interpretation war falsch. zum einen überschätzt K62 wegen $\z_n\to-\infty$ in $(r/L)^\z_n$ die small-scale intermittency ($r/L<1$!), zum anderen gibt es zuviele $\Stot<0$, K62 überschätzt also die intermittency der realisierungen (sagt intermittent, obwohl gar nicht intermittent, weil in K62 mehr intermittenz gewöhnt ist).}\\
In contrast to random cascade models, the FT can be satisfied by modifying the K62 correction parameter. However, to satisfy the FT, we need a value of $\d_1=6.48$, which is far too out compared to the predicted value $\d_1\simeq0.19$, ruling out to consider corrections of the involved quantities. Also, $\d_1=6.48$ substantially tightens the condition $|u(r)|\!>\!\d_1r$\remark{, such that only \todo{wieviele??} contribute to the FT}.\\
In any case, the inconsistency that $\Dg(u,r)<0$ for $|u(r)|\!<\!\d_1r$ remains. Motivated by the constant offset observed in practically all experimentally estimated $\Dg(u,r)$, a modification of the diffusion coefficient in the form\linebreak $\Dg(u,r)=\d_0a_2-\d_1a_2 u+a_2u^2/r$ seems promising. The only consistent value for $\d_0$ that ensures $\Dg(u,r)\geq0$ turns out to be $\d_0=\d_1^2/4$, since for $\d_0<\d_1^2/4$ is $\Dg(u,r)<0$ still possible, and for $\d_0>\d_1^2/4$ we encounter a complex valued EP $\Sm[\uc]$. However, the resulting FT becomes singular for too large fluctuations of $u(r)$.\\

For completeness we report that the second\;law\,-\,like equation takes the form
\begin{align}
	\lla R[\uc]\,\ln\frac{u_L\,(u_L-\d\,L)^{\,\nu-1}}{u_r\,(u_r-\d\,r)^{\,\nu-1}}\rra \;\geq\; \lla \ln\frac{p_r(u_r)}{p_L(u_L)}\rra = -\DS \;,
\end{align}
but refrain from attempting an interpretation.\\

As a last comment, we mention that the above procedure used to obtain the form $\Sm[\uc]=R[\uc]-\Dp$ as in (\ref{eq:eq:FTs_Yak_SmRDp}) can be generalised to arbitrary polynomial forms of $F(u,r)$ and $D(u,r)$. This generalised procedure is similar to the manipulations of (\ref{eq:FTs_multifrac_phi_powser}) for the multifractal model, only that the expansion is in $1/u$ instead of $u$, and the coefficients $f_k$ and $d_k$ may have arbitrary $r$-dependencies.\remark{und wird sogar einfacher!}

\cleardoublepage
\section{Asymptotic analysis} \label{s_turbulence_asymp}
In this chapter we will apply the asymptotic analysis developed in chapter \ref{s_asymp} to the Markovian description of fully developed turbulence. This analysis will include the examination of realisations giving rise to extreme values of $\Sm$, the asymptotic form of $p(u,r)$ and the resulting asymptotic scaling exponents $\z_n$.\\
The results presented in this chapter are rather preliminary, in the sense that they were obtained only recently.

\subsection{Realisations of extreme entropy} \label{ss_turbulence_asymp_extreme}
In the previous chapters we found that small-scale intermittency is formally related to negative EP $\Stot[\uc]<0$. From the discussion of the second\;law\,-\,like inequality resulting from the K62 model, (\ref{eq:turbFTs_K62_secondlaw}), followed that the EP $\Sm[\uc]$ is equivalent to the log-multiplier of velocity increments of the turbulent cascade, and negative $\Sm[\uc]$ imply positive multipliers which signify an inverse turbulent cascade. This interpretation of negative $\Sm[\uc]$ is not as clear for more sophisticated models than K62, however, it is interesting to examine the most likely realisations that give rise to extreme values of $\Sm[\uc]$. These realisations are nothing else than the solutions of the Euler-Lagrange equations (ELE) deriving from the path integral representation of the respective MP.\\
We can readily transfer the ELE obtained in chapter (\ref{eq:asymp_ELE}) to the case of velocity increments $u(r)$,
\begin{subequations} \label{eq:tasymp_ELE_constr}
  \begin{align}
  \begin{split}
    \dddu &= \mfrac{\ddu \+ F}{2D}\big(D'(\ddu \- F)+2\ddD\big) \\
     &\hspace{70pt}+FF'-\ddF+2DJ''+2Dik(\pt_u\-\dd_r\pt_{\ddu})w \;,
  \end{split}  \label{eq:tasymp_ELE_constr_ode}\\
		0&=\mfrac{\ddu_0+F_L}{2D_L}-ik\pt_{\ddu}w_L+\L'_L \;,  \label{eq:tasymp_ELE_constr_BC0} \\
		0&=\mfrac{\ddu_\l+F_\l}{2D_\l}-ik\pt_{\ddu}w_\l \;. \label{eq:tasymp_ELE_constr_BC1}
	\end{align}
\end{subequations}
Here, in analogy to $x(t)$, primes denote partial derivatives with respect to $u$ and dots with respect to $r$, and we dropped the arguments of $u_r$, $F(u_r,r)$, $D(u_r,r)$, $w(u_r,\ddu_r,r)$ and derivatives, and indices denote evaluation at scales $r$, $L$ and $\l$ respectively.\\
Substitution of the constraint 
\begin{subequations} \label{eq:tasymp_ELE_constraint}
  \begin{align}
		\Sm[\uc] &= \int_{L}^{\l} w(u_r,\dot u_r,r) \di r \;, \\
		w(u,\dot u,r) &= \dot u\,\frac{F(u,r)}{D(u,r)} \;,
	\end{align}
\end{subequations}
and simplifications yield
\begin{subequations} \label{eq:tasymp_ELE_Sm}
  \begin{align}
  \begin{split}
    \dddu &= \mfrac{1}{2}\mfrac{D}{D'}(\ddu^2-F^2) + \ddu\mfrac{\dot D}{D} + F'F + 2DJ' - (1\-2ik)\mfrac{\dot F D - F\dot D}{D}
    \end{split}  \label{eq:tasymp_ELE_ode}\\
		0&=\mfrac{\ddu_0+F_L}{2D_L}-ik\mfrac{F_L}{D_L}+\L'_L \;,  \label{eq:tasymp_ELE_BC0} \\
		0&=\mfrac{\ddu_\l+F_\l}{2D_\l}-ik\mfrac{F_\l}{D_\l} \;. \label{eq:tasymp_ELE_BC1}
	\end{align}
\end{subequations}
\vspace{1pt}\\

\begin{figure}[t] 
	\begin{center}
		\includegraphics[width=0.75\textwidth]{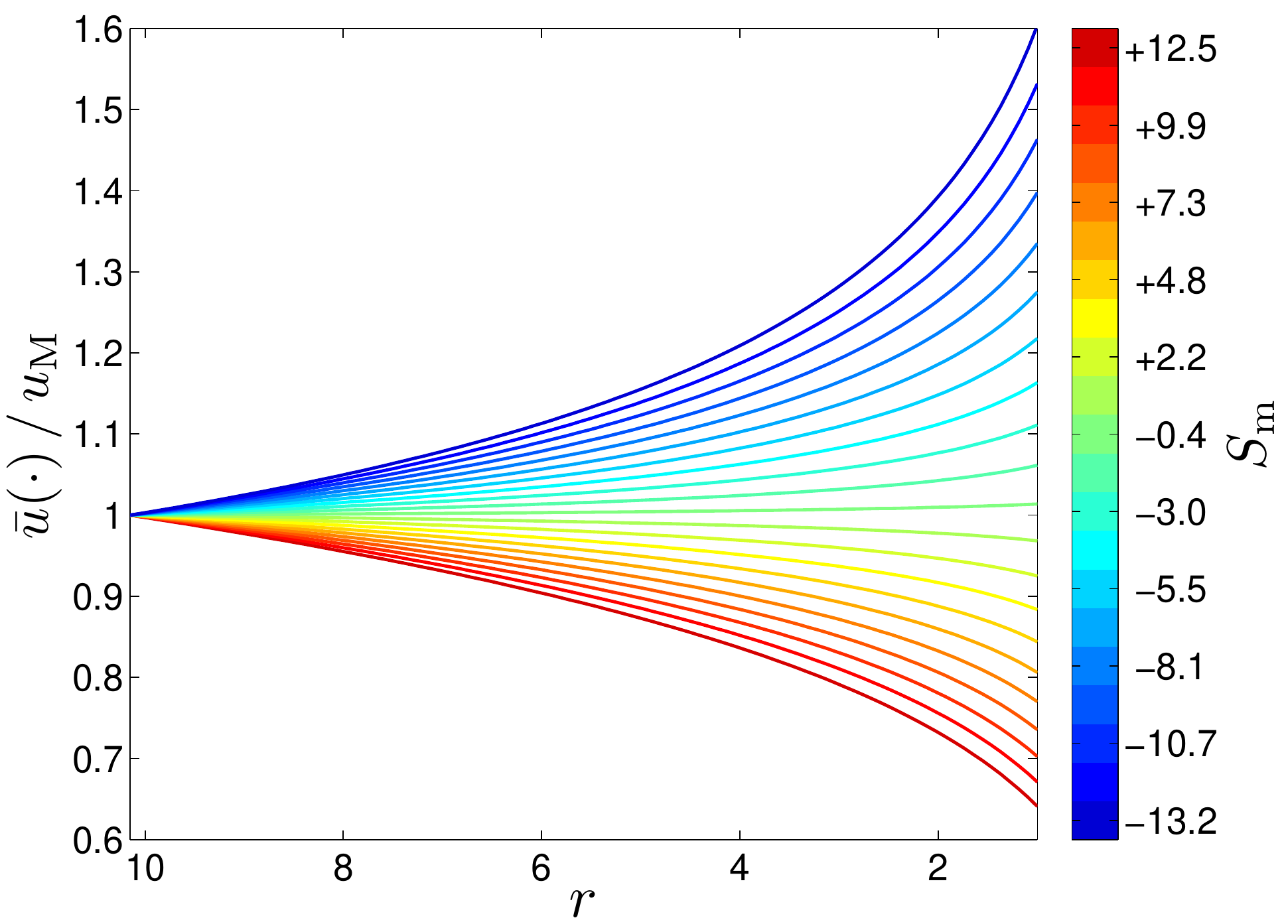}
		\caption{\label{ff:Sm_trajectories_K62}\figtxt{Most likely realisations of the K62 process for various values of EP $\Sm$. More specific, the shown trajectories are the solutions (\ref{eq:tasymp_optfluc_Sm_K62}) of the ELE (\ref{eq:tasymp_ELE_Sm_K62}) subject to the constraint (\ref{eq:tasymp_ELE_constraint}) for the stochastic process defined by (\ref{eq:tasymp_K62_SDE}).}}		
	\end{center}
\end{figure}

Let us consider as an analytically solvable example the K62 coefficients from (\ref{eq:D1D2_K62}),
\begin{alignat}{2}
	&F(u,r) = - a\,\frac{u}{r} &\sep& D(u,r) = b\,\frac{u}{r^2} \;,\nn
	&a = \frac{3+2\mu}{9} &\sep& b = \frac{\mu}{18} \;, \label{eq:tasymp_K62_SDE}
\end{alignat}
for which the ELE becomes
\begin{subequations} \label{eq:tasymp_ELE_Sm_K62}
  \begin{align}
		\dddu(r) &= \frac{\ddu(r)}{u(r)} - \frac{\ddu(r)}{r} \;, \\
		0 &= \ddu_LL + a(2ik-1)u_L + 2bu_L^2\L'(u_L) \;, \\
		0 &= \ddu_\l\l + a(2ik-1)u_\l \;.
	\end{align}
\end{subequations}
The solutions of this ELE are power laws,
\begin{align}
	\bu(r;k) = u_\mr{M} \Big(\frac{r}{L}\Big)^{\displaystyle a(1-2ik)} \;,
\end{align}
where $u_\mr{M}$ denotes the mode of the initial PDF $p_L(u_L)$. To obtain the optimal fluctuations $\bu(r;\Sm)$ as a function of the EP $\Sm$, we adjust $k$ such that $\bu(r;k)$ satisfies the constraint
\begin{align}
	\Sm(k) &= \Sm[\bu(r;k)] = -a\nu(1-2ik)\ln\frac{\l}{L} \nn
	\Ra\quad ik(\Sm) &= \frac{1}{2}+\frac{\Sm}{2a\nu\ln(\l/L)} \;.
\end{align}
The optimal fluctuations hence read
\begin{align} \label{eq:tasymp_optfluc_Sm_K62}
	\bu(r;\Sm) = u_\mr{M} \Big(\frac{r}{L}\Big)^{\mfrac{\Sm}{\nu\ln(L/\l)}} \;,
\end{align}
in which the exponent is the EP $\Sm$ scaled with a measure for the length of the inertial range. The optimal fluctuations are depicted for various values of $\Sm$ in figure \ref{ff:Sm_trajectories_K62}.\\
A few special cases are apparent.
\\
(i) For an initial PDF $p_L(u_L)$ with zero mean and zero skewness it is\linebreak $u_\mr{M}=0$ and hence $\bu(r;\Sm)\equiv0$. In other words, the most likely (and average\remark{weil ohne skewness eben symmetrisch, und dann mean=mode}) realisation in a K62 process without initial skewness will always be identical zero, regardless the value of EP we impose. This uncoupling of $\bu(r;\Sm)$ and $\Sm$ for vanishing skewness hints that the formal EP $\Sm$ indeed has bearings with the energy production in the cascade.\\
(ii) The most likely realisation for $\Sm=0$ is $\bu(r;\Sm)\equiv u_\mr{M}$, identical to the cases in which no turbulence generation takes place, i.e. $L=0$, or in which the inertial range extends to zero scale or is of infinite length, i.e. $\l=0$ or $L\to\infty$ respectively.\\
The connection to the log-multiplier $\ln H=\ln(u(r)/u_L)$, discussed in the context of the K62 second law (\ref{eq:turbFTs_K62_secondlaw}), is established by writing
\begin{align}
	\ln\,\frac{\bu(r;\Sm)}{u_\mr{M}} = \mfrac{\Sm}{\nu\ln(L/\l)}\,\ln\,\frac{r}{L} \;,
\end{align}
in which the sign of $\Sm$ obviously determines whether $H<1$ or $H>1$, signifying a direct or inverse cascade process.\\

Let us now see how the optimal fluctuations $\bu(r;\Sm)$ look like for the experimentally estimated drift and diffusion (\ref{eq:D1D2_JFM_D1D2}) from \cite{Renner2001}, used in \cite{Nickelsen2013} (chapter \ref{s_turbulence_FT}).\\
Solving the resulting ELE analytically is hopeless and we refrain from writing down the explicit equations. Instead we employed the relaxation method \texttt{bvp4c} implemented in the numerical computing environment \texttt{MatLab}. The relaxation algorithm takes an initial guess for the solution and adapts this guess iteratively\remark{three-stage Lobatto IIIa formula} to satisfy the linear set of equations that arises from the boundary value problem to be solved. For small values of $\Sm$ we used the K62 solution (\ref{eq:tasymp_optfluc_Sm_K62}) as the initial guess, for large values of $\Sm$ we used an already obtained solution for a similar value of $\Sm$.
\begin{figure}[t] 
	\begin{center}
		\includegraphics[width=0.75\textwidth]{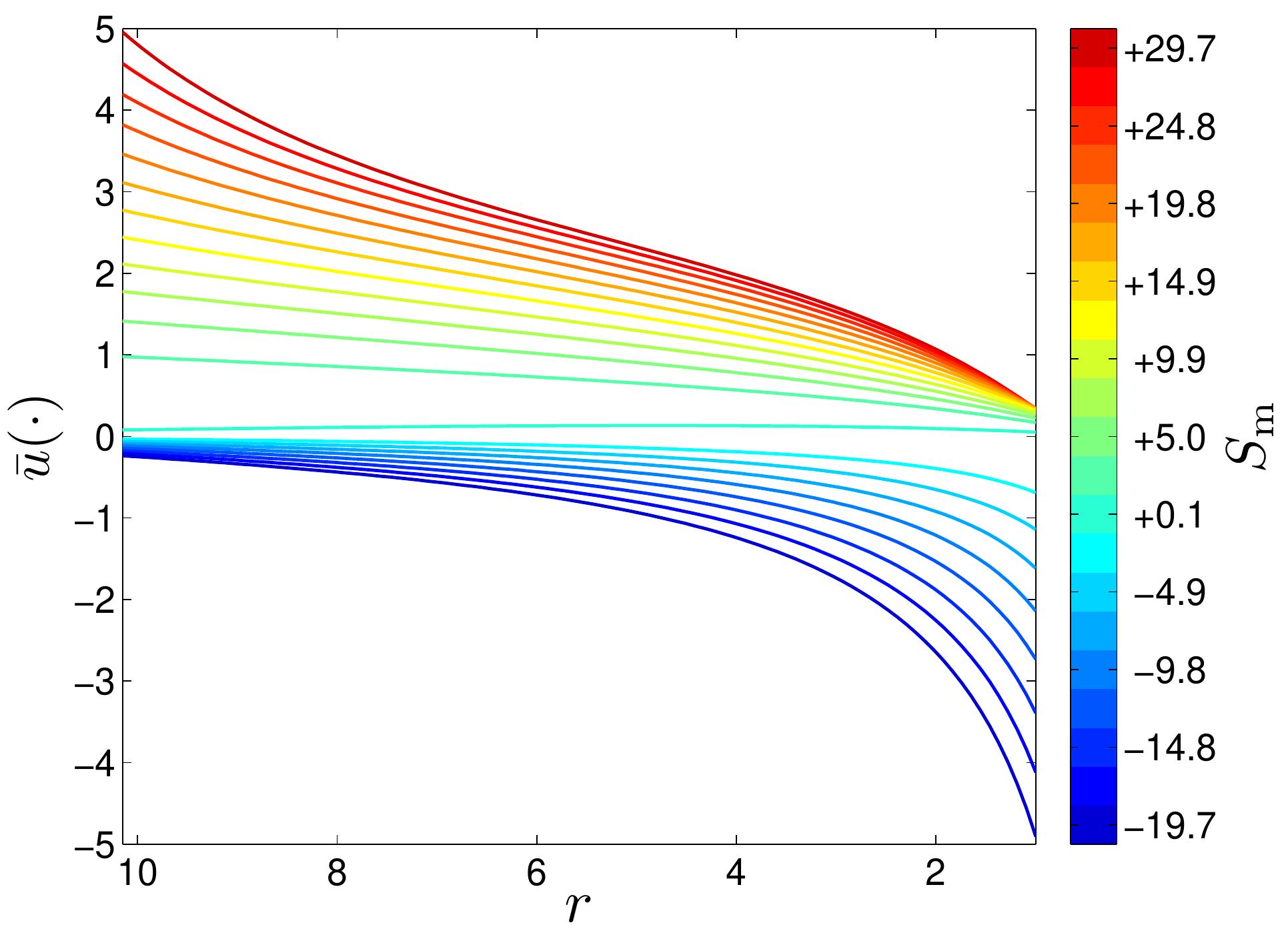}
		\caption{\label{ff:Sm_trajectories_JFM}\figtxt{Most likely realisations of the stochastic process defined by drift and diffusion estimated in \cite{Renner2001} from experimental data for various values of EP $\Sm$. The realisations are obtained from solving the ELE (\ref{eq:tasymp_ELE_Sm}) with \texttt{bvp4c} in \texttt{MatLab}.}}
	\end{center}	
\end{figure}\\
The solutions $\bu(r;\Sm)$ of the ELE for the experimentally estimated drift and diffusion using \texttt{bvp4c} are depicted in figure \ref{ff:Sm_trajectories_JFM}. It is apparent that $\bu(r;\Sm)$ shows qualitatively the same behaviour as for the K62 case, which may not prove but indicate that the interpretation of $\Sm$ within the theoretical K62 model is also valid in realistic turbulent flows. Nevertheless, a striking difference is that for positive $\Sm$ the initial values of the optimal fluctuations clearly deviate from the mode of the initial distribution.\\
A peculiarity arises in solving the ELE numerically: The sign of $\bu(r;\Sm)$ is prone to the initial guess. By taking as an initial guess the K62 solution (\ref{eq:tasymp_optfluc_Sm_K62}), we obtained the solutions shown in figure \ref{ff:Sm_trajectories_JFM}. But by reversing the sign of the initial guess, also the solutions $\bu(r;\Sm)$ reverse their sign, see for example figure \ref{sf:Sm_trajectories_JFM2_pos}. Hence, the numerical solution of the ELE appears to be bistable, and consequently, the path-probability $P[\bu(r;\Sm)]$ is bimodal. By comparing the values for the action $\SS[\bu(r;\Sm)]$ for entirely positive and entirely negative solutions in figure \ref{sf:Sm_trajectories_JFM2_action}, we find that in the case of negative $\Sm$ both solutions are balanced, whereas for positive $\Sm$, the negative solutions tend to be more likely. The {\it average} realisations $\uc$ giving rise to extreme values of EP, as discussed in \cite{Nickelsen2013} (chapter \ref{ss_turbulence_PRL}), hence have contributions from both the negative and the positive modes of $P[\bu(r;\Sm)]$.
\begin{figure}[t] 
  \subfloat[][\figsubtxt{Positive solutions $\bu(r;\Sm)$.}]{\label{sf:Sm_trajectories_JFM2_pos}
	\includegraphics[width=0.5\textwidth]{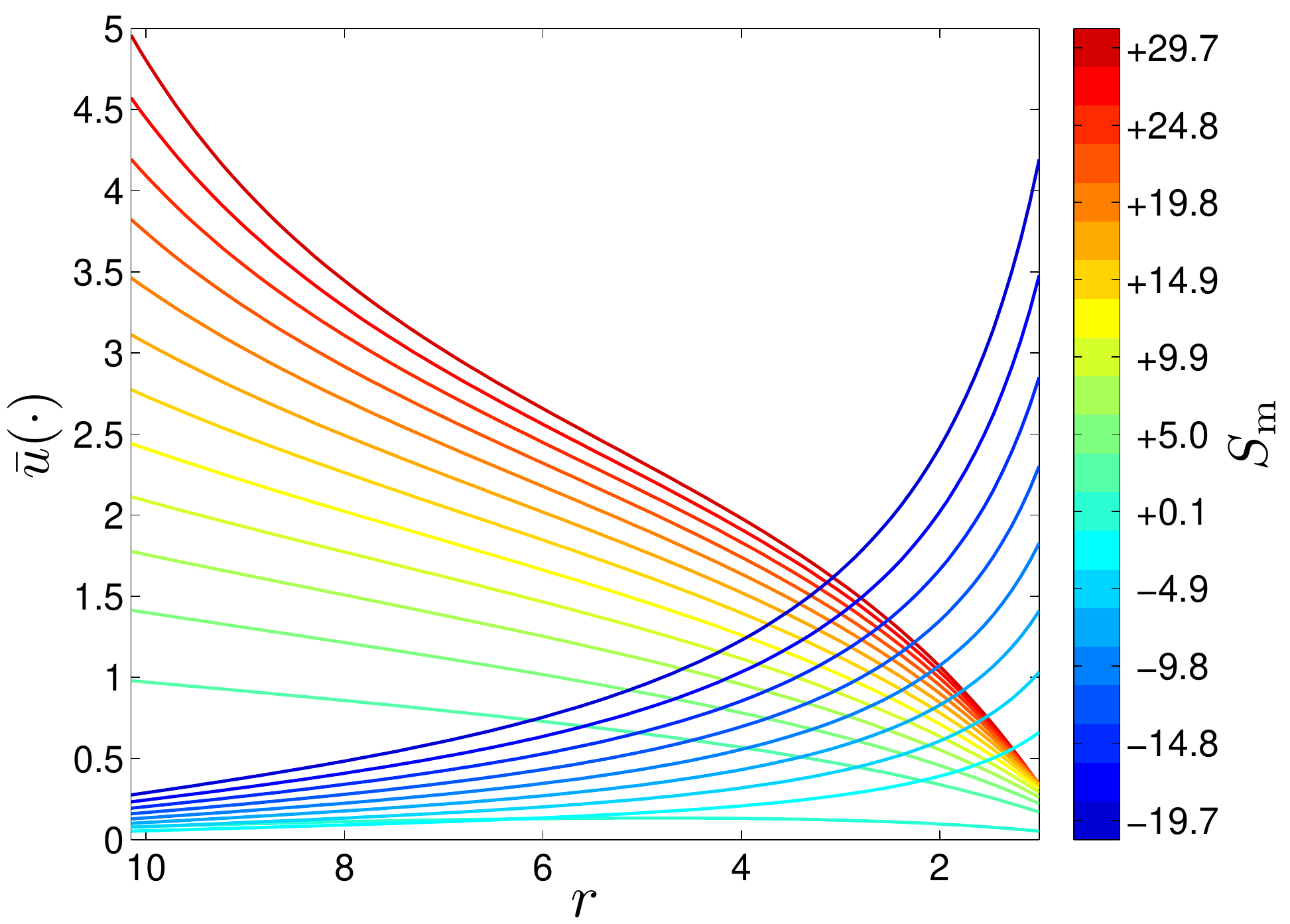}}
  \subfloat[][\figsubtxt{The action $\SS[\buc]$ for positive and negative solutions $\bu(r;\Sm)$ as function of $\Sm$.}]{\label{sf:Sm_trajectories_JFM2_action}
	\includegraphics[width=0.45\textwidth]{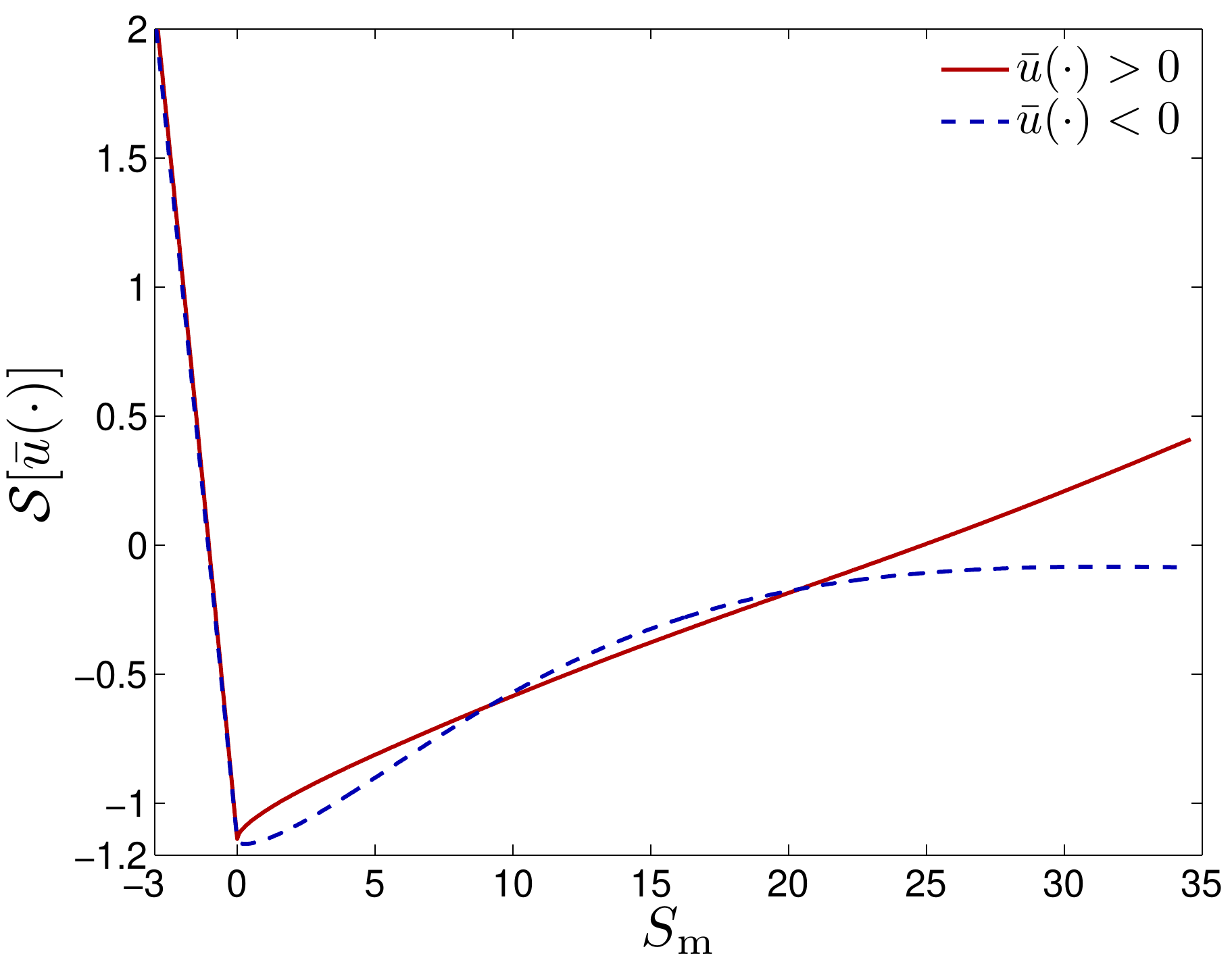}}
  \caption{\label{ff:Sm_trajectories_JFM2}\figtxt{The ELE (\ref{eq:tasymp_ELE_Sm}) is numerically bistable. Depending on the initial guess, the solutions of the ELE using \texttt{bvp4c} are positive or negative, for positive and negative values of $\Sm$ separately. In (a) we show an example of only positive solutions, the negative solutions are barely distinguishable from the mirrored positive solutions. In (b) we assess the probability $P[\buc]=\exp\big[\!-\!\SS[\buc]\,\big]$ of positive and negative solutions by means of the action $\SS[\buc]$.}}
\end{figure}\\
The finding that for positive values of $\Sm$ the negative solutions are more likely is in accord with the four-fifths law, $\lla u(r)^3\rra\eq-\frac{4}{5}\beps r\!<\!0$. The balance of probability of positive and negative solutions for negative values of $\Sm$, suggesting a vanishing skewness in the realisations $\{u(r);\,\Sm\!<\!0\}$, remains curious. We should also keep in mind, that by comparing only the probabilities $P[\bu(r;\Sm)]$ evaluated at the {\it optimal} fluctuation $\buc$, we obtain no information about the fluctuations in the vicinity of $\buc$. We therefore refrain from a further interpretation and leave it for future studies.
\remark{According to the four-fifths law, $\lla u(r)^3\rra=-\frac{4}{5}\beps r$, an ensemble of inverse turbulent cascade should display a positive skewness. Computing  $\lla u(r,\Sm)^3\rra/r$ instead of $\lla u(r,\Sm)\rra$ should therefore reveal whether $\Sm<0$ indeed corresponds to an inverse cascade. \todo{hatte ich mal gemacht, kam glaub ich nichts definitives bei raus.} Wahrscheinlich zeigt $\Sm<0$ nur ne kombi aus extremen negativen und positiven $\lla u(r,\Sm)\rra$ an, müsste mal wirklich die verteilung anschauen. In K62 ist es ja auch so, für symmetrisches $p_L$ überlagern sich beide fälle genau gleich, nur die leicht negative skewness lässt den negativen teil überwiegen.}
\remark{wenn ich mir die resultierende asymptotik für $p(\Sm)$ anschaue, dann sehe ich tatsächlich dass linker schwanz gut passt (bräuchte nur faktor zwei wenn ich pef hätte), während der reichte teil nicht hunderprozentig passt. da müsste man wohl die beiträge der negativen und die positiven $\bu(r)$ entsprechend gewichtet berücksichtigen.}

\subsection[Asymptotic ${p(u,r)}$ and ${\z_n}$]{Asymptotic $\bs{p(u,r)}$ and $\bs{\z_n}$} \label{ss_turbulence_asymp_tails}
In the previous section, we used the ELE to calculate the most likely realisations giving rise to preset values of $\Sm$. Instead of imposing the constraint with respect to $\Sm$, we can also impose the boundary condition that $u(r)$ takes a certain value $u^\ast$ at scale $r^\ast$ with $L>r^\ast>\l$.\\
In this case, the ELE (\ref{eq:tasymp_ELE_constr}) reads
\begin{subequations} \label{eq:tasymp_ELE_pur}
  \begin{align}
    \dddu &= \mfrac{\ddu \+ F}{2D}\big(D'(\ddu \- F)+2\ddD\big)+FF'-\ddF+2DJ'' \;, \label{eq:tasymp_ELE_pur_ode}\\
		0&=\mfrac{\ddu_0+F_L}{2D_L}+\L'_L \;,  \label{eq:tasymp_ELE_pur_BC0} \\
		u(r^\ast)&=u^\ast \;. \label{eq:tasymp_ELE_pur_BC1}
	\end{align}
\end{subequations}
Solving the above ELE for a certain value of $u^\ast$ and $r^\ast$ yields the asymptotics
\begin{align} \label{eq:tasymp_pur_JFM}
	p(u^\ast,r^\ast) \sim \exp\big[\-\SS[u(r;\,u^\ast,r^\ast]\,\big] \;.
\end{align}
Hence, by varying $u^\ast$ and $r^\ast$ we obtain an approximation of $p(u,r)$ which is asymptotically exact in the limit of $|u|\to\infty$, if we assume that infinite values for $u$ are singularly rare. \\

\begin{figure}[!t] 
  \subfloat[][\figsubtxt{Asymptotic and analytic solution, zoomed out.}]{\label{sf:tasymp_K62_zout}
	\includegraphics[width=0.49\textwidth]{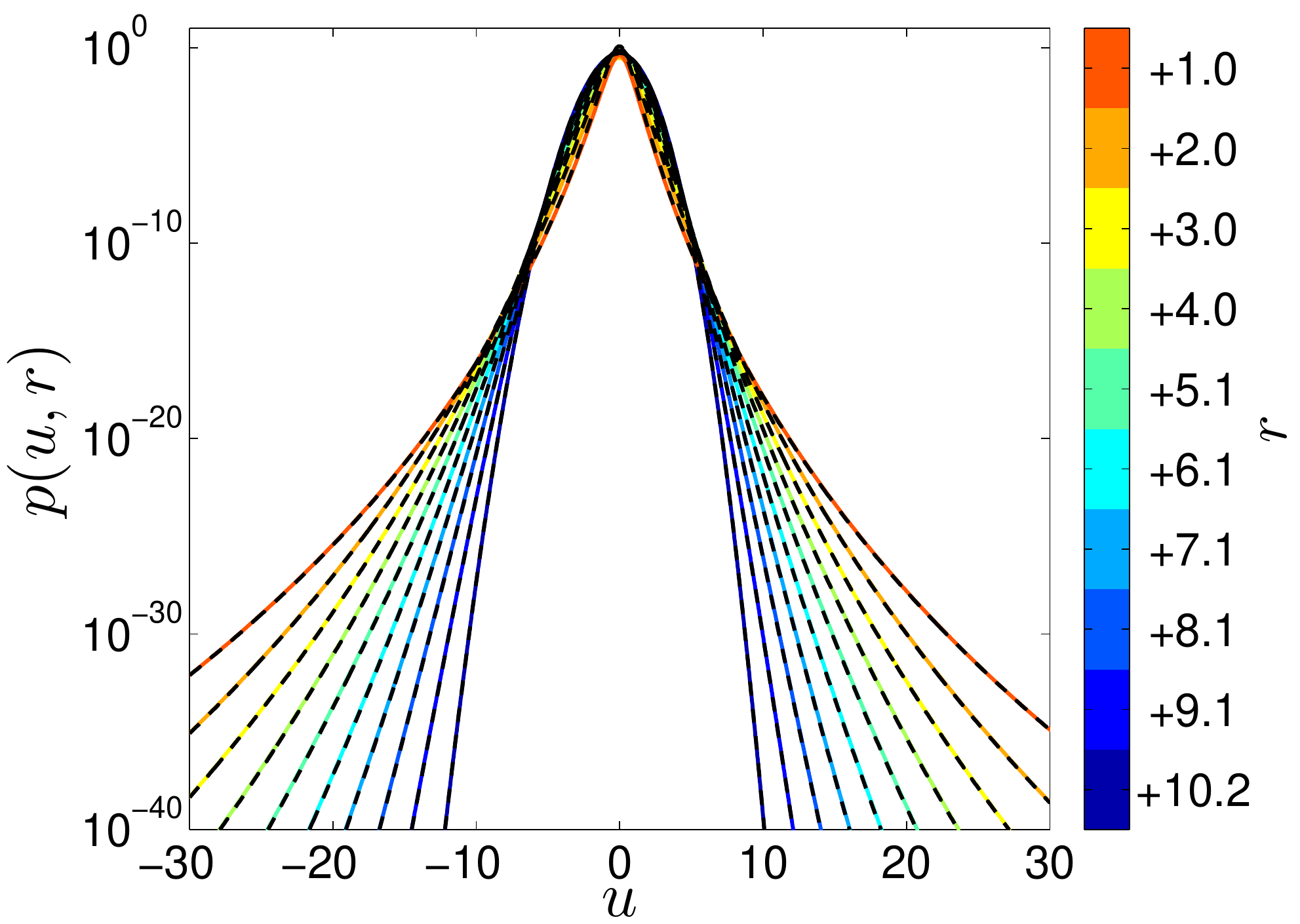}}
  \subfloat[][\figsubtxt{Asymptotic and analytic solution, zoomed in.}]{\label{sf:tasymp_K62_zin}
	\includegraphics[width=0.49\textwidth]{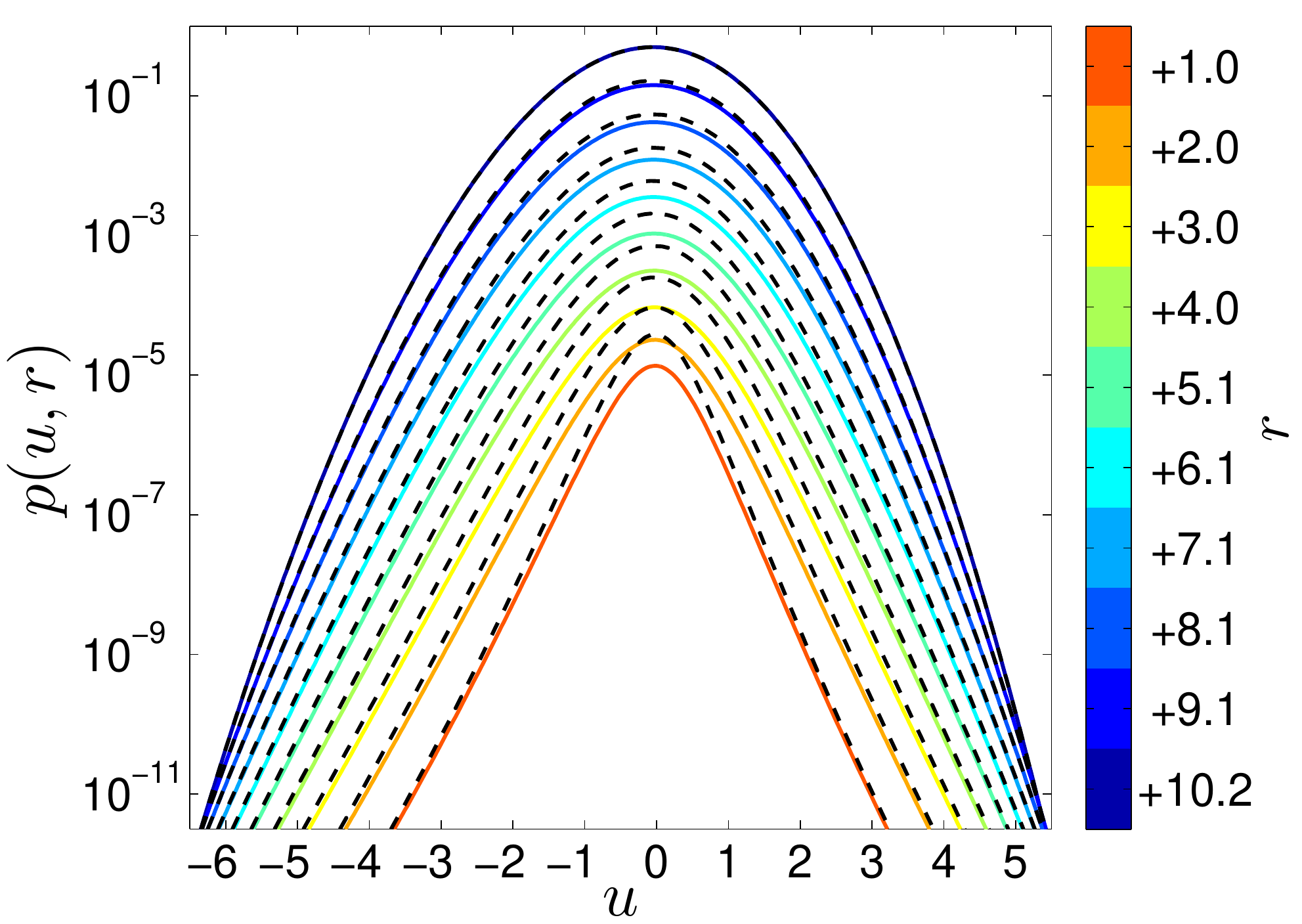}}
  \caption{\label{ff:tasymp_K62}\figtxt{Asymptotics of $p(u,r)$ for the K62 process defined by (\ref{eq:tasymp_K62_SDE}). The coloured lines are the asymptotic solutions given by (\ref{eq:tasymp_pur_K62}) for $\mu=0.25$, $L=10.15$ and $\l=1$ and various values of scale $r$. The dashed lines are the analytic solutions given in (\ref{eq:D1D2_K62_FPE_sol}). The pre-exponential factor of the asymptotics is obtained from fits to the analytic solution in the far tails of $p(u,r)$.}}
\end{figure}

Taking again the K62 model as an analytic accessible example, we obtain from (\ref{eq:tasymp_ELE_Sm_K62}) the ELE
\begin{subequations} \label{eq:tasymp_ELE_pur_K62}
  \begin{align}
		\dddu(r) &= \frac{\ddu(r)}{u(r)} - \frac{\ddu(r)}{r} \;, \\
		0 &= \ddu_LL - au_L + 2bu_L^2\L'(u_L) \;, \\
		u(r^\ast)&=u^\ast \;.
	\end{align}
\end{subequations}
Note that the ODE remains the same since $F(u,r)/D(u,r)$ is not a function of scale $r$. The solutions are again power laws,
\begin{align} \label{eq:tasymp_optfluc_pur_K62}
	u(r) \eq u^\ast\,\Big(\frac{r}{r^\ast}\Big)^{\displaystyle c} \;,
\end{align}
where the exponent $c$ has to be determined from
\begin{align}
	\begin{split}
		0&=(a\-c)\,p_L\big(u_L(c)\big) + 2b\,u_L(c)\,p'_L\big(u_L(c)\big) \nn 
		u_L(c)&=u^\ast\,\Big(\frac{L}{r^\ast}\Big)^{\displaystyle c}  \;.
	\end{split}
\end{align}
Substitution of (\ref{eq:tasymp_optfluc_pur_K62}) into the action $S[\uc]$ yields the asymptotics
\begin{align}
	p(u,r) &\sim \Big(\mfrac{L}{r}\Big)^{\mscale[1.1]{\frac{(c-a)^2}{4b}+\frac{b-a}{2}} }	\;,
\end{align}
or, with $a=(3+2\mu)/9$ and $b=\mu/18$ substituted from (\ref{eq:D1D2_K62}),
\begin{align} \label{eq:tasymp_pur_K62}
	p(u,r) &\sim \Big(\mfrac{L}{r}\Big)^{\mscale[1.1]{\frac{(3c-1-2\mu/3)^2}{2\mu}-\frac{2+\mu}{12}}}	\;.
\end{align}
In figure \ref{ff:tasymp_K62}, we compare the asymptotic solution with the exact solution from (\ref{eq:D1D2_K62_FPE_sol}), in which we used $\mu=0.25$, $L=10.15$ and $\l=1$, and we took $p_L(u_L)$ from fitting a skew-normal distribution to the experimentally determined result using again the data from \cite{Renner2001} as in the previous section. The agreement between the measured and fitted $p_L(u_L)$ can be judged from \ref{sf:tasymp_exper_pur} for the largest scale $r\eq L$. From figure \ref{ff:tasymp_K62} it is apparent that the asymptotics of $p(u,r)$ coincides with the analytical result for values of $u$ that are barely accessible in a comparable experiment.\footnote{Such as the experiment in \cite{Renner2001} with matching $L=10.15$, $\l=1$ and $p_L(u_L)$.}\\ 
The asymptotics in (\ref{eq:tasymp_pur_K62}) only includes the exponential factor, a constant pre-exponential factor has been obtained from fitting the asymptotic $p(u,r)$ to the analytical solution (\ref{eq:D1D2_K62_FPE_sol}) for values of $u$ that are sufficiently far in the tails to ensure the validity of the asymptotic assumption. For a more realistic process, where the analytic solution is not known, the determination of a constant pre-exponential factor requires an interval of $u$-values in which both the asymptotics is valid and the histogram determined from an experiment or simulation approximates $p(u,r)$ sufficiently well.
\begin{figure}[!t]
	\begin{center}
		\includegraphics[width=0.7\textwidth]{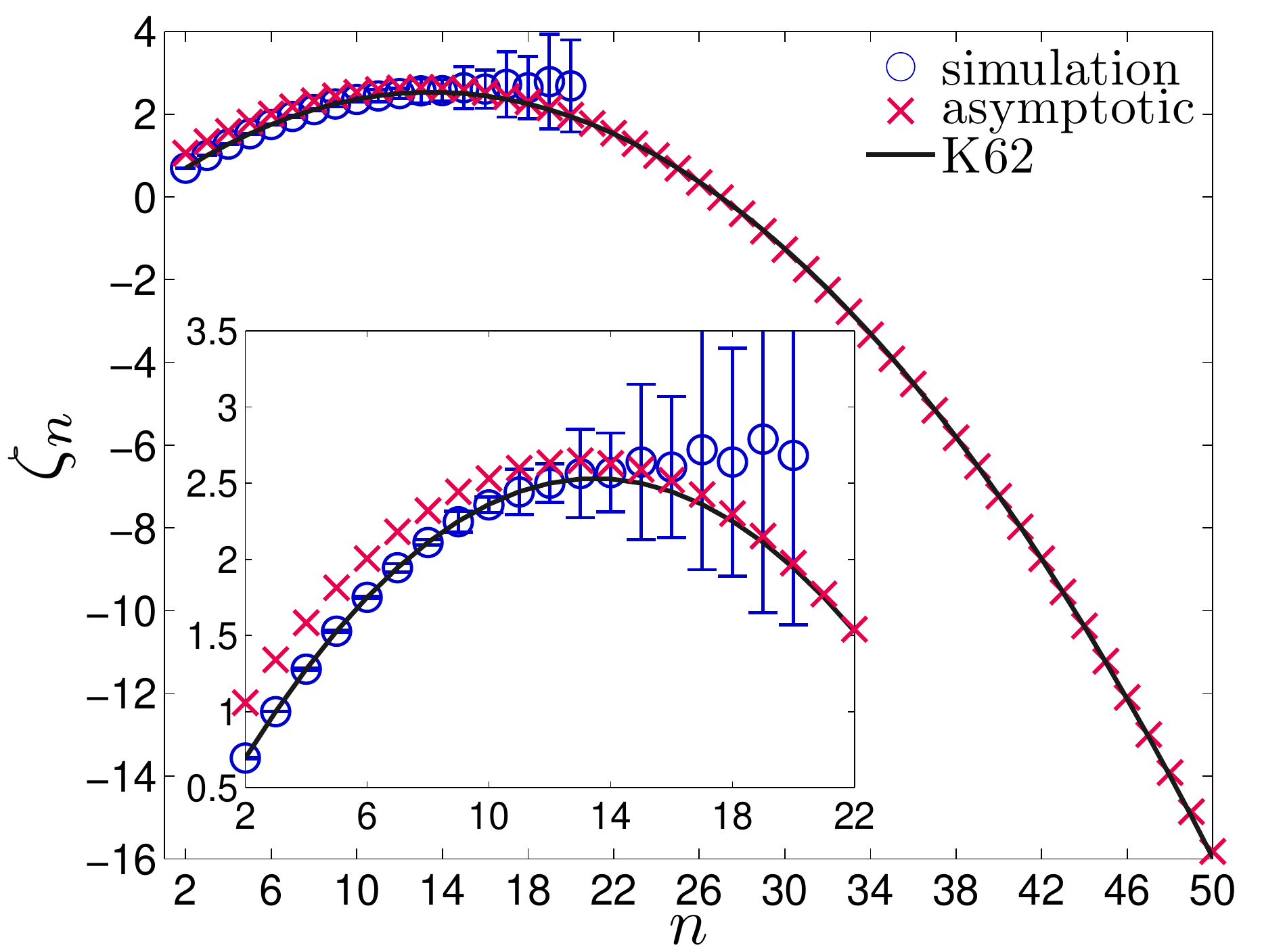}
		\caption{\label{ff:tasymp_zn_K62}\figtxt{Asymptotic scaling exponents for the K62 process for $\mu=0.25$. The asymptotic solutions shown in figure \ref{ff:tasymp_K62} imply structure functions $\Str^n(r)$, from which asymptotic scaling exponents have been derived, shown as red crosses. The scaling exponents determined from an ensemble of $10^7$ realisations $u(r)$, generated from the analytical solution of the K62 SDE in (\ref{eq:D1D2_K62_SDE_sol}), are included as blue circles, the exact scaling law (\ref{eq:FDT_K62}) for the K62 model is added as a black line. The scaling exponents are obtained by fitting the ESS formula (\ref{eq:FDT_ESS_Sn}), the error bars are the standard deviations of the $\z_n$ determined from $10^3$ blocks of $10^4$ realisations $u(r)$.}}
	\end{center}
\end{figure}\\

The asymptotic $p(u,r)$ can be used to determine the scaling exponents $\z_n$ of the corresponding structure functions asymptotically. To this end, the structure functions are computed by
\begin{align}
	\Str^n(r) = \int u^n\,p(u,r) \di u \;,
\end{align}
and, according to the ESS procedure explained after (\ref{eq:FDT_ESS_Sn}), $\z_n$ is obtained as the slope of a linear fit of $\ln S^n(r)$ as function of $\ln S^3(r)$.\\
In figure \ref{ff:tasymp_zn_K62}, we demonstrate that by using the asymptotics of $p(u,r)$ from figure \ref{ff:tasymp_K62}, the scaling exponents resulting from the proposed procedure are excellent in a range of $n=20$ to $n=50$. For values $n<20$, the determination of the moments rests on the centre of $p(u,r)$ for which the asymptotic solution is not accurate, for $n>50$ the range in $u$ covered by the asymptotic solution is not sufficient\footnote{The part of $p(u,r)$ being relevant for structure functions $\Str^n(r)$ beyond order $n=50$ is so far in the tails, that the asymptotics of $p(u,r)$ exceeds the standard machine accuracy of MatLab. A limitation that can be overcome without much effort.}.\\

Instead of using K62 as the analytical accessible example, it is interesting to see how the method performs for the experimental data from \cite{Renner2001} as used in the previous section.\\
We take again the drift and diffusion coefficients $\Dfg(u,r)$ experimentally estimated in \cite{Renner2001}. By substituting the numerical solutions of the ELE obtained by using \texttt{bvp4c} in \texttt{MatLab} into (\ref{eq:tasymp_pur_JFM}), we yield the asymptotics of $p(u,r)$ defined by the Markov process extracted from the experimental data. In figure \ref{sf:tasymp_exper_pur}, we compare the asymptotic $p(u,r)$ with the experimental result, for which we took for $p_L(u_L)$ again the skew-normal distribution fitted to the experimental data.\\
In determining a pre-exponential factor, we face the problem explicated above for the K62 example: The experimentally determined $p(u,r)$ does not reach sufficiently far into the tail of $p(u,r)$ where the asymptotic solution is valid. As a preliminary work-around, we determined the 10-th order structure function $\Str^{10}(r)$ from both the experimental data and the asymptotics, and adjusted the pre-exponential factor such that $\Str^{10}(r)$ is identical for both. The $10$-th order structure function was chosen as it proved to be the largest order $n$ for which $\Str^n(r)$ is reliable.\\

We stress that the estimation of $\Dfg(u,r)$ can only use the available experimental measurements covering the range of $u$-values as apparent from figure \ref{sf:tasymp_exper_pur}. Consequently, by taking these $\Dfg(u,r)$ to determine the asymptotic solution of the FPE, we extrapolate the measured $p(u,r)$ to higher values of $|u|$ under the assumption that the form of the FPE defined by $\Dfg(u,r)$ governs the evolution of $p(u,r)$; very much like assuming that a random variable is Gaussian distributed and predicting the probability of unobservable realisations by estimating mean and variance from the available data. It is therefore arguable to expect that by determining the asymptotic solution of the FPE, we also obtain the true asymptotic behaviour of the underlying experiment.\\
Nevertheless, for completeness, we show the resulting asymptotic scaling exponents in figure \ref{sf:tasymp_exper_zn} which are clearly unsatisfactory. We impute this unsatisfactory result to the uncertain determination of the pre-exponential factor.\\
Indeed, premature results give hope that an explicit determination of the pre-exponential factor as demonstrated in section \ref{ss_asymp_application_toy} improves the situation considerably for two reasons. First, we do not have to determine the pre-exponential factor from fits, and therefore do not rely on experimentally determined $p(u,r)$. And second, since we also achieve the $u$-dependency of the pre-exponential factor, the range of $u$ values for which the asymptotics is valid should enlarge. Efforts to determine the pre-exponential factor have to be left for future studies.
\begin{figure}[!t] 
  \subfloat[][\figsubtxt{Asymptotic $p(u,r)$ together with experimental result.}]{\label{sf:tasymp_exper_pur}
	\includegraphics[width=0.52\textwidth]{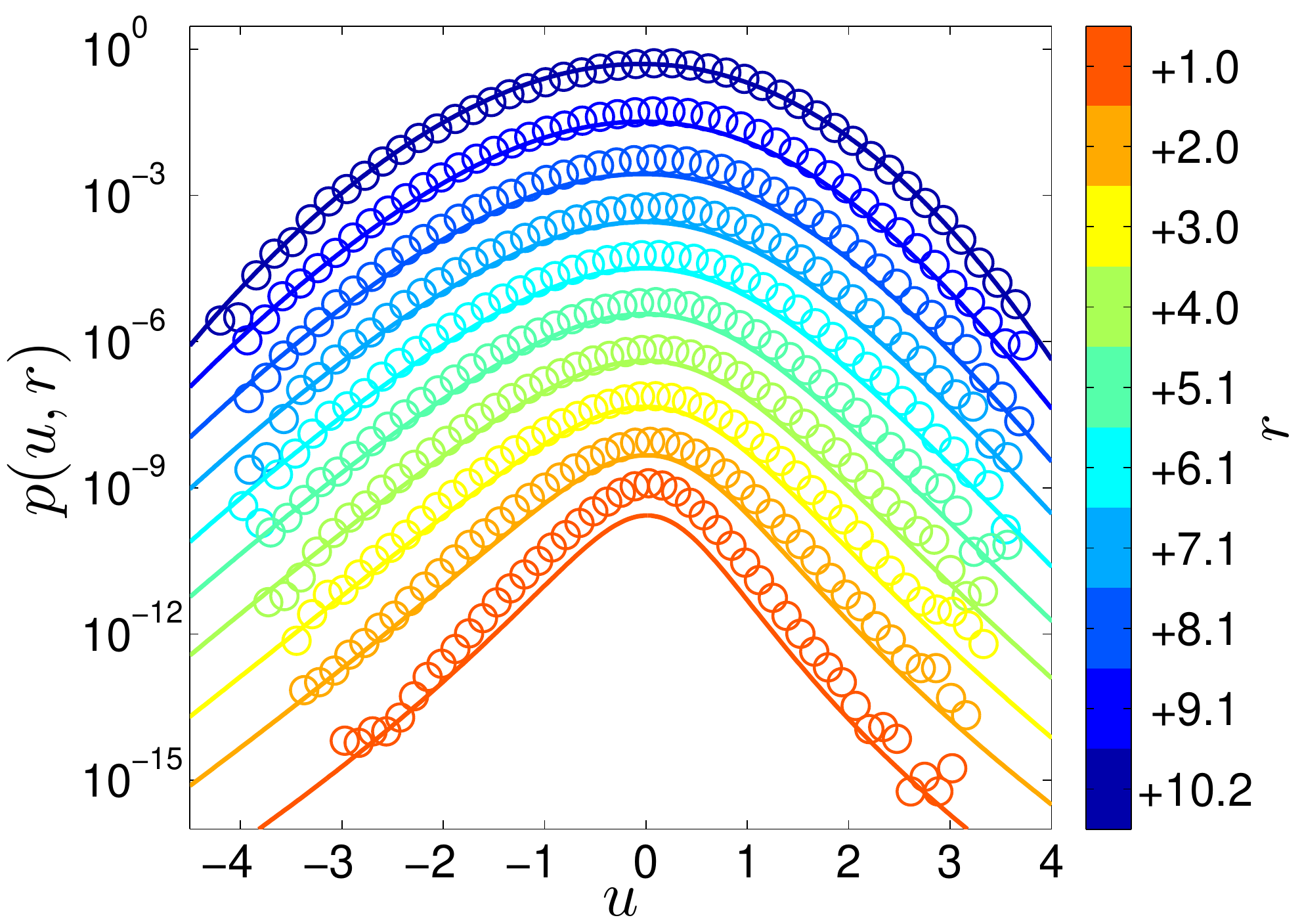}}
  \subfloat[][\figsubtxt{Asymptotic scaling exponents in comparison.}]{\label{sf:tasymp_exper_zn}
	\includegraphics[width=0.4785\textwidth]{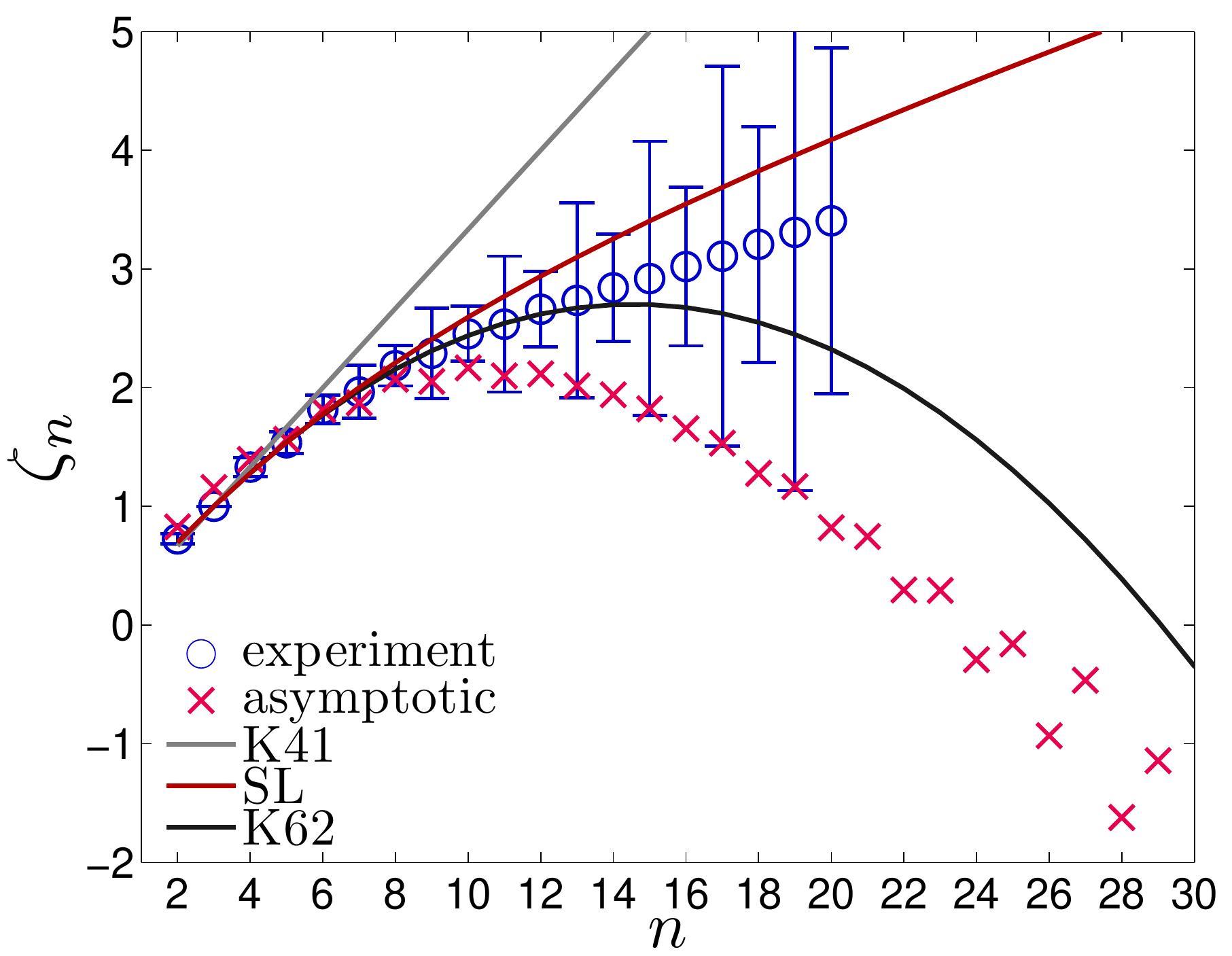}}
  \caption{\label{ff:tasymp_exper}\figtxt{Asymptotics of $p(u,r)$ for the Markov process defined by the experimentally estimated drift and diffusion (\ref{eq:D1D2_JFM_D1D2}). The coloured lines in (a) are the asymptotic solutions given by (\ref{eq:tasymp_pur_JFM}) for various values of scale $r$, which derive from the numeric solutions of the corresponding ELE (\ref{eq:tasymp_ELE_pur}) obtained with \texttt{bvp4c}. The coloured symbols are the corresponding histograms determined from the experimental data which was used in \cite{Renner2001} to estimate the drift and diffusion coefficients used here. The asymptotic solutions imply structure functions $\Str^n(r)$, from which asymptotic scaling exponents have been derived, shown as red crosses in (b). In addition, scaling exponents determined from the exponential data and some theoretical predictions are included into (b), please refer to figure \ref{ff:zetan_ESS}. The pre-exponential factor of the asymptotics is obtained from fitting $\Str^{10}(r)$ to the experimental result.}}
\end{figure}\\

In closing, we mention that instead of using the asymptotics of $p(u,r)$ to determine high order structure functions, the moment equation of the corresponding FPE can be used to set up an iteration procedure.\\
The general form of drift and diffusion estimated in \cite{Renner2001} from experimental data reads
\begin{subequations}
  \begin{alignat}{9}
		&\Df(u,r) =\;& -&a_0(r) \;&-\;& a_1(r)\,u \;&+\;& a_2(r)\,u^2 \;&-\;& a_3(r)\,u^3 \;& \;,   \\
		&\Dg(u,r) =\;&  &b_0(r) \;&-\;& b_1(r)\,u \;&+\;& b_2(r)\,u^2 &. & & 
	\end{alignat}
\end{subequations}
Employing the moment equation (\ref{eq:FPE_moments}) and solving for the highest order structure function yields
\begin{align}\label{eq:tasympt_mom-it}
		\Str^{n\+4} = &\frac{1}{a_3(r)}\Big[ (n\+1)b_0(r)\Str^{n}(r) + \big(a_0\+(n\+1)b_1(r)\big)\Str^{n\+1}(r)\\
		&\quad+ \big(a_1(r)\+(n\+1)b_2(r)\big)\Str^{n\+2}(r) + a_2(r)\Str^{n\+3}(r) + \mfrac{1}{n\+2}\pt_r\Str^{n\+2}(r)\Big]\;. \nonumber
\end{align}
The above formula yields the $(n\+4)$\,-\,th order structure function and needs as input the three next smaller structure functions.\\
One way to initialise the iteration is using $\Str^1(r)\equiv1$ and taking\linebreak $\Str^2(r)=c_2r^{\z_2}$ and $\Str^3(r)=c_3r^{\z_3}$ with $c_{2,3}$ and $\z_{2,3}$ fitted to the experimental data, arguing that a scaling law is a good approximation for the second and third order structure functions, and the derivative $\pt_r\Str^{2}$ can be calculated analytically. In this case, we have nested equations of depth $n-4$ to determine the $n$-th order structure function. Another possibility is to take $\Str^{1}(r)$, $\Str^{2}(r)$ and $\Str^{3}(r)$  directly from measurements or simulations of the SDE, compute $\pt_r\Str^{2}$ numerically and iterate to higher order structure functions by means of numerical $\Str^n(r)$.\\
In any case, after very few steps, the iteration results into structure functions that depart drastically from the experimentally determined ones and become divergent. We hence face a highly sensitive dependency on the initial conditions as characteristic for chaotic systems. In addition, taking simpler forms of drift and diffusion reveals that dropping the odd term $b_1(r)\,u$ of the diffusion coefficient, responsible for the skewness of the dynamics\remark{die even terms in drift haben vernachlässigbaren einfluss, steht auch in JFM} and correlation between the additive and multiplicative noise (cf. (\ref{eq:D1D2_JFM_SDE_two-xi})), the iteration procedure improves in comparison with according simulation results.\\
We leave the implications of these observations for further studies.

\cleardoublepage
\section{Closing discussion}
We close the part on turbulence with a summarising discussion.
\subsection{Classification}
The Markov representation is a unifying framework for a variety of established theories of turbulence, allowing a classification of these theories in terms of Markov processes.\\
The central turbulence theory is the K62 model. The K62 model may be viewed as the harmonic oscillator of the Markovian description of turbulence, in the sense that it is fully analytical solvable, captures the main features of turbulence, and from an alteration of the K62 process more realistic models of turbulence arise. Depending on the kind of alteration, the generalisations of the K62 model belong to different branches of turbulence research.\\

The {\bf K62 model}, with scaling exponents $\z_n=n/3-\mu/18n(n-3)$, is characterised by drift $\Df(u,r)\propto u/r$ and diffusion $\Dg(u,r)\propto u^2/r$. The fluctuation theorem associated with K62 involves only boundary terms. Considering only the deterministic process defined by $\Df(u,r)$, we retrieve the trivial case of K41, $\z_n=n/3$. We now give an overview of the emerging models of turbulence by various alterations of K62.\\

{\bf (i)} A linear drift $\Df(u,r)=a(r)u$ and quadratic diffusion $\Dg(u,r)=b(r)u^2$ for arbitrary $a(r)$ and $b(r)$ represent {\bf log-normal random cascade models}, that is to say, the random variable $u$ is described by a log-normal distribution. The explicit form of $a(r)$ and $b(r)$ determines the variance $\s^2(r)$ of the log-normal distribution, the associated fluctuation theorem receives an integral contribution.\\
Only for $a(r)\propto1/r$ and $b(r)\propto1/r$, retrieving the K62 model, it is $\s^2(r)\sim\ln(r)$ and the moments of $u(r)$ exhibit a scaling law.\\
By altering $a(r)$ and $b(r)$ but keeping the $r$-dependency equal, that is $a(r)/b(r)=\const$, we abandon a scaling law, but the boundary form of the K62 fluctuation theorem is retained. The random cascade process for $a(r)/b(r)=\const$ has been termed {\bf continuously self-similar}, and the stochastic processes becomes a stationary process.\\
For the special choice $a(r)\sim b(r)\sim\pt_r\ln S^m(r)$ the moments of $u(r)$ scale as $\big(S^m(r)\big)^{\z_n}$, being the conjecture of {\bf extended self-similarity}, where $m=3$ is the most common choice.\\
A shortcoming of log-normal random cascade models is the vanishing {\bf skewness} in the statistics of $u(r)$. In terms of Markov processes, a skewness can principally be added to the dynamics\footnote{Initialising the process with a asymmetric $p_L(u_L)$ introduces a decaying skewness to the process, even for a dynamics defined by $\Dk(u,r)$ being invariant under $u\mapsto -u$.} by, e.g., introducing an additive noise term coupled to the existing multiplicative noise term in the stochastic differential equation, or by augmenting the Fokker-Planck equation with a suitable third Kramers-Moyal coefficient $\Dth(u,r)$.\\

{\bf (ii)} Instead of keeping the $u$ dependency and altering the $r$ dependency, we may also keep $\Dfg(u,r)\propto 1/r$ but allow an arbitrary combination of powers in $u$. For this choice of $\Dfg(u,r)$, the moments of $u(r)$ become a superposition of scaling laws with distinct linear exponents. Allocating the scaling laws with fractal subspaces of $\mathbb{R}^3$, we obtain the essence of {\bf multifractal models} of turbulence. The explicit correspondence to multifractal models, however, remains to be established. The associated fluctuation theorem is again of boundary form, involving the moments of $u(r)$.\\

{\bf (iii)} Keeping the K62 drift but adding to the K62 diffusion a term linear in $u$, that is $\Dg(u,r)\sim \b u+bu^2/r$, we obtain the {\it continuous} approximation of {\bf Yakhot's model} of turbulence.\\
The full equivalence to Yakhot's model is achieved by taking also the higher Kramers-Moyal coefficients into account, $\Dk(u,r)\sim \b_k u^{k-1}+b_ku^k/r$ for $k\geq2$. The stochastic process is now a {\it discontinuous} process.\\
The coefficient $\b_k$ accounts for effects of energy injection. Neglecting that effect by setting $\b_k=0$, the Kramers-Moyal expansion implies {\bf Yakhot's scaling law} $\z_n=n/3\,(B+3)/(B+n)$.\\
An expansion of Yakhot's scaling law to third order in the anomaly parameter $\d_2=\z_2\,-2/3$ recovers the scaling exponents suggested by {\bf L'vov and Procaccia} being cubic in $n$. Cubic scaling exponents correspond to a Kramers-Moyal expansion truncated after the third term.\\

{\bf (iv)} The K62 process is composed of a deterministic component and a Gaussian process. Substituting the Gaussian process with a jump process, where the moments of the jump density are $\Ak(u,r)\propto (bu)^k/r$ with $b=(2/3)^{1/3}-1$, defines a jump process being equivalent to the {\bf log-Poisson random cascade model}. This model exhibits scaling laws with the scaling exponents $\z_n=n/9+2-(2/3)^{n/3}$ derived by {\bf She and Leveque}.\\

{\bf (v)} Drift and diffusion coefficients can also be obtained directly from {\bf experimental measurements}. Generally, $\Df(u,r)$ and $\Dg(u,r)$ are of similar form as in the K62 process, but $\Dg(u,r)$ involves two extra terms, being constant and linear in $u$, and each term in $\Dfg(u,r)$ is a different function of $r$. The stationary distribution at scale $r=L$ defined by $\Dfg(u,L)$ generally differs from the measured initial distribution $p_L(u)$ of the process. Accordingly, non-equilibrium is imposed by both, initialising the process off equilibrium, and driving the process. \todo{Plot von D1 und D2 für K62, Castaing, Yakhot und JFM?}

\subsection{Interpretation}
\remark{das Vasen-Bild: $\ln(L/r$ ist die Anzahl der Brüche, und $u(r)$ ist die größe der Bruchstücke. Eine zerdepperte Vase macht eine Realisierung $u(r)$. Wir haben viele Vasen der Größe $u_L$, wobei wir $u_L$ aus ner Gauß-verteilung ziehen. Entropie wird bei den Fällen konsumiert, wo größere Bruchstücke bei rauskommen als die typische Vase selbst war.}
The thermodynamic interpretation of Markov processes discussed in the first part provokes a similar fashioned interpretation of the Markov representation of fully developed turbulence.\\

Starting point is again the K62 process. Being a stationary process in logarithmic scale $\ln(L/r)$, the K62 process is a relaxation process from an initial distribution $p_L(u_L)\neq\d(u_L)$. Consequently, non-equilibrium can only be imposed by initialising the process off equilibrium. In the limit $\ln(L/r)\to\infty$, all moments of $u(r)$ will be zero and we are left with a free laminar flow. The average tendency of the K62 process towards small scales is therefore a decrease of fluctuations.\\
The second law like inequality discussed in section \ref{ss_turbulence_PRL} confirms indeed that the average log-multiplier $\lla H_r\rra=\lla\ln(u(r)/u_L\rra$ is bounded from above by the difference in Shannon entropy $\D S(r)$ between final and initial state of the turbulent cascade. The Shannon entropy difference $\D S(r)$ is found to be negative for all scales and monotonically decreasing with decreasing $r$, implying that strictly $\lla u(r)\rra<\lla u_L\rra$, in particular towards the end of the cascade. However, by assessing the multiplier statistics, we have seen that instances with $H_r>1$ remain possible throughout the inertial range, the maximum value of $H_r$ being approximately constant.\remark{reference to inertial range since $p_L(u_L)$ and $\mu$ were taken from a real flow for which the inertial range was $L=10\l$.}\remark{In a second law like inequality, the average entropy $\lla\Sm\rra$ is bounded from below by $-1/\nu\D S(r)$ with $\D S(r)$ being the difference in Shannon entropy between final and initial state of the turbulent cascade. The Shannon entropy difference $\D S(r)$ is found to be negative for all scales and monotonically decreasing with decreasing $r$, implying that strictly $\lla\Sm\rra>0$. The K62 process therefore may be interpreted as a process where the turbulent structures consume no entropy from some sort of reservoir.(entropy consumption des systems ist ja erlaubt, also $\Sm<0$, wird ja ausm reservoir entnommen. darf eben nur noch größer sein als die difference zwischen end- und anfangszustand, weil die entropie ja nur 'gespeichert' wird und nicht wirklich vernichtet wird.) Since $\Sm<0$ corresponds to velocity fluctuations $u(r)$ exceeding the typical fluctuations $u_L$ they evolved from, the reservoir appears to provide additional velocity fluctuations to the cascade. This is in accord with the construction of the K62 process as a random cascade process.(da hatten wir ja auch einfach geschwindigkeits fluktuationen hinzugefügt.)}\\
The convergence of the K62 fluctuation theorem is dependent on realisations $u_r>u_L$, corresponding to spatial velocity fluctuations which, contrary to the average tendency, build up while evolving towards smaller scales. Since $\Sm=\nu\ln(u_L/u_\l)$ (with $\nu=(6+4\mu)/\mu\approx27$), these increasing spatial velocity fluctuations correspond formally to a consumption of entropy, $\Sm<0$, but with $\Sm$ bearing no reference to thermodynamic entropy production. Nevertheless, since $\Sm<0$ corresponds to velocity fluctuations $u(r)$ exceeding the typical fluctuations $u_L$ they evolved from, it is tempting to think of some kind of reservoir that provides velocity fluctuations throughout the cascade, in accord with the construction of the K62 process as a random cascade process in section \ref{ss_MAR_stochasticprocess}. Accepting such a perception of a reservoir, the form of the diffusion coefficient, $\Dg\sim u^2$, suggests that in a cascade process the turbulent energy $u(r)^2$ is the analog of the thermal energy $\kB T$ in a thermodynamic process. But this perception bears an inconsistency, since $u(r)^2$ depends on the realisation $u(r)$ itself and can therefore not be regarded as an equilibrium property of an ideal reservoir. It is safer to think of $\Dg\sim u^2$ as the result of accumulated randomness towards the end of the cascade.\\
In any case, realisations $\uc$ with $u(r)>u_L$ are characterised by\linebreak $\Sm[\uc]<0$. Such realisations might be the result of mingling cascades, where one cascade was initialised by such a large fluctuation that $u(r)$ is still larger than the large-scale fluctuation $u_L$ of a second cascade. Another possibility is that one strong small-scale fluctuation arises from an aggregation of many small-scale structures, corresponding to an inverse cascade with multipliers larger than one. The fact that strong small-scale fluctuations are significantly more frequent than equally strong large-scale fluctuations, cf. figure \ref{ff:sm-sc-intm}, indicates that an aggregation of small-scale structures is the predominant mechanism that causes $u(r)>u_L$, and we claim that $\Sm[\uc]<0$ is a signature of a concurrent inverse cascade in addition to the direct cascade.\remark{oder sollte man sich hier eher ne peitsche vorstellen? dh, die fluktuationen werden gerade deswegen so stark, weil sie so klein werden, ähnlich wie ne pirrouette?}\\

In stochastic thermodynamics, $\Sm[\xc]<0$ accounts for collisions between fluid molecules and particles in which the particle is kicked primarily from one side, rectifying thermal noise into directed motion. The exact dynamics of such collisions is not resolved by the stochastic process, taking place below the Markov-Einstein time scale $\tME$. What is the analog mechanism in a turbulent cascade?\\
In constructing the K62 process as a random cascade process in section \ref{ss_MAR_stochasticprocess}, we coarse-grained the exact dynamics of the break-up of eddies as a Gaussian noise accounting only for the outcome of the break-up, in close analogy to consider the collision of a Brownian particle with fluid molecules only as a random change in particle velocity. Accordingly, we expect the mechanism that is responsible for $\Sm[\uc]<0$, implying $u(r)>u_L$, to take place during the break-up of eddies.\remark{dh, für skalenschritte $\D r<\D\rME\approx\l$, also zwischen einem kasakdenschritt, denn $\D\rME$ is ja gerade ein kaskadenschritt. aber was hat dann $\rME\approx\l$ zu bedeuten?} The decrease of scale $r$ after a break-up is given by $\rME\approx\l$, with $\l$ being the Taylor microscale indicating the scale below which molecular friction is prevailing. It is therefore tempting to speculate about a connection between $\Sm[\uc]<0$ and molecular friction, such as $\Sm[\uc]<0$ being the hallmark of velocity fluctuations which build up in the dissipative range, enter the inertial range from below and manifest themselves as intermittent fluctuations at small scales, $u(r)>u_L$.\remark{da $u_{\lchar}\approx u_L$ nun mal die großen fluktuationen sind, und die kleinen $u_\l$ sich erst noch durch die kaskade entwickeln müssen, müssen gemessene kleine fluktuationen $u_\l>u_L$ aus einem prozess hervorgegangen sein, der fluktuationen verstärkt hat. denkbar wäre eine überlagerung (=zusammenschluss) von gleichgerichteten kleinen strukturen, entspräche einer inversen kaskade. oder eben ein stau bei der dissipation, weil die dissipative range bei größer werdendem $\Rey$ kleiner wird, so dass diese verdichtung von kleinen wirbeln ein zusammenschluss von wirbeln begünstigt.} The underlying mechanism might be that the rate of energy dissipation is smaller than the rate of energy transfer into the dissipative range, provoking occasional aggregations of accumulated small turbulent structures to intermittent fluctuations, characterised by instances of {\it negative} energy transfer rates  which ensures that on average we retain the balance between energy dissipation and energy transfer. For high Reynolds numbers, implying a smaller dissipative range, this mechanism would intensify. In the picture of multifractal models we may say that less space-filling turbulent structures towards the end of the cascade create a bottleneck of energy transfer. However, these considerations are clearly speculative, but might offer new perspectives worth pursuing.\\

The above considerations were mainly based on the K62 model. The Markov process defined by drift and diffusion estimated from experimental data exhibits qualitatively the same characteristics as the K62 process, in particular the correspondence between $\Sm[\uc]<0$ and realisations $u(r)$ which increase towards smaller scales, as discussed in sections \ref{ss_turbulence_PRL} and \ref{ss_turbulence_asymp_extreme}. We therefore assume that the above considerations also apply qualitatively to real turbulent flows.\\
The failure of K62 in modelling the extreme value statistics of $u(r)$, demonstrated by the break-down of the K62 fluctuation theorem when applied to experimental data, however, suggests that the details of the K62 process need to be modified, cf. the discussion of modifications in the previous section. Such a modification may be to substitute the Gaussian noise with Poissonian noise, leading to the significantly more successful scaling exponents derived by She and Leveque. Another modification is to combine various cascade processes taking place in fractal subspaces as in multifractal models.\remark{Or, as a last example, modify the variance and mean of the Gaussian noise in the K62 process.}\\
An important modification is the inclusion of a skewness, i.e. breaking the invariance with respect to $u\mapsto-u$. The K62 model and related random cascade models do not include a skewness in $u$: The trajectories $u(r)$ generated by solving the respective stochastic differential equations will hence be negative and positive to equal shares, provided the initial distribution $p_L(u_L)$ is symmetric. This symmetry corresponds to reversible processes in the turbulent cascade, as a balanced occurence of positive and negative $u(r)$ imply that instances of positive energy transfer are as frequent as instances of negative energy transfer.\\
In contrast, the stochastic process defined by drift and diffusion obtained from experimental data does exhibit a skewness. Indeed, restricting the realisations $\uc$ to a positive production of associated entropy,\linebreak $\Sm[\uc]>0$, the asymptotic analysis of section \ref{ss_turbulence_asymp_extreme} revealed a slight unbalance in favour of {\it negative}, decreasing $u(r)$, relating negative skewness with a decay of spatial velocity fluctuations, in compliance with the average tendency. Realisations $\uc$ giving rise to a consumption of entropy, $\Sm[\uc]<0$, on the other hand, exhibit an increase of fluctuations towards small scales, and a mere balance between negative and positive $u(r)$. The similar analysis of experimentally measured realisations $u(r)$ presented in \cite{Nickelsen2013} (section \ref{ss_turbulence_PRL}) confirmed that decreasing $u(r)$ are characterised by $\Sm+\D S>0$ and increasing $u(r)$ by $\Sm+\D S<0$, where the Shannon entropy $\D S$ can be shown to have a negligible influence, as is the case for the K62 process.\\

In conclusion, a consumption of entropy $\Sm$ signifies fluctuations $u(r)$ that contribute to small-scale intermittency, bearing reference to instances of inverse cascades. The relation to transfer and dissipation of energy in the cascade and a negative skewness in the statistics of $u(r)$ is intriguing but remains unclear. If a model of turbulence accounts for the correct balance between direct and inverse cascades coexisting in a realistic turbulent flow, the fluctuation theorem resulting from the Markov representation of that model will be satisfied for experimental data.

\subsection{Possible future studies}
The results presented in this thesis offer a couple of starting points for future studies.\\

We have demonstrated the equivalence of Markov processes to many established models of turbulence. Besides this theoretical legitimation of using Markov processes to approach turbulence, we have also given a survey of experimental evidence that the turbulent cascade is indeed Markovian. The natural next step is, instead of finding the Markovian counterpart of {\it existing} models of turbulence, to use the gathered knowledge on the Markov description of turbulence and construct a theoretical model on the very level of Markov processes. A starting point may be exact relations of velocity increments as the K\'{a}rm\'{a}n-Howarth-Kolmogorov relation (\ref{eq:FDT_KHK}), or to specify the Markov representation of multifractal models (\ref{eq:D1D2_power_multifrac}) to consider subspaces with inverse cascades, or to adapt other existing Markov representations in order to better reproduce experimental results.\\
The aim of such efforts, instead finding the amplitudes and exponents of scaling laws, would be to find a form of drift and diffusion coefficients, and even a jump density, to capture the universal features of fully developed turbulence. In doing so, the comparison with the presented forms of MPs in chapter \ref{s_MAR}, the convenient experimental verification using fluctuation theorems demonstrated in chapter \ref{s_turbulence_FT} and the asymptotic analysis discussed in chapter (\ref{s_turbulence_asymp}) may be of assistance.\\

Apart from being a test equation for possible drift and diffusion coefficients (and even jump densities), the fluctuation theorem is for itself an intriguing object to focus on. The implications of scale reversal to the direction of the energy cascade, for instance, is an open question, which also leaves the interpretation of the associated entropy production to be of rather speculative nature.\\
Instead of scale reversal, other involutions are possible, such as a reversal of velocity direction, $u\mapsto-u$, bearing reference to time-reversal. This involution is of interest, as it reverses the skewness of velocity increments, and along with the skewness also the energy transfer in the turbulent cascade. The entropy in the fluctuation theorem building on this involution therefore directly targets the irreversibility of energy transfer. Unfortunately, or interestingly, the linear term in the diffusion coefficient, which proved to be the important term of the skewness, provokes a divergent behaviour of the entropy, spoiling the convergence of its fluctuation theorem. The reason for this divergence could not be clarified satisfactorily.\\
Other courses of action may be a fluctuation theorem in two dimensions. A promising starting point is the Markov analysis in $u(r)$ and $\eps(r)$ by Renner et al., which we discussed at the end of section \ref{ss_MAR_estimateD1D2}. Or two sources of noise acting on different scales as in (\ref{eq:D1D2_JFM_SDE_two-xi}). Both could allow an analysis of non-equilibrium steady states in the very dynamics of $u(r)$ or $\eps(r)$, involving the housekeeping entropy production to maintain the steady states and the non-negative excess entropy for transitions between steady states. These two entropy productions and the total entropy production obey each a separate fluctuation theorem, implying three faces of the second law, as discussed at the end of section \ref{ss_td-interpration_FTs}.\\

An intricate point in the Markov analysis of experimental data is the justification to truncate the Kramers-Moyal expansion after the second term and obtain a Fokker-Planck equation, as it relies on proving that exactly $\Dfr(u,r)\equiv0$. In fact, due to experimental and statistical limitations, it is only possible to show $\Dfr(u,r)\approx0$. Accordingly, instead of asking {\it how small} $\Dfr(u,r)$ is, the more appropriate question is to ask {\it how likely} is $\Dfr(u,r)\equiv0$. A Baysian analysis of the hypothesis ``$\Dfr(u,r)\equiv0$'' is promising to tackle that question.\\

The analysis of optimal fluctuations $u(r)$ giving rise to extreme entropy productions given in section \ref{ss_turbulence_asymp_tails} raises the question how the skewness of $u(r)$ is respected in the Markov representation. The Euler-Lagrange equation for other Markov representations than the K62 process and the experimentally extracted Markov process should therefore be interesting.\\
To allow a more profound analysis, the fluctuation determinant for multiplicative noise, determined for additive noise in \cite{Nickelsen2011}, would account for the vicinity of the optimal fluctuations $u(r)$.\\
A computation of the fluctuation determinant for multiplicative noise which determines the pre-exponential factor would also improve the asymptotics of $p(u,r)$, as discussed in section \ref{ss_turbulence_asymp_tails}, significantly.\\

Other open issues are to determine the jump density of the log-Poisson model, (\ref{eq:D1D2_bK62_SL_expo_Psik}), and to establish the explicit corresponds between multifractal models and the power series form of drift and diffusion coefficients, (\ref{eq:D1D2_power_multifrac}).

\chapter*{Conclusions}
\addcontentsline{toc}{chapter}{Conclusions}
Markov processes are well suited to capture both, thermodynamic processes of nanoscopic systems under non-equilibrium conditions, and cascade processes in fully developed turbulence. In contrast to macroscopic systems which are dominated by the {\it average} tendency of their microscopic dynamics, the properties of nanoscopic systems are characterised by {\it rare} stochastic realisations of their dynamics.\\
In closing this thesis, we relate the main statements of stochastic thermodynamics with the main results discussed for fully developed turbulence.\\

A paradigm to study stochastic thermodynamics is the Brownian motion of tiny particles suspended in a fluid. But many other systems are well described in the framework of stochastic thermodynamics. Closely related to Brownian motion are processes in biological cells, in particular the functioning of molecular motors. Benefiting from the understanding of molecular motors, the engineering of nanoscopic devices is on the rise, revealing an intimate relation between information and entropy. Stochastic thermodynamics is well suited to define the various working principles of a nanoscopic device. Other examples for applications of stochastic thermodynamics are active particles, nanoscale measurements and many more.\\
Regarding fully developed turbulence, a variety of established theories find a representation as a Markov process. A central role takes the K62 theory by Kolmogorov. Altering and augmenting the Markov representation of the K62 model leads to extended self-similarity, random cascade models, models resulting from field theoretic approaches and multifractal models. In addition, it is possible to capture the features of a turbulent flow by extracting a Markov process directly from experimental data.\\

The intuitive formulation of Brownian motion in terms of stochastic differential equations allows to study the effects of inhomogeneous media and temperature gradients.\\
The unified representation of the various established theories of fully developed turbulence as Markov processes is convenient for a contrasting juxtaposition of these models. In addition, combining features of models and incorporating refinements to overcome inconsistencies become feasible tasks.\\

An advantage of stochastic thermodynamics is its applicability to single realisations of the system under consideration. Notions of heat and work allow a formulation of the first law on the level of individual trajectories. The second law, being an inequality for entropy production, are stated more precisely as fluctuation theorems, being equalities. Both the first law and the fluctuation theorems hold arbitrarily far from equilibrium.\\
The Markov representation of models of turbulence address individual probes of the spatial structures of a turbulent flow, and, vice versa, allow to generate artificial data that carry the exact properties specified by the model under consideration. Formally, the Markov representation associates production of entropy to individual probes of spatial structures which in turn satisfy a fluctuation theorem. Validating the fluctuation theorem for extracted Markov processes from experimental data is an on-the-fly criterion for the quality of the extraction.\\

The entropy production of individual trajectories of nanoscopic systems is not restricted by the second law, only the average entropy production is bound to be positive. The rare individual trajectories that {\it consume} entropy are contrary to the average tendency of the dynamics and as such of particular interest. Molecular motors, for instance, are capable of rectifying thermal noise to perform useful work. Possible implications for nanoscopic devices are intriguing.\\
The average tendency of spatial velocity fluctuations in fully developed turbulence is to diminish towards small scales. This decrease of fluctuations is associated with a break-up of turbulent structures and a production of entropy. Despite the rather macroscopic dimensions of a turbulent flow, small-scale velocity fluctuations, exceeding in magnitude the large-scale fluctuations and associated with a consumption of entropy, are a predominant phenomenon in a turbulent flow, a phenomenon which is known as small-scale intermittency. The Markov representation suggests that these intermittent small-scale fluctuations result from an aggregation of turbulent structures.\\

Fluctuation theorems are valid arbitrarily far from equilibrium, but nevertheless involve equilibrium state variables. This connection allows the recovery of equilibrium information from non-equilibrium measurements. A prominent application that exploits this property of fluctuation theorems is the determination of free-energy profiles from force-spectroscopy of single biomolecules.\\
Fluctuation theorems are prone to the balance between entropy producing and entropy consuming fluctuations. In fully developed turbulence, velocity fluctuations associated with entropy consumption are responsible for small-scale intermittency. Accordingly, fluctuation theorem arising from the Markov representation of a model of turbulence will only hold for experimental measurements if this model precisely assesses the occurrence of intermittent small-scale fluctuations. The K62 theory is an example where the fluctuation theorem fails an experimental test, whereas the fluctuation theorem for the Markov process extracted from experimental measurements holds for these measurements. The latter finding validates that continuous Markov processes are well suited to capture small-scale intermittency, calling for a meaningful Markov theory of fully developed turbulence.\\

A reliable recovery of equilibrium information from non-equilibrium measurements is dependent on capturing the statistics of entropy consuming events. Typically, these events are also the rare events of the non-equilibrium process, hampering the recovery of equilibrium information. The path integral formulation of continuous Markov process is suitable to develop an asymptotic method which allows to examine these rare events. The asymptotic method also quantifies the probability of the rare event, improving the recovery of equilibrium information from non-equilibrium measurements.\\
In the context of Markov representations of fully developed turbulence, the asymptotic method can be used to determine the probability of extreme intermittent small-scale fluctuations of velocity and asymptotically extrapolate the scaling exponents.\\

Overall, the Markov representation of cascade processes constitutes a consistent approach to fully developed turbulence. The Markov processes representing the two Kolmogorov theories arise naturally from applying the Markov assumption to the cascade of turbulent structures. The results of the developing field of stochastic thermodynamics provide valuable tools for the analysis of the Markov representations. Accordingly, the Markov analysis of fully developed turbulence is a promising endeavour, from which new insights regarding the mechanism small-scale intermittency and the irreversibility of energy transfer can be expected, and therefore deserves further study.

\cleardoublepage
\pagebreak
\bibliographystyle{mybst_arxiv}
\bibliography{dissrefs_arxiv}

\begin{appendix}

\newpage
\quad
\cleardoublepage
\chapter{Technical details of continuous MPs}
\cleardoublepage
\section{\Ito~calculus} \label{AA_Itocalc}
The material presented in this chapter is adapted from part 4 of the book by Gardiner \cite{Gardiner2009}.\\
In this appendix, we provide a quick introduction to the calculus of stochastic differential equations (SDEs). The simplest form of a SDE is the Wiener process $W(t)$ which is a sequence of Gaussian random number with zero mean and variance $t$. As for infinitesimal $t$ also the variance is infinitesimal small, the Wiener process is continuous but can be shown to be non-differentiable. On the other hand, integration of a random function $\xi(t)$ with zero mean and correlation function $\lla\xi(t-t')\rra=\d(t-t')$ reproduces the Wiener process $W(t)$. In that sense, the Gaussian white noise $\xi(t)$ is the non-existing derivative of $W(t)$,
\begin{align}
	\dd W(t) = W(t+\dd t) - W(t) = \xi(t)\dd t \;.
\end{align}
Despite this pathology, $\xi(t)$ can be meaningfully combined with differential equations to form SDEs, constituting in many cases a simpler approach to stochastic processes than equivalent probabilistic descriptions. However, we have to keep in mind that $\xi(t)$ is only rigorously defined under an integral, very much like the $\d$-function.\\

The general form of a SDE is
\begin{align} \label{eq:A1_generalSDE}
	\ddx(t) = f\big(x(t),t\big) + g\big(x(t),t\big)\,\xi(t) \sep x(t=t_0)=x_0 \;.
\end{align}
The solution of this SDE is the stochastic variable $x(t)$ which can in principle be obtained by integrating
\begin{align}
	x(t) = x_0 + \int_{t_0}^{t} f\big(x(\t),\t\big) \di\t + \int_{t_0}^{t} g\big(x(\t),\t\big) \di W(\t) \;.
\end{align}
As the integrands are stochastic, the occurring {\it stochastic integrals} are not of Riemann type. For Riemann integrals, applying lower sum, upper sum, trapezoidal rule or other suitable discretisations, the value of the integral will always be the same in the continuous limit.\\
This is not the case for stochastic integrals, with the consequence that SDEs are only well defined if along with the SDE also the rule of discretisation is specified. The mathematically simplest rule is the {\it pre-point} rule, where the integrand is evaluated at the beginning of each discretisation interval. The SDE is then called {\it \Ito~SDE}. The drawback with the \Ito~convention is that the rules of calculus need to be modified. Ordinary calculus can be used if the {\it mid-point} rule is used, for which the integrand is evaluated at the centre of each discretisation interval. In this case, the SDE is referred to as {\it Stratonovich SDE}.\footnote{More specific, the Stratonovich stochastic integral takes the average of beginning and end of the discretisation intervals, and is therefore only {\it related} to the mid-point rule (\cite{Gardiner2009} p.\,96). In the limit of infinitesimal discretisation intervals, however, this distinction becomes irrelevant.} The various discretisation rules will be formalised in appendix (\ref{AA_SDE2FPE}).\\

The modified calculus arising from the \Ito~interpretation of a SDE is called {\it \Ito~calculus}. Building on the \Ito~convention of stochastic integrals, it can be shown that $\dd W^2=\dd t$\remark{displacement of a Brownian particle is not proportional to the elapsed time, but rather to its square root} and $\dd W^{n+2}=0$ for $n\geq1$, which is the starting point of \Ito~calculus.\\
With regard to SDEs, the most important implication of \Ito~calculus is the {\it \Ito~lemma}. For a stochastic variable $x(t)$, the \Ito~lemma states that the total differential of a function $f\big(x(t),t\big)$ reads
\begin{align} \label{eq:A1_Ito-lemma}
	\dd f\big(x(t),t\big) = f'\big(x(t),t\big)\dd x + \tfrac{1}{2}f''\big(x(t),t\big)\dd W(t)^2 + \dot f\big(x(t),t\big)\dd t \;.
\end{align}
Since $\dd W(t)=\sqrt{\dd t}$, the \Ito~lemma collects all terms of order $\dd t$.\\
We exemplify the use of \Ito's~lemma (\ref{eq:A1_Ito-lemma}) by changing the variable $x(t)$ to $y\big(x(t)\big)$ and get the formula
\begin{align}
	\dd y(x) &= y'(x)\dd x + \tfrac{1}{2}y''(x)\dd x^2 \nn
	&= y'(x)\big[f(x,t)\dd t + g(x,t)\dd W(t)\big] + \tfrac{1}{2}y''(x)g(x,t)^2\dd W(t)^2 \nn
	\Ra\;\dot y\big(x(t)\big) &= y'(x)\big[f(x,t) + g(x,t)\xi(t)\big] + \tfrac{1}{2}y''(x)g(x,t)^2 \;, \label{eq:A1_Ito-formula}
\end{align}
where we used \Ito's~lemma (\ref{eq:A1_Ito-lemma}) in the first line and substituted $\dd x(t)$ from the general form of the SDE in (\ref{eq:A1_generalSDE}) in the second line. This prescription is known as {\it \Ito's~formula} for change of variables and is often used to solve SDEs.\\

Let us consider {\it geometric Brownian motion} (GBM) with constant drift as an example. The SDE reads
\begin{align} \label{eq:A1_GBM_Ito}
	\dot x(t) = a(t)x(t) + \sqrt{2b(t)}x(t)\xi(t) \sep \text{It\'{o}} \;.
\end{align}
This SDE can be solved using the transformation $y(x)=\ln x$. 
We identify $f(x,t)=a(t)x$ and $g(x,t)=\sqrt{2b(t)}x$ and get for $y(t)$ the SDE
\begin{align}
	\dot y\big(x(t)\big) &= y'\big(x(t)\big)\big[a(t)x(t) + \sqrt{2b(t)}x(t)\xi(t)\big] + y''\big(x(t)\big)b(t)x(t)^2 \nn
	&= a(t) + \sqrt{2b(t)}\xi(t) - b(t) \nn
	\Ra\;\dot y(t) &= \big(a(t)\-b(t)\big) + \sqrt{2b(t)}\xi(t) \;. \label{eq:A1_GMB_Ito-formula}
\end{align}
Note that by making use of \Ito's~formula, we now have the drift \mbox{$a(t)\-b(t)$} with the noise induced component $-b(t)$.\\
The above SDE can readily be integrated,
\begin{align} \label{eq:A1_GBM_Ito_ysol}
	y(t) = y(t_0) + A(t) - B(t) + \sqrt{2b(t)}Z(t) \;,
\end{align}
where 
\begin{subequations}
  \begin{align}
		A(t) = \int_{t_0}^{t}a(t')\di t' \;,\\
		B(t) = \int_{t_0}^{t}b(t')\di t' \;,\\
	\end{align}
\end{subequations}
and
\begin{align} \label{eq:A1_Z}
	Z(t) = \int_{t_0}^{t}\xi(t')\di t'
\end{align}
is a Gaussian random variable with zero mean and variance $t$, as we see from
\begin{subequations} \label{eq:A1_Z_properties}
  \begin{align}
		\lla Z(t) \rra &= \int\limits_{t_0}^{t}\lla\xi(t')\rra\di t' = 0 \;,\\
		\lla Z(t)^2 \rra &= \int\limits_{t_0}^{t}\!\!\int\limits_{t_0}^{t}\lla\xi(t')\xi(t'')\rra\di t'\dd t'' \nn
		&= \int\limits_{t_0}^{t}\!\!\int\limits_{t_0}^{t}\d(t'-t'')\di t'\dd t'' = \int\limits_{t_0}^{t}\dd t' =  t \;.
	\end{align}
\end{subequations}
Note that $\lla Z(\dd t)^2 \rra=\dd t$ in retrospect confirms $\dd W(t)^2=\dd t$ as used above. The random variable $y(t)$ is therefore also a Gaussian random variable with mean $A(t)\-B(t)$ and variance $2B(t)$. Accordingly, $x(t)$ is a log-normal random variable
\begin{align} \label{eq:A1_GBM_Ito_xsol}
	x(t) = x_0\exp\big[A(t)-B(t)+\sqrt{2B(t)}Z(t)\big] \;.
\end{align}
Its PDF reads
\begin{align} \label{eq:A1_GBM_Ito_psol}
	p(x,t) = \frac{1}{x\sqrt{4\pi B(t)}}\exp\bigg[-\frac{\big(\ln(x/x_0)-(A(t)\-B(t))\big)^2}{4B(t)}\bigg] \;.
\end{align}
The moments of a log-normal distribution all exist and read in this case
\begin{align} \label{eq:A1_GBM_Ito_momsol}
	\lla x(t)^n\rra = \lla x_0^n\rra\exp\Big[ \big(A(t)\-B(t)\big)\,n + B(t)\,n^2 \Big] \;.
\end{align}
In standard GBM, the coefficients $a(r)$ and $b(r)$ are constants, and\linebreak $A(t)-B(t)$ becomes $(a-b)(t-t_0)$. The ratio $g\dfns a/b$ is known as exponential growth constant\footnote{also, the exponential growth rate $g\dfns (a/b)/(t-t_0)$}, see \cite{Peters2013} for a discussion. Taking an arbitrary initial distribution $p_0(x_0)\neq\d(x_0)$, for $g>1$, all moments of $x(t)$ diverge exponentially, and for $g<1$, all moments decay exponentially. Hence, for $g<1$, the stationary solution is $\pst(x)=\d(x)$ for which all trajectories get stuck at $x=0$.\\
Note that interpretation of the SDE (\ref{eq:A1_GBM_Ito}) in Stratonovich convention would lead to the mean $A(t)$ instead of $A(t)\-B(t)$, as then ordinary calculus applies and the last term in the first line of (\ref{eq:A1_GMB_Ito-formula}) does not contribute.

\cleardoublepage
\section{Transformation of time in a SDE} \label{AA_timetrafo}
We want to transform time in the SDE
\begin{align} \label{eq:A2_SDE}
	\dot x(t) = f(x,t) + g(x,t)\xi(t) \;.
\end{align}
Integration from $t$ to $t+\Dt$, where $\Dt$ is a infinitesimal time-step, yields
\begin{align} \label{eq:A2_SDE_int}
	x(t+\Dt) &= x(t) + f_\a(x,t)\Dt + \int_{t}^{t+\Dt}g(x,t)\xi(t)\di t \nn
	&= x(t) + f_\a(x,t)\Dt + g_\a(x,t)Z(\Dt) \;.
\end{align}
Here we have used the mean-value theorem for integration, where the index $\a$ indicates evaluation at time $t+\a\Dt$ with $0\geq\a\geq1$, and we have used the Gaussian random variable $Z(\Dt)$ defined in (\ref{eq:A1_Z}) with zero mean and variance $\Dt$ from (\ref{eq:A1_Z_properties}).\\
In fact, the mean-value theorem theorem is not defined for a stochastic integral. The consequence that we applied it anyway is that in the limit $\Dt\to0$ the value chosen for $\a$ does still matter. The choice of $\a$ is equivalent to the choice of discretisation of the SDE, i.e. $\a=0$ for pre-point and $\a=1/2$ for mid-point. Indeed, we show in appendix \ref{AA_SDE2FPE} that the PDF $p(x,t)$ obeys the Fokker-Planck equation (FPE)
\begin{align} \label{eq:A2_FPE}
	\dot p(x,t) = \big[-\pt_x\big(f(x,t)+\a g'(x,t)g(x,t)\big) + \tfrac{1}{2}\pt_x^2g(x,t)^2 \big] p(x,t) \;,
\end{align}
where the value of $\a$ still enters.\\

\renewcommand{\tx}{\tilde{x}}%
We return to the question of time transformation. \\
Suppose we want transform time from $t$ to $s=a(t)$. For convenience, we denote $b(s)\dfns a^{-1}(s)$. Using $\dd s = \dot a(t)\dd t$, that is $\pt_t=\dot a\big(b(s)\big)\pt_s$, the SDE reads
\begin{align} \label{eq:A2_SDE_trans1}
	\dot a\big(b(s)\big)\dot\tx(s) = \tf(\tx,s) + \tg(\tx,s)\xi\big(b(s)\big) \;,
\end{align}
where $\tx(s)\dfns x(b(s))$, $\tf(\tx,s)\dfns f\big(\tx,b(s)\big)$ and $\tg(\tx,s)\dfns g\big(\tx,b(s)\big)$ are now evaluated for the new time $s$.\\
The noise $\xi\big(b(s)\big)$ needs special attention if we require that the properties of the stochastic integral $\tZ(\D s)$ of new noise $\txi(s)$ are of the form (\ref{eq:A1_Z_properties}), that is in particular $\big\langle \tZ(\D s)^2\big\rangle=\D s$. This requirement is met if we use
\begin{align} \label{eq:A2_SDE_trans_xi}
	\txi(s) = \frac{\xi\big(b(s)\big)}{\sqrt{\dot a\big(b(s)\big)}} \;,
\end{align}
since then we have consistently
\begin{align}
	\dd s &= \big(\txi(s)\dd s\big)^2 = \frac{\big[\,\xi\big(b(s)\big)\dd s\,\big]^2}{\dot a\big(b(s)\big)} \nn
	&= \frac{\big[\,\xi(t)\dot a(t)\dd t\,\big]^2}{\dot a(t)} = \dot a(t)\big[\,\xi(t)\dd t\,\big]^2 = \dot a(t) \dd t \;, \label{eq:A2_SDE_trans_Z_elegant}
\end{align}
where we used $[\xi(t)\dd t]^2\eq\dd t$.\\
With an equivalent\remark{equivalent, weil wir in obiger begründung eigentlich auch integriert haben, aber ein infinitesimales stück $\d\ t$ wodurch das integral zu ner multiplikation mit $\dd t$ wird.} but more tedious calculation we can also explicitly show that
\begin{align}
	\big\langle \tZ(\D s)^2 \big\rangle &= \int\limits_{s}^{s+\D s}\int\limits_{s}^{s+\D s}\lla\txi(s')\txi(s'')\rra\di s'\dd s'' \nn
	&= \int\limits_{s}^{s+\D s}\int\limits_{s}^{s+\D s}\frac{1}{\sqrt{\dot a\big(b(s')\big)\,\dot a\big(b(s'')\big)}} \lla\xi\big(b(s')\big)\xi\big(b(s'')\big)\rra\di s'\dd s'' \nn
	&= \int\limits_{b(s)}^{b(s+\D s)}\int\limits_{b(s)}^{b(s+\D s)}\frac{\dot a(t')\,\dot a(t'')}{\sqrt{\dot a(t')\,\dot a(t'')}}\lla\xi(t')\xi(t'')\rra\di t'\dd t'' \nn
	&= \int\limits_{b(s)}^{b(s+\D s)}\int\limits_{b(s)}^{b(s+\D s)}\dot a(t')\,\d(t'-t'')\di t'\dd t'' \nn
	&= \int\limits_{b(s)}^{b(s+\D s)}\dot a(t'')\dd t'' = a\big(b(s\+\D s)\big) - a\big(b(s)\big) = \D s \;. \label{eq:A2_SDE_trans_Z}
\end{align}
\remark{es sollte doch auch möglich sein zu zeigen, dass dann $\txi(s)$ wieder normal $\d$-korreliert ist...}Note that instead of $\big[\xi(t)\dd t\big]^2=\dd t$ in the derivation (\ref{eq:A2_SDE_trans_Z_elegant}), we used here that $\langle\xi(t)\xi(t')\rangle=\d(t-t')$.\\
Substitution of (\ref{eq:A2_SDE_trans_xi}) into (\ref{eq:A2_SDE}) yields the transformed SDE 
\begin{align} \label{eq:A2_SDE_trans2}
	\dot\tx(s) = \frac{\tf(\tx,s)}{\dot a\big(b(s)\big)} + \frac{\tg(\tx,s)}{\sqrt{\dot a\big(b(s)\big)}}\,\txi(s) \;.
\end{align}
We reason from $\big\langle\tZ(\D s)^2\big\rangle=\D s$ that also $\big\langle\txi(s)\txi(s')\big\rangle=\d(s-s')$ and obtain by comparison with (\ref{eq:A2_FPE}) the equivalent FPE 
\begin{align} \label{eq:A2_FPE_trans}
	\dot{\tilde p}(x,s) = \bigg[-\pt_x\bigg(\frac{\tf(x,s)}{\dot a\big(b(s)\big)}+\a\,\frac{\tg'(x,s)\tg(x,s)}{\dot a\big(b(s)\big)}\bigg) + \frac{1}{2}\pt_x^2\,\frac{\tg(x,s)^2}{\dot a\big(b(s)\big)} \bigg]\,\tilde p(x,s) 
\end{align}
for the PDF $\tilde p(x,s)\dfns p(x,b(s))$. This FPE results also from applying the same time transformation directly to (\ref{eq:A2_FPE}) as it should be.\remark{siehe diss p.1-6}\\

Alternatively, we could also require that in the SDE with transformed time, (\ref{eq:A2_SDE_trans1}), the noise must have the correlation function
\begin{align} \label{eq:A2_SDE_trans_xi_corr}
	\lla\,\xi\big(b(s)\big)\,\xi\big(b(s')\big)\,\rra = \sqrt{\dot a\big(b(s)\big)}\,\d(s-s')
\end{align}
in order to have a Gaussian random variable $Z(\D s)$ with variance $\D s$ in the integrated SDE, cf. (\ref{eq:A2_SDE_int}), and to retain the equivalence to a FPE as between (\ref{eq:A2_SDE}) and (\ref{eq:A2_FPE}).\remark{das machen immer renner und co (nämlich $-r\pt_r$), schreiben aber nicht wie dann die korrelationsfunktion ihres rauschens aussieht (nämlich $-r\d(r-r')$), siehe auch diss p.14ab}\\

As an example, let us again consider GBM with constant drift as in (\ref{eq:A1_GBM_Ito}), but now with the coefficients $a(t)=a/t$ and $b(t)=b/t$. The SDE reads
\begin{align} \label{eq:A2_GBM}
	\dot x(t) = \frac{a}{t}\,x(t) + \sqrt{\frac{2b}{t}}\,x(t)\,\xi(t) \sep \text{$\a$-point} \;.
\end{align}
with $\d$-correlated $\xi(t)$. By noting $f(x,t)=a/t\,x$ and $g(x,t)=\sqrt{2b/t}\,x$, the equivalent FPE from (\ref{eq:A2_FPE}) follows as
\begin{align} \label{eq:A2_FPE_GBM1}
	\dot p(x,t) = \Big[-\pt_x\Big(\mfrac{a}{t}\,x+\a \mfrac{2b}{t}\,x\Big) + \pt_x^2\mfrac{b}{t}\,x^2\Big] p(x,t) \;,
\end{align}
or, by multiplying with $t$,
\begin{align} \label{eq:A2_FPE_GBM2}
	t\,\dot p(x,t) = \big[-(a +2\a b)\pt_x\,x + b\pt_x^2\,x^2\big] p(x,t) \;.
\end{align}
Transformation to logarithmic time, i.e. $s=a(t)=\ln t$, $t=b(s)=\exp s$ and $\dot a(t)=1/t$, yields the FPE
\begin{align} \label{eq:A2_FPE_GBM_trans}
	\dot{\tilde p}(x,s) = \big[-(a +2\a b)\pt_x\,x + b\pt_x^2\,x^2\big] \tilde p(x,s)
\end{align}
with now time independent coefficients.\\
The equivalent SDE is accordingly 
\begin{align} \label{eq:A2_GBM_trans}
	\dot \tx(s) = a\,x(t) + \sqrt{2b}\,\tx(s)\,\txi(s) \sep \text{$\a$-point} \;,
\end{align}
where $\txi(s)$ is again $\d$-correlated.\\
On the other hand, if we multiply the SDE with $t$, we obtain 
\begin{align} \label{eq:A2_GBM2}
	t\,\dot x(t) = a\,x(t) + \sqrt{2b}\,x(t)\,\sqrt{t}\xi(t) \sep \text{$\a$-point} \;,
\end{align}
and see that transformation to $s=\ln t$ does not only involve $t\pt_t=\pt_s$ but also $\txi(s)=\sqrt{b(s)}\xi(b(s))=\sqrt{t}\xi(t)$ in order to reproduce (\ref{eq:A2_GBM_trans}) and $\big\langle\txi(s)\txi(s')\big\rangle=\d(s-s')$.

\cleardoublepage
\section{Derivation of FPE from SDE} \label{AA_SDE2FPE}
The material presented in this chapter is adapted from the article by Lau and Lubensky \cite{LauLubensky07PRE} and part 4 of the book by Gardiner \cite{Gardiner2009}.\\
MPs $X(t)$ in one continuous degree of freedom can be modelled by stochastic differential equations (SDEs) of the form
\begin{align} \label{eq:A3_SDE_noalpha}
	\ddx_t=f(x_t,t)+g(x_t,t)\,\xi(t) \sep x(t=t_0)=x_0
\end{align}
where $x_t=x(t)$ and with the Gaussian noise $\xi(t)$ defined by
\begin{align} \label{eq:A3_SDE_xi}
	\lla\xi(t)\rra=0 \sep \lla\xi(t)\xi(t')\rra=\d(t-t') \;.
\end{align}
The noise $\xi(t)$ itself is discontinuous, but its time-integral
\begin{align}
	Z(\Dt) = \int_{t}^{t+\Dt} \xi(t') \di t'
\end{align}
is continuous in $\Dt$. The random variable $Z(\Dt)$ is again Gaussian distributed with mean and variance given by
\begin{subequations} \label{eq:A3_Z-is-gauss}
  \begin{align}
		\lla Z(\Dt) \rra &= \int_t^{t+\Dt}\lla\eta(t')\rra\di t' = 0 \;, \label{eq:A3_Z-is-gauss_mean} \\
		\lla Z(\Dt)^2 \rra &= \int_t^{t+\Dt}\int_t^{t+\Dt}\lla\xi(t'')\xi(t')\rra\di t'' \di t' \nn
	&= \int_t^{t+\Dt}\int_t^{t+\Dt'}\delta(t''-t')\di t''\di t' = \int_t^{t+\Dt}\di t' = \Dt \;. \label{eq:A3_Z-is-gauss_variance}
	\end{align}
\end{subequations}

Attempting an integration of the SDE (\ref{eq:A3_SDE_noalpha}) results into
\begin{subequations}
  \begin{align}
		\int_{t}^{t+\Dt} \ddx(t') \di t' &= x(t+\Dt)-x(t) \nn
		&= \int_{t}^{t+\Dt}f(x_{t'},t')\di t' + \int_{t}^{t+\Dt}g(x_{t'},t')\,\xi(t') \di t' \label{eq:A3_SDEint_pre} \\
		&\stackrel{?}{\simeq} f(x_{t},t)\Dt + g(x_{t},t)\,Z(\Dt)\;,  \label{eq:A3_SDEint_noalpha}
	\end{align}
\end{subequations}
where we encounter in the second line {\it stochastic integrals}, and the third line is preliminary, as we will see now.\\ Consider the second stochastic integral involving $\xi(t)$,
\begin{align} \label{eq:A3_def_J_stochint}
  \JJ(t,\Dt) = \int_t^{t+\Dt}g\big(x(t'),t'\big)\,\xi(t')\di t' \;.
\end{align}
We apply the mean-value theorem and write
\begin{align} \label{eq:A3_J_MVT}
  \JJ(t,\Dt) = g\big(x(t_\a),t_\a\big)\int_t^{t+\Dt}\xi(t')\di t' \;,
\end{align}
where $t_\a=t+\a\Dt$ with $0\leq\a\leq1$ and $x(t_\a)\simeq x(t)+\a\Dx$ with $\Dx = x(t+\Dt)-x(t)$.\\
In a Riemann integral\remark{dann wenn integrand stetig, also eigentlich ist das integral mit $f$ ein Riemann integral}, the choice of $\a$ has no effect as soon as the limit $\Dt\to0$ is taken. But as $\xi(t)$ is discontinuous, we can not omit $\a$ from the integration of the SDE arguing that the effect of choosing $\a$ will be negligible for a sufficient small value of $\Dt$. We therefore refine the integration of the SDE given in (\ref{eq:A3_SDEint_noalpha}) by writing
\begin{align}
	x(t+\Dt) \simeq x(t) + f(x_t+\a\Dx,t_\a)\Dt + g(x_t+\a\Dx,t_\a)\,Z(\Dt) \label{eq:A3_SDEint_alpha}
\end{align}
and will now explore the effect of $\a$ on the statistics of $x(t+\Dt)$.\\

First we develop an integration scheme.\\
We require that the integration scheme has to be in first order of $\Dt$, which is not yet accomplished in (\ref{eq:A3_SDEint_alpha}), since we know from (\ref{eq:A3_Z-is-gauss_variance}) that $Z(\Dt)=\OO(\sqrt{\Dt})$.\\
To get (\ref{eq:A3_SDEint_alpha}) up to order $\Dt$, we expand $f(x+\a\Dx,t_\a)$ and $g(x+\a\Dx,t_\a)$ in $\Dx$ to sufficient order and substitute recursively $\Dx$ from (\ref{eq:A3_SDEint_alpha}), 
\begin{subequations} \label{eq:A3_SDE_fg_expansions_alpha}
  \begin{align}
		\hspace*{-20pt}f(x_t\+\a\Dx,t_\a) &= f(x_t,t_\a) + \OO\big(\Dt^{\frac{1}{2}}\big) \\
		g(x_t\+\a\Dx,t_\a) &= g(x_t,t_\a) + g'(x_t,t_\a)\a\Dx + \OO(\Dt) \nn
		&= g(x_t,t_\a) + \a g'(x_t,t_\a)g(x_t,t_\a)Z(\Dt) + \OO\big(\!\Dt^{\frac{3}{2}}\big)\!\!\!
	\end{align}
\end{subequations}
where we used that $\Dx=\OO(\sqrt{\Dt})$. Going with these two expansions back into (\ref{eq:A3_SDEint_alpha}), we yield the desired numeric integration scheme of linear order in $\Dt$
\begin{align} \label{eq:A3_SDE_num_alpha}
  x(t\+\Dt) = x_t + f(x_t,t)\Dt + g(x_t,t)Z(\Dt) + \a g'(x_t,t)g(x_t,t)Z(\Dt)^2 \;.
\end{align}
Here we have dropped the index on $t_\a$, since the stochastic integrals are only stochastic in $x(t)$ and $\xi(t)$, and we assume that $f(x,t)$ and $g(x,t)$ are differentiable in $t$.\remark{im limes $\Dt\to0$ geht $f(t+\a\Dt)$ linear gegen $f(t)$ (und nicht sublinear).}\remark{which is known as the Milstein scheme for $\a=\frac{1}{2}$ and with spurious drift correction. see englisch wikipedia on Milstein method (or LauLubsensky, or google.books for KloedenPlaten), there is the therm $1/2gg'(Z^2\-\Dt)$ and the ``$-\!\Dt$'' creates the spurious drift correction), and generalises the common Euler-Maruyama scheme... (see Gardiner, eq(15.4.9) and case a) on p.411)}\\ 
As we see from the expansions in (\ref{eq:A3_SDE_fg_expansions_alpha}), the choice of $\a$ amounts to choosing the point of evaluation in discretising the SDE. That is to say, for instance, $\a=0$ corresponds to the pre-point rule and $\a=1/2$ to the mid-point rule. In other words, if the SDE (\ref{eq:A3_SDE_noalpha}) is defined as an \Ito~SDE, we choose in (\ref{eq:A3_SDE_num_alpha}) $\a=0$ to solve the SDE numerically, and for the Stratonovich interpretation we choose $\a=1/2$.\\

Having in place the integrated SDE (\ref{eq:A3_SDE_num_alpha}) in linear order of $\Dt$, we can derive from it the evolution equation for the PDF $p(x,t)$, known as Fokker-Planck equation (FPE). To this end, we consider an observable $A\big(x(t)\big)$. We fix $x(t)$ and take $A\big(x(t\+\Dt)\big)$ as a random variable due to the Gaussian variable $Z(\Dt)$. The average of $A\big(x(t\+\Dt)\big)$ is therefore conditioned on the value of $x(t)$ and can be written in terms of the conditional PDF $p(x\+\Dx,t\+\Dt|x,t)$,
\begin{align} \label{eq:A3_Acondavg0}
	\lla A\big(x(t\+\Dt)\big)\rra = \int p(x\+\Dx,t\+\Dt|x,t)A(x\+\Dx)  \di(x\+\Dx) \;.
\end{align}
On the other hand, we can expand $\lla A\big(x(t\+\Dt)\big)\rra$ in $\Dt$,
\begin{align} \label{eq:A3_Acondavg1}
	  \lla A\big(x(t\+\Dt)\big)\rra &\simeq \lla A\big(x(t)+\ddx(t)\Dt\big)\rra\nn 
	  &\simeq A\big(x(t)\big) + A'\big(x(t)\big)\lla\dot x(t)\Dt\rra \nn
		&\qquad+ \tfrac{1}{2} A''\big(x(t)\big)\lla\dot x(t)^2\Dt^2\rra +\OO(\Dt^3) \;.
\end{align}
For the two averages follow by substituting the integrated SDE (\ref{eq:A3_SDE_num_alpha}),
\begin{subequations}
  \begin{align}
		\lla\dot x(t)\Dt\rra &\simeq \lla x(t+\Dt) - x(t) \rra \nn
		&\simeq f(x_t,t)\Dt + g(x_t,t)\lla Z(\Dt)\rra + \a g'(x_t,t)g(x_t,t)\lla Z(\Dt)^2\rra \nn
		&= f(x_t,t)\Dt + \a g'(x_t,t)g(x_t,t)\Dt  \\
		\lla(\dot x(t)\Dt)^2\rra &\simeq g(x_t,t)^2\lla Z(\Dt)^2\rra \nn
		&= g(x_t,t)^2\Dt \;,
	\end{align}
\end{subequations}
where we kept only terms of linear order in $\Dt$.\\
Substituting these averages back into (\ref{eq:A3_Acondavg1}) yields
\begin{align} \label{eq:A3_Acondavg2}
		\lla A\big(x(t\+\Dt)\big)\rra &\simeq A\big(x(t)\big)+ A'\big(x(t)\big)\big[f(x_t,t) + \a g'(x_t,t)g(x_t,t)\big]\,\Dt \nn
		&\hspace{56pt} + \tfrac{1}{2}A''\big(x(t)\big)g(x_t,t)^2\,\Dt \;.
\end{align}
Note that for $\a=0$ the above formula also derives from \Ito's~formula in (\ref{eq:A1_Ito-formula}) for the change of variable from $x(t)$ to $A\big(x(t)\big)$.\\
The expansion (\ref{eq:A3_Acondavg2}), involving only the observable $A\big(x(t)\big)$, can be passed on to an equation for $p(x,t)$ by substituting (\ref{eq:A3_Acondavg0}) for the l.h.s., and taking the average with respect to $x$ at time $t$,
\begin{align}
	&\int A(x\+\Dx)\,p(x\+\Dx,t\+\Dt)  \di(x\+\Dx) - \int A(x)\,p(x,t) \di x \nn
	&\qquad = \Dt\cdot\int\Big[A'(x)\big[f(x_t,t) + \a g'(x_t,t)g(x_t,t)\big] \nn
	&\hspace{80pt} + \tfrac{1}{2}A''(x)g(x_t,t)^2\Big]\,p(x,t) \di x \;.
\end{align}
Integration by parts and discarding the boundary terms due to the normalisation condition of $p(x,t)$ yields
\begin{align}
	&\int A(x\+\Dx)\,p(x\+\Dx,t\+\Dt)  \di(x\+\Dx) - \int A(x)\,p(x,t) \di x \nn
	&\qquad = \Dt\cdot\int A(x)\,\Big[-\pt_x\big[f(x_t,t) + \a g'(x_t,t)g(x_t,t)\big] \nn
	&\hspace{120pt} + \tfrac{1}{2}\pt_x^2g(x_t,t)^2\Big]\,p(x,t) \di x \;.
\end{align}
Since $A(x)$ is arbitrary, and $(x\+\Dx)$ is just an integration variable, we can read off in differential form
\begin{align}
	&\frac{p(x,t\+\Dt) - p(x,t)}{\Dt} = \Big[-\pt_x\big[f(x_t,t) + \a g'(x_t,t)g(x_t,t)\big]  \nn
	&\hspace{160pt} + \tfrac{1}{2}\pt_x^2g(x_t,t)^2\Big]\,p(x,t)
\end{align}
which in the continuous limit $\Dt\to0$ finally becomes the FPE
\begin{align} \label{eq:A3_FPE_alpha}
	\hspace*{-7pt}\ddp(x,t) = \Big[\-\pt_x\big[f(x_t,t) \+ \a g'(x_t,t)g(x_t,t)\big] + \tfrac{1}{2}\pt_x^2g(x_t,t)^2\Big]\,p(x,t) \;.
\end{align}
Note that the FPE depends on $\a$.\\

The FPE needs to be completed by the initial PDF $p_0(x)$ for the time $t_0<t$ from which the values $x_0$ are sampled. The solution of the FPE is then the PDF $p(x,t)$ for $x$ at time $t$. For the choice $p_0(x)=\d(x-x_0)$ the solution of the FPE will be the conditional PDF $p(x,t|x_0,t_0)$, from which by
\begin{align} \label{eq:A3_int_condPDF}
	p(x,t) = \int p(x,t|x_0,t_0)\,p_0(x_0) \di x_0
\end{align}
the solution of the FPE with arbitrary initial PDF $p_0(x_0)$ can be obtained. In that sense, $p(x,t|x_0,t_0)$ is the Green's function of the FPE and obeys the same FPE as $p(x,t)$,
\begin{align} \label{eq:A3_FPE_alpha_cond}
	&\ddp(x,t|x_0,t_0) = \Big[-\pt_x\big[f(x_t,t) + \a g'(x_t,t)g(x_t,t)\big] \nn
	&\hspace{85pt}+ \tfrac{1}{2}\pt_x^2g(x_t,t)^2\Big]\,p(x,t|x_0,t_0) \;,
\end{align}
but without specifying an initial condition.\\

A set of values for $x$ that follows $p(x,t)$ can in principle be obtained by integrating the SDE (\ref{eq:A3_SDE_noalpha}) using the scheme (\ref{eq:A3_SDE_num_alpha}) for a certain value of $\a$ and in the limit $\Dt\to0$.\footnote{or, if possible, by solving the SDE explicitly using $\a$-calculus.} However, in numeric integrations, only finite $\Dt$ are possible, and using the integration scheme (\ref{eq:A3_SDE_num_alpha}) will remain an approximation. Higher order integrations schemes exist, but in the majority of stochastic processes the scheme (\ref{eq:A3_SDE_num_alpha}) is sufficient.\remark{We demonstrate in figure xxx the performance of the scheme (\ref{eq:A3_SDE_num_alpha}) for GBM with ... and choices for $\a$.
\begin{figure}[t] 
  \subfloat[][\figsubtxt{bla}]{\label{sf:num-scheme}
	\includegraphics[width=0.45\textwidth]{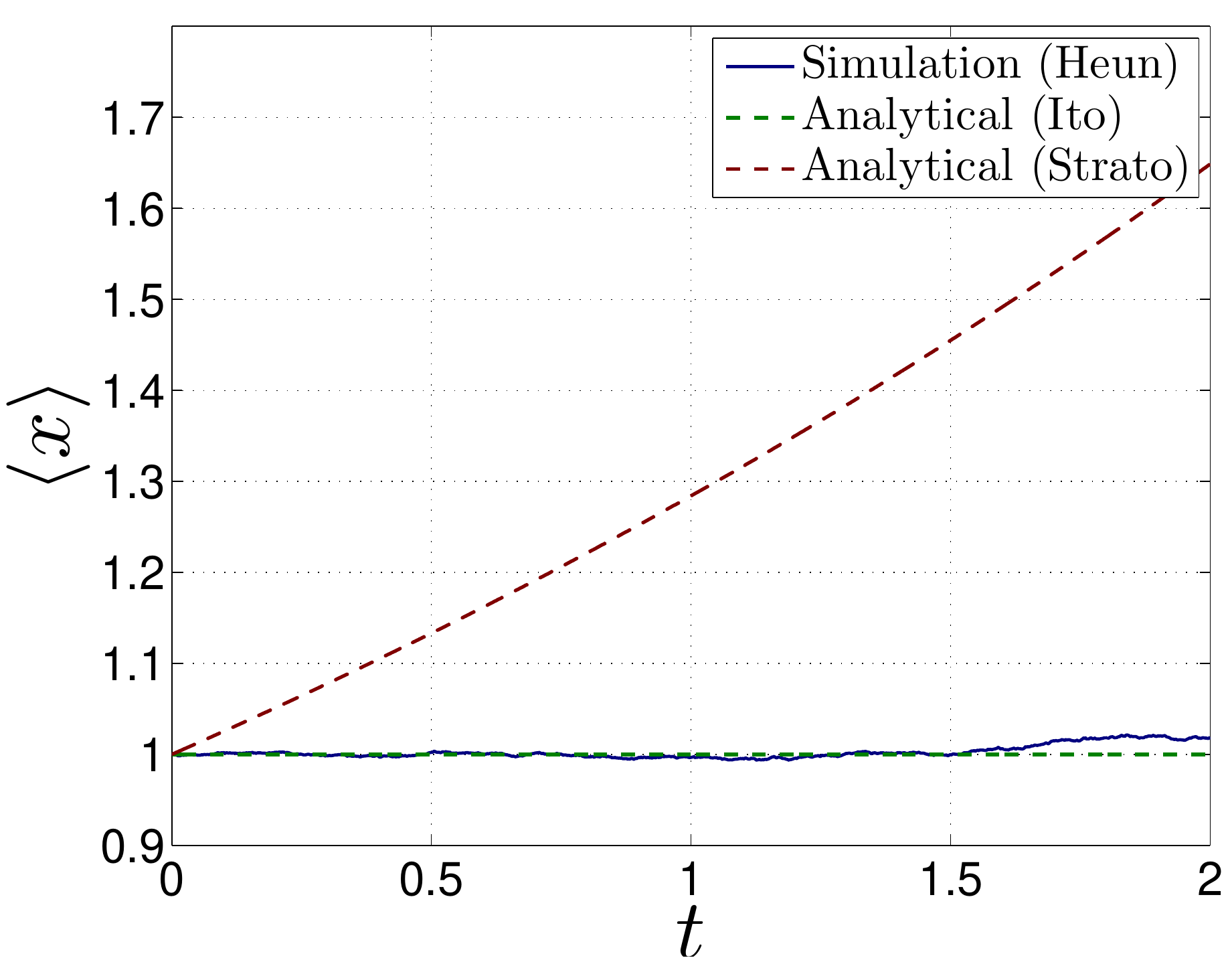}} \hfil
  \subfloat[][\figsubtxt{bla}]{\label{sf:num-scheme}
	\includegraphics[width=0.45\textwidth]{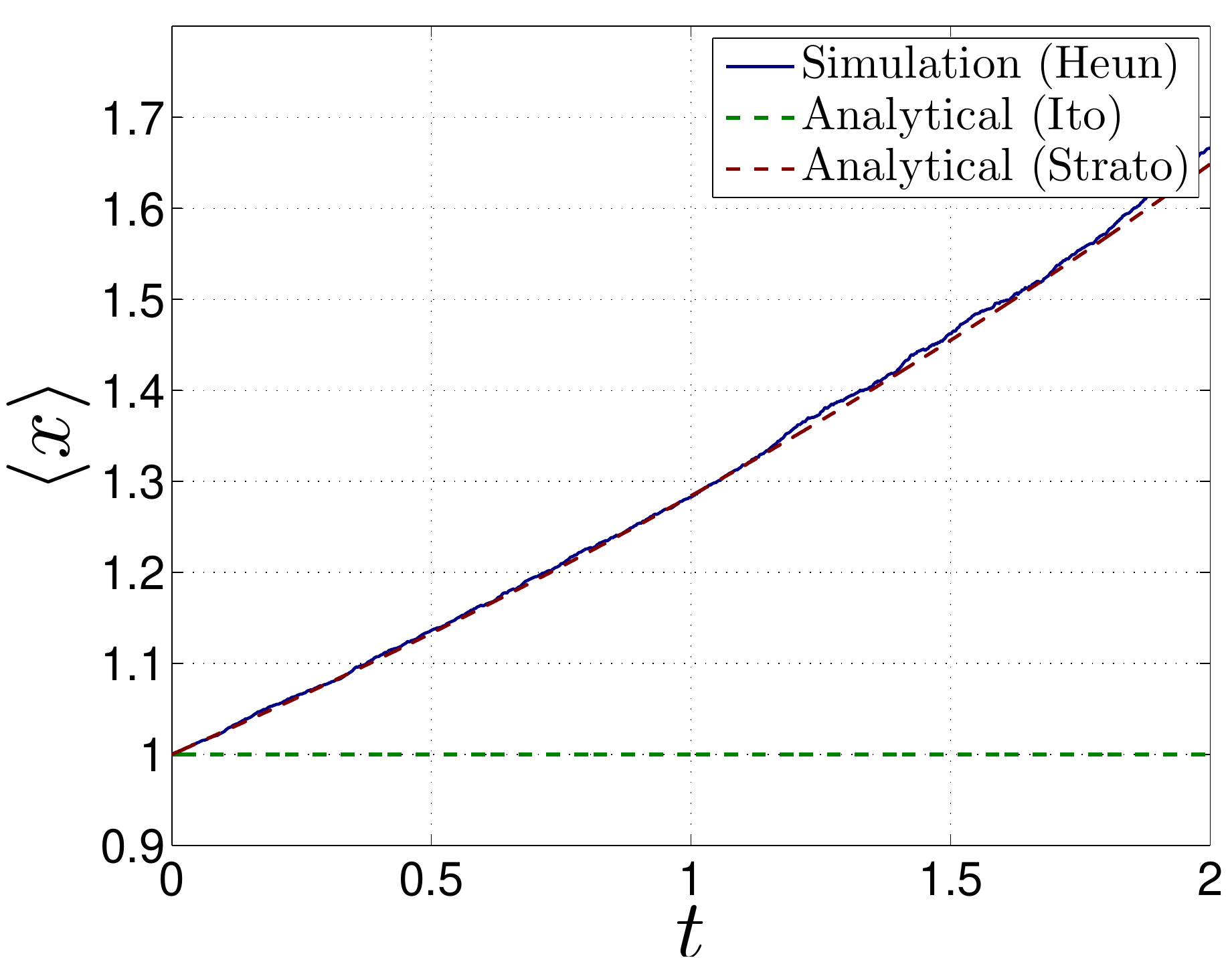}} \\
  \subfloat[][\figsubtxt{bla}]{\label{sf:num-scheme}
	\includegraphics[width=0.45\textwidth]{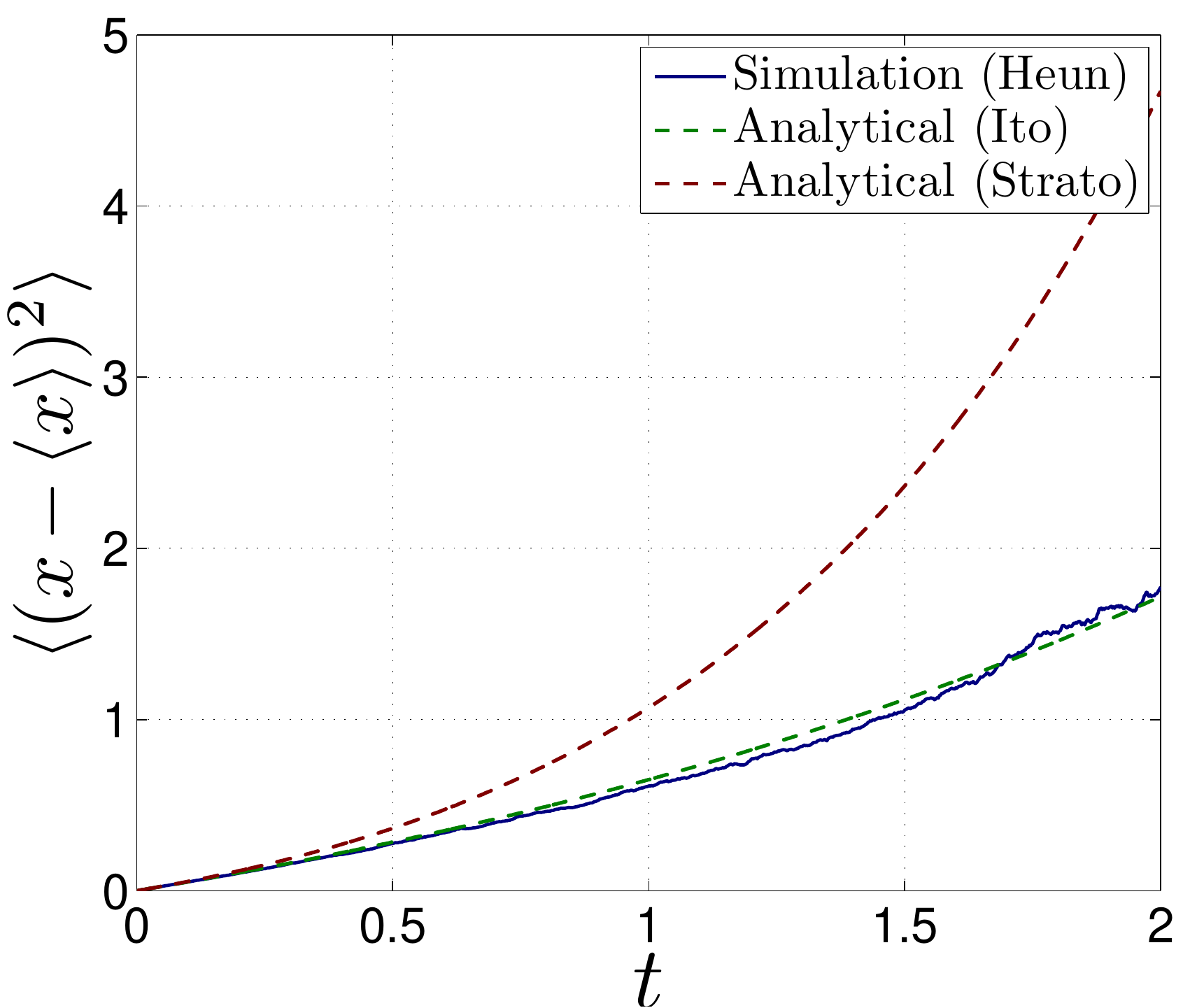}} \hfil
  \subfloat[][\figsubtxt{bla}]{\label{sf:num-scheme}
	\includegraphics[width=0.45\textwidth]{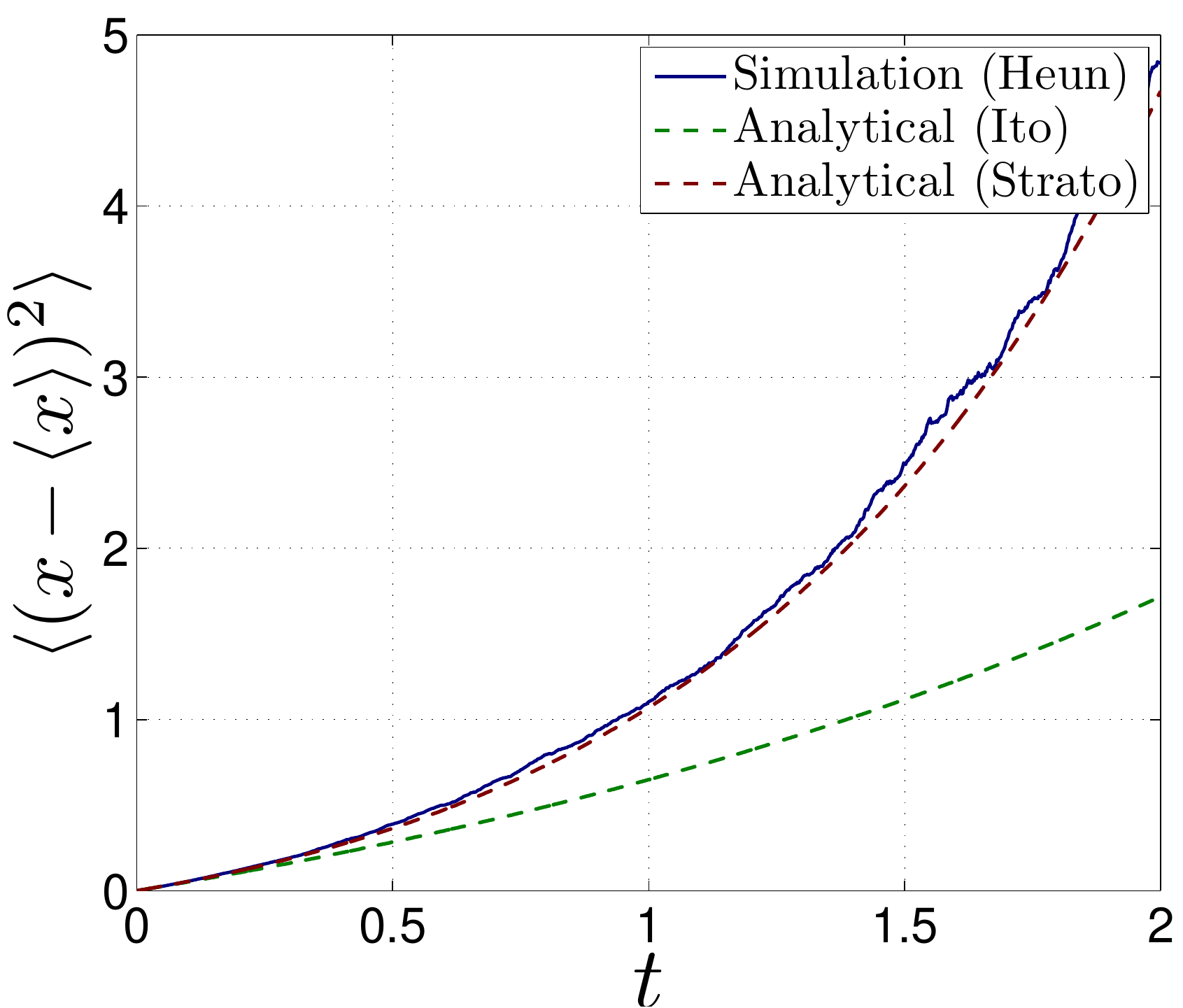}}
  \caption{\figtxt{bla}}
  \label{ff:num-scheme}
\end{figure}}\\

We note that Lau and Lubensky use in \cite{LauLubensky07PRE} the representation of a PDF as the expectation of the $\d$-function,
\begin{align}
	p(x,t|x_0,t_0) &= \int p(x')\,\d\big(x'-x(t;x_0,t_0)\big) \di x' \nn
	&= \lla\d\big(x-x(t;x_0,t_0)\big)\rra_{p(x)} \;, \label{eq:A3_PDF_delta-fct}
\end{align}
where the parametric dependency of $p(x,t|x_0,t_0)$ has been passed on to the random variable $x(t;x_0,t_0)$.\\
Our derivation of the FPE becomes equivalent to the derivation of Lau and Lubensky by substituting $A\big(x(t)\big)=\d\big(x-x(t;x_0,t_0)\big)$. 

\cleardoublepage
\section{Derivation of WPI from SDE} \label{AA_SDE2WPI}
The material presented in this chapter follows closely the article by Lau and Lubensky \cite{LauLubensky07PRE}.\\
The aim is to introduce the Wiener path integral (WPI) representation of continuous MPs. A WPI is used to express the conditional PDF as 
\begin{align} \label{eq:A4_WPI_first}
  p(x_t,t|x_0,t_0) &= \int\limits_{(x_0,t_0)}^{(x_t,t)} \Di\xc\,P[\xc|x_0] \;.
\end{align}
Here, $P[\xc|x_0]$ is the probability density functional of the path $\xc$, and $\Di\xc$ is the integration measure in function space. The path integral above is the sum of the probabilities of all paths $\xc$ that connect a given initial value $x_0$ at time $t_0$ with $x_t$ at time $t$. The notation $\xc$ emphasises that the whole trajectory is considered, and not, as $x(t')$ might suggest, the value of $x(t)$ at a single time $t'$. Instead of probability density functional we will also call $P[\xc|x_0]$ {\it path probability}.\\
The WPI above is still only symbolic. The most direct way to concretise the path integral is to discretise time as $t_i=t_0+i\Dt$ where $\Dt$ is the time step, and also discretise the path $\xc$ as the sequence $(x_1,x_2,\dots,x_N)$, where $x_1=x_0$, $x_N=x_t$, $N=(t-t_0)/\Dt$ and $x_i=x(t_i)$. Each $x_i$ at time $t_i$ is a random variable and hence follows a PDF $p_i(x_i,t_i)$.\\
The Markov assumption states that the PDF only depends on the most recent event, and we can write\remark{nicht übersehen, da ist ein $i$ als index a $p_i(x_i,t_i)$, ist also die marginalisierte PDF und daher für jeden Zeitpunkt ne andere. dank markov können wir aber die $i$ abhängigkeit allein auf das nächst-zurückliegende $x_{i-1}$ zurückführen.}
\begin{subequations}
  \begin{align}
		p_i(x_i,t_i) &= p(x_i,t_i|x_{i-1},t_{i-1}) \quad \forall\; 2\leq i\leq N \;, \\
		p_1(x_1,t_1) &= p_0(x_1) \;.
	\end{align}
\end{subequations}
Hence, the MPs is completely determined by the {\it propagator}\linebreak $p(x_{i+1},t_{i+1}|x_i,t_i)$ and the initial PDF $p_0(x_1)$.\\
As a direct consequence of the Chapman-Kolmogorov relation (\ref{eq:CKR}), the solution of the conditional FPE can be written as the integration of the product of all propagators with respect to all intermediate variables,
\begin{align} \label{eq:A4_WPI_prototype}
	p(x_t,t|x_0,t_0) = \int\left[\prod\limits_{i=2}^{N-1}\dd x_i\right]\;\prod\limits_{i=1}^{N-1}\,p(x_{i+1},t_{i+1}|x_i,t_i) \;.
\end{align}
The above formula can be viewed as the prototype of a WPI, where we identify the prototypical path probability and integration measure as
\begin{subequations} \label{eq:A4_WPI_protos}
  \begin{align}
		P[\xc|x_0] &\sim \prod\limits_{i=1}^{N-1}\,p(x_{i+1},t_{i+1}|x_i,t_i) \;, \label{eq:A4_WPI_protos_P}\\
		\Di\xc &\sim \prod\limits_{i=2}^{N-1}\dd x_i \;. \label{eq:A4_WPI_protos_D}
	\end{align}
\end{subequations}
The propagator is equivalent to Green's function of the FPE, cf. (\ref{eq:A3_int_condPDF}), which can be obtained from the FPE (\ref{eq:A3_FPE_alpha_cond}) for small $\Dt$ (\cite{Risken89} p.\,73),
\begin{align}
	p(x_{i+1},t_i+\Dt|x_i,t_i) &\simeq \frac{1}{\sqrt{2\pi\Dt g(x_i,t_i)^2}} \label{eq:A4_propa_Risken} \\
	&\qquad\times\exp\bigg[\frac{x_{i+1}\-x_i-f(x_i,t_i) \- \a g'(x_i,t_i)g(x_i,t_i)}{2\Dt\,g(x_i,t_i)^2}\bigg] \;. \nonumber
\end{align}
However, the interpretation of the resulting stochastic integral for the functional $P[\xc|x_0]$ is ambiguous.\\
In the following, we will instead build on the SDE discretised in $\a$-point, (\ref{eq:A3_SDEint_alpha}), to yield a definition of $P[\xc|x_0]$ and $\Di\xc$ in terms of the discretisation introduced above.\\

The essence of the derivation is to transform from the random variable $Z(\Dt)$, of which we know the statistical properties from (\ref{eq:A3_Z-is-gauss}), to the random variable $x_{i+1}$. The resulting PDF for $x_{i+1}$ is then the propagator $p(x_{i+1},t_{i+1}|x_i,t_i)$ used above.\\
We formulate the mentioned transformation of probability as
\begin{align}
	\NN(Z)\,\dd Z = p(x_{i+1},t_{i+1}|x_i,t_i)\,\dd x_{i+1} \;,
\end{align}
where $\NN(Z)$ is the PDF of $Z$. To calculate the Jacobian, we need $Z$ as a function of $x_{i+1}$. At this point, the underlying SDE enters the calculation by solving the discretised SDE (\ref{eq:A3_SDEint_alpha}) for $Z$,
\begin{align} \label{eq:A4_def_h}
	h(x_{i+1};\,t_{i+1},x_i,t_i)\dfn\frac{x_{i+1}-x_i-\fa\,\Dt}{\ga}=Z \;.
\end{align}
Here, we take $\fa$ and $\ga$ to be evaluated in $\a$-point, that is
\begin{align} \label{eq:A4_def_faga}
	\fa \dfn f(x_i+\a\Dx,t_i+\a\Dt) \sep \ga \dfn g(x_i+\a\Dx,t_i+\a\Dt) \;.
\end{align}
The transformation of probability then takes the form
\begin{align}
	\hspace*{-7pt} p(x_{i+1},t_{i+1}|x_i,t_i) &= \bigg|\frac{\pt h(x_{i+1};\,t_{i+1},x_i,t_i)}{\pt x_{i+1}}\bigg|\,\NN\big(h(x_{i+1};\,t_{i+1},x_i,t_i)\big).
\end{align}
The next step is to determine the two factors above which will form the propagator $p(x_{i+1},t_{i+1}|x_i,t_i)$.\\
The first factor, using (\ref{eq:A3_SDEint_alpha}), is readily determined to be
\begin{align} \label{eq:A4_propa1}
	\Big|\frac{\pt h(x_{i+1};\,t_{i+1},x_i,t_i)}{\pt x_{i+1}}\Big| = \frac{1-\a \fax\Dt}{\ga} - \a \gax\frac{x_{i+1}-x_i-\fa\Dt}{\ga^2}.
\end{align}
The second factor is best determined by writing the $\NN(Z)$ as the expectation of a $\d$-function
\begin{align} \label{eq:A4_Jacob}
	\NN(h(x_{i+1};\,t_{i+1},x_i,t_i)) = \lla \d\big(Z'-h(x_{i+1};\,t_{i+1},x_i,t_i)\big) \rra_{\NN(Z')} \;,
\end{align}
in the same fashion as in (\ref{eq:A3_PDF_delta-fct}). Using the Fourier representation of the $\d$-function, we obtain\remark{ich darf ja vorzeichen des arguments der $\d$-fkt umkehren, entspricht trafo $k\mapsto-k$ im integral.}
\begin{align} \label{eq:A4_d(h-Z)}
	\d\big(h(x_{i+1};\,t_{i+1},x_i,t_i)-Z'\big) = \frac{1}{2\pi}\int&\exp\big[ik\,h(x_{i+1};\,t_{i+1},x_i,t_i)\big] \nn
	&\;\times\exp\big[-ik\,Z'\big]\di k \;.
\end{align}
The average $\lla\dots\rra_{\NN(Z')}$ only affects the factor in the second line.\\
At this point, the statistical properties stated in (\ref{eq:A3_Z-is-gauss}) are taken into account by substituting for $\NN(Z)$ a Gaussian distribution with zero mean and variance $\Dt$ and performing the resulting Gaussian integral,
\begin{align} \label{eq:A4_Zexp}
	\lla\int\exp\big[-ik\,Z'\big]\di k\rra_{\NN(Z')} &= \int\NN(Z')\,\int \exp\big[-ik\,Z'\big]\di k \di Z' \nn
	&=\frac{1}{\sqrt{2\pi\Dt}}\int\int\exp\left[-\frac{Z'^2}{2\Dt}-ikZ'\right]\di Z'\di k \nn
  &=\int\exp\Big[-\mfrac{1}{2}\,k^2\Dt\Big]\di k \;.
\end{align}
Combining (\ref{eq:A4_propa1}) with (\ref{eq:A4_Jacob}) and (\ref{eq:A4_d(h-Z)}), (\ref{eq:A4_def_h}), (\ref{eq:A4_Zexp}) and writing $\Dx=x_{i+1}-x_i$, we obtain
\begin{align} \label{eq:A4_propa2}
	p(x_{i+1},t_{i+1}|x_i,t_i) &= 
    \frac{1}{2\pi \ga}\int\left[1-\a \fax\Dt - \a \gax\,\frac{\Dx-\fa\Dt}{\ga}\right] \nn
  &\hspace{40pt}\times\exp\left[ik\,\frac{\Dx-\fa\Dt}{\ga}-\mfrac{1}{2}k^2\Dt \right]\di k\,,
\end{align}
which is already the desired propagator, but not in the desired exponential form (\ref{eq:A4_propa_Risken}). \\
We could in principle write the prefactor in the rectangular brackets as $\exp\ln[\dots]$ and expand $\ln[\dots]$ to second order to collect also the terms $\Dx^2=\OO(\Dt)$. This procedure, however, is considered inconsistent with the concept of path integrals (\cite{LauLubensky07PRE} p.\,8).\\
We therefore rewrite the problematic term, that is the one that involves $\Dx$, as the derivative of the exponential function,
\begin{align}
	&-\a \gax\int\left[\frac{\Dx-\fa\Dt}{\ga}\right]\,\exp\left[ik\,\frac{\Dx-\fa\Dt}{\ga}-\mfrac{1}{2}k^2\Dt \right]\di k \nn
  =&-\a \gax\int\exp\left[-\mfrac{1}{2}k^2\Dt\right](-i\pt_{k})\exp\left[ik\,\frac{\Dx-\fa\Dt}{\ga}\right]\di k \nn
  =&\,\a ik \gax\Dt\int \exp\left[ik\,\frac{\Dx-\fa\Dt}{\ga}-\mfrac{1}{2}k^2\Dt \right]\di k \;,
\end{align}
where the last line follows by integration by parts and discarding the boundary terms\remark{$\sim\exp[-k^2]$}. The remaining prefactor does not involve $\Dx$ and can be included into the exponential function by writing it in the form $\exp\ln[\dots]$ and expanding $\ln[\dots]$ to linear order in $\Dt$,
\begin{align}
	p(x_{i+1},t_{i+1}|x_i,t_i) 
	&= \frac{1}{2\pi \ga}\int\exp\bigg[ik\,\frac{\Dx-\fa\Dt}{\ga}-\mfrac{1}{2}k^2\Dt\bigg] \nn
  &\hspace{120pt} \times \left(1-\a f'_i\Dt+ik \gax\Dt\right)\di k \nonumber\\[10pt]	
  &=\frac{1}{2\pi \ga}\int\exp\bigg[ik\,\frac{\Dx-\fa\Dt}{\ga}-\mfrac{1}{2}k^2\Dt \nn
  &\hspace{120pt} + \ln\big[1+\a\big(ik\gax - \fax\big)\Dt\big]\bigg]\di k \nonumber\\[10pt]	
  &=\frac{1}{2\pi \ga}\int\exp\bigg[ik\,\frac{\Dx-\fa\Dt}{\ga}-\mfrac{1}{2}k^2\Dt \nn
  &\hspace{120pt} + \a(ik\gax - \fax)\Dt \bigg]\di k \nonumber\\[10pt]	
  &=\frac{1}{2\pi \ga}\int\exp\bigg[-k^2\frac{\Dt}{2}+
		k\frac{i\Dt}{\ga}\bigg(\frac{\Dx}{\Dt}\-\fa\Dt\+\a \gax \ga\bigg) \nn
  &\hspace{120pt} - \a \fax\Dt \bigg]\di k \;. \label{eq:WPI_HS_discrete}
\end{align}
The resulting Gaussian integral can readily be calculated and we finally end up with\footnote{Defining $y_i\dfns k/\ga$ instead of performing the Gaussian integration yields the so called Hubbard-Stratonovich transform of a WPI, cf. \cite{LauLubensky07PRE}, which has important applications in field theories.}
\begin{align}
	p(x_{i+1},t_{i+1}|x_i,t_i) &= \frac{\sqrt{2\pi}}{2\pi \ga\sqrt{\Dt}}\exp\left[-\frac
    {\frac{\a \fax\Dt^2}{2}+\frac{\Dt^2}{4\ga^2}\left(\frac{\Dx}{\Dt}-\fa+\a \gax \ga\right)^2}
    {\frac{\Dt}{2}}\right] \nn
  &=\frac{1}{\sqrt{2\pi\Dt \ga^2}}\exp\left[
    -\frac{\Dt}{2\ga^2}\bigg(\frac{\Dx}{\Dt}\-\fa\+\a \ga \gax\bigg)^2\-\a\Dt \fax\right] \label{eq:A4_propa3}
\end{align}
which is of the expected form in (\ref{eq:A4_propa_Risken}), only that we now have the Jacobian $a\Dt \fax$, and $f(x,t)$ and $g(x,t)$ have to be evaluated in $\a$-point. Note that for pre-point, $\a=0$, both forms are equivalent.\\
Having the propagator in place, we can go back to our prototype of a WPI, (\ref{eq:A4_WPI_prototype}), and obtain after substituting (\ref{eq:A4_propa3}),
\begin{align}
	p(x_t,t|x_0,t_0) &= \int\left[\prod\limits_{i=2}^{N-1}\dd x_i\right]\;\prod\limits_{i=1}^{N-1}\,p(x_{i+1},t_{i+1}|x_i,t_i) \nonumber \\[10pt]
  &= \int\left[\prod\limits_{i=2}^{N-1}\dd x_i\right]\;\prod\limits_{i=1}^{N-1}\frac{1}{\sqrt{2\pi g_i^2\Dt}} \nn
  &\hspace{30pt}\times\exp\bigg[-\frac{\Dt}{2\ga^2}\bigg(\frac{x_{i+1}\-x_i}{\Dt}
		-\fa+\a \ga \gax\bigg)^2 \nn
		&\hspace{80pt}-\a\Dt \fax\bigg] \nonumber \\[10pt]
  &= \frac{1}{\sqrt{2\pi g_N^2\Dt}}\int\left[\prod\limits_{i=2}^{N-1}\frac{\di x_i}{\sqrt{2\pi g_i^2\Dt}}\right] \\
	&\hspace{30pt}\times\exp\bigg[-\sum\limits_{i=1}^{N-1}\frac{\Dt}{2\ga^2}\bigg(\frac{x_{i+1}\-x_i}{\Dt}
		-\fa+\a \ga \gax\bigg)^2\nn
		&\hspace{80pt}-\a\Dt \fax\bigg] \;. \nonumber
\end{align}
We are now able to substitute our prototype of the integration measure, (\ref{eq:A4_WPI_protos_D}), by the definition
\begin{align} \label{eq:A4_def_WPI_Di}
  \int\Di\xc \dfn \lim\limits_{\stackrel{N\to\infty}{\Dt\to0}} \frac{1}{\sqrt{2\pi \Dt g_N^2}}\int\left[\prod\limits_{i=2}^{N-1}\frac{\di x_i}{\sqrt{2\pi g_i^2\Dt}}\right] \;,
\end{align}
and the prototype of the path-probability in (\ref{eq:A4_WPI_protos_P}) by
\begin{align} \label{eq:A4_def_WPI_P}
	P[\xc|x_0] \dfn \lim\limits_{\stackrel{N\to\infty}{\Dt\to0}}\; \exp{\Big[-\Dt\sum\limits_{i=1}^{N-1} s_i(x_i,\,x_{i+1})\Big]}
\end{align}
with
\begin{align} \label{eq:A4_def_WPI_s}
	s_i(x_i,\,x_{i+1}) \dfn \frac{1}{2\ga^2}\bigg(\frac{x_{i+1}\-x_i}{\Dt}
		+\fa+\a \ga \gax\bigg)^2-\a \fax \;,
\end{align}
In this form, the continuous limit for the path probability can be explicitly performed and we obtain
\begin{align} \label{eq:A4_def_WPI_P_conti}
	P[\xc|x_0] = \exp\big[-\SS[\xc]\,\big]
\end{align}
with the action functional
\begin{subequations} \label{eq:A4_def_WPI_Ss_conti}
  \begin{align}
		\SS[\xc] &\dfn \int_{t_0}^{t} s\big(x(\t),\ddx(\t),\t\big) \di \t \;, \label{eq:A4_def_WPI_S_conti} \\
		s(x,\ddx,\t) &\dfn \frac{\big[\ddx - f(x,\t) + \a g'(x,\t)g(x,\t)\big]^2}{2g(x,\t)^2}+\a f'(x,\t) \;. \label{eq:A4_def_WPI_s_conti}
	\end{align}
\end{subequations}
Note that the integral in (\ref{eq:A4_def_WPI_S_conti}) is a stochastic integral, and for discretisation, $f(x,t)$ and $g(x,t)$ have to be discretised in $\a$-point as defined in (\ref{eq:A4_def_faga}).\\

In closing, we remark that from this form of the WPI, the corresponding FPE can be retrieved \cite{LauLubensky07PRE}, which brings us full circle with regard to the three equivalent formulations of continuous MPs.

\cleardoublepage
\section{Overview of SDE, FPE and WPI} \label{AA_overview}
We provide an overview of three equivalent descriptions of continuous MPs (MPs), being stochastic differential equations (SDEs), the Fokker-Planck equation (FPE) and Wiener path integrals (WPI), along with a matching first order numerical integration scheme (NUM).\\
We distinguish the two cases where the MP is initially given by a SDE, or where the MP is initially given by a FPE, and, respectively, the equivalent other forms are desired.\\
In the same fashion, we also provide the equivalent SDEs in \Ito~and Stratonovich convention.\\

\paragraph{SDE} Suppose a continuous MP is given by a SDE defined in $\a$-convention ($\a=0$ for \Ito~and $\a=1/2$ for Stratonovich), then are equivalent
\begin{subequations} \label{eq:A5_overviewSDE}
  \begin{align}
		\mr{(SDE)} \quad& \ddx_t = f(x_t,t) + g(x_t,t)\,\xi(t) \sep \a\text{-point}  \label{eq:A5_overviewSDE_SDE}\\[10pt]
		\begin{split}
		\mr{(NUM)} \quad& x(t\+\Dt) = x_t + f(x_t,t)\Dt + g(x_t,t)Z(\Dt)   \\
		           \quad& \hspace{110pt} + \a g'(x_t,t)g(x_t,t)Z(\Dt)^2
		\end{split} \label{eq:A5_overviewSDE_NUM}\\[10pt]
		\begin{split}
		\mr{(FPE)} \quad& \ddp(x,t) = \Big[\-\pt_x\big[f(x,t) \+ \a g'(x,t)g(x,t)\big] \\
		           \quad& \hspace{150pt} + \tfrac{1}{2}\pt_x^2g(x,t)^2\Big]\,p(x,t)
		\end{split} \label{eq:A5_overviewSDE_FPE}\\[10pt]
		\mr{(WPI)} \quad& \SS[\xc] = \int_{t_0}^{t}\frac{\big[\ddx - f(x_\t,\t) + \a g'(x_\t,\t)g(x_\t,\t)\big]^2}{2g(x_\t,\t)^2} \label{eq:A5_overviewSDE_WPI}\\[5pt]
		            \quad& \hspace{140pt} +\a f'(x_\t,\t) \di\t	\sep \a\text{-point} \;. \nonumber
	\end{align}
\end{subequations}
The index $t$ denotes that $t$ is to be taken as argument. The Gaussian noise $\xi$ has zero mean and is $\d$-correlated, $\lla\xi(t-t')\rra=\d(t-t')$, and according to (\ref{eq:A3_Z-is-gauss}), $Z(\Dt)$ is a Gaussian random variable with zero mean and variance $\Dt$. The $\a$-point discretisation is defined in (\ref{eq:A4_def_faga}).\\
According to van Kampen (\cite{Kampen1981,vanKampen2007}), when deterministic and stochastic influences have distinguished sources, the SDE is to be taken for\linebreak $\a=1/2$, in order to retain a clear-cut interpretation of $f(x_t,t)$ as force and $g(x_t,t)\xi(t)$ as noise. Otherwise, the choice of $\a$ is a matter of taste. \newpage

\paragraph{FPE in \Ito-form} Suppose a continuous MP is given by a FPE in terms of drift and diffusion, 
\begin{subequations} \label{eq:A5_overview_D1D2}
  \begin{align}
		\Df(x,t) &= f(x,t) + \a g'(x,t)g(x,t)\;, \\
		\Dg(x,t) &= \mfrac{1}{2}g(x,t)^2 \;,
	\end{align}
\end{subequations}
then are equivalent
\begin{subequations} \label{eq:A5_overviewFPE-D12}
  \begin{align}
		\mr{(FPE)} \quad& \ddp(x,t) =  \big[-\pt_x\Df(x,t) + \pt_x^2\Dg(x,t)\big]\,p(x,t) \label{eq:A5_overviewFPE-D12_FPE} \\[10pt]
		\begin{split}
		\mr{(SDE)}  \quad& \ddx_t = \Df(x,t) - \a\Dgx(x_t,t) \\
							 \quad& \hspace{100pt} + \sqrt{2\Dg(x_t,t)}\,\xi(t) \sep \a\text{-point} 
		\end{split}\label{eq:A5_overviewFPE-D12_LE}\\[10pt]
		\begin{split}
		\mr{(NUM)} \quad& x(t\+\Dt) = x(t)  + \big[\Df(x,t) + (1\-2\a)\Dgx(x_t,t)\big]\,\Dt \\
		           \quad& \hspace{60pt} + \sqrt{2\Dg(x_t,t)}\,Z(\Dt) + \a \Dgx(x_t,t)\,Z(\Dt)^{\,2}
		\end{split} \label{eq:A5_overviewFPE-D12_NUM}\\[10pt]
		\mr{(WPI)} \quad& S[\xc] = \int\limits_{t_0}^{t} \frac{\big(\ddx_\t - \Df(x,t) + 2(\a\-1)\Dgx(x_\t,\t)\big)^2}{4\Dg(x_\t,\t)}  \nn
							 \quad& \hspace{60pt} + \a(\Dfx(x_\t,\t) - \Dgx(x_\t,\t)) \label{eq:A5_overviewFPE-D12_WPI} \\
							 \quad& \hspace{80pt} + \a(1\-\a)\Dgxx(x_\t,\t) \di \t	\sep \a\text{-point} \;. \nonumber
	\end{align}
\end{subequations}
Again, the index $t$ denotes that $t$ is to be taken as argument. The Gaussian noise $\xi$ has zero mean and is $\d$-correlated, $\lla\xi(t-t')\rra=\d(t-t')$, and according to (\ref{eq:A3_Z-is-gauss}), $Z(\Dt)$ is a Gaussian random variable with zero mean and variance $\Dt$. The $\a$-point discretisation is defined in (\ref{eq:A4_def_faga}).\\
Note that the SDE in this forms defines the same dynamics regardless the choice of $\a$. \newpage

\paragraph{FPE in Stratonovich-form} The FPE is often given in the so-called Stratonovich form, that is in terms of force and diffusion,
\begin{subequations} \label{eq:A5_overview_FD}
  \begin{align}
  \begin{split}
		F(x,t) &= \Df(x,t) - \Dgx(x,t) \\
		&= f(x,t) + (\a-1)g'(x,t)g(x,t) \;,
	\end{split} \\
  \begin{split}
		D(x,t) &= \Dg(x,t) = \mfrac{1}{2}g(x,t)^2 \;.
	\end{split}
	\end{align}
\end{subequations}
For convenience, we also give the table of equivalent LE and WPI for this case,
\begin{subequations} \label{eq:A5_overviewFPE-FD}
  \begin{align}
		\mr{(FPE)} \quad& \ddp(x,t) = \big[-\pt_xF(x,t) + \pt_xD(x,t)\pt_x\big]\,p(x,t) \label{eq:A5_overviewFPE-FD_FPE} \\[10pt]
		\begin{split}
		\mr{(SDE)}  \quad& \ddx_t = F(x_t,t) + (1-\a)D'(x_t,t) \\
							 \quad& \hspace{100pt} + \sqrt{2D(x_t,t)}\,\xi(t) \sep \a\text{-point} 
		\end{split}\label{eq:A5_overviewFPE-FD_SDE}\\[10pt]
		\begin{split}
		\mr{(NUM)} \quad& x(t\+\Dt) = x(t)  + \big[F(x_t,t) + (1\-\a)D'(x_t,t)\big]\,\Dt \\
		           \quad& \hspace{90pt} + \sqrt{2D(x_t,t)}\,Z(\Dt) + \a D'(x_t,t)\,Z(\Dt)^{\,2}
		\end{split} \label{eq:A5_overviewFPE-FD_NUM}\\[10pt]
		\mr{(WPI)} \quad& S[\xc] = \int\limits_{t_0}^{t} \frac{\big(\ddx_\t - F(x_\t,\t) + (2\a\-1)D'(x_\t,\t)\big)^2}{4D(x_\t,\t)} \label{eq:A5_overviewFP-FDE_WPI} \\
							 \quad& \hspace{60pt} + \a F'(x_\t,\t) + \a(1\-\a)D''(x_\t,\t) \di \t	\sep \a\text{-point} \;. \nonumber
	\end{align}
\end{subequations}
Again, the index $t$ denotes that $t$ is to be taken as argument. The Gaussian noise $\xi$ has zero mean and is $\d$-correlated, $\lla\xi(t-t')\rra=\d(t-t')$, and according to (\ref{eq:A3_Z-is-gauss}), $Z(\Dt)$ is a Gaussian random variable with zero mean and variance $\Dt$. The $\a$-point discretisation is defined in (\ref{eq:A4_def_faga}).\\
Note that the SDE in this forms defines the same dynamics regardless the choice of $\a$. \newpage

\paragraph{\Ito~and Stratonovich SDE} The two predominantly used interpretations of SDEs are those of \Ito~($\a=0$) and Stratonovich ($\a=1/2$). It often is convenient, to switch from a \Ito~SDE to a Stratonovich one and vice versa. Again, $\xi$ is a Gaussian noise with zero mean and correlation $\lla\xi(t-t')\rra=\d(t-t')$. See (\ref{eq:A4_def_faga}) for the definition of discretisation in $\a$-point.\\[20pt]
The key observation is that a SDE of the form\remark{diss p.28}
\begin{align} \label{eq:A5_alpha}
	\ddx_t = f(x_t,t) - \a g'(x_t,t)g(x_t,t) + g(x_t,t)\,\xi(t) \sep \a\text{-point} \;,
\end{align}
defines the same dynamics for arbitrary $\a$ with $0\leq\a\leq1$.
\vspace{20pt}\\
For the \Ito~SDE
\begin{subequations} \label{eq:A5_pre2mid}
\begin{align}
	\ddx_t = f(x_t,t) + g(x_t,t)\,\xi(t) \sep \text{pre-point} \;,
\end{align}
the equivalent Stratonovich SDE reads
\begin{align}
	\ddx_t = f(x_t,t) - \mfrac{1}{2}g'(x_t,t)  + g(x_t,t)\,\xi(t) \sep \text{mid-point} \;.
\end{align}
\end{subequations}
\vspace{20pt}\\
For the Stratonovich SDE
\begin{subequations} \label{eq:A5_mid2pre}
\begin{align}
	\ddx_t = f(x_t,t) + g(x_t,t)\,\xi(t) \sep \text{mid-point} \;,
\end{align}
the equivalent \Ito~SDE reads
\begin{align}
	\ddx_t = f(x_t,t) + \mfrac{1}{2}g'(x_t,t)  + g(x_t,t)\,\xi(t) \sep \text{pre-point} \;.
\end{align}
\end{subequations}
\vspace{20pt}\\
In general, for a SDE defined in $\a$-point
\begin{subequations} \label{eq:A5_a2a}
\begin{align}
	\ddx_t = f(x,t) + g(x_t,t)\,\xi(t) \sep \a\text{-point}
\end{align}
the equivalent SDE in $\tilde\a$-point reads
\begin{align}
	\ddx_t = f(x,t) + (\a\-\tilde\a)g'(x,t)g(x,t)  + g(x_t,t)\,\xi(t) \sep \tilde\a\text{-point}
\end{align}
\end{subequations}

\end{appendix}

\end{document}